UNIVERSITY OF MONTENEGRO

FACULTY OF ELECTRICAL ENGINEERING

MSc Andjela Draganic

# ANALYSIS OF NON-STATIONARY MULTICOMPONENT SIGNALS WITH A FOCUS ON THE COMPRESSIVE SENSING APPROACH

**- PhD thesis -**

// ACKNOWLEDGEMENT

My heartfelt appreciation goes to my supervisor, Prof. Dr. Srdjan Stankovic, for giving me the opportunity to work at the Laboratory for Multimedia, Faculty of Electrical Engineering in Podgorica, for the given trust, the enormous patience, professional help and encouragement he provided me during the work on the master and doctoral thesis. I am grateful for the numerous ideas that have inspired the publication of several scientific papers on which the thesis is based upon. I am also very grateful for the time he devoted reading the thesis and for the numerous comments and suggestions that contributed to its significant improvement.

Special thanks to Prof. Dr. Irena Orovic for the time she devoted to me during the entire period of my work at the Laboratory for Multimedia, for the encouragement and help during the whole period of work on the master and doctoral thesis. I am grateful for the numerous discussions and suggestions she provided to me which helped in writing scientific papers and overcoming the obstacles I encountered while working on the thesis.

I thank my friends for the patience and support they provided me during the PhD studies.

I owe a special gratitude to my family for understanding and support that greatly eased the work on this thesis.

Andjela Draganic

# INFORMATION ON DOCTORAL DISSERTATION

Abstract:

The characterization of multicomponent signals with particular emphasis on musical and communication signals is one of the problems studied in the dissertation. In order to provide an efficient analysis of the multicomponent signals, the possibility to separate signal components is observed. The procedure for decomposition and classification of the signal components whose energy and physical characteristics differ in the time-frequency domain is proposed in this work. A special focus in the dissertation is on the application of the compressive sensing approach in multicomponent signals. The compressive sensing method becomes popular in the field of signal processing until recently, and its application in various fields can increase the acquisition and transmission speed, reduce the complexity of devices, and reduce energy consumption. The procedure that applies the compressive sensing in the classification of the wireless communication signals is proposed. The algorithms for reconstruction of the compressive sensed signals are intensively developing, and therefore special emphasis in the dissertation is devoted to the hardware implementation of one of the algorithms for sparse signal reconstruction.

# THESIS OVERVIEW

One of the challenges in signal processing is the analysis of multicomponent signals of variable spectral content. Time-frequency representation of these signals can provide an efficient analysis if a suitable cross-terms free time-frequency distribution is used. A large number of time-frequency distributions is defined, and which one will be used depends on the signal nature and the application's specific requirements. This work deals with the musical signals and signals that appear in communication systems, whose components' energy and physical characteristics in the time-frequency domain are different. Having in mind multicomponent nature of those signals, distributions that provide cross terms free representation, i.e. the S-method and the complex-time distributions are used for their analysis.

A special attention is paid to the analyzing the possibility to separate the signal components in order to analyze each component individually. A decomposition and classification procedure in communications' signal is proposed, where different disturbances are frequent, as well as interferences among the signal components belonging to different standards. Bearing in mind that standard approaches for component classification do not provide satisfactory results if the components' energy differs, the focus was on solving this problem. An iterative procedure for separating signal components based on the eigenvalue and eigenvectors decomposition is defined. The signal classification can be made having in mind that the components, belonging to different standards in wireless communications, have different physical characteristics. The classification is made based on the parameters of the components, estimated by visualizing the separated components in the time-frequency domain. The proposed procedure does not require a priori knowledge of the number of signal components or their frequency positions.

The constant tendency for faster signal processing and transmission nowadays leads to a new approach in signal acquisition according to the compressive sensing principles. Compared to the conventional approach in signal acquisition, that is based on the Sampling theorem, the compressive sensing approach has many advantages. The Sampling theorem requires a signal to be sampled with a frequency at least two times higher than the maximal signal frequency, in order to faithfully represent and reconstruct the signal. Sampling in such way may produce a large number of signal samples, which may complicate samples transmission and storage. Therefore, a compression is a necessary step after the acquisition in a conventional approach. The compressive sensing approach allows simultaneous acquisition and signal compression. However, certain conditions have to be satisfied related to the signal nature and acquisition procedure, in order to successfully apply the compressive sensing approach. Those conditions are met in a large number of real applications, which opens up numerous possibilities for using this approach in signal reconstruction from a small set of acquired coefficients. Therefore, a special attention is devoted to the hardware realization of the algorithm for reconstruction of signals that have a compact representation in a transformation domain. Modifications of the original algorithm are proposed in order to reduce the complexity of the system. The system is scalable and allows matrix dimensions to be changed in accordance with the changes in the number of available signal samples and the number of frequency components of the signal.

# Content

# SADRŽAJ

# Lista slika

# Lista tabela

# Spisak skraćenica

| Simbol | Značenje |
|---|---|
| BP | Basis Pursuit |
| BPDN | Basis Pursuit Denoising |
| BPSK | Binary phase-shift keying |
| CoSaMP | Compressive Sampling Matching Pursuit |
| CS | Komprimovano (kompresivno) očitavanje |
| DFT | Diskretna Fourier-ova transformacija |
| DPSK | Differential phase-shift keying |
| DSSS | Direct Sequence Spread Spectrum |
| EVD | Dekompozicija na sopstvene vrijednosti |
| FHSS | Frequency Hopping Spread Spectrum |
| HT | Hermitska transformacija |
| IF | Trenutna frekvencija |
| IHT | Iterative Hard Thresholding |
| IRS | Iterative re-weighted/shrinkage |
| ISM | Industrial, Scientific and Medical |
| IST | Iterative Soft Thresholding |
| LASSO | Least Absolute Shrinkage and Selection Operator |
| LARS | Least Angle Regression |
| ML | Maximum-likelihood |
| MP | Matching Pursuit |
| MRI | Magnetic Resonance Imaging |
| MUSIC | MUltiple SIgnal Classification |
| OMP | Orhogonal Matching Pursuit |
| *pdf* | Funkcija gustine vjerovatnoće |
| SFAR 2D | Simple and Fast Algorithm for Reconstruction of 2D signals |
| SNR | Odnos signal šum |
| SPEC | Spektrogram |
| STFT | Kratkotrajna Fourier-ova transformacija |
| SVD | Dekompozicija na singularne vrijednosti |
| TF | Vremensko-frekvencijska distribucija |
| WD | Wigner-ova distribucija |
| 2D FT | Dvodimenziona Fourier-ova transformacija |

# Uvod

Analiza multikomponentnih signala promjenljivog spektralnog sadržaja predstavlja izazov u oblasti obrade signala. U svrhu efikasne analize ovakvih signala, koristi se njihovo vremensko-frekvencijsko predstavljanje, sa ciljem visoko-rezolucione reprezentacije trenutne frekvencije posmatranih signala (ili njihovih pojedinačnih komponenti). Izbor odgovarajuće vremensko-frekvencijske distribucije zavisi kako od tipa signala tako i od zahtjeva konkretne aplikacije. Česta, neželjena pojava u vremensko-frekvencijskom predstavljanju multikomponentnih signala jesu kros-članovi. Kako su u ovom radu analizirani multikomponentni muzički i komunikacioni signali, to su za njihovu analizu korišćene distribucije koje obezbjeđuju predstavu bez kros članova, a fokus je bio na primjeni S-metoda, kao i na distribucijama sa kompleksnim argumentom vremena.

U cilju efikasnije analize multikomponentnih signala, predložene su procedure za odvajanje komponenti signala, bazirane na sopstvenim vrijednostima i vektorima. Procedure su definisane korišćenjem vremensko-frekvencijskih reprezentacija, prilagođenim radu sa multikomponentnim signalima, sa primjenama u muzičkim i signalima u bežičnim komunikacijama. Procedure su efikasne u radu sa signalima kod kojih su komponente jednakih energija, kao i u radu sa onim signalima čije su komponente različitih energija.

U bežičnim komunikacijama odvajanje komponenti signala je od posebnog interesa za njihovu klasifikaciju, naročito kod signala koji pripadaju interferirajućim standardima u ovim komunikacijama – Bluetooth i IEEE 802.11b standardu. Osim klasifikacije komponenti, kod ovih signala je testirana mogućnost primjene novog pristupa u akviziciji signala – kompresivnog odabiranja. Cilj je redukcija količine podataka koja se šalje komunikacionim kanalom, a da se na prijemnoj strani i dalje ima kompletna informacija o signalu. Uslov je da signal ima konciznu, tj. kompaktnu predstavu u nekom domenu.

Kod kompresivnog očitavanja odbirci signala se uzimaju slučajnim rasporedom, a ne u ekvidistantnim vremenskim intervalima kao što je to bio slučaj sa akvizicijom u skladu sa Teoremom o odabiranju. To znači da se prilikom akvizicije ne uzimaju sve vrijednosti, nego samo određeni, mali skup koeficijenata. Iz tako malog skupa prikupljenih odbiraka rekonstruiše se kompletan signal, primjenom rekonstrukcionih algoritama. Ovakav pristup ima brojne prednosti u poređenju sa dosadašnjim načinom odabiranja signala, u smislu smanjenja potrošnje energije, povećanja brzine rada prilikom akvizicije i smanjenja memorijskih zahtjeva.

Imajući u vidu popularnost pomenutog pristupa, posebna pažnja u radu je posvećena i primjeni komprimovanog očitavanja i novih rekonstrukcionih algoritama na signale u komunikacijama kao i na muzičke signale. Testirani su novi domeni i signali koji imaju kompaktnu predstavu u tim domenima. Razmatrani su jednodimenzioni i dvodimenzioni slučajevi. Posmatrane su različite vremensko-frekvencijske reprezentacije kao domeni u kojima signali imaju kompaktnu predstavu: kratkotrajna Fourier-ova transformacija, Wigner-ova distribucija, S-metod, distribucije iz Cohen-ove klase. Imajući u vidu da, pri radu sa visoko nestacionarnim signalima standardne vremensko-frekvencijske reprezentacije ne daju zadovoljavajuće rezultate (u smislu tačnosti procjene trenutne frekvencije), analizirane su distribucije sa kompleksnim argumentom vremena. Posmatrana je mogućnost kombinovanja ovih distribucija sa pristupom kompresivnog odabiranja. Definisan je 2D pristup zasnovan na kompresivnom odabiranju, korišćenjem ambiguity funkcije četvrtog reda. Imajući u vidu da distribucija sa kompleksnim argumentom vremena koncentriše energiju signala oko trenutne frekvencije u vremensko-frekvencijskoj ravni, te da je u ovom domenu obezbjeđeno kompaktno predstavljanje, odbirci se uzimaju iz ambiguity funkcije i to slučajnim rasporedom kako bi se zadovoljilo svojstvo inkoherentnosti. Analizirana je mogućnost rekonstrukcije vremensko-frekvencijske distribucije iz malog skupa ambiguity koeficijenata. Zatim je testirana tačnost estimacije trenutne frekvencije iz dobijene vremensko-frekvencijske reprezentacije. Posmatrani su slučajevi sa različitim brojem ambiguity mjerenja. Procedura je testirana na radarskim signalima. Za signale sa šumom definisana je robustna procedura, zasnovana na L-estimacionom pristupu. Umjesto dosadašnje prakse usrednjavanja koeficijenata signala, preostalih nakon primjene L-estimacije, uklonjeni koeficijenti zahvaćeni šumom se mogu uspješno rekonstruisati primjenom tehnika kompresivnog odabiranja. Pokazano je da se, osim na radarske, procedura može uspješno primijeniti i na muzičke signale.

Struktura rada je sljedeća: U prvoj glavi je dat opis multikomponentnih muzičkih signala i signala u komunikacijama koji su analizirani kroz rad. Matematičke transformacije i vremensko-frekvencijske distribucije korišćene za analizu ovih signala, date su u istoj glavi. U drugoj glavi opisana je procedura dekompozicije harmonijskih (muzičkih) i neharmonijskih signala, bazirana na sopstvenim vrijednostima. Treća glava sadrži uvod u princip kompresivnog odabiranja, matematičku formulaciju kao i rekonstrukcione algoritme upotrijebljene u radu. U ovoj glavi date su neke od aplikacija kompresivnog odabiranja. U četvrtoj glavi opisan je 2D pristup estimaciji trenutne frekvencije nestacionarnih signala, sa primjenom kompresivnog odabiranja na ambiguity funkciju i distribuciju sa kompleksnim argumentom vremena. Peta glava sadrži prijedlog procedure za klasifikaciju signala u bežičnim komunikacijama, koja je zasnovana na sopstvenim vektorima i kompresivnom odabiranju. U šestoj glavi dat je prijedlog hardverske arhitekture algoritma za rekonstrukciju kompresivno očitavanih signala. U cilju pojednostavljenja hardverske implementacije, u ovoj glavi predložene su adekvatne modifikacije algoritma.

# 1. Multikomponentni muzički i komunikacioni signali i matematičke transformacije za njihovu analizu i obradu

U analizi i obradi signala koriste se brojne matematičke metode, prilagođene različitim tipovima signala [1]-[5]. Često se koriste Fourier-ova analiza, analiza u vremensko-frekvencijskom domenu, *wavelet* analiza i slično. Signali mogu biti deterministički ukoliko su jednoznačno opisani nekom funkcijom ili pravilom [2], [4]. Ako signali uzimaju vrijednosti iz skupa mogućih vrijednosti sa određenom vjerovatnoćom, onda se nazivaju stohastičkim i za njihov opis se koristi teorija vjerovatnoće [4].

U ovom radu bavićemo se analizom kako determinističkih, tako i signala koji se javljaju u realnim aplikacijama. Fokus je analiza multikomponentnih signala, na prvom mjestu onih signala koji se koriste u komunikacijama, kao i muzičkih signala. U Poglavljima 1.1 i 1.2 biće više riječi o obje vrste signala, kao i tehnikama za njihovu analizu.

## 1.1. Signali u komunikacijama - *Spread Spectrum* modulacijske tehnike

U zavisnosti od zahtjeva sistema u pogledu brzine prenosa podataka i potrošnje energije, definišu se različiti standardi u bežičnim komunikacijama. Poželjne osobine u svim postojećim komunikacionim standardima su niska potrošnja energije i brz prenos podataka. Takođe, obezbjeđivanje sigurnog prenosa informacija je od velikog značaja u bežičnim tehnologijma. U tom cilju koriste se modulacione tehnike koje se baziraju na širenju spektra – tzv. *spread spectrum* tehnike [6]-[13]. Ove tehnike obezbjeđuju robusnost na smetnje, kao što su inter-simbolske interferencije (ISI) i različiti tipovi šumova [6]. U ovom radu smo se fokusirali na signale koji koriste *Industrial, Scientific and Medical* (ISM) opseg frekvencija, tj. opseg frekvencija od 2.4 GHz do 2.4835 GHz [7]. Iz dana u dan sve je više korisnika ISM frekvencijskog opsega, pa je samim tim veća i vjerovatnoća pojave smetnji u sistemu usljed prisustva raznih tipova signala. Upravo to je jedan od osnovnih razloga zbog koga su posmatrani ovi signali. Cilj je prevazilaženje problema pojave smetnji u sistemima definisanim ovim standardima. U literaturi se može naći opis mnogobrojnih tehnika razvijenih sa ciljem redukcije smetnji, kao što su tehnike bazirane na proširenju spektra signala.

Tehnike proširenja spektra - *Spread spectrum* tehnike [7], [9]-[11] proširuju frekvencijski spektar signala korišćenjem koda koji je jedinstven za svakog korisnika i nekorelisan sa posmatranim signalom. Kao rezultat, dobija se mnogo širi propusni opseg u poređenju sa propusnim opsegom nemodulisanog signala.

Postoji više tipova *spread spectrum* modulacija. Često korišćena dva tipa modulacija su *direct sequence spread spectrum* (DSSS) i *frequency hopping spread spectrum* (FHSS) modulacije [7], [9]-[13]. Ove dvije modulacijske tehnike su od posebnog interesa u ovom radu, jer pripadaju interferirajućim standardima u bežičnim komunikacijama. Kao što je napomenuto, problem interferencije signala koji pripadaju ovim standardima obrađivan je u literaturi, a u ovom radu biće predložen pristup za klasifikaciju signala koji je zasnovan na odvajanju komponenti signala i posmatranju svake komponente pojedinačno. Pristup je baziran na dekompoziciji na sopstvene vrijednosti i mjerenju koncentracije u transformacionim domenima.

### 1.1.1. DSSS i FHSS modulacije

Prva modulacijska tehnika koju smo posmatrali u ovom radu, DSSS modulacija, koristi se u IEEE 802.11b standardu za bežični LAN [11]. Standard IEEE 802.11 [11] radi u nelicenciranom, 2.4 GHz ISM opsegu frekvencija i koristi jedan od 13 preklapajućih, 22 MHz širokih kanala. Svaki kanal se karakteriše svojom centralnom frekvencijom, a rastojanje između susjednih centralnih frekvencija je 5 MHz. Standard omogućava prenos signala na udaljenosti od 10 do 100 metara, uz brzinu prenosa oko 11 Mbps.

Modulacijska tehnika korišćena u IEEE 802.11b standardu bazirana je na promjenama faze signala koji prenosi informaciju. Promjene faze signala definisane su pseudoslučajnom sekvencom. Modulisani, DSSS signal, sa binarno faznom modulacijom (*binary phase-shift keying* - BPSK) ili diferencijalnom faznom modulacijom (*differential phase-shift keying* - DPSK) može se matematički predstaviti sljedećom relacijom [11]:

$$m(t) = b(t)c(t) = Ad(t)c(t)\cos(2\pi f_c t + \theta), \tag{1.1}$$

gdje je $A$ amplituda signala, $d(t)$ je modulacija signala, $c(t)$ je pseudoslučajna sekvenca za proširenje spektra signala, $f_c$ je frekvencija nosioca a $\theta$ je faza signala u trenutku $t$=0. Bez znanja pseudoslučajne sekvence, na prijemnoj strani, signal ne može biti uspješno demodulisan. Ako širinu spektra signala $d(t)$ označimo sa $B_d$, a širinu spektra signala $c(t)$ označimo sa $B_c$, rezultujući signal $m(t)$ imaće širinu spektra $B_c >> B_d$.

Pseudoslučajna sekvenca, koja ima za cilj proširenje spektra signala, je oblika [11]:

$$c(t) = \sum_{i=-\infty}^{\infty} c_i \psi(t - iT_c), \tag{1.2}$$

gdje $c_i = \pm 1$, i:

$$\psi(t) = \begin{cases} 1, & 0 \le t < T_C \\ 0, & \text{za ostalo } t \end{cases}. \tag{1.3}$$

Na prijemu, signal se ponovo množi sa sekvencom $c(t)$. Ako je zadovoljeno da je $\psi(t)$ pravougaona funkcija, onda važi da je $c(t)$=±1 i $c^2(t)$=1, pa je množenjem signala sa $c(t)$ na prijemu moguće dobiti polazni signal $b(t)=Ad(t)\cos(2\pi f_C t+\theta)$ [11]. Šeme predajnika i primjenika DSSS modulacije su prikazane na slici 1.1, dok je primjer DSSS signala dat na slici 1.2.

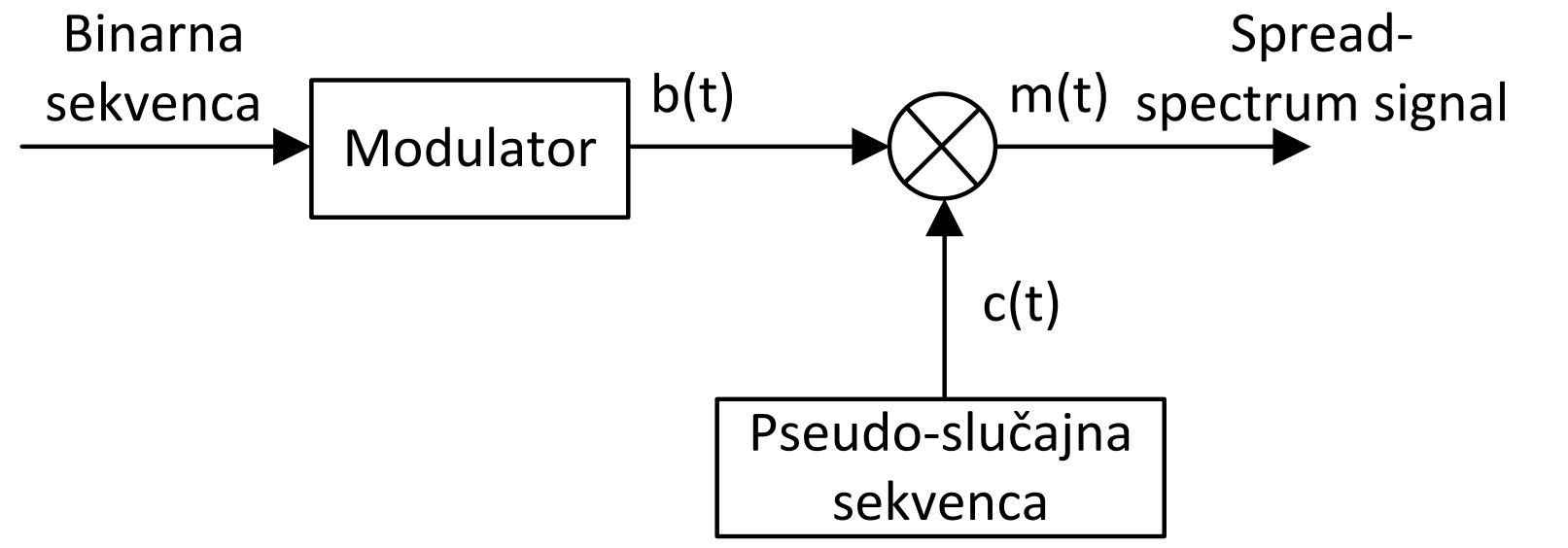

**a)**

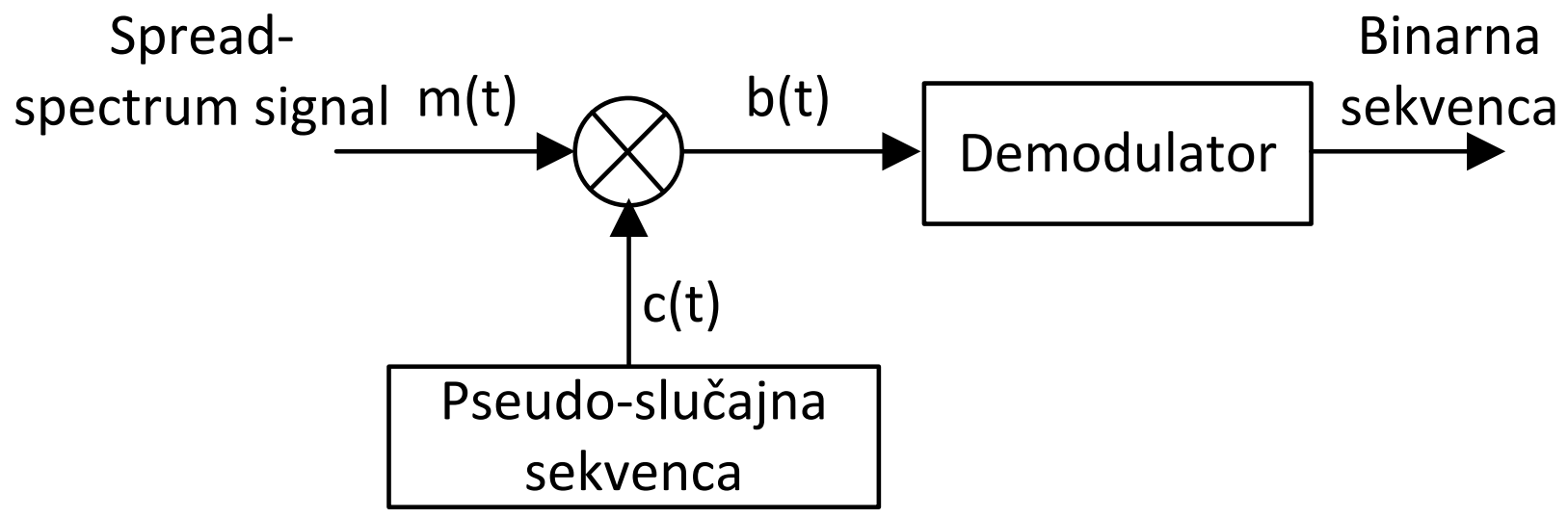

**b)**

**Slika 1.1: a) DSSS predajnik; b) DSSS prijemnik**

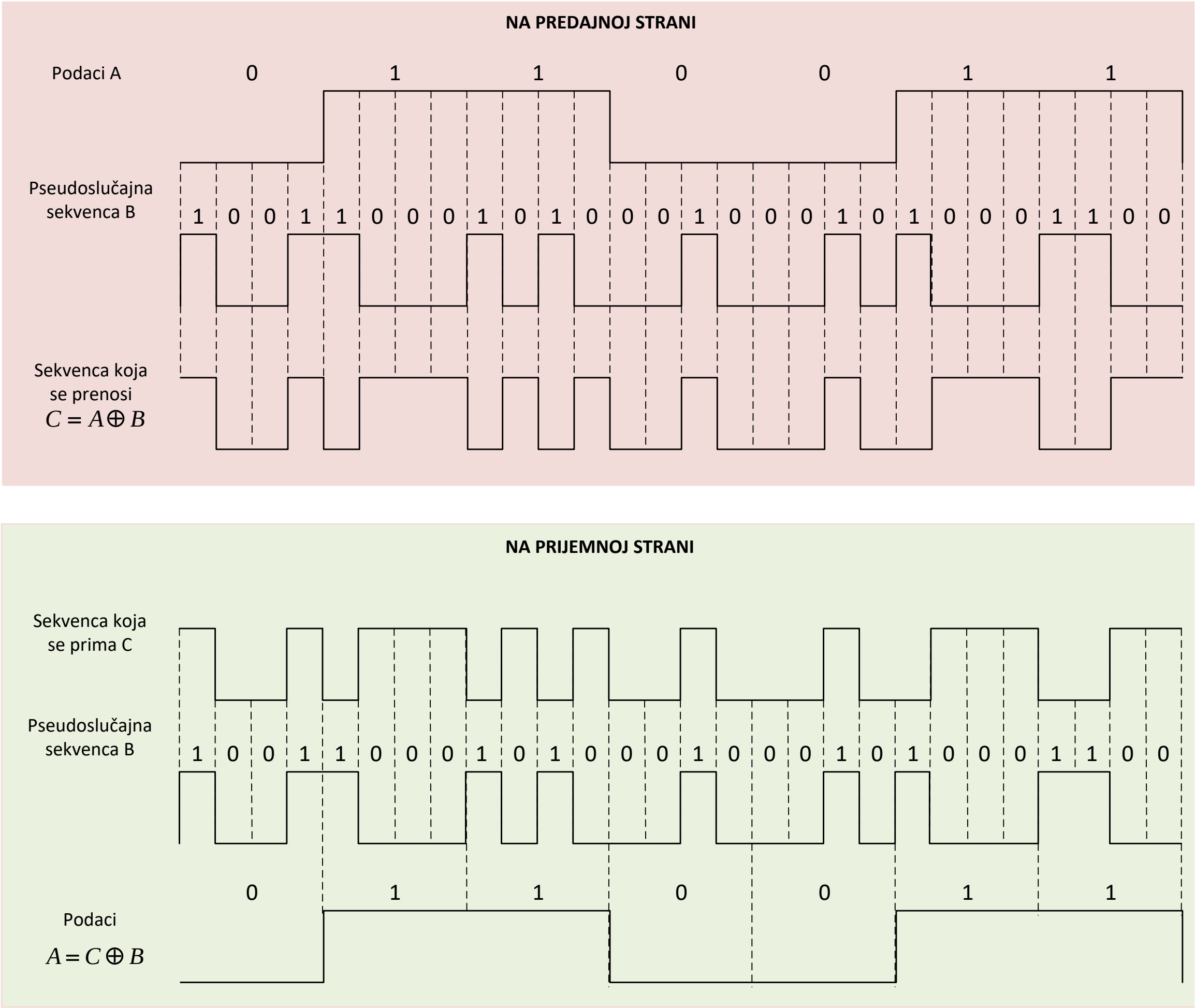

**Slika 1.2: Primjer DSSS signala**

Druga modulacijska tehnika, FHSS [7], [9]-[12], koristi se u Bluetooth standardu i zasnovana je takođe na pomjeranju frekvencije nosioca na pseudoslučajan način. Pseudoslučajna sekvenca u FHSS modulaciji definiše frekvencije na kojima se pojavljuju djelovi signala. Sekvenca frekvencija na kojima se pojavljuju djelovi signala naziva se *frequency-hopping pattern*.

Frekvencije zauzimaju određeni opseg koji se naziva *hopping band* i koji ima *M* frekvencijskih kanala. Vremenski interval između dva hop-a je hop interval. Osim vremenskog intervala, frekvencija hop-a i njegovo trajanje predstavljaju karakteristike FHSS signala. Predajnik i prijemnik FHSS signala prikazani su na slici 1.3.

U cilju uspješne demodulacije signala, pseudoslučajna sekvenca mora biti poznata kako predajniku, tako i prijemniku. Postoje dvije vrste FHSS modulacije:

- Spora FHSS modulacija,
- Brza FHSS modulacija.

Ako se uzme da se frekvencija nosioca mijenja svakih $T_C$ sekundi, a da je trajanje simbola $T_S$, onda važi da je $T_C \geq T_S$ kod spore FHSS, a $T_C < T_S$ kod brze FHSS modulacije. Standard koji koristi FHSS modulaciju, Bluetooth standard, je komunikacijski protokol čiji signali djeluju u ISM opsegu frekvencija. Bluetooth tehnika koristi radio talase koji omogućavaju komunikaciju na rastojanjima do 10 m, a frekvencijski opseg ovog standarda se dijeli u više nivoa, tačnije 79 frekvencija.

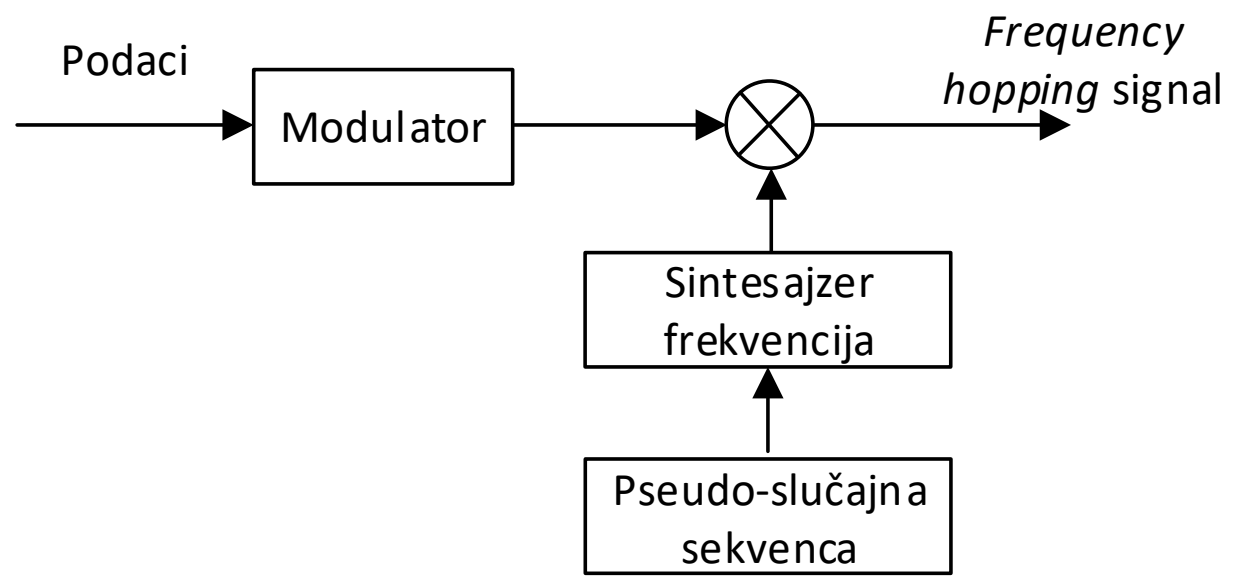

**a)**

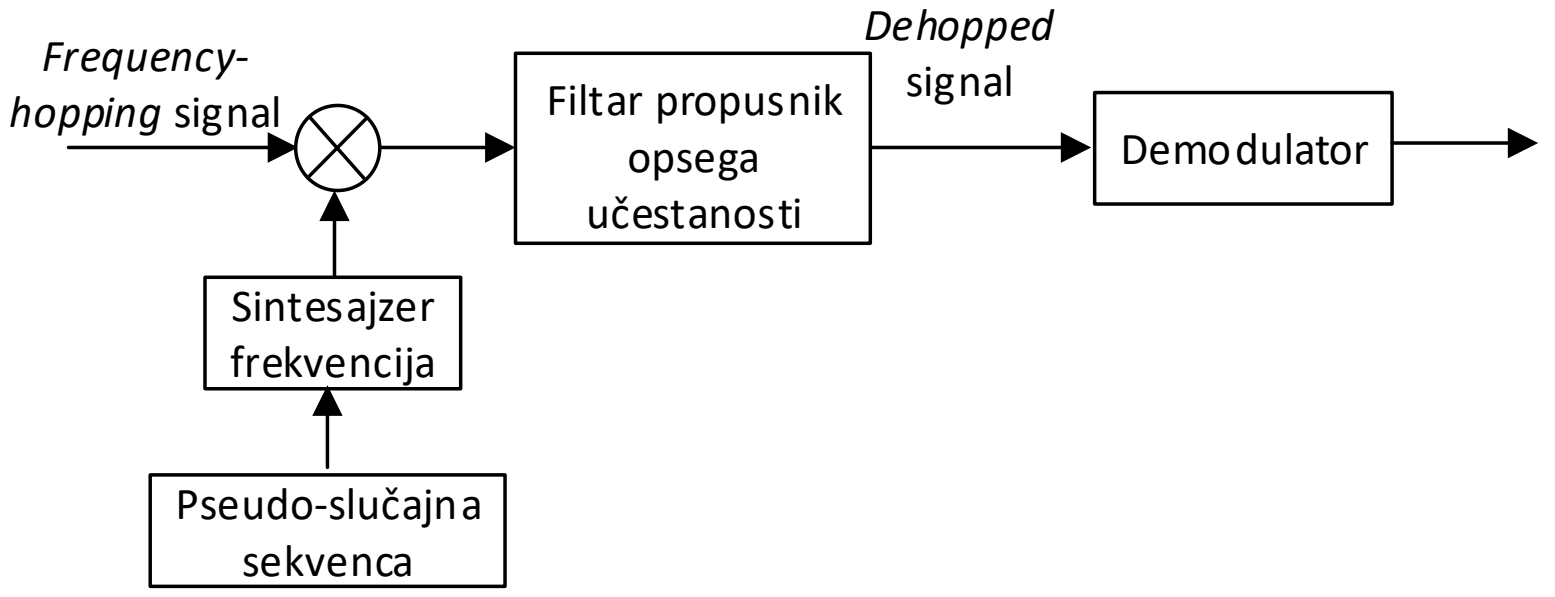

**b)**

**Slika 1.3: a) Predajnik i b) prijemnik FHSS sistema**

Postoje brojni pristupi za analizu signala u bežičnim komunikacijama. Signali se mogu analizirati npr. u vremenskom, frekvencijskom, vremensko-frekvencijskom domenu. Za karakterizaciju FHSS i DSSS signala moguće je koristiti vremensko-frekvencijsku analizu [6], [7], [9]. Estimacija parametara komponenti FHSS i DSSS signala može se vršiti na osnovu njihove vremensko-frekvencijske reprezentacije [7]. Ona je značajna za identifikaciju bežičnog standarda, kao i identifikaciju načina rada u komunikacionim sistemima.

## 1.2. Muzički signali

U fizičkom pogledu, zvuk predstavlja talasne fluktuacije oko vibrirajućeg materijala [15]. Osnovne karakteristike zvuka su brzina propagacije, frekvencija i nivo zvučnog pritiska. U ovom radu posmatran je jedan tip audio signala, tj. muzički signali, nastali sviranjem na različitim instrumentima. Muzički signali su zvučni signali koje karakteriše frekvencija od 20 Hz do 20 KHz [15]. Zvuk kod muzičkih instrumenta može da se generiše na više načina. Vibracije žice kod žičanih instrumenata (violine, violončela, klavira) ili vibracije vazduha kod duvačkih instrumenata (flaute ili klarineta) rezultiraju pojavom zvučnih talasa. Vibracije žice, vazduha ili membrana na jednoj frekvenciji, posljedično izazivaju vibracije na čitavom skupu frekvencija koje su cjelobrojni umnožak osnovne (fundamentalne) frekvencije. Ovi umnošci osnovne frekvencije nazivaju se harmonicima.

Muzički signali se mogu analizirati u različitim domenima: vremenskom, frekvencijskom, vremensko-frekvencijskom, i slično. Analiza frekvencijskog sadržaja kao i amplituda muzičkih signala tema je mnogih radova [16]-[21]. Spektar muzičkih signala sastavljen je od većeg broja sinusoidalnih komponenti koje se nazivaju harmonicima.

Procedura odvajanja komponenti signala koja je predložena u radu, primijenjena je nad muzičkim signalima. Odvajanjem komponenti muzičkog signala i posmatranjem svake komponente pojedinačno obezbjeđuje se efikasnija analiza istog. U literaturi postoje i druge tehnike za analizu muzičkih signala. Brza tehnika harmonijskog prilagođavanja - *Fast Harmonic Matching Pursuit*, primijenjena je na audio signale u [17], sa ciljem njihove aproksimacije kombinacijom od $M$ harmonijskih struktura. Odvajanje harmonijskih komponenti može se vršiti i filtriranjem u spektralnom domenu, ali u tom slučaju frekvencije svakog harmonika treba da budu unaprijed poznate [16].

## 1.3. Transformacije za analizu signala

### 1.3.1. Fourier-ova transformacija

Fourier-ova transformacija je često korišćena matematička transformacija u obradi signala. Ona pruža informacije o frekvencijama koje postoje u signalu, kao i informacije

o amplitudama komponenti posmatranog signala, za razliku od predstave signala u vremenskom domenu koja daje informacije o promjenama amplitude signala tokom vremena. Frekvencijski sadržaj signala naziva se spektar signala.

Fourier-ova transformacija signala $x(t)$ definiše se na sljedeći način [2], [4], [15]:

$$X(\omega)=\int_{-\infty}^{\infty} x(t)e^{-j\omega t}dt\,. \tag{1.4}$$

Vremenski oblik signala se može dobiti korišćenjem inverzne Fourier-ove transformacije, koja je definisana sljedećom relacijom:

$$x(t)=\frac{1}{2\pi}\int_{-\infty}^{\infty} X(\omega)e^{j\omega t}d\omega\,. \tag{1.5}$$

Prethodne relacije se mogu zapisati u matričnom obliku. Za direktnu Fourier-ovu transformaciju važi relacija:

$$\mathbf{X}=\mathbf{\Psi x}\ , \tag{1.6}$$

gdje je **X** vektor koeficijenata Fourier-ove transformacije. Vremenski oblik signala, vektor **x**, dobija se primjenom inverzne transformacije:

$$\mathbf{x}=\mathbf{\Psi}^{-1}\mathbf{X}\,. \tag{1.7}$$

Matrice **Ψ** i $\mathbf{\Psi}^{-1}$ su matrice direktne i inverzne Fourier-ove transformacije, definisane na sljedeći način (za signal dužine $N$ odbiraka) [15]:

$$\mathbf{\Psi}=\begin{bmatrix} 1 & 1 & 1 & 1 & \dots & 1 \\ 1 & e^{-j\frac{2\pi}{N}} & e^{-2j\frac{2\pi}{N}} & e^{-3j\frac{2\pi}{N}} & \dots & e^{-(N-1)j\frac{2\pi}{N}} \\ 1 & e^{-2j\frac{2\pi}{N}} & e^{-4j\frac{2\pi}{N}} & e^{-6j\frac{2\pi}{N}} & \dots & e^{-(N-1)2j\frac{2\pi}{N}} \\ \dots & \dots & \dots & \dots & \dots & \dots \\ 1 & e^{-(N-2)j\frac{2\pi}{N}} & e^{-2(N-2)j\frac{2\pi}{21}} & e^{-3(N-2)j\frac{2\pi}{N}} & \dots & e^{-(N-1)(N-2)j\frac{2\pi}{N}} \\ 1 & e^{-(N-1)j\frac{2\pi}{N}} & e^{-2(N-1)j\frac{2\pi}{21}} & e^{-3(N-1)j\frac{2\pi}{21}} & \dots & e^{-(N-1)(N-1)j\frac{2\pi}{N}} \end{bmatrix}$$

$$\mathbf{\Psi}^{-1}=\frac{1}{N}\begin{bmatrix} 1 & 1 & 1 & 1 & \dots & 1 \\ 1 & e^{j\frac{2\pi}{N}} & e^{2j\frac{2\pi}{N}} & e^{3j\frac{2\pi}{N}} & \dots & e^{(N-1)j\frac{2\pi}{N}} \\ 1 & e^{2j\frac{2\pi}{N}} & e^{4j\frac{2\pi}{N}} & e^{6j\frac{2\pi}{N}} & \dots & e^{(N-1)2j\frac{2\pi}{N}} \\ \dots & \dots & \dots & \dots & \dots & \dots \\ 1 & e^{(N-2)j\frac{2\pi}{N}} & e^{2(N-2)j\frac{2\pi}{N}} & e^{3(N-2)j\frac{2\pi}{N}} & \dots & e^{(N-1)(N-2)j\frac{2\pi}{N}} \\ 1 & e^{(N-1)j\frac{2\pi}{N}} & e^{2(N-1)j\frac{2\pi}{N}} & e^{3(N-1)j\frac{2\pi}{N}} & \dots & e^{(N-1)(N-1)j\frac{2\pi}{N}} \end{bmatrix}. \tag{1.8}$$

Analiza u Fourier-ovom domenu ima nedostatke u radu sa nestacionarnim signalima [2]-[4]. Naime, ovom analizom moguće je odrediti frekvencije na kojima se javljaju komponente signala, ali ne i vrijeme njihovog pojavljivanja i vremensko trajanje komponenti. U cilju prevažilaženja ovih nedostataka Fourier-ove transformacije uvedena je vremensko-frekvencijska analiza [15]-[33]. Ona omogućava analizu signala po vremenu i po frekvenciji istovremeno, pruža informacije o frekvencijama i vremenu pojavljivanja komponenti signala a omogućava i jasno razlikovanje multikomponentnih od monokomponentnih signala. To je nama od posebnog značaja imajući u vidu da su u radu posmatrani multikomponenti signali. Opis nekih često korišćenih vremensko-frekvencijskih distribucija, kao i distribucija koje su korišćene u radu, dat je u Poglavlju 1.4.

U praktičnim aplikacijama neophodno je koristiti diskretnu formu Fourier-ove transformacije, koja je definisana na sljedeći način [2], [15]:

$$DFT(k)=\sum_{n=0}^{N-1} x(n)e^{-j\frac{2\pi}{N}nk}, \tag{1.9}$$

dok je inverzni oblik Fourier-ove transformacije definisan kao:

$$x(n)=\frac{1}{N}\sum_{k=0}^{N-1} DFT(k)e^{j\frac{2\pi}{N}nk}. \tag{1.10}$$

### 1.3.2. Hermitska transformacija

Hermitska transformacija je još jedna često korišćena transformacija u obradi signala, a koju ćemo koristiti i za analizu signala posmatranih u ovom radu [15], [34]-[39]. Hermitska transformacija ima primjenu u brojnim aplikacijama u obradi signala, prvenstveno zbog sljedećih osobina: Hermitske bazne funkcije obezbjeđuju dobru lokalizaciju signala i u vremenskom i u transformacionom domenu. Zbog toga Hermitska transformacija nalazi primjenu u kompresiji slike, kompjuterskoj tomografiji, biomedicinskim aplikacijama i slično.

Hermitske funkcije mogu se definisati korišćenjem Hermitskih polinoma. Hermitski polinom *n*-tog reda se definiše kao [15]:

$$H_n(t)=(-1)^n e^{t^2}\frac{d^n(e^{-t^2})}{dt^n},\ n=0,1,2,\ldots , \tag{1.11}$$

U nastavku su dati oblici za prva 4 Hermitska polinoma, dobijena na osnovu relacije (1.11):

$$\begin{aligned}&H_0(t)=1,\ H_1(t)=2t,\ H_2(t)=4t^2-2,\\&H_3(t)=8t^3-12t,\ H_4(t)=16t^4-48t^2+12\end{aligned}. \tag{1.12}$$

Kontinualne Hermitske funkcije sada se mogu definisati na sljedeći način:

$$\psi_n(t)=\frac{e^{-t^2}H_n(t)}{\sqrt{2^n n!\sqrt{\pi}}}. \tag{1.13}$$

Hermitski polinomi zadovoljavaju sljedeću rekurzivnu relaciju:

$$\begin{aligned}&H_n(t)=2tH_{n-1}(t)-2(t-1)H_{n-2}(t),\\&H_0(t)=1,\ H_1(t)=2t.\end{aligned} \tag{1.14}$$

Hermitske funkcije reda $n$, gdje je $n\geq 2$, opisane su sljedećim relacijama [15]:

$$\begin{aligned}&\psi_0(t)=\frac{1}{\sqrt[4]{\pi}}e^{-\frac{t^2}{2}},\ \psi_1(t)=\frac{\sqrt{2}t}{\sqrt[4]{\pi}}e^{-\frac{t^2}{2}},\\&\psi_n(t)=t\sqrt{\frac{2}{n}}\psi_{n-1}(t)-\sqrt{\frac{n-1}{n}}\psi_{n-2}(t),\end{aligned} \tag{1.15}$$

I Hermitski polinomi i Hermitske funkcije zadovoljavaju osobinu ortogonalnosti, i formiraju ortogonalnu bazu za predstavljanje signala. Signal $x(t)$ se može predstaviti korišćenjem Hermitskih funkcija, u sljedećem obliku [15]:

$$x(t)=\sum_{n=0}^{\infty}K(n)\psi_n(t). \tag{1.16}$$

Prethodna relacija predstavlja Hermitski razvoj, a koeficijenti Hermitskog razvoja se računaju u skladu sa sljedećom relacijom:

$$K(n)=\int_{-\infty}^{\infty}x(t)\psi_n(t)dt. \tag{1.17}$$

U praksi se radi sa diskretnim signalima dužine $N$, i sa konačnim brojem Hermitskih funkcija $P$. Zbog toga se posmatrani signal mora odabrati na određeni način. Kada radimo sa Hermitskom transformacijom, signal $x(t)$ je potrebno odabrati u tačkama koje odgovaraju nulama Hermitskog polinoma $N$-tog reda, a kao rezultat dobija se diskretni signal $x(d)$. I Hermitske funkcije takođe treba da budu odabrane u nulama Hermitskog polinoma $N$-tog reda. Diskretna Hermitska transformacija se zatim definiše kao:

$$K(n)=\sum_{d=0}^{N-1}x(d)\psi_n(d), \tag{1.18}$$

i predstavlja aproksimaciju originalnog signala dobijenu korišćenjem $P$ Hermitskih funkcija ($P \leq N$). Koeficijenti Hermitske transformacije obično se računaju pomoću Gauss-Hermite kvadraturne aproksimacije:

$$K(n) \approx \frac{1}{N}\sum_{d=0}^{N-1} \mu_{N-1}^{n}(d)x(d), \tag{1.19}$$

gdje su konstante $\mu$ definisane na sljedeći način:

$$\mu_{N-1}^{n}(d) = \frac{\psi_n(d)}{\left[\psi_{N-1}(d)\right]^2}. \tag{1.20}$$

U matričnoj formi, Hermitska transformacija se može predstaviti kao [15]:

$$K = \mathbf{H}s,$$

$$\begin{bmatrix} K(0) \\ K(1) \\ \dots \\ K(P-1) \end{bmatrix} = \frac{1}{N}\begin{bmatrix} \frac{\psi_0(0)}{\left(\psi_{P-1}(0)\right)^2} & \dots & \frac{\psi_0(N-1)}{\left(\psi_{P-1}(N-1)\right)^2} \\ \frac{\psi_1(0)}{\left(\psi_{P-1}(0)\right)^2} & \dots & \frac{\psi_1(N-1)}{\left(\psi_{P-1}(N-1)\right)^2} \\ \dots & \dots & \dots \\ \frac{\psi_{P-1}(0)}{\left(\psi_{p-1}(0)\right)^2} & \dots & \frac{\psi_{P-1}(N-1)}{\left(\psi_{p-1}(N-1)\right)^2} \end{bmatrix}\begin{bmatrix} x(0) \\ x(1) \\ \dots \\ x(N-1) \end{bmatrix}, \tag{1.21}$$

gdje je **K** vektor Hermitskih koeficijenata, **H** je Hermitska transformaciona matrica, a **x** je signal. Inverzna formulacija Hermitske transformacije je data sa:

$$\mathbf{x} = \mathbf{\Psi K},$$

$$\mathbf{\Psi} = H^{-1} = \begin{bmatrix} \psi_0(1) & & \cdots & \psi_0(N) \\ \psi_1(1) & & \cdots & \psi_1(N) \\ \vdots & \vdots & \ddots & \vdots \\ \psi_{P-1}(1) & & \cdots & \psi_{P-1}(N) \end{bmatrix} \tag{1.22}$$

gdje $\mathbf{\Psi}$ predstavlja matricu inverzne Hermitske transformacije.

## 1.4. Vremensko-frekvencijska analiza signala

Realni signali razlikuju se po svojoj prirodi: stacionarnosti, broju komponenti u signalu i slično. Stoga se koriste različiti metodi za obradu i analizu različitih tipova signala. Vremensko-frekvencijska analiza je posebno pogodna za analizu nestacionarnih signala koji imaju brze promjene spektralnog sadržaja.

Zavisno od aplikacije i tipa signala definisan je veliki broj vremensko-frekvencijskih distribucija. Dakle, izbor vremensko-frekvencijske distribucije koja se primjenjuje zavisi od vrste signala. Idealna vremensko-frekvencijska distribucija koncentriše energiju oko trenutne frekvencije (*instantaneous frequency* - IF) u vremensko-frekvencijskoj ravni [2]-[4], [15], [40]-[48]. Za signal oblika $Ae^{j\phi(t)}$, gdje $A$ predstavlja amplitudu signala, trenutna frekvencija se definiše kao prvi izvod faze signala $\phi(t)$: $IF(t)=\frac{1}{2\pi}\frac{d\phi(t)}{dt}$. Idealna vremensko-frekvencijska reprezentacija za ovakav signal treba da bude u sljedećem obliku:

$$TF(t,\omega)=2\pi A^2\delta(\omega-\phi^{'}(t)). \tag{1.23}$$

Međutim, u mnogim realnim slučajevima postoji rasipanje energije oko trenutne frekvencije. Ono nastaje kao posljedica uticaja prozora i viših izvoda faze (drugog, trećeg, itd.) i zavisi kako od vrste signala, tako i od primijenjene vremensko-frekvencijske distribucije. Realna forma vremensko-frekvencijske distribucije, definisana relacijom:

$$TF(t,\omega)=2\pi A^2\delta(\omega-\phi^{'}(t))*_{\omega}\mathcal{F}\left\{w(t)\right\}*_{\omega}\mathcal{F}\left\{e^{jS(t,\tau)}\right\}, \tag{1.24}$$

uključuje i faktor rasipanja $S(t,\tau)$. Funkcija $w(t)$ u relaciji (1.24) predstavlja funkciju prozora, $\mathcal{F}$ je operator Fourier-ove transformacije, a $*_{\omega}$ označava konvoluciju u frekvencijskom domenu.

Vremensko-frekvencijska analiza se koristi i u analizi multikomponentnih signala. Važno je napomenuti da je pojam trenutna frekvencija definisan samo za monokomponentne signale. Kod multikomponentnih signala, trenutna frekvencija se odnosi na trenutnu frekvenciju pojedinačnih komponenti signala.

U Poglavljima 1.4.1 i 1.4.2 biće dat osvrt na često korišćene vremensko-frekvencijske distribucije, kao i na distribucije koje se koriste za analizu multikomponentnih signala. Fokus će biti na linearnim distribucijama, na nekim od kvadratnih vremensko-frekvencijskih distribucija, a biće objašnjene i distribucije višeg reda, prilagođene radu sa visoko nestacionarnim signalima, koji će biti predmet analize u ovom radu.

### 1.4.1. Linearne vremensko-frekvencijske transformacije

Neke od često korišćenih linearnih vremensko-frekvencijskih reprezentacija su kratkotrajna Fourier-ova transformacija (*Short Time Fourier Transform* - STFT), polinomijalna Fourier-ova transformacija, *wavelet* transformacije, i slično [2], [22], [23]. Linearne vremensko-frekvencijske transformacije imaju osobinu da je transformacija linearne kombinacije signala jednaka linearnoj kombinaciji transformacija [4], [15]. Za signal oblika $x(t)=c_1x_1(t)+...+c_Px_P(t)$ linearna vremensko-frekvencijska reprezentacija definisana je na sljedeći način:

$$TF\{x(t)\}=c_1X_1(t,\omega)+...+c_PX_P(t,\omega),$$

odnosno (1.25)

$$TF\{x(t)\}=\sum_{i=1}^{P}TF\{x_i(t)\},$$

gdje je $X_P(t,\omega)$ vremensko-frekvencijska transformacija signala $x_P(t)$. Iz prethodne relacije može se zaključiti da je vremensko-frekvencijska reprezentacija multikomponentnog signala $x(t)$ jednaka sumi vremensko-frekvencijskih reprezentacija pojedinačnih komponenti signala.

### *Kratkotrajna Fourier-ova transformacija - STFT*

Kratkotrajna Fourier-ova transformacija (STFT) je Fourier-ova transformacija signala odsječenog prozorom i ubraja se u grupu linearnih transformacija. Izračunavanjem lokalnog spektra svakog uprozorenog dijela signala, dobija se ukupni vremensko-frekvencijski prikaz. Kratkotrajna Fourier-ova transformacija signala $x(t)$, uz korišćenje funkcije prozora $w(\tau)$, definisana je na sljedeći način [2], [4], [15]:

$$STFT(t,\omega)=\int_{-\infty}^{\infty}x(t+\tau)w(\tau)e^{-j\omega\tau}d\tau. \quad (1.26)$$

Kvadratna verzija STFT-a se naziva spektrogram (SPEC) i definisan je kao:

$$SPEC(t,\omega)=\left|STFT(t,\omega)\right|^2=\left|\int_{-\infty}^{\infty}x(t+\tau)w(\tau)e^{-j\omega\tau}d\tau\right|^2. \quad (1.27)$$

Dakle, spektrogram spada u grupu kvadratnih transformacija. Rezolucija signala, odnosno trenutne frekvencije u vremensko-frekvencijskoj ravni zavisi od širine prozora $w(\tau)$. Faktor rasipanja dobija se razvojem faze signala u Taylor-ov red, i kod SPEC-a sadrži sve više izvode faze signala:

$$S(t,\tau)=\sum_{i=2}^{\infty}\phi^{(i)}(t)\tau^i/i!\ . \quad (1.28)$$

### 1.4.2. Neke od kvadratnih vremensko-frekvencijskih distribucija

***Wigner-ova distribucija***

U cilju poboljšanja rezolucije STFT-a uvedene su kvadratne vremensko-frekvencijske distribucije. Osim već pomenutog spektrograma, često korišćena kvadratna distribucija je Wigner-ova distribucija (WD) [2], [4], [24]:

$$WD(t,\omega)=\int_{-\infty}^{\infty} x\left(t+\frac{\tau}{2}\right)x^{*}\left(t-\frac{\tau}{2}\right)e^{-j\omega\tau}d\tau\,. \qquad (1.29)$$

Wigner-ova distribucija idealno koncentriše linearno frekvencijski modulisane signale (faktor rasipanja je jednak nuli). Međutim, ova distribucija dovodi do pojave tzv. unutrašnjih interferencija (*inner-interferences*) kod signala sa nelinearnim promjenama trenutne frekvencije. U takvim slučajevima, faktor rasipanja sadrži neparne izvode faze i definisan je relacijom:

$$S(t,\tau)=\sum_{i=1}^{\infty}\phi^{(2i+1)}(t)\frac{(\tau/2)^{2i+1}}{(2i+1)!}\,. \qquad (1.30)$$

U slučaju multikomponentnih signala, WD pored unutrašnjih interferencija unosi i neželjene, kros-članove, koji se javljaju na aritmetičkoj sredini između svaka dva člana signala i posljedica su kvadratne prirode WD-a. Na primjer, WD signala $x(t)=x_1(t)+x_2(t)$, ima oblik:

$$WD(t,\omega)=WD_{x_1x_1}(t,\omega)+WD_{x_2x_2}(t,\omega)+WD_{x_1x_2}(t,\omega)+WD_{x_2x_1}(t,\omega), \qquad (1.31)$$

gdje $WD_{x_1x_1}(t,\omega)$ i $WD_{x_2x_2}(t,\omega)$ predstavljaju WD korisnih komponenti signala, dok $WD_{x_1x_2}(t,\omega)$ i $WD_{x_2x_1}(t,\omega)$ predstavljaju WD-ove nastale od kros-komponenti.

U cilju prevazilaženja nedostataka STFT-a i WD-a, uvedeni su S-metod, distribucije iz Cohen-ove klase kao i distribucije sa kompleksnim argumentom vremena.

***S-metod***

S-metod je kvadratna vremensko-frekvencijska distribucija, opisana sljedećom relacijom:

$$SM(t,\omega)=\frac{1}{2\pi}\int_{-\infty}^{\infty}P(\theta)STFT\left(t,\omega+\frac{\theta}{2}\right)STFT^{*}\left(t,\omega-\frac{\theta}{2}\right)d\theta\,, \tag{1.32}$$

gdje je $P(\theta)$ funkcija prozora u frekvencijskom domenu [2], [4], [15]. Kao i WD, i S-metod se odlikuje dobrom koncentracijom komponenti signala u vremensko-frekvencijskoj ravni, boljom u poređenju sa STFT-om i SPEC-om [2], [4], [15], [20], [25]. Takođe, u velikom broju slučajeva S-metod prevazilazi nedostatak WD-a koji se odnosi na kros-članove, tako što ih redukuje ili potpuno eliminiše. Wigner-ova distribucija i spektrogram predstavljaju specijalne slučajeve S-metoda, koji se dobijaju za funkcije prozora $P(\theta)=1$ (u ovom slučaju dobija se pseudo Wigner-ova distribucija), i $P(\theta)=2\pi\delta(\theta)$ (za ovu funkciju prozora dobija se spektrogram). U numeričkim realizacijama koristi se diskretna forma S-metoda, opisana relacijom [15]:

$$SM(n,k)=\sum_{i=-L_d}^{L_d}P(i)STFT(n,k+i)STFT^{*}(n,k-i), \tag{1.33}$$

gdje je sa $L_d$ označena širina prozora u frekvencijskom domenu [2], [4], [15] a koja ne bi trebalo da bude veća od polovine rastojanja između dva auto-člana signala. U suprotnom, doći će do pojave kros-članova u S-metodu.

S-metod ne zadovoljava marginalne uslove. Za razliku od S-metoda, distribucije iz Cohen-ove klase mogu da zadovoljavaju ove uslove, pa su pogodnije za primjenu u slučajevima kada se zahtijeva zadovoljenje ovih uslova.

***Ambiguity funkcija i Cohen-ova klasa distribucija***

Cohen-ova klasa distribucija [2], [20], [22], [40] takođe pripada kvadratnim distribucijama. Postoji veliki broj distribucija iz ove klase, u zavisnosti od jezgra koje se koristi, a od kojeg zavise i osobine te distribucije. Tako razlikujemo distribucije zasnovane na Choi-Williams-ovom, Gausov-om, Zhao-Atlas-Marks-ovom, Born-Jordan-ovom jezgru i druge [20], [22], [23], [40]. Parametri jezgra se biraju tako da obezbijede što je moguće bolju koncentraciju komponenti signala, a istovremeno da redukuju kros komponente.

Cohen-ova klasa distribucija može da se definiše polazeći od Wigner-ove distribucije, odnosno od ambiguity funkcije koja predstavlja dvodimenzionalnu Fourier-ovu transformaciju Wigner-ove distribucije i definisana je sljedećom relacijom:

$$AF(\theta,\tau)=\mathcal{F}_{t,\omega}\{WD(t,\omega)\}=\int_{-\infty}^{\infty}x\left(t+\frac{\tau}{2}\right)x^{*}\left(t-\frac{\tau}{2}\right)e^{-j\theta t}dt. \qquad (1.34)$$

Prilikom definisanja distribucija iz Cohen-ove klase polazi se od pretpostavke da su auto-članovi signala u ambiguity domenu locirani u blizini koordinatnog početka i osa $\theta$ i $\tau$, dok su kros članovi dislocirani od osa. Korišćenjem jezgra niskopropusnog tipa moguće je redukovati kros članove. Cohen-ova klasa distribucija se generalno može opisati sljedećom relacijom:

$$CD(t,\omega)=\frac{1}{2\pi}\int_{-\infty}^{\infty}\int_{-\infty}^{\infty}K(\theta,\tau)AF(\theta,\tau)e^{-j\theta t-j\omega\tau}d\tau d\theta, \qquad (1.35)$$

gdje je $K(\theta,\tau)$ dvodimenziona funkcija jezgra. Adekvatnim odabirom funkcije jezgra i podešavanjem njegovih parametara, moguće je redukovati ili u potpunosti eliminisati kros-članove. Međutim, jezgro utiče na koncentraciju članova signala u vremensko-frekvencijskoj ravni i u većini slučajeva je pogoršava. Zbog toga uvijek postoji kompromis između redukcije kros-komponenti i očuvanja dobre koncentracije komponenti signala.

Choi-Williams distribucija dobija se filtriranjem ambiguity funkcije eksponencijalnim jezgrom oblika $e^{-\theta^2\tau^2/\sigma^2}$, gdje je širina jezgra kontrolisana parametrom σ.

Zhao-Atlas-Marks jezgro omogućava redukciju kros-članova u velikoj mjeri, uz istovremeno zadržavanje dobre frekvencijske rezolucije. Nalazi primjenu u filtriranju govornih signala, a koristi jezgro oblika:

$$l(\theta,\tau)=|\tau|\frac{\sin(\theta\tau/2)}{\theta\tau/2}w(\tau), \qquad (1.36)$$

gdje je $w(\tau)$ funkcija prozora.

Born-Jordan-ova distribucija je specijalan slučaj Zhao-Atlas-Marks distribucije, sa jezgrom oblika (za $|\tau|=1$ i $w(\tau)$), koje se podešava promjenom parametara $\theta$ i τ:

$$l(\theta,\tau)=\frac{\sin(\theta\tau/2)}{\theta\tau/2}. \qquad (1.37)$$

Gausovo jezgro opisanom je relacijom:

$$l(\theta,\tau)=e^{-(\theta^2/2\sigma_1{}^2+\tau^2/2\sigma_2{}^2)}, \tag{1.38}$$

a oblik ovog jezgra je moguće podešavati promjenom vrijednosti parametara $\sigma_1$ i $\sigma_2$. Izgled nekih od pomenutih jezgara dat je na slici 1.4.

Međutim, u slučaju signala sa brzo promjenljivom faznom funkcijom, distribucije iz Cohen-ove klase ne daju zadovoljavajuću tačnost pri estimaciji trenutne frekvencije komponenti signala. U tom slučaju moguće je koristiti distribucije sa kompleksnim argumentom vremena, koje obezbjeđuju visoku koncentraciju signala u vremensko-frekvencijskoj ravni, a kao posljedicu toga, obezbjeđuju tačniju estimaciju trenutne frekvencije pojedinačnih komponenti signala, u poređenju sa standardnim distribucijama.

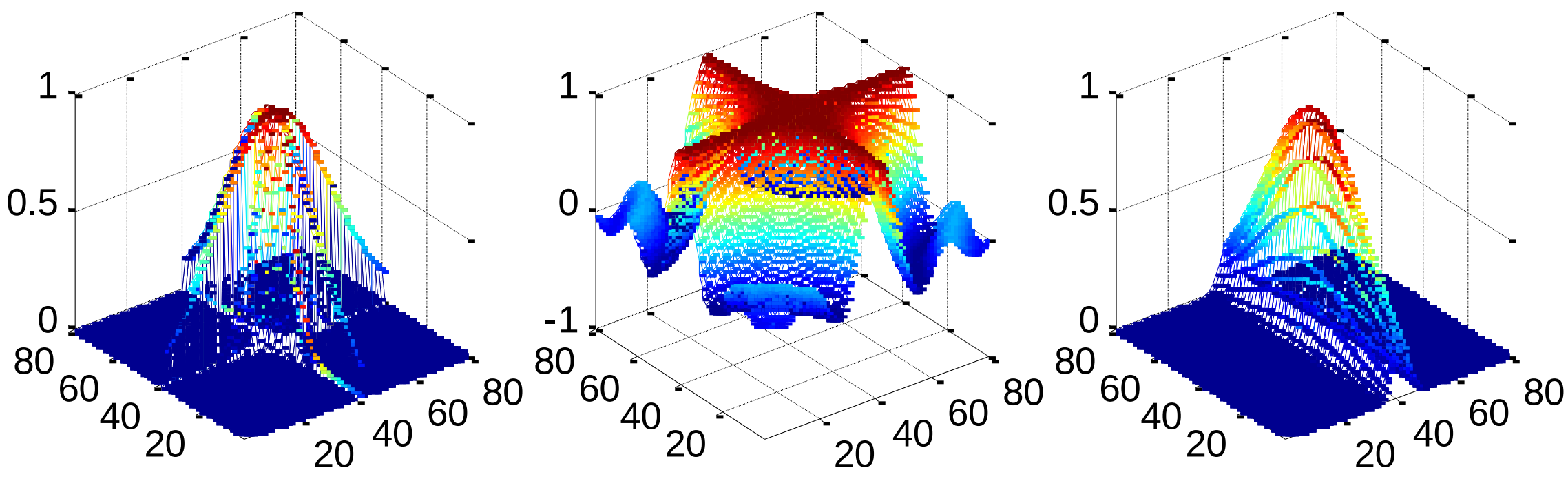

**Slika 1.4: Izgled a) Choi-Williams-ovog; b) Born Jordan-ovog i c) Gauss-ovog jezgra**

***Distribucije sa kompleksnim argumentom vremena***

Opšta forma distribucija sa kompleksnim argumentom vremena [2], [41], *N*-tog reda, može se definisati na sljedeći način:

$$CTD_N(t,\omega)=\int_{-\infty}^{\infty} R_N(t,\tau)e^{-j\omega\tau}d\tau, \tag{1.39}$$

gdje je sa $R_N(t,\tau)$ definisan moment signala sa kompleksnim argumentom, u skladu sa sljedećom relacijom:

$$R_N(t,\tau)=\prod_{k=0}^{N-1} x^{e^{-j\frac{2\pi k}{N}}}\left(t+\frac{\tau}{N}e^{j\frac{2\pi k}{N}}\right), \tag{1.40}$$

ili:

$$R_N(t,\tau)=\prod_{\substack{i=-N/2,\\ i\neq 0}}^{N/2} x\left(t+\frac{\tau}{Nsign(i)(a_i+jb_i)}\right)^{sign(i)(a_i+jb_i)}, \tag{1.41}$$

gdje $N$ predstavlja paran broj koji odgovara redu distribucije, a $a_i$ i $b_i$ su tačke na jediničnom krugu u kompleksnoj ravni. Distribucija $N$-tog reda sa kompleksnim argumentom vremena definisana je kao [42]:

$$\begin{aligned} CTD_N(t,\omega) &= \int_{-\infty}^{\infty} R_N(t,\tau)e^{-j\omega\tau}d\tau = \\ &= \int_{-\infty}^{\infty} \prod_{\substack{i=-N/2,\\ i\neq 0}}^{N/2} x\left(t+\frac{1}{N}\frac{\tau}{sign(i)(a_i+jb_i)}\right)^{sign(i)(a_i+jb_i)} e^{-j\omega\tau}d\tau. \end{aligned} \tag{1.42}$$

Faktor rasipanja je oblika:

$$\begin{aligned} S(t,\tau) &= \phi^{(3)}(t)\frac{\tau^3}{16\cdot 3!(a+jb)^2}+\phi^{(5)}(t)\frac{\tau^5}{256\cdot 5!(a+jb)^4}+ \\ &+\phi^{(7)}(t)\frac{\tau^7}{4096\cdot 7!(a+jb)^6}+.... \end{aligned} \tag{1.43}$$

Polazeći od momenta sa kompleksnim argumentom vremena moguće je definisati realni i kompleksni dio ambiguity funkcije. Ambiguity funkcija za moment signala sa kompleksnim argumentom definisana je sljedećom relacijom [42]:

$$AF_{ct}(\theta,\tau)=\int_{-\infty}^{\infty} R_{cN}(t,\tau)e^{-j\theta t}dt, \tag{1.44}$$

gdje je $R_{cN}(t,\tau)$ imaginarni dio momenta $R_N(t,\tau)$, dok je ambiguity funkcija za moment signala sa realnim argumentom definisana kao [42]:

$$AF_{rt}(\theta,\tau)=\int_{-\infty}^{\infty} R_{rN}(t,\tau)e^{-j\theta t}dt=\int_{-\infty}^{\infty} x\left(t+\frac{\tau}{N}\right)x^*\left(t-\frac{\tau}{N}\right)e^{-j\theta t}dt, \tag{1.45}$$

gdje je $R_{rN}(t,\tau)$ realni dio momenta $R_N(t,\tau)$. Obje ove funkcije filtriraju se funkcijom jezgra $K(\theta,\tau)$:

$$\begin{aligned} AF_{rt}^K(\theta,\tau) &= K(\theta,\tau)AF_{rt}(\theta,\tau), \\ AF_{ct}^K(\theta,\tau) &= K(\theta,\tau)AF_{ct}(\theta,\tau). \end{aligned} \tag{1.46}$$

Konvolucija ove dvije funkcije sa funkcijom prozora $W(\varepsilon)$ daje rezultujuću ambiguity funkciju [42]:

$$AF_{CTD}(\theta,\tau)=\int_{-\infty}^{\infty}\int_{-\infty}^{\infty}\int_{-\infty}^{\infty}W(\varepsilon)e^{-j\varepsilon\tau_1}e^{j\varepsilon(\tau-\tau_1)}AF_{rt}^{K}(\theta_1,\tau_1)AF_{ct}^{K}(\theta-\theta_1,\tau-\tau_1)d\tau_1 d\theta_1 d\varepsilon, \qquad (1.47)$$

Konačno, distribucija sa kompleksnim argumentom vremena dobija se u skladu sa sljedećom relacijom:

$$CTD(t,\omega)=\frac{1}{2\pi}\int_{-\infty}^{\infty}\int_{-\infty}^{\infty}AF_{CTD}(\theta,\tau)e^{j\theta t-j\omega\tau}d\tau d\theta. \qquad (1.48)$$

Adekvatnim izborom reda distribucije kompleksnog argumenta, može se smanjiti uticaj viših izvoda faze i poboljšati rezolucija u vremensko-frekvencijskoj ravni.

## 1.5. Robustne transformacije

Robustna statistika uvedena je sa ciljem efikasnije analize signala zahvaćenih impulsnim šumom, i daje fundamentalne principe za rješavanje problema u kojima se javljaju impulsne smetnje [49]-[52]. Standardne transformacije i vremensko-frekvencijske distribucije daju zadovoljavajuće rezultate kod signala zahvaćenih Gauss-ovim šumom. Međutim, ukoliko je signal zahvaćen impulsnim šumom, rješenje dobijeno primjenom standardnih transformacija ili vremensko-frekvencijskih distribucija neće biti zadovoljavajuće, pa se u tim slučajevima koriste drugi pristupi. Osnovni pristupi za estimaciju signala u zavisnosti od prisutnog šuma, su M-, L- i R- estimacioni pristupi [2], [15]. M-estimacioni pristup (*maximum likelihood estimate*) primjenjuje se u slučajevima kada su poznate statističke osobine šuma, ali je ovaj pristup osjetljiv na varijacije funkcije gustine vjerovatnoće šuma (*pdf*), i ne obezbjeđuje rješenje u zatvorenoj formi [2]. Zbog toga su uvedeni drugi pristupi, kao što su R- i L-estimacija. Ovi pristupi mogu dati zadovoljavajuće rješenje u prisustvu kombinacije Gausovog i impulsnog šuma [2].

### 1.5.1. L-estimacioni pristup

L-estimacioni pristup [2], [4], [15] može se primijeniti na diskretne transformacije signala, odbacivanjem dijela odbiraka zahvaćenih šumom i usrednjavanjem preostalih koeficijenata (koeficijenata signala koji su pomnoženi baznim funkcijama). Do sada je ovaj pristup primijenjivan uglavnom na Fourier-ovim baznim funkcijama, što je rezultiralo definisanjem robustne Fourier-ove transformacije [2], [4]. Međutim, ovaj pristup se koristi i u vremensko-frekvencijskoj analizi, pa tako u literaturi nalazimo robustne vremensko-frekvencijske distribucije kao npr. robustnu STFT, robustnu WD, robustni S-metod, robustnu formu distribucija sa kompleksnim argumentom vremena, [2], [4], [15], [20], [49]-[52] itd.

L-estimacioni pristup zasnovan je na α-trimovanom filtru. Uveden je zbog analize signala koji su zahvaćeni mješavinom različitih tipova šuma, ali se može efikasno primijeniti na signale zahvaćene samo impulsnim ili Gausovim šumom, gdje daje medijan formu ili formu srednje vrijednosti, kao specijalne slučajeve.

Opšti oblik transformacije u diskretnoj formi može se zapisati na sljedeći način:

$$K(p)=\sum_{n=0}^{N-1} x(n)\psi_p(n), \quad p\in[0,...,N-1], \tag{1.49}$$

gdje se pretpostavlja da je $s(n)$ signal zahvaćen šumom, tj. $x(n)=s(n)+v(n)$, a $v(n)$ predstavlja šum. Filtrirani transformacioni koeficijenti, $K(p)$, mogu se dobiti rješavanjem optimizacionog problema, tj. minimizacijom funkcije gubitaka $L(e)$:

$$K(p)=\arg\min_{\mu}\sum_{n=0}^{N-1} L(e)=\arg\min_{\mu}\sum_{n=0}^{N-1} L\left(K(p)-\mu\right). \tag{1.50}$$

gdje je sa $K(p)$ označena transformacija signala zahvaćenog šumom. Na osnovu pristupa ML pristupa (ML-*maximum likelihood*) [2], [15], za poznatu funkciju gustine vjerovatnoće *pdf*, $p_v(e)$, forma funkcije gubitaka se određuje na osnovu sljedeće relacije:

$$L(e)=-\log[p_v(e)]. \tag{1.51}$$

Ukoliko je signal zahvaćen Gausovim šumom, koristi se filtar srednje vrijednosti. Naime, za $p_v(e)\sim e^{-|e|^2}$, funkcija gubitaka oblika $L(e)=|e|^2$ je ML estimator za ovaj tip šuma. U skladu sa ML pristupom, u slučaju impulsnih šumova (npr. Laplace-ovog, Cauchy-ijevog ili α-stabilnog šuma), funkcija gubitaka treba da bude u obliku $L(e)=|e|$. Međutim, ML pristup je osjetljiv na varijacije *pdf* funkcije šuma. Takođe, u praktičnim aplikacijama signali su obično zahvaćeni kombinacijom impulsnog i Gausovog šuma, pa u takvim slučajevima Huber-ova teorija estimacije rješenje bazira na L-estimacionom pristupu.

L-estimacija neke transformacije $K(p)$ definiše se na sljedeći način [2]:

$$K_L(p)=\sum_{i=0}^{N-1} l_i x_s(i), \tag{1.52}$$

gdje parameter $N$ definiše dužinu signala, a $l_i$ su koeficijenti filtra. Sortirane vrijednosti signala pomnožene sa baznim funkcijama $\psi_p(n)$ date su u vektoru $x_s$, koji je dobijen na sljedeći način:

$$x_s(n)=sort\left\{x(n)\psi_p(n),\ \ n\in[0,N-1]\right\}. \tag{1.53}$$

Vrijednosti koeficijenata filtra $l_i$ se mogu dobiti na sljedeći način (za parno $N$):

$$l_i=\begin{cases}\dfrac{1}{4a+N(1-2a)}, & i\in[(N-2)a,a(2-N)+N-1],\\ 0, & \text{za ostale vrijednosti parametra } i,\end{cases} \tag{1.54}$$

i izvedeni su po analogiji sa $\alpha$-trimovanim filtrom. Standardna forma transformacije dobija se za $a$=0, dok se medijan forma dobija za $a$=0.5.

### 1.5.2. Robustne forme distribucija zasnovanih na ambiguity domenu

U četvrtoj glavi korišćena je robustna forma distribucije sa kompleksnim argumentom vremena, bazirana na robustnoj formi ambiguity funkcije. Robustna forma ambiguity funkcije dobija se rješavanjem optimizacionog problema [20]:

$$AF(p,n)=\arg\min_{\varepsilon}\sum_{k=-M/2}^{M/2-1} L(e(p,n,k))=\arg\min_{\varepsilon}\sum_{k=-M/2}^{M/2-1} L\left(x(k+n)x^*(k-n)e^{-j\frac{2\pi}{M}kp}-\varepsilon\right). \tag{1.55}$$

L-forma ambiguity funkcije se može definisati na sljedeći način:

$$AF_L(p,n)=\sum_{i=-M/2}^{M/2-1} l_i(r_i^s(p,n)+j\cdot i_i^s(p,n)), \tag{1.56}$$

gdje su $r_i^s$ i $i_i^s$ sortirani elementi od $R(p,n)$ i $I(p,n)$, tj.:

$$\begin{aligned} r_i\in R(p,n)&=\left\{\mathrm{Re}\left[x(k+n)x^*(k-n)e^{-j\frac{2\pi}{M}kp}\right], n\in\left[-\frac{M}{2},\frac{M}{2}\right)\right\},\\ i_i\in I(p,n)&=\left\{\mathrm{Im}\left[x(k+n)x^*(k-n)e^{-j\frac{2\pi}{M}kp}\right], n\in\left[-\frac{M}{2},\frac{M}{2}\right)\right\}.\end{aligned} \tag{1.57}$$

Koeficijenti $l_i$ su definisani relacijom (1.54). Robustna forme distribucije sa kompleksnim argumentom vremena *RCTD* se, dakle, može definisati na sljedeći način:

$$RCTD(t,\omega)=\frac{1}{2\pi}\int_{-\infty}^{\infty}\int_{-\infty}^{\infty}AF_{CTD}^{R}(\theta,\tau)e^{j\theta t-j\omega\tau}d\tau d\theta\,, \tag{1.58}$$

gdje je sa $AF_{CTD}^{R}(\theta,\tau)$ označena robustna forma ambiguity funkcije.

# 2. Dekompozicija na sopstvene vrijednosti i sopstvene vektore i njena primjena u muzičkim signalima

Dekompozicija na sopstvene (*Eigenvalue Decomposition* - EVD) ili dekompozicija na singularne (*Singular Value Decomposition* - SVD) vrijednosti i vektore predstavlja metod koji transformiše originalne, korelisane promjenljive, u skup nekorelisanih varijabli [53]-[57]. Singularni i sopstveni vektori imaju brojne primjene u obradi signala. Koriste se za karakterizaciju signala i njihovih komponenti. Primijenjivi su na vremensko-frekvencijske distribucije u cilju izdvajanja nekih specifičnih osobina signala koji se kasnije mogu koristiti za njihovu karakterizaciju. U tom cilju mogu se koristiti kako sopstvene/singularne vrijednosti, tako i sopstveni/singularni vektori, naročito vektori koji odgovaraju najvećim sopstvenim/singularnim vrijednostima [15], [16], [53], [55]-[57]. Dekompozicija na sopstvene ili singularne vektore se može koristiti u redukciji šuma a kombinuje se i sa neuralnim mrežama u cilju klasifikacije EEG signala.

U ovom radu korišćena je EVD, a upotrijebljena je u analizi multikomponentnih signala u bežičnim komunikacijama kao i za analizu muzičkih signala. Naime, analiza multikomponentnih signala često zahtijeva razdvajanje komponenti i analizu svake komponente zasebno. Odvajanje komponenti signala može se vršiti u vremenskom domenu (ako komponente signala imaju konačno trajanje i ako se intervali u kojima se komponente pojavljuju ne preklapaju), ili u frekvencijskom domenu (propuštanjem signala kroz odgovarajući filtar, pod uslovom da komponente zauzimaju različite frekvencijske opsege). Ukoliko nijesu poznate frekvencije komponenti signala, analiza se može sprovesti korišćenjem vremensko-frekvencijskih distribucija, koje se mogu kombinovati sa EVD/SVD-om u cilju uspješnog razdvajanja komponenti signala.

Imajući u vidu da je EVD specijalan slučaj SVD-a, prvo će biti uvedena SVD. Dekompozicija na singularne vrijednosti, SVD, za posmatranu matricu **S**, definisana je na sljedeći način [15], [16], [20]:

$$\mathbf{S} = \mathbf{U}\mathbf{\Sigma}\mathbf{V}^{\mathbf{T}}, \tag{2.1}$$

gdje $\mathbf{\Sigma}$ predstavlja dijagonalnu matricu singularnih vrijednosti, a vrijednosti su sortirane u opadajućem redu duž glavne dijagonale. Matrice **U** i **V** su ortonormalne matrice čije kolone predstavljaju lijeve i desne singularne vektore. Ako je matrica **S** pravougaona matrica dimenzija $M \times N$ ($M > N$), onda matrice koje se dobijaju dekompozicijom imaju sljedeće dimenzije: **U** je dimenzija $M\times M$, **Σ** je dimenzija $M\times N$ a matrica **V** ima veličinu $N\times N$. Ovo važi u slučaju potpune SVD. U cilju uštede memorijskih resursa, može se računati nepotpuna SVD, i to na sljedeći način:

- računanjem samo $N$ kolona matrice **U** (tj. matrica $\mathbf{U}_{M\times M}$ sada ima dimenzije $\mathbf{U}_{M\times N}$);
- računanjem samo $N$ vrsta matrice **Σ** (tj. matrica $\mathbf{\Sigma}_{M\times N}$ sada ima dimenzije $\mathbf{\Sigma}_{N\times N}$).

Ukoliko je matrica **S** kvadratna matrica dimenzija $N\times N$, dekompozicija se svodi na dekompoziciju na sopstvene vrijednosti:

$$\mathbf{S} = \mathbf{U\Sigma U}^{\mathbf{T}}, \tag{2.2}$$

gdje su $\mathbf{U}_{N\times N}$ i $\mathbf{\Sigma}_{N\times N}$ matrice sopstvenih vektora i sopstvenih vrijednosti, respektivno.

Kao što je napomenuto, u ovom radu koristili smo EVD za razdvajanje komponenti signala iz autokorelacione funkcije. Autokorelaciona funkcija, na koju se primijenjuje EVD, formira se na osnovu odgovarajuće vremensko-frekvencijske distribucije. Vremensko-frekvencijska distribucija se bira tako da obezbjeđuje dobru koncentraciju komponenti signala u vremensko-frekvencijskoj ravni a eliminiše uticaj kros-članova.

## 2.1. Razdvajanje komponenti signala korišćenjem dekompozicije na sopstvene vrijednosti i vremensko-frekvencijske analize

Polazeći od izraza za inverznu Wigner-ovu distribuciju, $c$-ta komponenta nekog multikomponentnog signala se može predstaviti na sljedeći način [15], [20], [53]:

$$x_c(n+m)x_c^*(n-m) = \frac{1}{N+1}\sum_{k=-N/2}^{N/2} WD_c(n,k)e^{j\frac{2\pi}{N+1}k2m}. \tag{2.3}$$

Zamjenom $n+m=p$ i $n-m=q$, dobija se:

$$x_c(p)x_c^*(q) = \frac{1}{N+1}\sum_{k=-N/2}^{N/2} WD_c\left(\frac{p+q}{2},k\right)e^{j\frac{2\pi}{N+1}(p-q)k}. \tag{2.4}$$

Lijeva strana prethodne relacije odgovara autokorelacionoj matrici:

$$R_c(p,q) = x_c(p)x_c^*(q), \tag{2.5}$$

gdje je $x_c(p)$ vektor kolona čiji elementi odgovaraju elementima signala, a $x_c^*(q)$ je vektor vrsta sa kompleksno konjugovanim vrijednostima. Za $M$ komponenti signala relacija (2.5) postaje:

$$\sum_{c=1}^{M} R_c\left(p,q\right) = \frac{1}{N+1} \sum_{k=-N/2}^{N/2} \sum_{c=1}^{M} WD_c\left(\frac{p+q}{2},k\right) e^{j\frac{2\pi}{N+1}(p-q)k}. \tag{2.6}$$

Imajući u vidu da je S-metod multikomponentnog signala $x(n)$, gdje je $x(n) = \sum_{c=1}^{M} x_c(n)$, jednak sumi Wigner-ovih distribucija pojedinačnih komponenti, tj. $SM(n,k) = \sum_{c=1}^{M} WD_c\left(n,k\right)$, onda se relacija (2.6) može napisati u sljedećom obliku:

$$\sum_{c=1}^{M} R_c\left(p,q\right) = \frac{1}{N+1} \sum_{k=-N/2}^{N/2} SM\left(\frac{p+q}{2},k\right) e^{j\frac{2\pi}{N+1}(p-q)k}. \tag{2.7}$$

Ako se na autokorelacionu matricu, dobijenu prethodnom relacijom, primijeni SVD dobija se sljedeća relacija:

$$R(p,q) = \sum_{c=1}^{M} R_c(p,q) = \mathbf{U\Sigma V}^T. \tag{2.8}$$

Posmatran je slučaj kada je vremensko-frekvencijska distribucija predstavljena kvadratnom matricom. Tada je autokorelaciona funkcija $R(p,q)$ kvadratna matrica koja je simetrična u odnosu na glavnu dijagonalu, pa važi da su lijeva i desna matrica sopstvenih vektora jednake, tj. $\mathbf{U} = \mathbf{V}$. Matricu singularnih vrijednosti označićemo sa lambda, tj. $\mathbf{\Sigma} = \mathbf{\Lambda}$.

Problem dekompozicije svodi se na računanje sopstvenih vektora, pri čemu svaki sopstveni vektor dominantno odgovara jednoj komponenti signala. Drugim riječima, sopstveni vektori su jednaki komponentama signala sa tačnošću do fazne konstante [57]. Autokorelaciona matrica $R(p,q)$ dekomponuje se u skladu sa sljedećom relacijom:

$$R(p,q) = \sum_{c=1}^{M} R_c(p,q) = \sum_{j=1}^{M} \lambda_j u_j(n) u^*{}_j(n). \tag{2.9}$$

gdje $\lambda_j$ predstavljuju sopstvene vrijednosti a $u_j(n)$ sopstvene vektore autokorelacione matrice $\mathbf{R}$. Dobijeni sopstveni vektori odgovaraju komponentama signala, kojih je $M$, dok sopstvene vrijednosti odgovaraju energijama komponenti signala.

## 2.2. Dekompozicija muzičkih signala

Postoje brojni pristupi u analizi muzičkih signala, polazeći od analize u frekvencijskom domenu, do detaljnijih pristupa koji koriste vremensko-frekvencijske predstave signala [14]-[21]. U poređenju sa analizom u frekvencijskom domenu, pristupi zasnovani na vremensko-frekvencijskom predstavljanju su efikasniji prilikom praćenja promjena u muzičkom signalu. Analiza u vremensko-frekvencijskom domenu može pomoći u razlikovanju pojedinih muzičkih instrumenata na kojima je ton odsviran [21]. U ovom radu je predložena procedura razdvajanja komponenti muzičkog signala, zasnovana na vremensko-frekvencijskoj analizi. Razdvajanjem komponenti signala moguća je preciznija analiza a moguće su i određene obrade muzičkih tonova.

Procedura zasnovana na dekompoziciji na sopstvene vektore ne zahtijeva bilo kakvo predznanje o signalu, pa se efikasno može primijeniti i na harmonijske i na neharmonijske signale. Imajući u vidu da energije komponenti muzičkih signala variraju u zavisnosti od frekvencije na kojoj se komponenta nalazi, u radu je predloženo iterativno rješenje. Predloženi algoritam je prikazan na slici 2.1. Dekompozicija je zasnovana na više iteracija, a u svakoj iteraciji odvajaju se komponente čije su energije približno jednake [15], [20].

Ako se sa $x(t)$ označi signal koji se dekomponuje, onda se njegova Fourier-ova transformacija može označiti sa $X(\omega)$. Parametar $K$ označava broj komponenti koje se izdvajaju u jednoj iteraciji.

Prvi korak algoritma je računanje S-metoda za čitav signal, i primjena dekompozicije na dobijenu matricu:

$$R(p,q)=\frac{1}{N+1}\sum_{k=-N/2}^{N/2} SM\left(\frac{p+q}{2},k\right)e^{j\frac{2\pi}{N+1}(p-q)k}=\sum_{j=1}^{K}\lambda_j u_j(n)u^*_j(n) \ . \qquad (2.10)$$

Imajući u vidu da je spektar signala simetričan, broj izdvojenih komponenti biće jednak $K/2$. U sljedećem koraku, S-metod se računa za svaki sopstveni vektor $u_j$ koji odgovara $j$-oj komponenti signala, u skladu sa sljedećom relacijom:

$$SM_j(n,k)=\sum_{l=-L_j}^{L_j} STFT_j(n,k+l)STFT_j^*(n,k-l), \qquad (2.11)$$

gdje parameter $j$ ima vrijednosti $j$=1,2,…,$K$/2. Centralna frekvencija $j$-te komponente signala u prvoj iteraciji određuje se u skladu sa sljedećom relacijom:

$$\omega_j = \underset{k}{argmax}\{SM_j(n,k)\}. \tag{2.12}$$

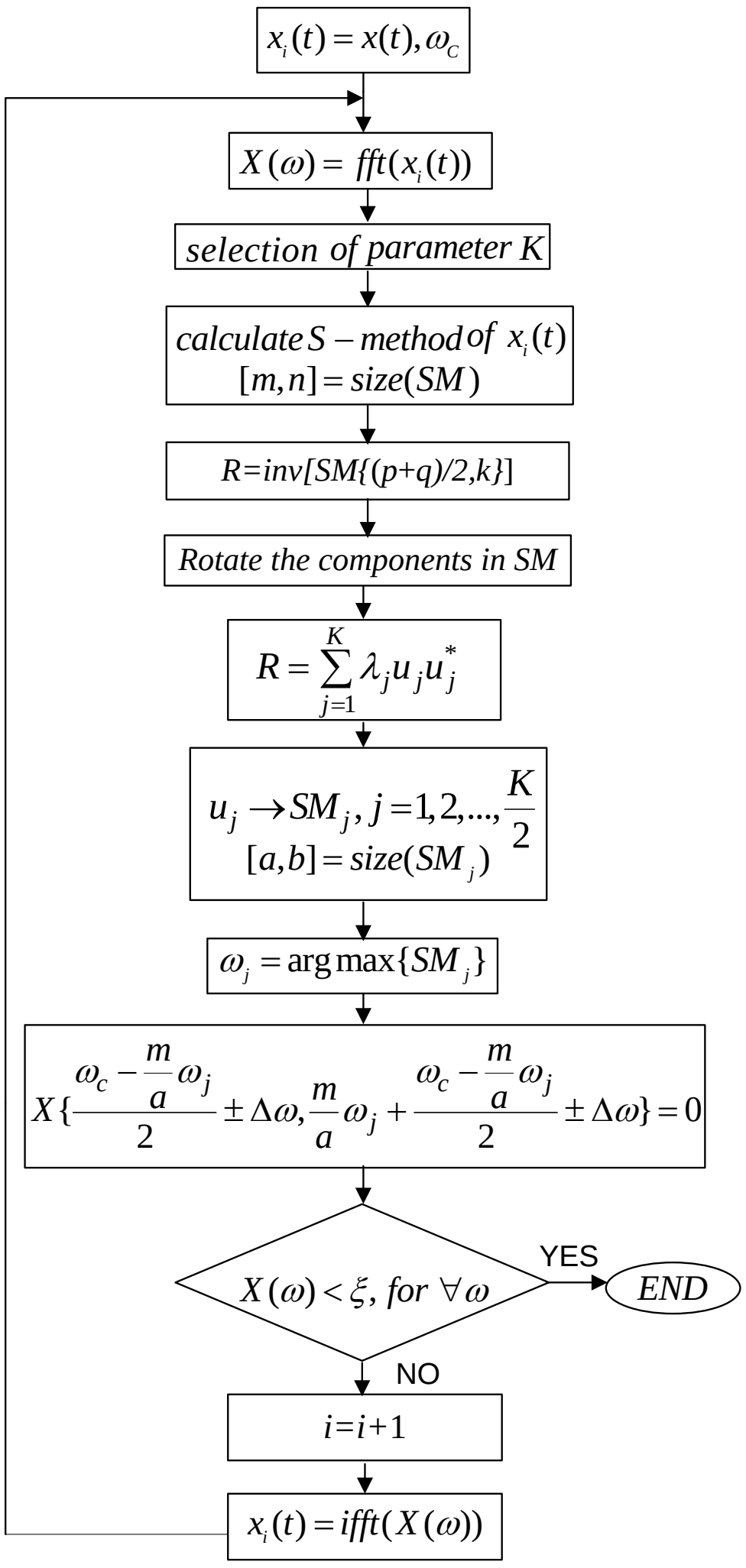

**Slika 2.1: Algoritam dekompozicije muzičkih signala**

Nakon identifikacije pozicija odvojenih komponenti (tj. njihovih centralnih frekvencija), odgovarajući region frekvencija u Fourier-ovoj transformaciji postavlja se na nulu [15], [16], [20]:

$$X\{\omega\} = \begin{cases} 0, \text{ ako važi } \omega \in \left\{ \left( \frac{\omega_c}{2} - \frac{m}{2a}\omega_j \right) \pm \Delta\omega, \ \frac{m}{a}\omega_j + \left( \frac{\omega_c}{2} - \frac{m}{2a}\omega_j \right) \pm \Delta\omega \right\}, \\ X\{\omega\}, \qquad \text{za ostalo } \omega. \end{cases} \tag{2.13}$$

Parametri $m$ i $a$ odgovaraju širinama prozora koji su korišćeni za računanje kratkotrajne Fourier-ove transformacije (STFT-a) za $SM$ i $SM_j$, respektivno. U cilju dobijanja signala za sljedeću iteraciju, računa se inverzna Fourier-ova transformacija u narednom koraku procedure:

$$x_{i+1}(t) = ifft(X(\omega)) \ . \tag{2.14}$$

Zatim se algoritam ponavlja, primijenjujući korake procedure na signal $x_{i+1}(t)$. Teorijski, procedura bi se završila kada bi se odvojile sve komponente signala, tj. kada bi se uklonio čitav frekvencijski sadržaj iz $X(\omega)$. U idealnom slučaju važilo bi da je tada $X(\omega)=0$. Međutim, u praksi je obično prisutan šum u signalu i teško je postići da je $X(\omega)=0$. Umjesto uslova $X(\omega)=0$, algoritam će se završiti kada je $X(\omega) < \xi$, gdje je $\xi$ prag određen empirijski i aproksimativno je jednak energiji šuma u Fourier-ovoj transformaciji signala $X(\omega)$.

### 2.2.1. Dekompozicija signala violine i flaute

Predložena procedura testirana je na signalima flaute i violine. Odabrana su ova dva signala za testiranje predloženog algoritma zbog primjetne razlike u njihovom spektralnom sadržaju. Naime, signal flaute ima značajno manji broj harmonika u poređenju sa posmatranim tonom violine [16], [20].

Signal falute je dužine 1000 odbiraka. S-metod je računat sa vrijednošću parametra $L$=6, u cilju postizanja dobre koncentracije komponenti signala u vremensko-frekvencijskoj ravni. Pojava kros-članova je izbjegnuta (imajući u vidu da su komponente relativno razmaknute u vremensko-frekvencijskoj ravni). S-metod signala flaute prikazan je na slici 2.2. Na slici je logaritamski prikaz, kako bi se uočile i komponente veoma malih energija, koje se u standardnom prikazu vremensko-frekvencijske reprezentacije ne bi mogle uočiti.

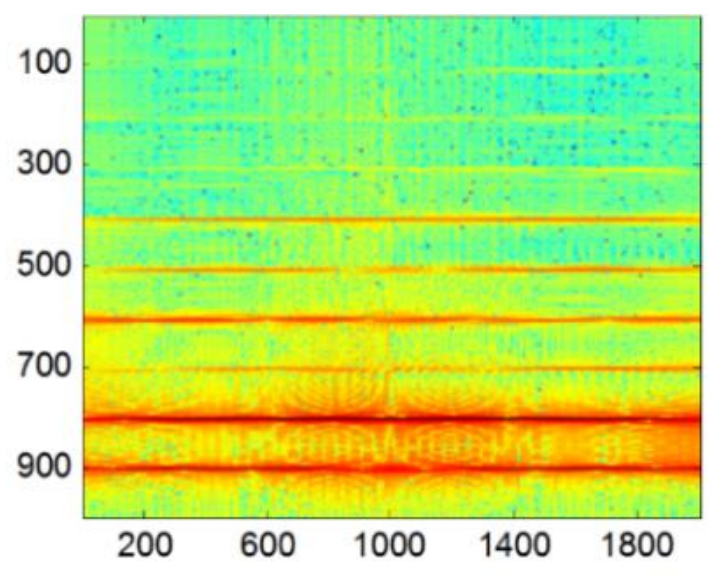

**Slika 2.2: S-metod signala flaute (horizontalna osa označava vrijeme, vertikalna osa označava frekvenciju)**

Komponente najvećih energija odvajaju se u prvoj iteraciji. U eksperimentima je odabrano da se odvajaju po dvije komponente u svakoj iteraciji. Drugim riječima, parametar *K* je 4 (imajući u vidu da jednoj komponenti, zbog simetričnosti spektra, odgovaraju dva sopstvena vektora). Za većinu signala moguće je odabrati veću vrijednost parametra *K*, međutim, u primjeru je uzeta optimalna vrijednost koja će raditi u većini slučajeva. Dakle, za sve iteracije uzeta je ista vrijednost *K*=4. Komponente koje su odvojene kroz 5 iteracija prikazane su na slici 2.3.

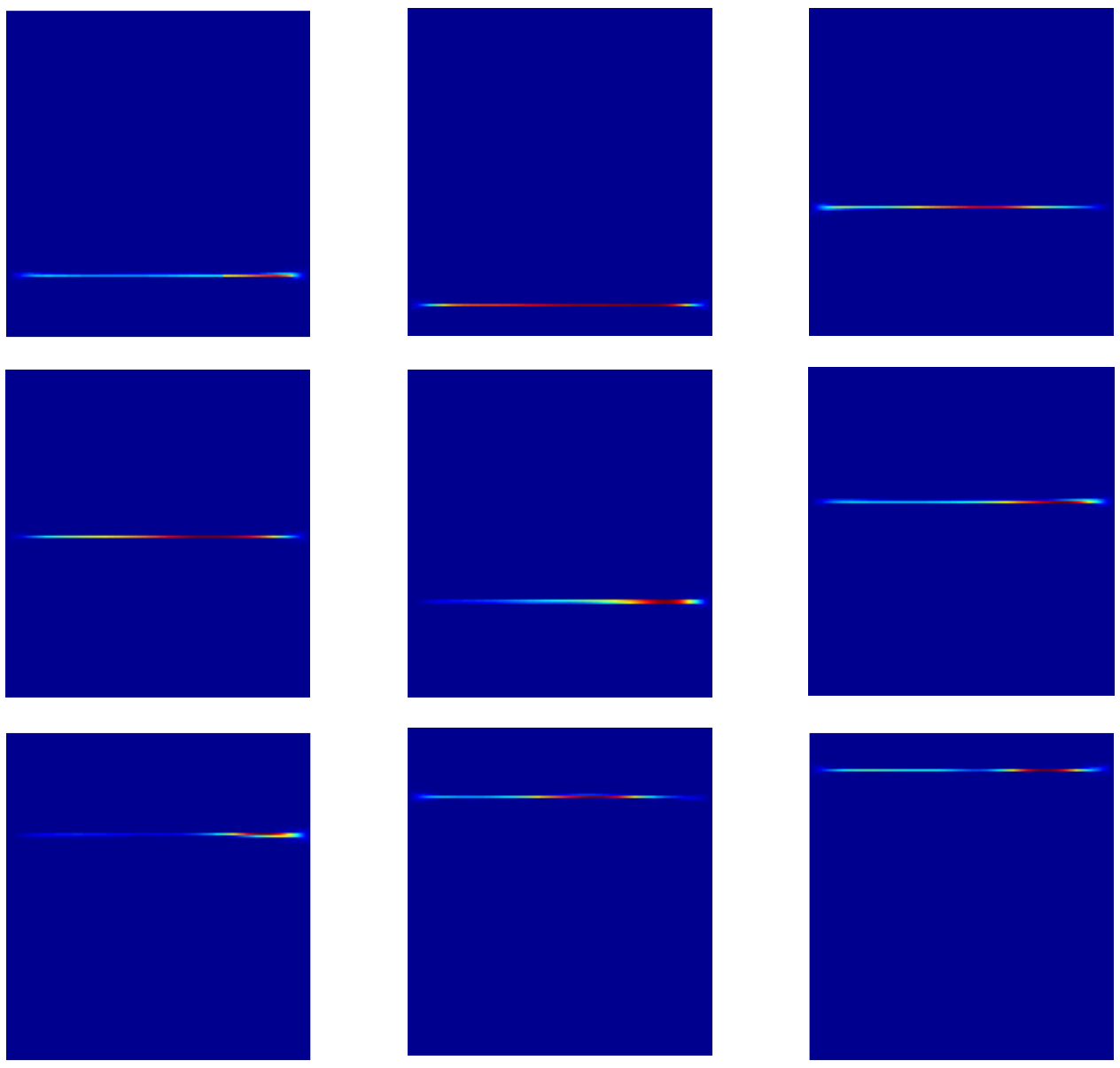

**Slika 2.3: Komponente signala flaute (na pozitivnim frekvencijama) koje su odvojene u 5 iteracija (horizontalna osa označava vrijeme, vertikalna osa označava frekvenciju)**

Predloženi algoritam testiran je i na kompleksnijim signalima, gdje je takođe pokazao uspješne rezultate, čak i u slučaju veoma bliskih komponenti signala u vremensko-frekvencijskoj ravni. Na primjer, posmatran je signal violine koji se sastoji od velikog broja harmonika na bliskim frekvencijama, kao što je prikazano na slici 2.4.

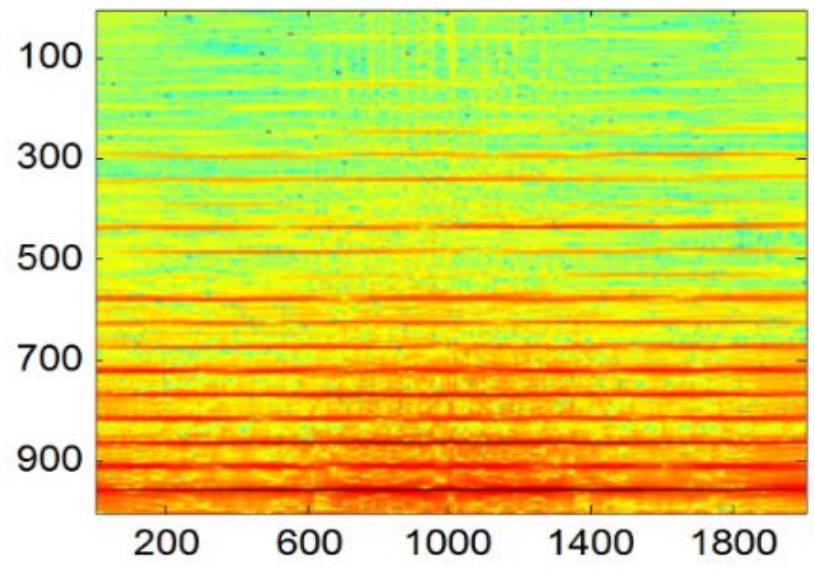

**Slika 2.4: S-metod signala violine (horizontalna osa označava vrijeme, vertikalna osa označava frekvenciju)**

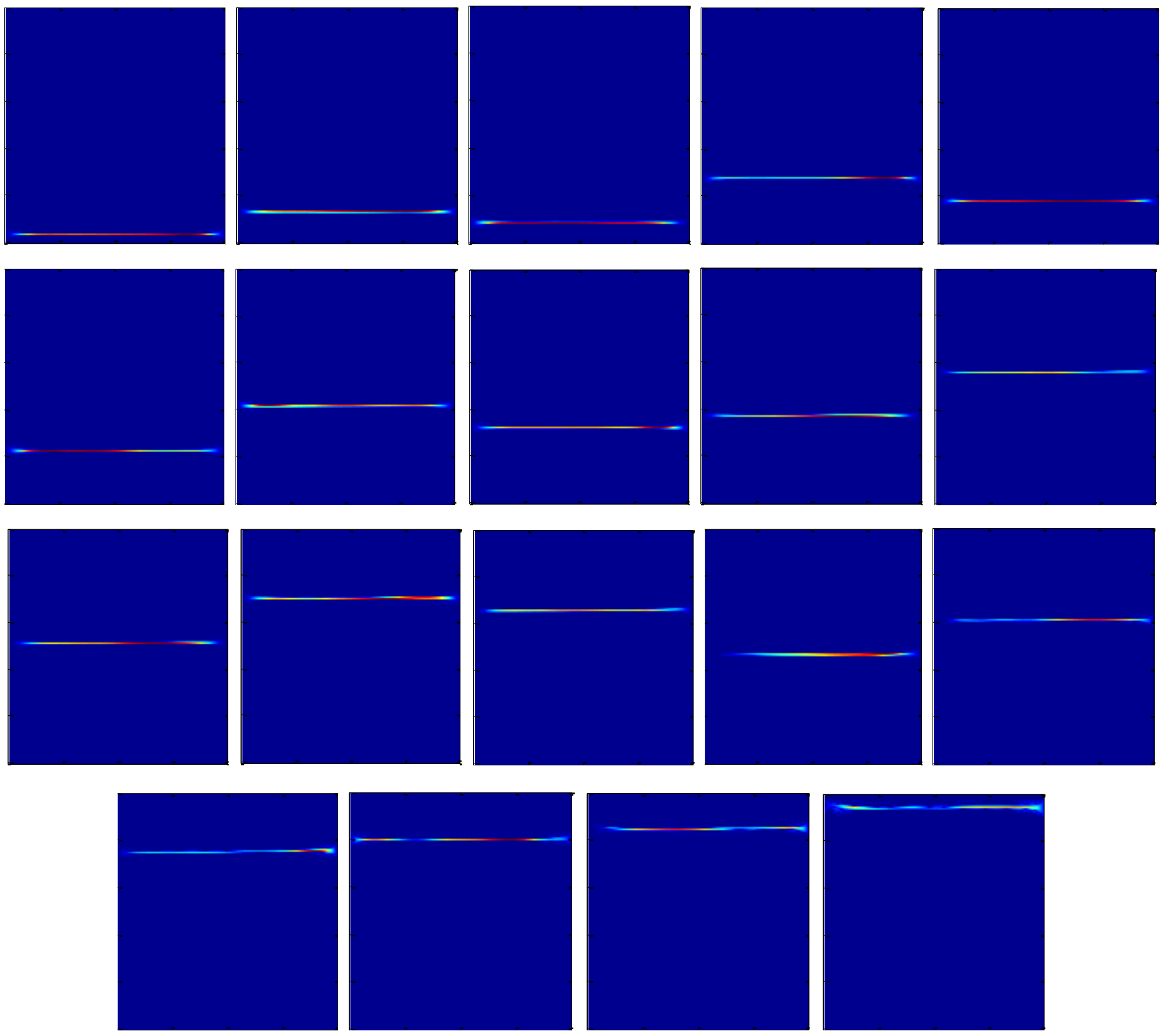

**Slika 2.5: Komponente signala violine, odvojene kroz 10 iteracija; Horizontalna osa označava vrijeme, vertikalna osa označava frekvenciju (prikazane su samo pozitivne frekvencije)**

Sa slike 2.4 se može vidjeti da je razmak između komponenti u vremensko-frekvencijskom domenu kod signala violine manji u poređenju sa razmakom komponenti kod signala flaute. Procedura dekompozicije primijenjena je na isti način i sa istim vrijednostima parametara kao i u prethodnom primjeru. Kao što se može vidjeti sa slike 2.5, svih 19 komponenti signala su uspješno odvojene kroz 10 iteracija.

### 2.2.2. Poređenje sa MUSIC algoritmom

Predloženi algoritam poređen je sa jednim od standardnih algoritama za harmonijsku analizu signala - MUltiple SIgnal Classification (MUSIC) algoritmom [16]. Algoritma MUSIC je primijenjen na audio signal, u cilju odvajanja komponenti signala. Posmatran je signal flaute iz prethodnog primjera. Pseudospektar signala, dobijen primjenom MUSIC algoritma, kao i spektar diskretne Fourier-ove transformacije prikazani su na slici 2.6. Može se primijetiti da pikovi u MUSIC pseudospektru ne odgovaraju DFT komponentama signala. U cilju primjene MUSIC algoritma, potrebno je specificirati broj *P* koji odgovara broju komponenti sinusoidalnog signala, što je u praksi često nepoznato.

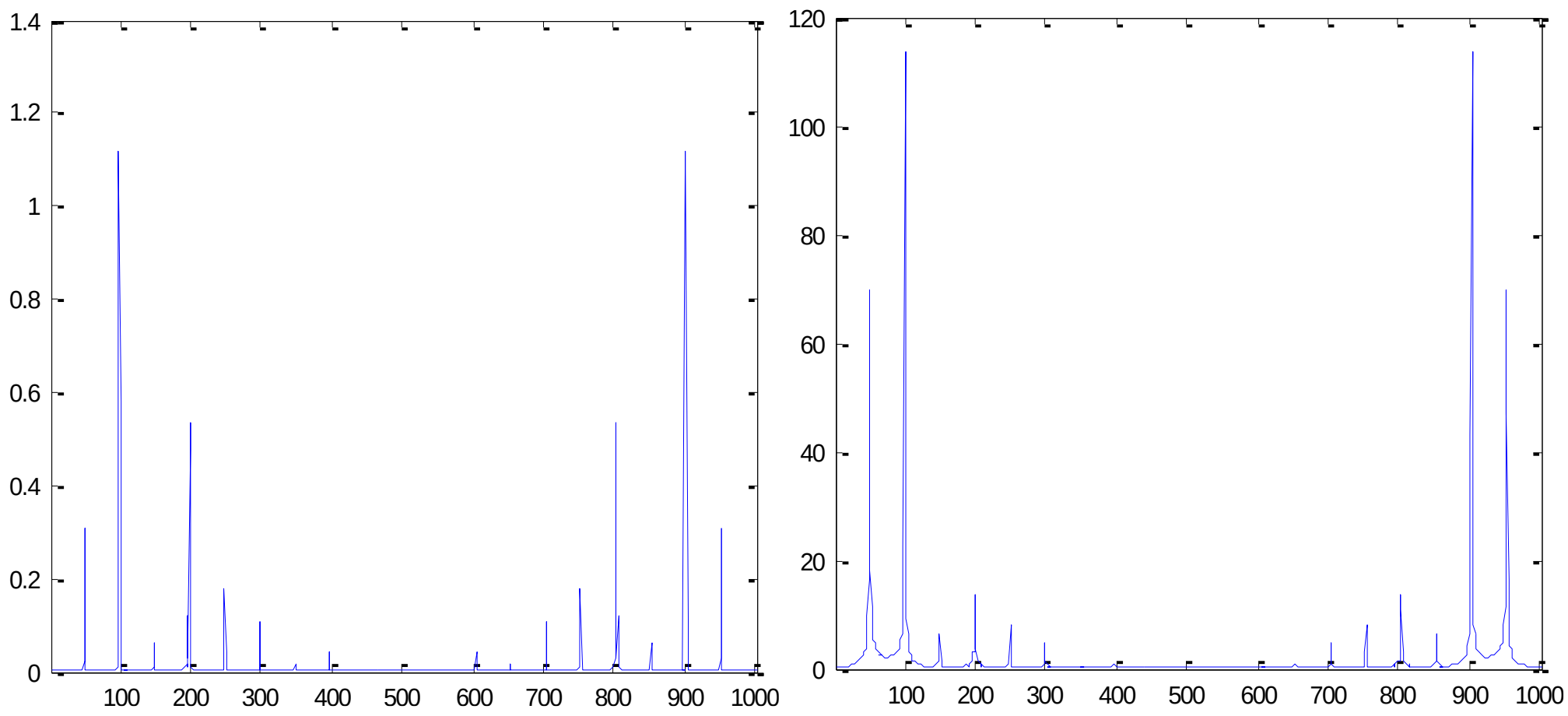

**a) b)**

**Slika 2.6: a) MUSIC pseudospektar, b) Diskretna Fourier-ova transformacija signala flaute**

Čak i da je broj komponenti signala poznat (npr., $P$=9 kod posmatranog signala flaute), odvajanje svih komponenti signala ovim algoritmom ne bi bilo moguće. Naime, prvih 6 sopstvenih vektora koji se dobijaju kao rezultat primjene MUSIC algoritma, odgovaraju komponentama signala (slika 2.7a). Ostatak sopstvenih vektora (eigenvektora) su, ili djelovi prvih 6 komponenti, ili pripradaju šumu (slika 2.7b), a imajući u vidu da se signal sastoji od ukupno 9 komponenti, zaključuje se da 3 komponente ovog signala ne mogu biti odvojene primjenom MUSIC algoritma. Performanse ovog algoritma se ne poboljšavaju niti u slučaju povećanja vrijednosti parametra *P*. Još jedan nedostatak ovog algoritma je loša rezolucija u vremensko-frekvencijskom domenu, koja umogome zavisi od širine prozora.

### 2.2.3. Primjer signala zahvaćenog šumom

Efikasnost predloženog algoritma testirana je i u prisustvu šuma [20]. Posmatrano je nekoliko slučajeva različite jačine šuma u signalu (tj. različitog odnosa signal-šum SNR). Pokazano je da dekompozicija uspješno odvaja komponente signala sve dok jačina šuma ne postane dominantna u odnosu na energiju nekih komponenti signala. Naime, tada se može desiti da komponente na visokim frekvencijama, zbog svoje niske energije, budu potpuno pokrivene šumom.

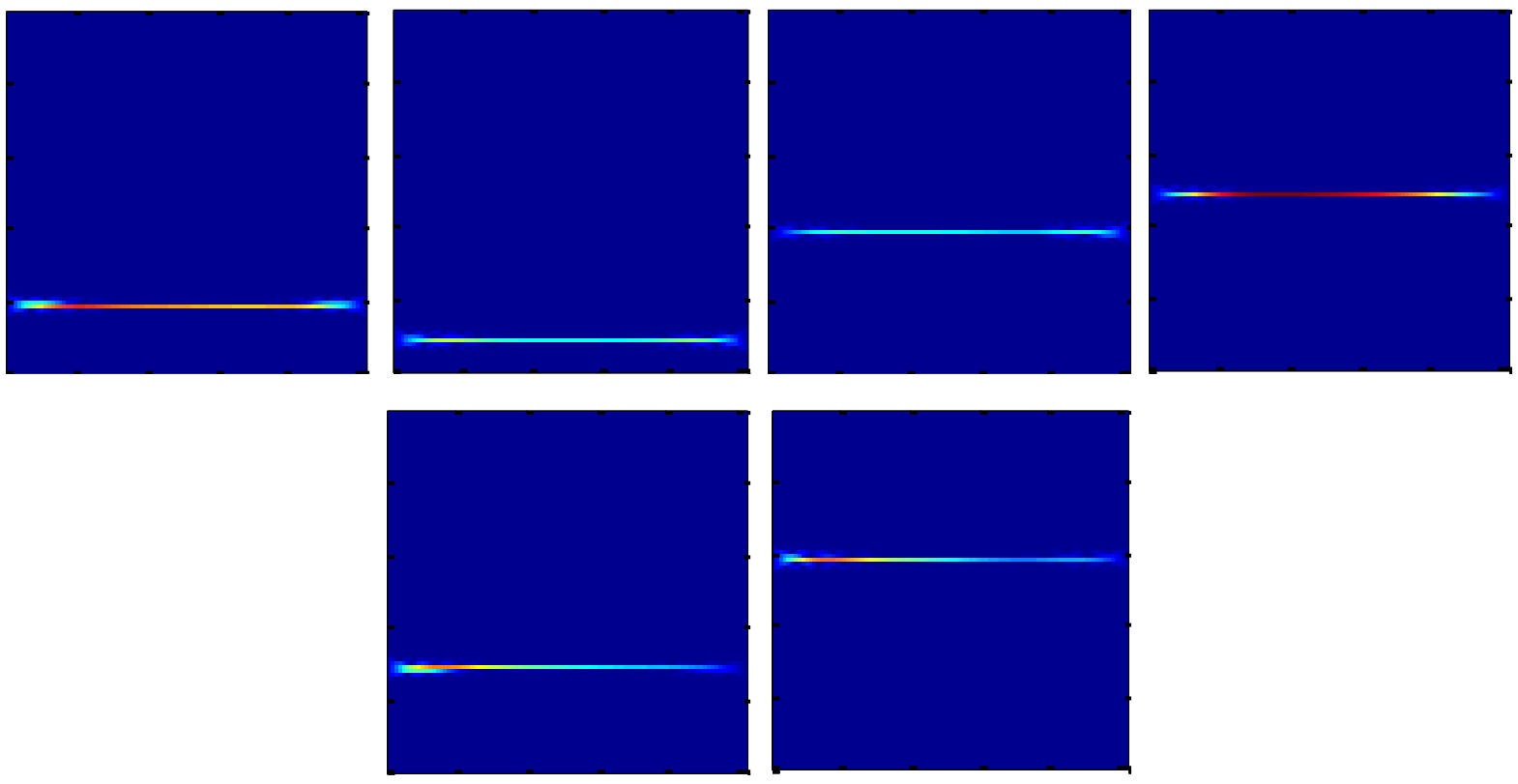

a)

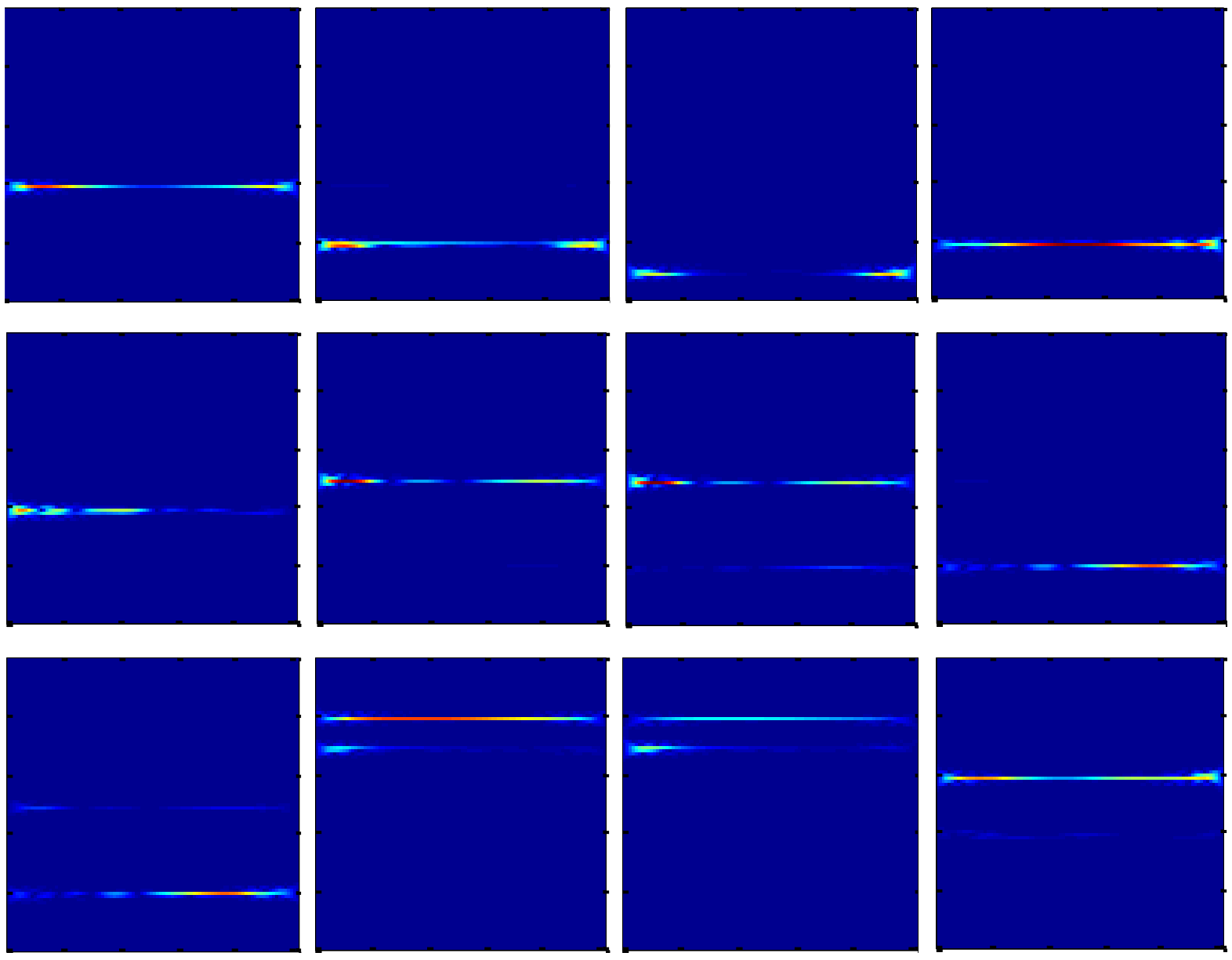

b)

**Slika 2.7: S-metod dobijen na bazi sopstvenih vektora odvojenih MUSIC algoritmom: a) odvojene komponente koje predstavljaju komponente signala, b) odvojene komponente koje predstavljaju djelove komponenti signala ili su šum**

Eksperimentalno je potvrđeno da predloženi algoritam uspješno odvaja komponente signala (čak i one na visokim frekvencijama) dok god je odnos signal-šum iznad 15 dB. Povećanjem intenziteta šuma više nije moguće razlikovati komponente niskih energija od šuma. Tada procedura razdvajanja komponenti ne radi. Primjeri dekompozicije signala flaute u prisustvu šuma, za različite vrijednosti SNR-a, prikazani su na slici 2.8, dok je broj uspješno odvojenih komponenti dat u tabeli 2-1.

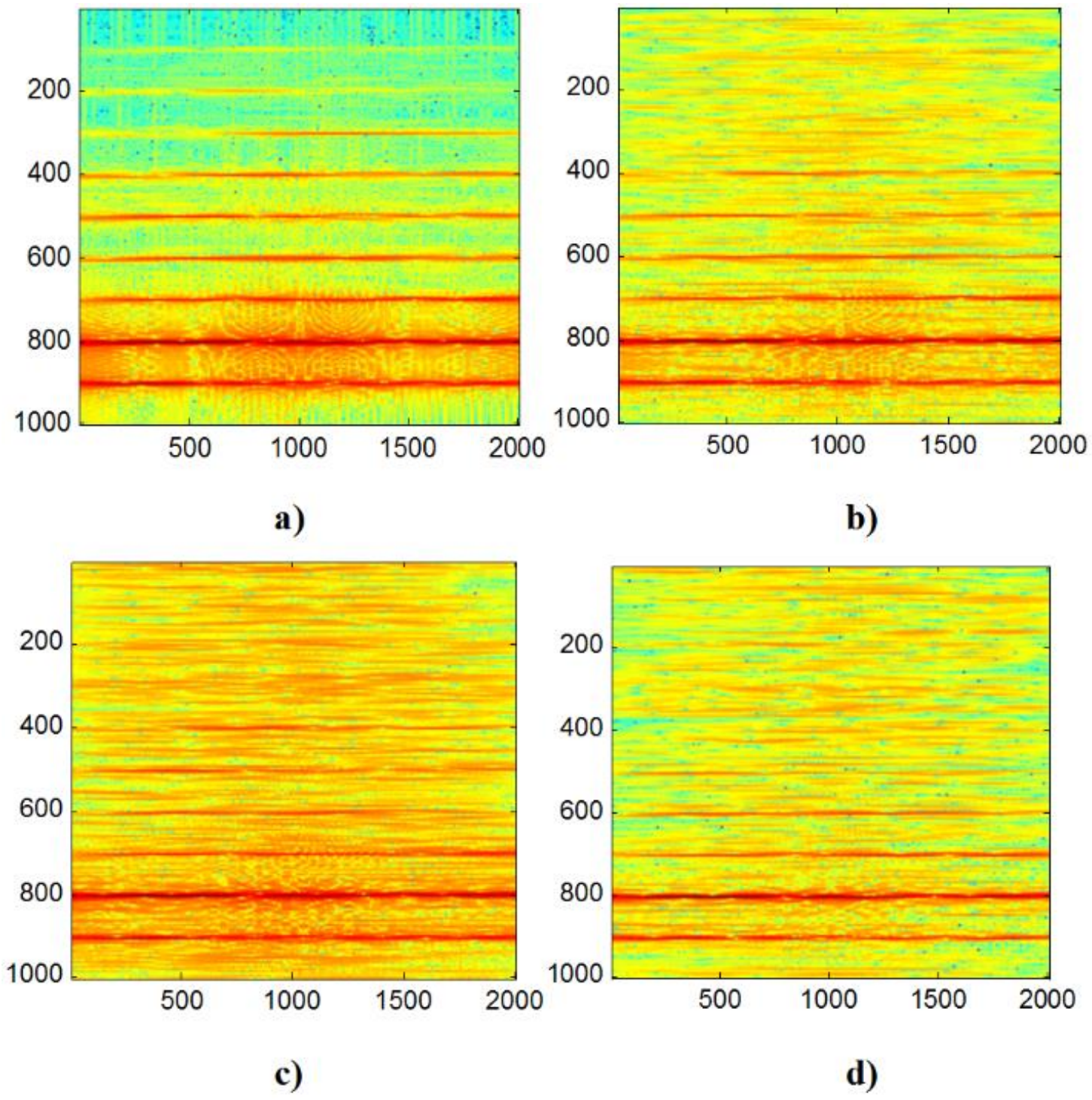

**Slika 2.8: a) Signal bez šuma; b) – d) Signal zahvaćen šumom: b) SNR=15dB, c) SNR=12 dB, d) SNR= 9 dB**

**Tabela 2-1. Broj odvojenih komponenti signala u zavisnosti od nivoa šuma**

| **Nivo šuma** | **Broj odvojenih komponenti (od ukupno 9 komponenti signala)** |
|---|---|
| SNR=9dB | 7 |
| SNR=12dB | 8 |
| SNR=14dB | 8 |
| SNR=15dB | 8 |
| SNR>15dB | 9 |

### 2.2.4. Performanse algoritma za dekompoziciju u slučaju neharmonijskog signala

Posmatrajmo primjer primjene algoritma za dekompoziciju u slučaju neharmonijskog signala. Signal se sastoji od 12 komponenti. Komponente su različitog trajanja, različitih energija, a različit je i frekvencijski razmak između susjednih komponenti signala.

Vremensko-frekvencijska reprezentacija neharmonijskog signala prikazana je na slici 2.9. Vrijednost parametra $L$ prilikom računanja S-metoda je odabrana da bude 6, a određeno je da se u svakoj iteraciji odvajaju po 2 komponente (tj. vrijednost parametra $K$ je 4, jer je signal realan, pa svakoj komponenti signala odgovaraju po 2 sopstvena vektora).

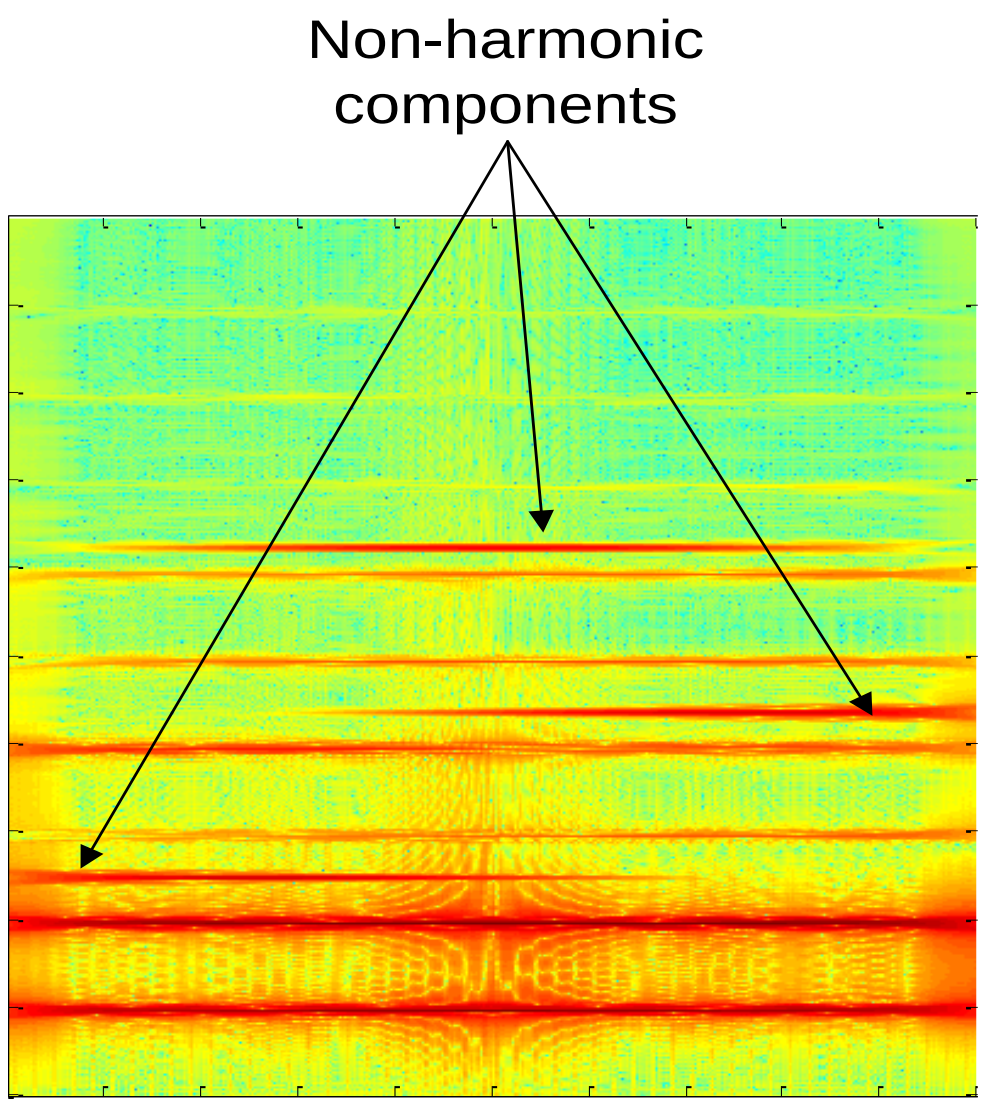

**Slika 2.9: S-metod posmatranog neharmonijskog signala (horizontalna osa označava vrijeme, vertikalna osa označava frekvenciju)**

Performanse predloženog algoritma su slične kao i performanse u dosada posmatranim slučajevima, što dokazuje da predloženi algoritam ne zahtijeva da signal koji se dekomponuje ima harmonijsku strukturu.

Odvojene komponente signala kroz iteracije prikazane su na slici 2.10.

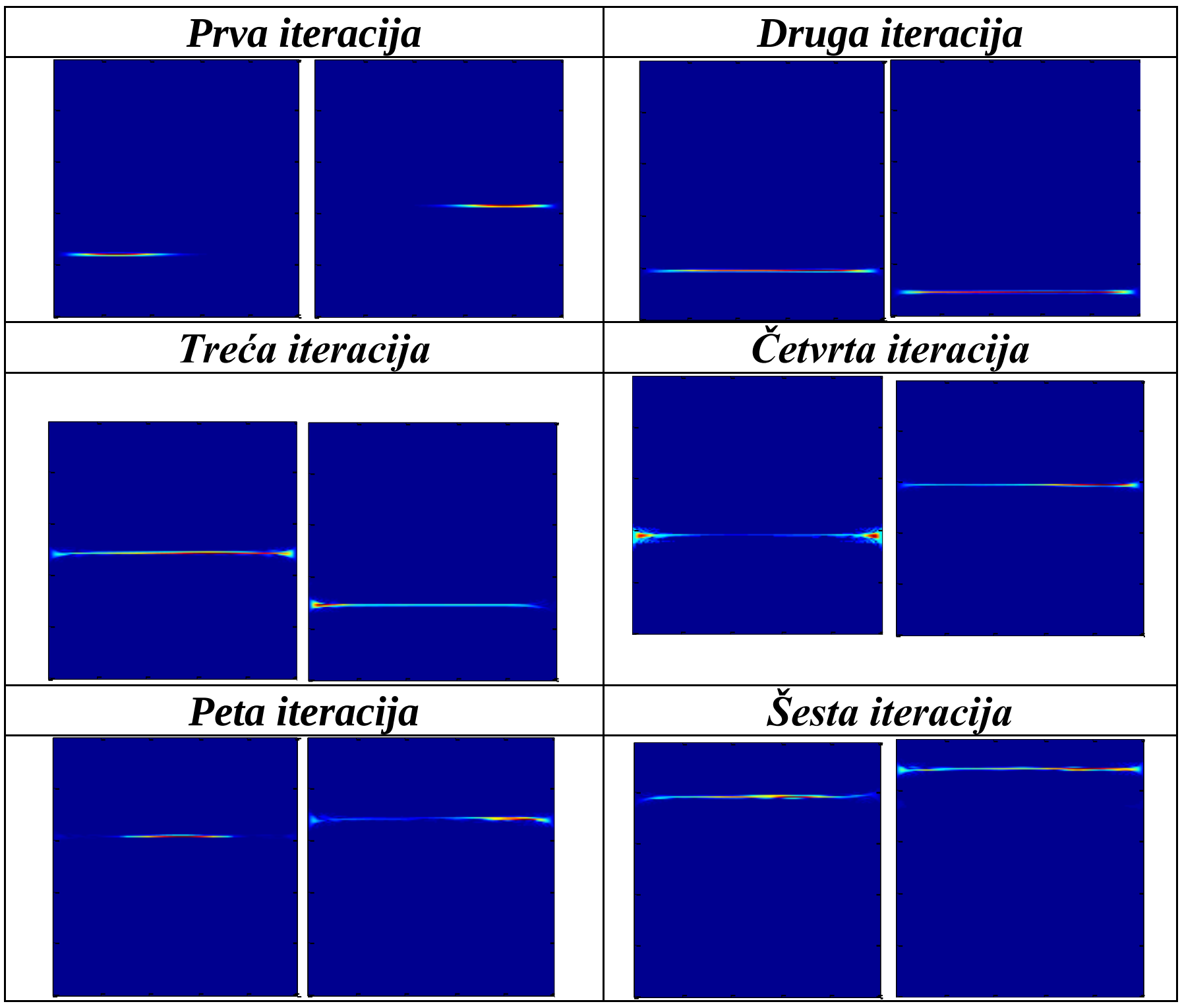

**Slika 2.10: Odvojene komponente neharmonijskog signala tokom šest iteracija (horizontalna osa - vrijeme, vertikalna osa - frekvencija)**

# 3. Compressive Sensing

## 3.1. Osnovni koncepti Compressive Sensing-a

Kontinualni signali, ograničenog propusnog opsega, prilikom odabiranja u skladu sa Shannon-Nyquist-ovom teoremom o odabiranju, mogu proizvesti ogromnu količinu podataka, koju treba dalje procesirati. Ovakav tradicionalan način odabiranja signala, koji zahtijeva akviziciju signala frekvencijom koja je najmanje dva puta veća od njegove maksimalne frekvencije, u nekim primjenama može biti neefikasan naročito u onim aplikacijama u kojima se javljaju visoko-frekventni signali. Takođe, mnoštvo senzora koji se koriste pri akviziciji ovakvih signala zahtijevaju veliku potrošnju energije.

Dalje, imajući u vidu da se većina signala koji se javljaju u realnim aplikacijama odlikuje velikim stepenom redudantnosti, javila se ideja o kompresiji sa gubicima. Kompresija sa gubicima bazirana je na odbacivanju određenog procenta odbiraka signala, iz domena u kojem signal ima konciznu predstavu, pod pretpostavkom da eliminisani odbirci neće biti od značaja prilikom prikaza signala.

Činjenica da većina signala koji se javljaju u realnim aplikacijama ima kompaktnu predstavu u nekom od transformacionih domena, nalazi svoje mjesto u teoriji kompresivnog očitavanja (*Compressive Sensing* - CS) [58]-[82]. Naime, jedna od ideja CS-a je kompresija pri akviziciji signala. Drugim riječima, CS nastoji da omogući direktnu akviziciju signala u komprimovanoj formi, prikupljanjem mnogo manje odbiraka u poređenju sa brojem odbiraka koji zahtijeva Teorema o odabiranju. Odabiranje signala u skladu sa CS-om, kao posljedicu ima nedostatak odbiraka u signalu. Osim namjernog izbjegavanja prikupljanja pojedinih odbiraka signala (što je rezultat načina akvizicije odbiraka), nedostajuće vrijednosti u signalu mogu nastati i kao posljedica procedura za otklanjanje šuma. U oba pomenuta slučaja nedostajuće odbirke signala treba rekonstruisati, a to se radi rekonstrukcionim algoritmima [15], [58]-[63], [90]-[96], o kojima će više riječi biti u Poglavlju 3.3.

Neki od koncepata koji se danas koriste u okviru CS pristupa, poznati su od ranije [85], [97], [98]. Na primjer, metod najmanjih kvadrata, baziran na minimizaciji normi, korišćen je od strane Claerbout-a i Muir-a još 1973. godine [97]. Minimizacija $\ell_1$-norme gradijenta

slike – tj. minimizacija totalne varijacije, predložena je 1990. godine od strane Rudin-a, Osher-a i Fatemi-ja [85], u cilju otklanjanja šuma iz slike. Ideja CS-a počinje intenzivnije da se razvija od momenta kada je pokazano da mali skup mjerenja može da obezbijedi tačnu rekonstrukciju signala.

U nastavku, akcenat je stavljen na često korišćene algoritme za rekonstrukciju signala razrijeđenog spektra [67]-[94], kao i na praktične aplikacije CS pristupa [95]-[171]. Važno je napomenuti da, kod CS pristupa, postoje specifični zahtjevi koji se odnose na signal i na mjerenja signala, i koji treba da budu ispunjeni u cilju uspješne rekonstrukcije. Jedan od uslova je kompaktna predstava signala u nekom domenu. Signal može imati kompaktnu predstavu u vremenskom, frekvencijskom, vremensko-frekvencijskom domenu [15], [50], [97], [118], [173]. Ovaj uslov je zadovoljen za većinu realnih signala. Drugi uslov je inkoherentnost između matrice mjerenja i bazne matrice domena u kom signal ima konciznu predstavu.

Postoji veliki broj aplikacija CS-a, od aplikacija u jednodimenzionim signalima do raznih aplikacija u obradi slike i videa. Neke od aplikacija su prilagođene radu u realnom vremenu. Kontinuiran razvoj u oblasti CS-a posljedica je potreba za smanjenjem kompleksnosti uređaja, povećanjem brzine akvizicije podataka i njihovog prenosa, kao i smanjenjem potrošnje energije. Imajući u vidu da se CS koristi za dobijanje informacije iz što je moguće manje dostupnih podataka, važno je istaći njegovu primjenu u medicini, a naročito u magnetnoj rezonansi (*Magnetic Resonance Imaging* – MRI) [104], [106], [107]. Smanjenjem broja koeficijenata neophodnih za dobijanje MR slike moguće je značajno smanjiti i vrijeme izloženosti pacijenta MR uređaju, čime se smanjuje negativan uticaj aparature za snimanje na pacijenta. Druga značajna primjena CS-a je u radarskim aplikacijama [115]-[120], [132]-[138]. Kompresivno očitavanje nalazi primjenu i u komunikacionim i mrežnim sistemima, bežičnim senzornim mrežama (*wireless sensor networks* - WSN), kognitivnim radio sistemima za ispitivanje spektra signala, itd. Neke od aplikacija opisane su u Poglavlju 3.4.

## 3.2. Matematička formulacija Compressive Sensing pristupa

Pretpostavimo signal $\mathbf{x}$, dužine $N$, koji ima kompaktnu predstavu u nekom transformacionom domenu (koji označavamo matricom direktne transformacije $\mathbf{\Psi}$, tj.

matricom inverzne transformacije $\mathbf{\Psi}^{-1}$). Vektor mjerenja **y** može se definisati na sljedeći način [15], [64]-[66]:

$$\mathbf{y} = \mathbf{\Phi\Psi}^{-1}\mathbf{X}, \tag{3.1}$$

gdje matrica **Φ** modeluje slučajno pododabiranje posmatranog signala. Ako je signal *K*-spars (*K*-kompaktan), to znači da je samo *K* od ukupno *N* koeficijenata iz transformacionog domena nenulto. Prilikom akvizicije odabrano je *M* od ukupno *N* odbiraka signala, a odbirci su smješteni u vektor mjerenja **y** (važi da je $M<<N$ i $M>2K$). Vektor **X** predstavlja vektor transformacionih koeficijenata, pa važi relacija:

$$\mathbf{X} = \mathbf{\Psi x}. \tag{3.2}$$

Različiti transformacioni domeni su u upotrebi: DFT domen [15], domen diskretne kosinusne transformacije (DCT) [15], [83], [163] *wavelet* domen, HT domen [87], [88], [112], vremensko-frekvencijski domeni [15], [118], [138], [172]-[177] itd. Osim koncizne predstave signala u nekom domenu, inkoherentnost je jedan od uslova koji mora biti ispunjen da bi bila obezbjeđena uspješna rekonstrukcija signala iz malog broja uzetih mjerenja. Naime, matrica mjerenja **Φ** treba da bude inkoherentna sa transformacionom matricom **Ψ**. Koherentnost između dvije matrice predstavlja najveću korelaciju između vektora bilo koje dvije vrste ili kolone matrica. Mjera korelacije definisana je na sljedeći način [15]:

$$\mu(\mathbf{\Phi},\mathbf{\Psi}) = \sqrt{N} \max_{k\geq 1, j\leq N} \left| \left\langle \mathbf{\Phi}_k, \mathbf{\Psi}_j \right\rangle \right|, \tag{3.3}$$

gdje je *N* dužina signala, $\mathbf{\Phi}_k$ i $\mathbf{\Psi}_j$ su vektor vrsta i vektor kolona matrica **Φ** i **Ψ**, respektivno. Koherencija uzima vrijednosti iz intervala:

$$1 \leq \mu(\mathbf{\Phi},\mathbf{\Psi}) \leq \sqrt{N}. \tag{3.4}$$

Vrijednost koherencije je veća ako su dvije matrice više korelisane, a kod CS pristupa poželjno je da je vrijednost koherencije što je moguće manja.

Sistem jednačina (3.1) se može napisati i na sljedeći način:

$$\mathbf{y}_{M\times 1} = \mathbf{A}_{M\times N}\mathbf{X}_{N\times 1}, \tag{3.5}$$

gdje je **A** CS matrica i važi $\mathbf{A} = \mathbf{\Phi\Psi}^{-1}$. Sistem je neodređen, imajući u vidu da sadrži *M* jednačina sa *N* nepoznatih. U cilju pronalaženja optimalnog rješenja sistema, koriste se optimizacione tehnike. Optimalno rješenje povezano je sa razrijeđenošću spektra signala – najkompaktnije dobijeno rješenje je ono koje je najoptimalnije.

Postoji veliki broj algoritama za rješavanje optimizacionih problema. Neki od njih su bazirani na konveksnim optimizacijama kao što su: *Basis pursuit*, *Dantzig selector*, i algoritmi zasnovani na gradijentu [15], [58]-[60]. Ovi algoritmi omogućavaju veliki stepen tačnosti prilikom rekonstrukcije, ali su numerički zahtjevni. Računski manje zahtjevni su tzv. *greedy* algoritmi. U tu grupu spadaju recimo *Matching Pursuit* i *Orthogonal Matching Pursuit* [62], [63], [67]-[69]. Takođe, algoritmi koji su bazirani na računanju praga, omogućavaju veliku tačnost rekonstrukcije, uz nisku računsku kompleksnost: *Iterative Hard Thresholding* - IHT, *Iterative Soft Thresholding* – IST, *Automated Threshold Based Iterative Solution*, *Adaptive Gradient-Based Algorithm*, [61], [63], [70], [73] itd.

## 3.3. CS algoritmi

Razrijeđenost spektra signala se može definisati kao broj nenultih koeficijenata u vektoru, u nekom domenu. Opisuje se $\ell_0$-normom [15], [58], [59], [107], koja je definisana na sljedeći način:

$$\|\mathbf{x}\|_{\ell_0} = \lim_{p\to 0}\sum_{i=1}^{N}|x_i|^p \ , \tag{3.6}$$

i predstavlja broj nenultih elemenata u vektoru **x**:

$$\|\mathbf{x}\|_{\ell_0} = \mathrm{card}\{\mathrm{supp}(\mathbf{x})\} \le K . \tag{3.7}$$

Ukoliko signal **x** ima kompaktnu predstavu u nekom transformacionom domenu, rješenje nedeterminisanog sistema jednačina (3.1) može biti svedeno na minimizaciju $\ell_0$-norme, tj.:

$$\min\|\mathbf{X}\|_{\ell_0} \ \text{ uz uslov } \ \mathbf{y} = \mathbf{AX}. \tag{3.8}$$

Računanje $\ell_0$-norme je računski veoma zahtjevno i korišćenje ove norme nije efikasno u radu sa signalima zahvaćenim šumom. Zbog toga se češće u praksi koristi $\ell_1$-norma. Optimizacioni problem baziran na $\ell_1$-normi svodi se na sljedeću relaciju [15], [58], [59], [60]:

$$\min\|\mathbf{X}\|_{\ell_1} \ \text{ uz uslov } \ \mathbf{y} = \mathbf{AX}. \tag{3.9}$$

U sljedećem poglavlju su opisani neki od često korišćenih algoritama za rekonstrukciju signala koji imaju razrijeđenu spektralnu predstavu.

### 3.3.1. Konveksne optimizacije

***Basis Pursuit i Basis Pursuit Denoising algoritmi***

Postoje različiti pristupi kojima se rješavaju konveksni problemi oblika (3.9), a često korišćeni algoritmi zasnovani na konveksnim optimizacijama su: *Basis Pursuit* (BP), *Basis Pursuit De-Noising* (BPDN), *Least Absolute Shrinkage and Selection Operator* (LASSO), *Least Angle Regression* (LARS), itd [58], [62], [63]. Pristup zasnovan na konveksnoj, $\ell_1$-minimizaciji, koji pruža skoro optimalno rješenje, definisan je relacijom (3.9) i poznat pod nazivom *Basis Pursuit* [15], [83]. *Basis Pursuit* se može rješavati tzv. *primal-dual interior point* metodom. Problem (3.9) se u slučaju realnih **y**, **A** i **X** može svesti na sljedeću relaciju [15], [83]:

$$\min_t \sum t, \text{ uz uslov } -t \le \mathbf{X} \le t,\ \mathbf{y} = \mathbf{AX}, \tag{3.10}$$

gdje je $t$ uvedeno sa ciljem da se izbjegne apsolutna vrijednost u relaciji $\|\mathbf{X}\|_{\ell_1} = \sum_{i=1}^{N} |\mathbf{X}_i|$ [15]. *Primal-dual interior point* metod je opisan Algoritmom 1. U slučaju mjerenja zahvaćenih šumom, tj. ako je vektor mjerenja dat kao **y**=**AX**+**n**, gdje **n** označava šum i važi da je $\|\mathbf{n}\|_{\ell_2} \le \varepsilon$, optimizacioni problem postaje *Basis Pusuit Denoising* (BPDN) i definisan je relacijom [15]:

$$\min \|\mathbf{X}\|_{\ell_1} \text{ uz uslov } \|\mathbf{y} - \mathbf{AX}\|_{\ell_2} \le \varepsilon. \tag{3.11}$$

Parametar $\Delta$ u Algoritmu 1 dobija se traženjem prvih izvoda od $\Lambda$ u zavisnosti od argumenata. Parametar $u$ računa se korišćenjem metoda pod nazivom *backtracking line search* [15], pa se, na primjer, nova vrijednost za **X** može dobiti kao **X**=**X**+$u$($\Delta$**X**).

**Algoritam 1: Primal-dual interior point metod**

- Za poznati vektor mjerenja **y**, postavi se relacija $\mathbf{X} = \mathbf{X_0} = \mathbf{A^T y}$ .
- Postaviti $t_0 = \gamma|\mathbf{X_0}| + \lambda \max\{|\mathbf{X_0}|\}$, gdje su parametri $\gamma$ i $\lambda$ korisnički definisani.
- Sljedeći korak je formiranje tzv. Lagrangian-ove funkcije:

$$\Lambda\left(\mathbf{X}, t, g, -\frac{1}{\mathbf{X}_0 - t_0}, \frac{1}{\mathbf{X}_0 + t_0}\right) = f(t) + g(\mathbf{AX} - \mathbf{y}) - \frac{\mathbf{X} - t}{\mathbf{X}_0 - t_0} - \frac{\mathbf{X} + t}{\mathbf{X}_0 + t_0}, \text{ gdje je}$$

$$g = -\mathbf{A}\left(\frac{-1}{\mathbf{X}_0 - t_0} + \frac{1}{-\mathbf{X}_0 - t_0}\right).$$

- Ažuriranje svakog od argumenata Lagrangian-ove funkcije za $\Delta$ i $u$.

### *Adaptivni gradijentni algoritam*

Adaptivni gradijenti algoritam za rekonstrukciju signala razrijeđenog spektra predložen je u [86]. Algoritam priprada grupi konveksnih optimizacija. Početne vrijednosti dostupnih odbiraka signala se iterativno mijenjaju za vrijednosti +Δ i –Δ, a istovremeno se mjeri poboljšanje koncentracije u domenu u kom je signal rijedak. Algoritam je prvobitno definisan za DFT domen, a kasnije i za Hermitski transformacioni domen [88]. Vrijednosti nedostajućih odbiraka se ažuriraju vrijednostima gradijentnog vektora. Gradijentni vektor se dobija kao razlika između $\ell_1$-normi vektora koji su promijenjeni za +Δ i –Δ.

**Algoritam 2: Adaptivni gradijentni algoritam**

**Ulaz**: skup pozicija dostupnih odbiraka signala, $\mathbf{\Omega_a}$ i skup pozicija nedostajućih odbiraka signala: $\mathbf{\Omega_m}=\mathbf{N}\backslash\mathbf{\Omega_a}$; vektor mjerenja **y**; u slučaju 1D signala *n*=*n*, a u slučaju 2D signala $n=(n_x,n_y)$

- Postaviti $\mathbf{y}^{(0)}(n) \leftarrow \mathbf{y}(n)$ i $k \leftarrow 0$
- Postaviti $\Delta \leftarrow \max\left|\mathbf{y}^{(0)}(n)\right|$
- ***repeat***
- Postaviti $\mathbf{y}_p(n) \leftarrow \mathbf{y}^{(k)}(n)$
- ***repeat***
- $k \leftarrow k+1$
- ***for*** $t \in \mathbf{N}$ ***do***
- ***if*** $t \in \mathbf{\Omega}_m$ ***then***

$$X^{+}(f) \leftarrow \Im\{\mathbf{y}^{(k)}(n)+\Delta\}, \qquad \left(f=f \text{ u 1D slučaju; } f=(f_1,f_2) \text{ u 2D slučaju}\right)$$

$$X^{-}(f) \leftarrow \Im\{\mathbf{y}^{(k)}(n)-\Delta\}, \qquad \left(\Im \text{ DFT} - \text{1D slučaj; 2D DFT} - \text{2D slučaj}\right)$$

$$\mathbf{\Gamma}^{(k)}(t) \leftarrow \left\|X^{+}\right\|_{\ell_1} - \left\|X^{-}\right\|_{\ell_1},$$

***else***

- $\mathbf{\Gamma}^{(k)}(t) \leftarrow 0$
- ***end if***
- $\mathbf{y}^{(k+1)}(t) \leftarrow \mathbf{y}^{(k)}(t) - \mathbf{\Gamma}^{(k)}(t)$
- ***end for***
- $\beta_k = \arccos \dfrac{\left\langle \mathbf{\Gamma}^{k-1}\mathbf{\Gamma}^{k} \right\rangle}{\left\|\mathbf{\Gamma}^{k-1}\right\|_2^2 \left\|\mathbf{\Gamma}^{k}\right\|_2^2}$
- ***until*** $\beta_k < 170^{\circ}$
- $\Delta \leftarrow \Delta/\sqrt{10}$
- $R = 10\log_{10}\left(\sum_{n\in\Omega_m}\left|\mathbf{y}_p(n)-\mathbf{y}^{(k)}(n)\right|^2 / \sum_{n\in\Omega_m}\left|\mathbf{y}^{(k)}(n)\right|^2\right)$ ***until*** $R < R_{\max}$ – zahtijevana preciz.
- ***return*** $\mathbf{y}^{(k)}(n)$

**Izlaz**: rekonstruisani signal $\mathbf{y}^{(k)}(n)$

Algoritam daje zadovoljavajuće rezultate čak i u slučajevima kada posmatrani signal nema kompaktnu predstavu u nekom domenu. Koraci adaptivnog gradijentnog algoritma sumirani su kroz Algoritam 2, i to za slučajeve 1D i 2D signala.

### 3.3.2. Greedy algoritmi

*Greedy* algoritmi predstavljaju drugu grupu često korišćenih algoritama za dobijanje optimalnog rješenja sistema jednačina (3.5).

Ovi algoritmi su računski manje zahtjevni u poređenju sa algoritmima zasnovanim na minimizaciji $\ell_1$-norme. Zbog toga su i brži, ali ujedno i manje precizni. Bazirani su na iterativnom traženju onih elemenata transformacione matrice (tj. rječnika) koji najbolje opisuju posmatrani signal. Najčešće korišćeni algoritmi koji pripadaju grupi *greedy* algoritama su *Matching Pursuit* (MP), *Orthogonal Matching Pursuit* (OMP), *Compressive Sampling Matching Pursuit* (CoSaMP), itd. Koraci OMP-a opisani su Algoritmom 3 [15], [62], [63].

**Algoritam 3: Orthogonal matching pursuit - OMP**

- **Ulaz:** CS matrica $\mathbf{A}=\mathbf{\Phi\Psi}$, vektor mjerenja $\mathbf{y}$
- Inicijalizacija promjenljivih:
  - početni ostatak $\mathbf{r}_0=\mathbf{y}$; početno rješenje $\mathbf{X}_0=0$; matrica izabranih atoma $\mathbf{\Upsilon}_0=[]$.
- Ponavljati sljedeće korake dok se ne dođe do kriterijuma za zaustavljanje algoritma:
  - $\omega_n=\arg\max_{i=1,\ldots,M}\left|\left\langle \mathbf{r}_{n-1},\mathbf{A}_i\right\rangle\right|$ - traženje kolone sa maksimalnom korelacijom
  - $\mathbf{\Upsilon}_n\leftarrow\left[\mathbf{\Upsilon}_{n-1}\ \ \mathbf{A}_{\omega_n}\right]$ - ažuriranje matrice odabranih atoma
  - $\mathbf{X}_n=\arg\min_{\mathbf{X}}\left\|\mathbf{r}_{n-1}-\mathbf{\Upsilon}_n\mathbf{X}_{n-1}\right\|_{\ell_2}^2$ - rješavanje metoda najmanjih kvadrata
  - $\mathbf{r}_n=\mathbf{r}_{n-1}-\mathbf{\Upsilon}_n\mathbf{X}_{n-1}$ - ažuriranje ostatka
  - $n=n+1$
- **Izlaz:** $\mathbf{X}_P$ i $\mathbf{r}_P$, gdje $P$ označava broj iteracija

### 3.3.3. Algoritmi bazirani na pragu

***Iterative hard thresholding i Iterative soft thresholding***

Algoritmi koji su zasnovani na pragu su generalno mnogo brži u poređenju sa algoritmima baziranim na konveksnim optimizacijama. To su iterativni algoritmi kod kojih se funkcija pragovanja može opisati sljedećom relacijom [15], [62], [63], [70]:

$$x_i = T_\varepsilon\left(f(X_{i-1})\right). \tag{3.12}$$

Funkcija pragovanja je označena kao $T_\varepsilon$, dok je $f$ funkcija kojom se modifikuje izlaz prethodne iteracije a **X** je vektor koji ima razrijeđenu predstavu u nekom domenu. Signal može biti rekonstruisan korišćenjem tzv. grubog (*hard*) ili finog (*soft*) pragovanja, pa postoje dvije vrste iterativnog algoritama pragovanja, a to su *iterative hard thresholding* (IHT) i *iterative soft thresholding* (IST).

IHT algoritam postavlja sve osim $K$ najvećih koeficijenata vektora **X** (najvećih u pogledu amplitude) na nulu. Funkcija grubog pragovanja $H_K$ definisana je na sljedeći način [62]:

$$H_K(\mathbf{X}) = \begin{cases} X_i, & |X_i| > \varepsilon \\ 0, & \text{van} \end{cases} \tag{3.13}$$

Parametar ε označava prag. Algoritam je sumiran Algoritmom 4 [62]. Funkcija finog pragovanja se primjenjuje na svaki elemenat u vektoru **X**, a definisana je kao:

$$S_\lambda(X_i) = \begin{cases} X_i - \lambda, & \mathbf{X} > \lambda \\ 0, & |X_i| < \lambda \\ X_i + \lambda, & \mathbf{X} < -\lambda \end{cases}. \tag{3.14}$$

**Algoritam 4: Iterative hard thresholding**

**Ulaz:** Broj komponenti signala $K$, transformaciona matrica $\mathbf{\Psi}$, matrica mjerenja $\mathbf{\Phi}$, CS matrica **A,** vektor mjerenja **y**

**Izlaz:** aproksimacija signala, X

$\mathbf{X}_0 \leftarrow 0$

***for*** $i$=1,…, dok se ne zadovolji kriterijum zaustavljanja ***do***

$\mathbf{X}_i \leftarrow H_K\left(\mathbf{X}_{i-1} + \mathbf{A}^T(\mathbf{y} - \mathbf{A}\mathbf{X}_{i-1})\right)$

***end for***

***return*** $\mathbf{X} \leftarrow \mathbf{X}_i$

### *Automatizovani algoritam zasnovan na pragovanju*

Neiterativna i iterativna rješenja za rekonstrukciju signala razrijeđenog spektra, takođe zasnovana na pragovanju, predložena su u [61].

Predložena rješenja bazirana su na modelu šuma koji se javlja kao posljedica nedostajućih odbiraka u signalu. Korišćenjem predefinisane vjerovatnoće greške $P$, računa se prag $T$ koji razdvaja komponente signala od spektralnog šuma koji se javlja u transformacionom domenu, a koji je posljedica nedostajućih odbiraka. Algoritam koristi DFT kao domen u kom signal ima kompaktnu predstavu, ali se procedura može primijeniti i korišćenjem drugih transformacionih domena. Ovaj pristup obezbjeđuje uspješnu rekonstrukciju signala kroz samo jednu iteraciju, ukoliko broj dostupnih odbiraka zadovoljava uslov definisan algoritmom.

Međutim, ukoliko je broj dostupnih odbiraka signala, $M$, veoma mali, koristi se iterativna verzija algoritma. Vrijednost praga $T$ računa se u svakoj iteraciji. Ako je ulaz algoritma vektor $\mathbf{y}$, koji se sastoji od $M$ dostupnih odbiraka signala, dužina signala je $N$, vektor pozicija dostupnih odbiraka signala je $\Omega_a=\{n_1,\ldots,n_M\}$, $\mathbf{\Psi}$ je transformaciona matrica, $\mathbf{\Phi}$ je matrica mjerenja, CS matrica je $\mathbf{A}=\mathbf{\Phi\Psi}$ i varijansa Gausovog šuma $\sigma_N$, onda se iterativna verzija algoritma može opisati koracima koji su sumirani u Algoritmu 5. Neiterativna verzija algoritma biće korišćena za prijedlog hardverskog rješenja i biće detaljnije opisana u Glavi 6.

**Algoritam 5: Automatizovani iterativni algoritam zasnovan na pragovanju**

**Ulaz:** $M$, $N$, $\mathbf{y}$, $\mathbf{\Omega_a}=\{n_1,\ldots,n_M\}$, $\mathbf{\Psi}$, $\mathbf{\Phi}$, $\mathbf{A}=\mathbf{\Phi\Psi}$, $\sigma_N$ .

- Postaviti $\mathbf{k}=\varnothing$;

***for*** $i$=1 : $i$=$i$+1: ***until*** dok se ne detektuju sve komponente

- Računanje varijanse: $\sigma^2 = M\dfrac{N-M}{N-1}\sum_{i=1}^{M}\dfrac{y(i)^2}{M}$;
- Računanje praga $T$, za datu vrijednost vjerovatnoće $P$: $T=\sqrt{-\sigma^2\log\left(1-P(T)^{1/N}\right)}$ ;
- Računanje početne DFT, $\mathbf{X}_i$: $\mathbf{X}_i=\mathbf{y}(\mathbf{\Phi\Psi}^{-1})$ ;
- Ažuriranje skupa $\mathbf{k}$: $\mathbf{k}=\mathbf{k}\cup\arg\left\{|\mathbf{X}_i|>T/N\right\}$ ;
- Računanje: $\mathbf{F}=\left(\mathbf{A}^H\mathbf{A}\right)^{-1}\mathbf{A}\mathbf{y}$; (CS matrica **A** sadrži vrste definisane skupom **k**, i $M$ kolona DFT matrice)
- Ažuriranje vektora $\mathbf{y}$, $\forall k\in\mathbf{k}$: $\mathbf{y}=\mathbf{x}(\mathbf{\Omega}_a)-X(k)e^{j\frac{2\pi k\mathbf{\Omega}_a}{N}}$ ;
- Ažuriranje vektora "početne" DFT **X**, u skladu sa ažuriranim vektorom **y**;
- Ažuriranje $A^2=\sum|\mathbf{y}|^2/M$ i $\sigma^2=A^2M\dfrac{N-M}{N-1}$ ;
- ***if*** $A^2<\sigma_N^2$ ***break*** ;

***end for***

### *Automatizovani algoritam zasnovan na pragovanju za rekonstrukciju 2D signala*

Automatizovani algoritam zasnovan na pragovanju [140], korišćen za rekonstrukciju jednodimenzionih, kompaktnih signala, generalizovan je za dvodimenzione signale koji imaju kompaktnu predstavu u domenu dvodimenzione Fourier-ove transformacije (2D DFT). On omogućava uspješnu rekonstrukciju signala, čak i u slučajevima ako je manje od 10% odbiraka signala dostupno.

Ako se sa $\mathbf{S_F}$ i $\mathbf{S_P}$ označe ukupni i parcijalni skup odbiraka signala, tj.:

$$\begin{aligned}\mathbf{S_F} &= \{S_F(x,y):\ x\in\{1,...,I\},\ y\in\{1,...,J\}\},\\ \mathbf{S_P} &= \{S_P(x,y):\ x\in\{1,...,M\},\ y\in\{1,...,L\}\}, \quad \mathbf{S_P}\subset\mathbf{S_F},\end{aligned} \tag{3.15}$$

gdje je $I\times J$ ukupan broj odbiraka signala, $M$ i $L$ predstavljaju broj odbiraka duž $x$ i $y$ pravaca respektivno, ($M<I$, $L<J$), onda se varijansa šuma koji nastaje kao posljedica nedostajućih odbiraka signala u slučaju 2D signala računa na sljedeći način:

$$\sigma^2 = \sum_{i=1}^{K} W_i^2 ML \frac{IJ-ML}{IJ-1}, \tag{3.16}$$

gdje je sa $W_i$ označena amplituda $i$-te komponente signala. Prethodna relacija važi za $K$ komponenti signala. Ako je vjerovatnoća da su sve DFT komponente na pozicijama zahvaćenim šumom ispod nivoa komponenti signala $P$=0.99, onda se prag koji razdvaja komponente koje pripadaju signalu računa kao:

$$T = \sigma\sqrt{-\log(1-P^{1/(IJ)})}\ . \tag{3.17}$$

Pretpostavimo da je skup pozicija dostupnih odbiraka signala označen sa $\Omega_a$, $\mathbf{\Omega}_a = \{(x_1,y_1),...,(x_M,y_L)\}$, sa $\mathbf{\Psi}^{-1}$ je označena inverzna 2D DFT matrica, $\mathbf{y}$ je vektor mjerenja i $\mathbf{k}$ vektor pozicija komponenti signala koje su iznad praga. Vektor početne DFT se računa korišćenjem sljedeće relacije:

$$\mathbf{Y} = \mathbf{\Psi}_{\Omega_a}\mathbf{y}, \tag{3.18}$$

gdje $\mathbf{\Psi}$ označava 2D DFT matricu. Matrica $\mathbf{\Psi}_{\Omega_a}$ sadrži one kolone originalne 2D DFT matrice koje su definisane skupom $\Omega_a$. Iterativna verzija algoritma sumirana je u Algoritmu 7 (*Simple and Fast Algorithm for Reconstruction of 2D signals* - SFAR 2D [154]). Algoritam je testiran na simuliranim radarskim slikama, u slučaju kada je samo 25% odbiraka signala dostupno. Dvodimenziona DFT od originalnog signala prikazana je na slici 3.1a. Komponente početne 2D DFT koje su iznad praga prikazane su na slici 3.1b, dok je rekonstruisana 2D DFT signala prikazana na slici 3.1c.

**Algoritam 7: Iterativna verzija automatizovanog algoritma zasnovanog na pragovanju za rekonstrukciju 2D signala (SFAR-2D algoritam)**

**Ulazi:** vektor mjerenja **y**, vektor pozicija dostupnih odbiraka signala $\Omega_a$, 2D DFT matrica i inverzna 2D DFT matrica $\boldsymbol{\Psi}$ i $\boldsymbol{\Psi}^{-1}$, respektivno; $\mathbf{k} = \varnothing$

- Računanje varijanse: $\sigma^2 = \sum_{i=1}^{K} W_i^2 ML \frac{IJ - ML}{IJ - 1}$, $W^2 = \sum_{i=1}^{K} W_i^2$ ;
- Računanje praga $T$ za datu vrijednost vjerovatnoće $P$=0.99;
- Računanje vektora početne DFT, **Y**, koji odgovara vektoru mjerenja **y**: $\mathbf{Y} = \boldsymbol{\Psi}_{\Omega_a}\mathbf{y}$ ;
- ***for*** $i$=1 : $i$=$i$+1: ***until*** sve komponete ne budu detektovane

  $\mathbf{k} = \mathbf{k} \cup \arg\{|\mathbf{Y}| > T\}$ ;

  Računanje $\mathbf{F} = \left(\mathbf{A}^H\mathbf{A}\right)^{-1}\mathbf{A}\mathbf{y}, \quad \mathbf{A} = \Psi^{-1}(\Omega_a, \mathbf{k})$ ;

  Ažuriranje **y**, **Y**, $W^2$ i $\sigma^2$: **y=y-AY(k)**; $\mathbf{Y} = \boldsymbol{\Psi}_{\Omega_a}\mathbf{y}$ ; $W^2 = \sum\left(|\mathbf{y}|^2 / (ML)\right)$ i

  $\sigma^2 = \sum_{i=1}^{K} W_i^2 ML \frac{IJ - ML}{IJ - 1}$ ;

  ***end for***

**Izlaz:** Rekonstruisana DFT **F**

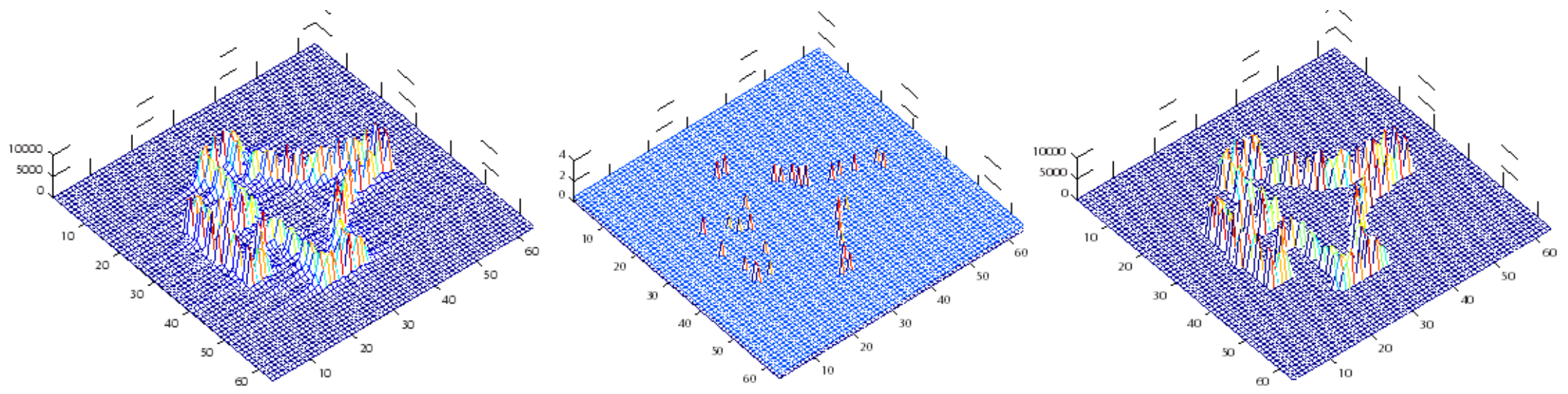

a) b) c)

**Slika 3.1: SFAR 2D algoritam primijenjen na simulirani radarski signal: a) 2D DFT originalnog signala, b) komponente 2D signala detektovane u jednom koraku – postavljanjem praga, c) 2D DFT rekonstruisanog signala**

## 3.4. CS aplikacije

Brojne su aplikacije CS pristupa, imajući u vidu da većina signala sa kojima se susrećemo u realnim situacijama ima kompaktnu predstavu u nekom domenu, a da se svojstvo inkoherentnosti može lako zadovoljiti, prilagođavanjem procedure akvizicije zahtjevima CS-a. Razvijeni su i hardverski uređaji zasnovani na principima CS-a [101]-[103]. Kompresivno očitavanje nalazi primjenu u multimedijalnim i medicinskim aplikacijama [103]-[114], radarima, zatim komunikacijama, [115]-[138], govornim i muzičkim signalima [139]-[142], podvodnim, akustičnim i linearno-frekvencijski modulisanim signalima [143]-[145]. Dalje, nalazi primjenu u rekonstrukciji slike [146]-[161], u video signalima [162]-[165], zaštiti digitalnih podataka [166]-[171] itd. Neke od aplikacija biće detaljnije opisane u narednim poglavljima.

### 3.4.1. Uređaji bazirani na CS-u

Postoji više tipova uređaja koji su zasnovani na principima CS-a. Grupa autora je u [101] predložila tzv. jednopikselnu kameru. U poređenju sa konvencionalnom kamerom, jednopikselna kamera je jednostavnija, manja i jeftinija. Analogno-informacioni konvertor (*analog-to-information converter* - AIC) predložen je u [103]. Konvertor je dizajniran tako da omogućava odabiranje u skladu sa Teoremom o odabiranju, ali takođe postoji i opcija za CS odabiranje signala. U radu sa signalima razrijeđenog spektra, ovaj dizajn omogućava povećanje energetske efikasnosti. Odabiranje po slučajnom rasporedu omogućeno je korišćenjem PN klok generatora.

### 3.4.2. CS u biomedicini

CS nalazi brojne primjere u raznim oblastima medicine, od kojih su naročito značajne primjene u snimanjima magnetnom rezonansom [104], [106], [107], elektroencefalografiji [109], elektrokardiografiji [110]-[113], elektrookulografiji, kao i elektromiografiji.

Metod za rekonstrukciju pododabranih MR slika korišćenjem homotropske aproksimacije $\ell_0$-norme predložen je u [107], i omogućava rekonstrukciju slika iz veoma malo dostupnih podataka. Metod je testiran kako na MR slikama, tako i na drugim medicinskim slikama.

Performanse različitih rekonstrukcionih algoritama, primijenjenih na signale u elektroencefalografiji, testirani su u [109]. Najbolji rezultati postignuti su primjenom *basis pursuit* algoritma.Testirana je efikasnost i *greedy* algoritama i pokazano je da *basis pursuit* daje bolje rezultate ali je računski zahtjevniji. Posmatrano je šest tipova rječnika (transformacija) i pokazano je da su B-Spline najpogodniji za CS prisup u ovim signalima.

CS pristup za brzo snimanje magnetnom rezonansom (tzv. *rapid MRI imaging*) predložen je u [160]. Kompaktna predstava MR slika u transformacionom domenu iskorišćena je u cilju smanjenja vremena koje je potrebno za skeniranje pacijenta, kao i za poboljšanje rezolucije posmatranih moždanih slika i angiograma.

TwIST (*Two-Step Iterative Shrinkage/Thresholding Algorithms for Image Restoration*) predložen u [161], je poboljšana verzija algoritma finog pragovanja – IST, a uveden je sa ciljem prevazilaženja problema spore konvergencije koji se javljao kod IST-a. Ovaj algoritam nalazi primjenu kako u rekonstrukciji MR slika, tako i u rekonstrukciji prirodnih slika.

### 3.4.3. Primjena u komunikacijama i radarima

Brojne su primjene CS-a u komunikacijama i radarima [116]-[138], naročito u bežičnim komunikacijama, širokopojasnim sistemima (*ultra wideband* - UWB) i ISAR (*Inverse synthetic aperture radar*) sistemima. Primjena CS-a u UWB komunikacijama opisana je u [125], gdje se koristi kompaktna predstava signala u vremenskom domenu. CS u distribuiranim radarskim mrežama, koje se sastoje od Wireless Local Area Network - WLAN rutera koji imaju ulogu prenosnika, opisan je u [126]. Kompresivno očitavanje se primenjuje u cilju smanjenja broja odbiraka signala koji se prenose prilikom centralne faze obrade, a u svrhu procjene položaja i brzine različitih meta. Predložena tehnika je u radu [126] poređena sa tzv. *matched filter* tehnikom, koja predstavlja tradicionalan metod za određivanje mete.

Primjena CS-a u rekonstrukciji osnovnog objekta kod ISAR signala predložena je u [132]. Signal ISAR se sastoji od dva dijela, osnovnog objekta i tzv. *micro-Doppler*-a prouzrokovanog brzo pokretnim djelovima mete, a segmenti signala se preklapaju u vremensko-frekvencijskoj ravni. Metod pronalazi i otklanja preklapajuće djelove, a zatim i rekonstruiše osnovni objekat. Odvajanje osnovnog objekta od *micro-Doppler*-a odrađeno je korišćenjem sortirane STFT (STFT je sortirana duž vremenske ose). Određeni procenat najjačih vrijednosti iz STFT-a je otklonjen, za svaku frekvenciju, a samim tim otklonjene su i komponente koje pripadaju *micro-Doppler*-u. Ostatak STFT-a predstavlja djelove osnovnog objekta ISAR signala. Odbijeni radarski signal nakon obrade i filtriranja, može se opisati relacijom:

$$x(t) = x_r(t) + x_m(t), \quad x_r(t) = \sum_{i=1}^{K} r_i e^{j\frac{2\pi f_{oi} t}{N}}, \tag{3.19}$$

gdje je $x_r(t)$ osnovni objekat, $x_m(t)$ *micro-Doppler*, $r_i$ označava amplitudu $i$-te komponente a $f_{oi}$ frekvenciju $i$-te komponente. Kratkotrajna Fourier-ova transformacija STFT signala $x(t)$ je: $S(t,f) = \sum_{l=0}^{L-1} x(t+l) e^{-j\frac{2\pi l f}{L}}$, ili $\mathbf{S}_L(t) = \mathbf{E}_L \mathbf{x}(t)$, gdje je $\mathbf{S}_L(t) = [S(t,0), \ldots, S(t,L-1)]^T$, $\mathbf{x}(t) = [x(t), x(t+1), \ldots, x(t+M-1)]^T$, $E(l,f) = e^{-j(2\pi f l/L)}$, a $\mathbf{E}_L$ je $L\times L$ DFT matrica. U analizi je korišćena STFT sa nepreklapajućim prozorima, računata sa korakom $L$, za vrijeme $t$, pa važi:

$$\mathbf{S} = [\mathbf{S}_L(0)^T, \mathbf{S}_L(M)^T, \ldots, \mathbf{S}_L(N-M)^T]^T = \mathbf{E}_{L,N}\mathbf{x}. \tag{3.20}$$

Skalarna vrijednost STFT-a za vremenski trenutak $t$ i frekvenciju $f$ označena je sa $S(t,f)$, $\mathbf{S}_L(t)$ je STFT vektor koji ima $L$ frekvencija u trenutku $t$, a $\mathbf{S}$ označava vektor svih STFT vrijednosti, za sve frekvencije $f$ i sve vremenske trenutke $t$. Matrica veličine $N\times N$, $\mathbf{E}_{L,N}$, i signal $\mathbf{x}$ su:

$$\mathbf{E}_{L,N} = \begin{bmatrix} \mathbf{E}_M & \mathbf{0}_M & \ldots & \mathbf{0}_M \\ \mathbf{0}_M & \mathbf{E}_M & \ldots & \mathbf{0}_M \\ \ldots & \ldots & \ldots & \ldots \\ \mathbf{0}_M & \mathbf{0}_M & \ldots & \mathbf{E}_M \end{bmatrix},$$
$$\mathbf{x} = [\boldsymbol{x}(0)^T, \boldsymbol{x}(M)^T, \ldots, \boldsymbol{x}(N-M)^T] = [x(0), x(1), \ldots, x(N-1)]^T. \tag{3.21}$$

Vektor $\mathbf{x}$ se može zapisati korišćenjem inverzne DFT matrice $\mathbf{E}_N^{-1}$ i DFT vektora $\mathbf{X}$, na sljedeći način:

$$\mathbf{x} = \mathbf{E}_N^{-1}\mathbf{X}, \quad \mathbf{S} = \mathbf{E}_{L,N}\mathbf{x} = \mathbf{E}_{L,N}\mathbf{E}_N^{-1}\mathbf{X} = \boldsymbol{\Psi}\mathbf{X}, \tag{3.22}$$

gdje je $\boldsymbol{\Psi}$ $N\times N$ transformaciona matrica. Nakon računanja STFT-a u skladu sa (3.22) primjenjuje se operacija sortiranja duž frekvencijske ose, za svaku frekvenciju: $s_{\text{sort}}(t,f) = \text{sort}\{S(t,f)\}$, $t=0, \ldots, L-1$. Sortira se u rastućem redosljedu i određeni procenat koeficijenata sa najnižim vrijednostima ($P$) i sa najvećim vrijednostima ($Q$) se otklanja iz sortirane STFT:

$$s_{cs}(f) = \{S_{SORT}(n,f), n = P, P+1, \ldots, L-Q\}, \tag{3.23}$$

gdje je $s_{CS}(f)$ vektor dostupnih STFT koeficijenata na frekvenciji $f$, dok je $\mathbf{S}_{CS}$ vektor STFT koeficijenata za svaku frekvenciju. Vektor svih dostupnih STFT koeficijenata je $\mathbf{S}_{CS}=\mathbf{\Psi}_{CS}\mathbf{X}$, a matrica $\mathbf{\Psi}_{CS}$ je formirana otklanjanjem vrsta originalne matrice koje odgovaraju otklonjenim STFT vrijednostima. Zatim se formira optimizacioni problem na sljedeći način:

$$\min \|\mathbf{X}\|_{\ell_1} \text{ uz uslov } \boldsymbol{S}_{CS} = \mathbf{\Psi}_{CS}\mathbf{X}. \tag{3.24}$$

Rezultati opisane procedure, primijenjeni na realni radarski signal, prikazani su na slici 3.2. Signal se sastoji od osnovnog objekta i tri reflektora koji rotiraju na frekvenciji ~ 60 RPM. Procenat uklonjenih koeficijenata iz sortirane STFT je 80%, a osnovni objekat je rekonstruisan od ostatka koeficijenata.

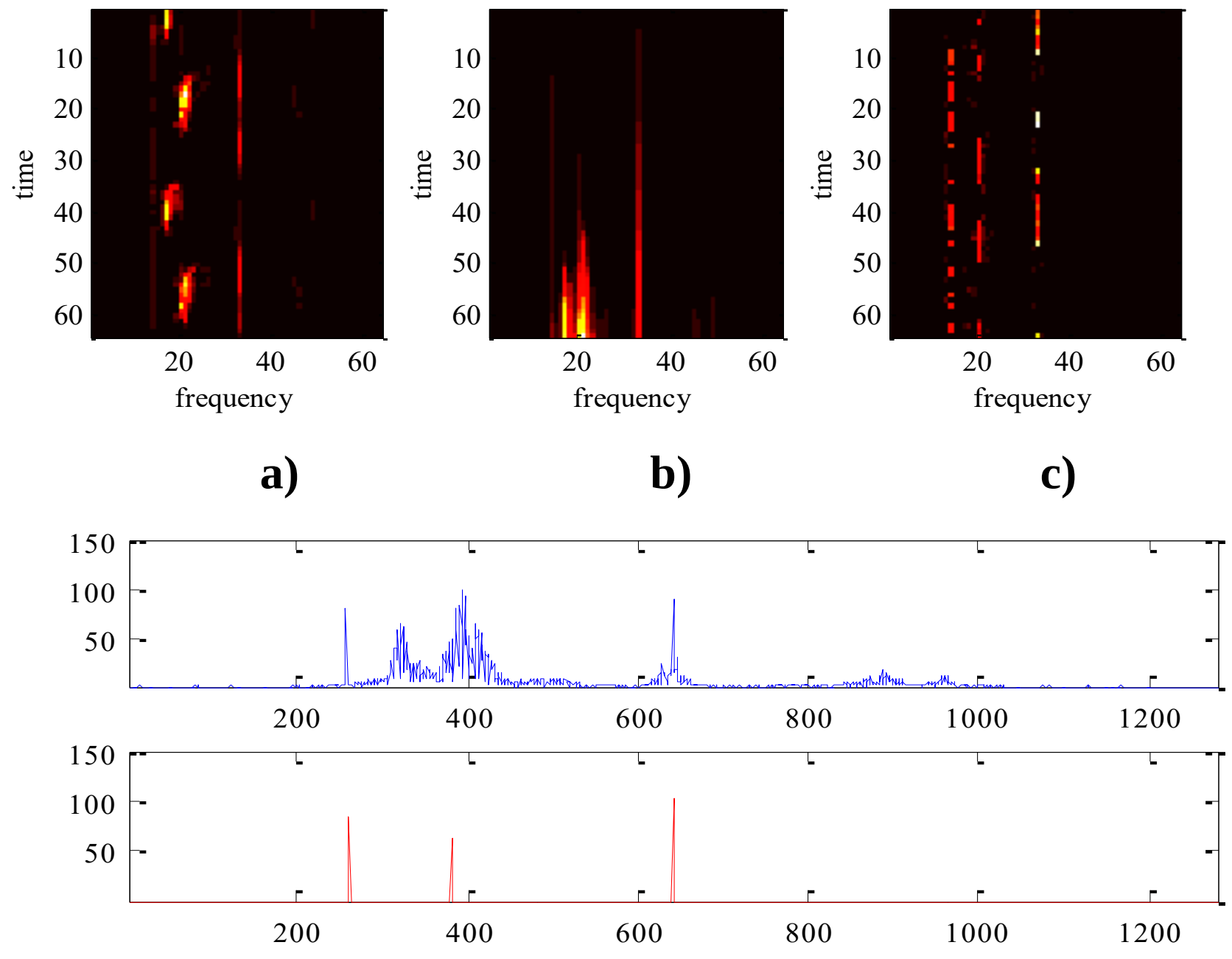

**d)**

**Slika 3.2: Realni radarski signal: a) STFT, b) sortirana STFT, c) preostali STFT koeficijenti nakon otklanjanja određenih regiona iz sortirane STFT, d) originalna DFT - plavo, i rekonstruisana DFT-crveno**

Ovaj pristup primijenjen je i u bežičnim komunikacijama [121], za razdvajanje signala koji pripadaju interferirajućim standardima u ovim komunikacijama - Bluetooth i IEEE 802.11b standardima. Predložena procedura je efikasna prilikom razdvajanja komponenti koje se preklapaju. U prvom koraku se odvajaju komponente IEEE 802.11b signala – iz sortirane STFT odbacuje se dio koeficijenata najmanjih vrijednosti koji pripadaju Bluetooth signalu. Ostatak STFT-a se rekonstruiše korišćenjem rekonstrukcionih algoritama, što rezultira rekonstrukcijom signala IEEE 802.11b. Kratkotrajnu Fourier-ovu transformaciju Bluetooth signala moguće je dobiti oduzimanjem rekonstruisane STFT od originalne, tj. polazne STFT, koja sadrži oba tipa signala.

Primjer signala koji se sastoji od Bluetooth i IEEE 802.11b signala, čije se komponente preklapaju u vremensko-frekvencijskoj ravni, prikazan je na slici 3.3.

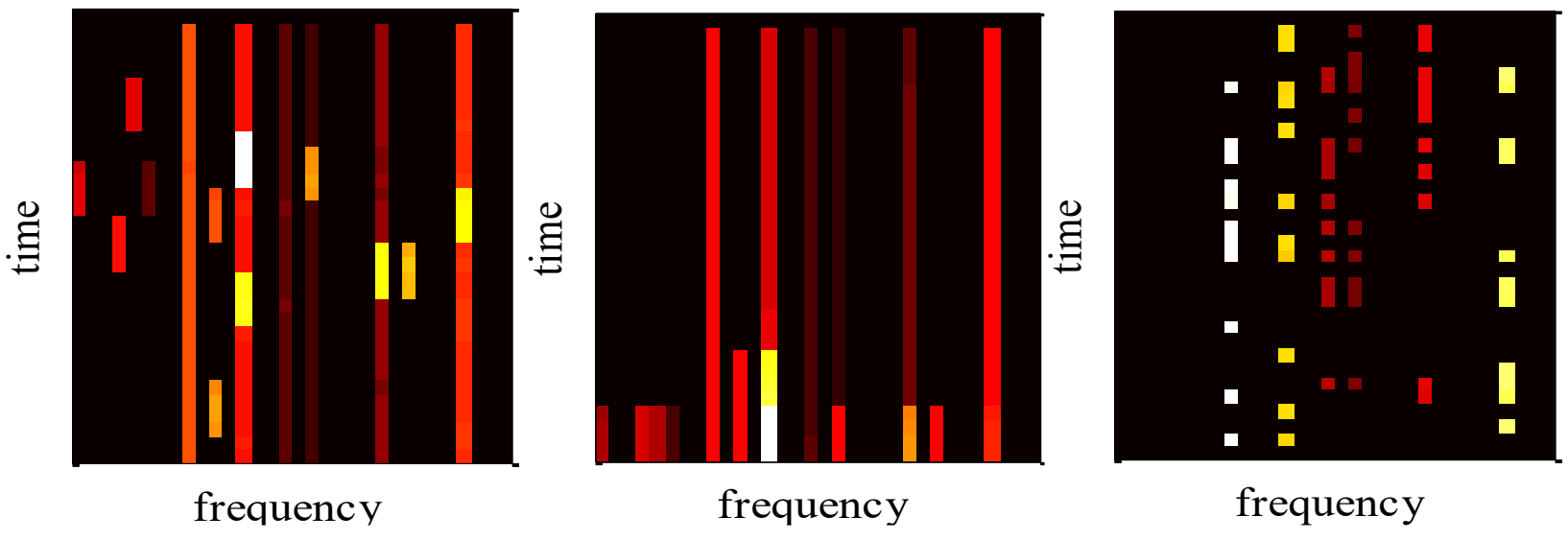

**a) b) c)**

**Slika 3.3: a) STFT originalnog signala, b) sortirana STFT, c) preostala STFT matrica, nakon odstranjivanja određenog procenta koeficijenata iz sortirane STFT**

Šest (dužih) komponenti pripada IEEE 802.11b signalu, a dvanaest kraćih komponenti pripada Bluetooth signalu. Korišćenjem opisane procedure, STFT je sortirana a zatim je određeni procenat najmanjih i najvećih koeficijenata otklonjen i STFT je rekonstruisana korišćenjem preostalih koeficijenata. Rekonstruisana STFT odgovara IEEE 802.11b signalu i prikazana je na slici 3.4a. Kratkotrajna Fourier-ova transformacija Bluetooth signala prikazana je na slici 3.4b.

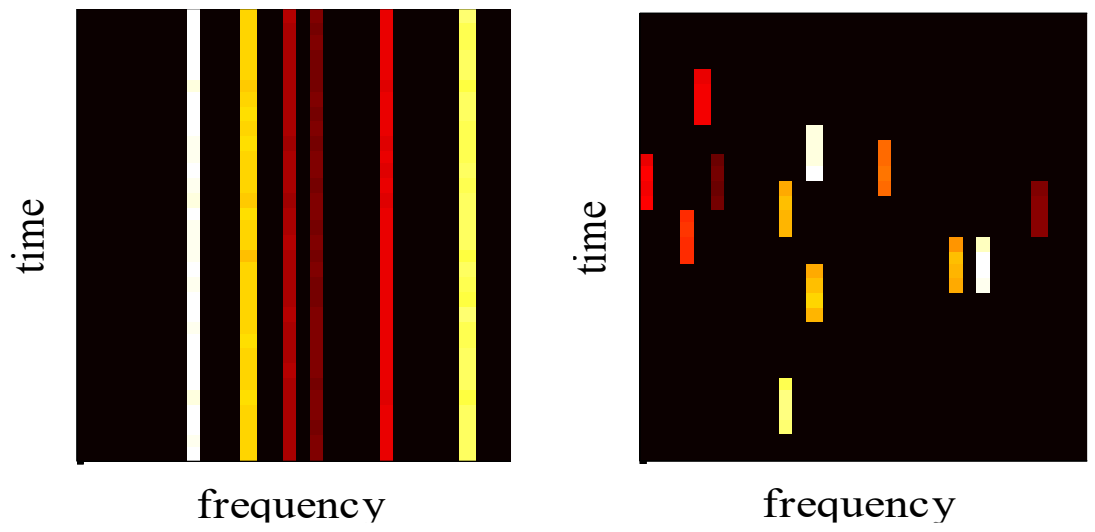

**a) b)**

**Slika 3.4: STFT a) IEEE 802.11b, b) Bluetooth signala**

Visoko-rezoluciona ISAR aplikacija predložena je u [122]. Pokazano je da predloženo rješenje, bazirano na CS-u, daje bolje rezultate u pogledu rezolucije slike od konvencionalnog, *range-Doppler* pristupa. Ako je signal definisan sljedećom relacijom:

$$x(t) = \sum_{k=1}^{K} W_k \rho\left(t/T_a\right) e^{-j2\pi f_k t} + \gamma(t), \tag{3.25}$$

gdje je $K$ broj najjačih centara, $\gamma(t)$ modeluje aditivni šum, $W_k$ je amplituda, $\rho(t/T_a)$ je jedinična pravougaona funkcija, $f_k$ je frekvencija nosioca a $T_a$ je vremensko trajanje *chirp* impulsa. Transformaciona i CS matrica označene su sa $\mathbf{\Psi}$ i $\mathbf{A}$, a signal se može zapisati i kao $x(t)=\mathbf{\Psi}\boldsymbol{\theta}+\gamma(t)$, gdje $\boldsymbol{\theta}$ označava vektor čije nenulte komponente odgovaraju kompleksnim amplitudama $K$ najjačih centara. Optimizacioni problem je definisan kao:

$$\min\left\|\hat{\boldsymbol{\theta}}\right\|_{\ell_1} \quad \text{uz uslov} \quad \left\|x(t)-\mathbf{A\Psi}\hat{\boldsymbol{\theta}}\right\|_{\ell_2} \leq \xi. \tag{3.26}$$

Nivo šuma označen je sa $\xi$ a vektor mjerenja sa $\mathbf{y}$, $\mathbf{y}=\mathbf{Ax}$, $\mathbf{A}_{M\times N}$ ($M<N$).

Rekonstrukcija ISAR slika, korišćenjem TwIST rekonstrukcionog algoritma predložena je u [134]. Rekonstrukcija je bazirana na metodu minimizacije gradijenta, koji je prilagođen rekonstrukciji signala (koji nemaju striktno konciznu predstavu niti u prostornom niti u transformacionom domenu, ali je gradijent slike rijedak i pogodan da se na njega primijene rekonstrukcioni algoritmi). Vektor mjerenja označen je sa $\mathbf{y}$: $\mathbf{y}=\mathbf{Af}+\mathbf{n}$, gdje je $\mathbf{A}$ matrica koja modeluje proces selekcije koeficijenta, $\mathbf{f}$ je signal koji treba estimirati a $\mathbf{n}$ je šum. U cilju estimiranja signala $\mathbf{f}$, formira se funkcija:

$$O(\mathbf{f})=\partial(\mathbf{y},\mathbf{Af})+\varepsilon R(\mathbf{f}), \tag{3.27}$$

gdje je $\partial(\mathbf{y},\mathbf{AF})$ funkcija koja modeluje razliku između $\mathbf{y}$ i $\mathbf{f}$, $\varepsilon$ je >0, $R(\mathbf{f})$ je regularizaciona funkcija (najčešće odgovara $\ell_1$ normi, ali se može definisati da minimizuje $\ell_0$ normu, $\ell_2$ normu, TV normu i slično). Ako minimizuje $\ell_1$ normu, onda je funkcija $O(\mathbf{f})$ definisana na sljedeći način:

$$O(\mathbf{f})=\left\|\mathbf{y}-\mathbf{Af}\right\|_{\ell_2}^2+\varepsilon\left\|\mathbf{f}\right\|_{\ell_1}. \tag{3.28}$$

Algoritam TwIST, korišćen za rješavanje optimizacionog problema, zasnovan je na *iterative shrinkage/thresholding* (IST) i *iterative re-weighted/shrinkage* (IRS) algoritmima. Rješenje se traži iterativnom procedurom, definisanom na sljedeći način:

$$\begin{aligned}&\mathbf{f}_1=\Gamma_\varepsilon(\mathbf{f}_0),\\&\mathbf{f}_{t+1}=(1-\alpha)\mathbf{f}_{t-1}+(\alpha-\beta)\mathbf{f}_t+\beta\Gamma_\varepsilon(\mathbf{f}_t),\ \ t\geq 1.\end{aligned} \tag{3.29}$$

Funkcija $\Gamma$ je definisana sljedećom relacijom:

$$\Gamma_\varepsilon(\mathbf{f})=\Upsilon_\varepsilon(\mathbf{f}+\mathbf{A}^T(\mathbf{y}-\mathbf{Af})), \tag{3.30}$$

a operator rašumljavanja $\Upsilon_\varepsilon$ je:

$$\Upsilon_\varepsilon=\arg\min_{\mathbf{f}}\frac{1}{\mu}\sum\nolimits_i\left\|D_i\mathbf{f}\right\|+\frac{1}{2}\left\|\mathbf{y}-\mathbf{Af}\right\|_{\ell_2}^2,\ \left\|D_i\mathbf{f}\right\|-\text{gradijentni operator}. \tag{3.31}$$

Procedura je testirana na realnoj radarskoj slici Mig25 (slika 3.5). Korišćena je slika dimenzija 64×64. Posmatrani su slučajevi kada je 33.4%, 38.5%, i 49.4% od ukupnog broja odbiraka dostupno.

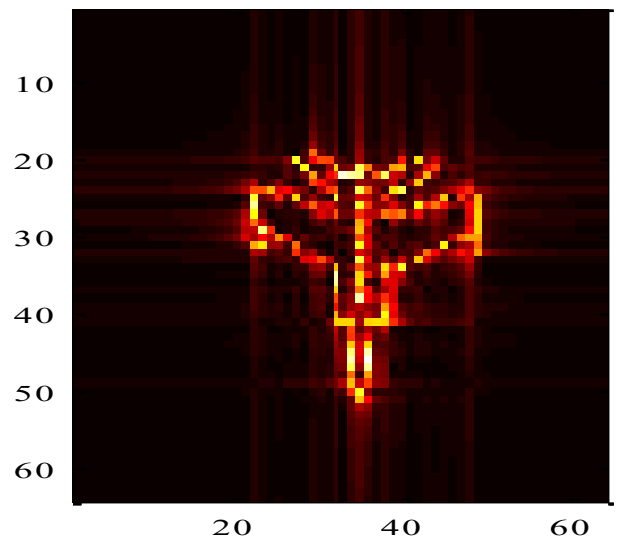

**Slika 3.5: ISAR slika Mig25**

Slike dobijene bez primjene rekonstrukcionog algoritma prikazane su na slikama 3.6 a), b) i c), za 33.4%, 38.5%, i 49.4% dostupnih koeficijenata, respektivno. Slike sadrže veliki procenat šuma, koji je posljedica nedostajućih odbiraka.

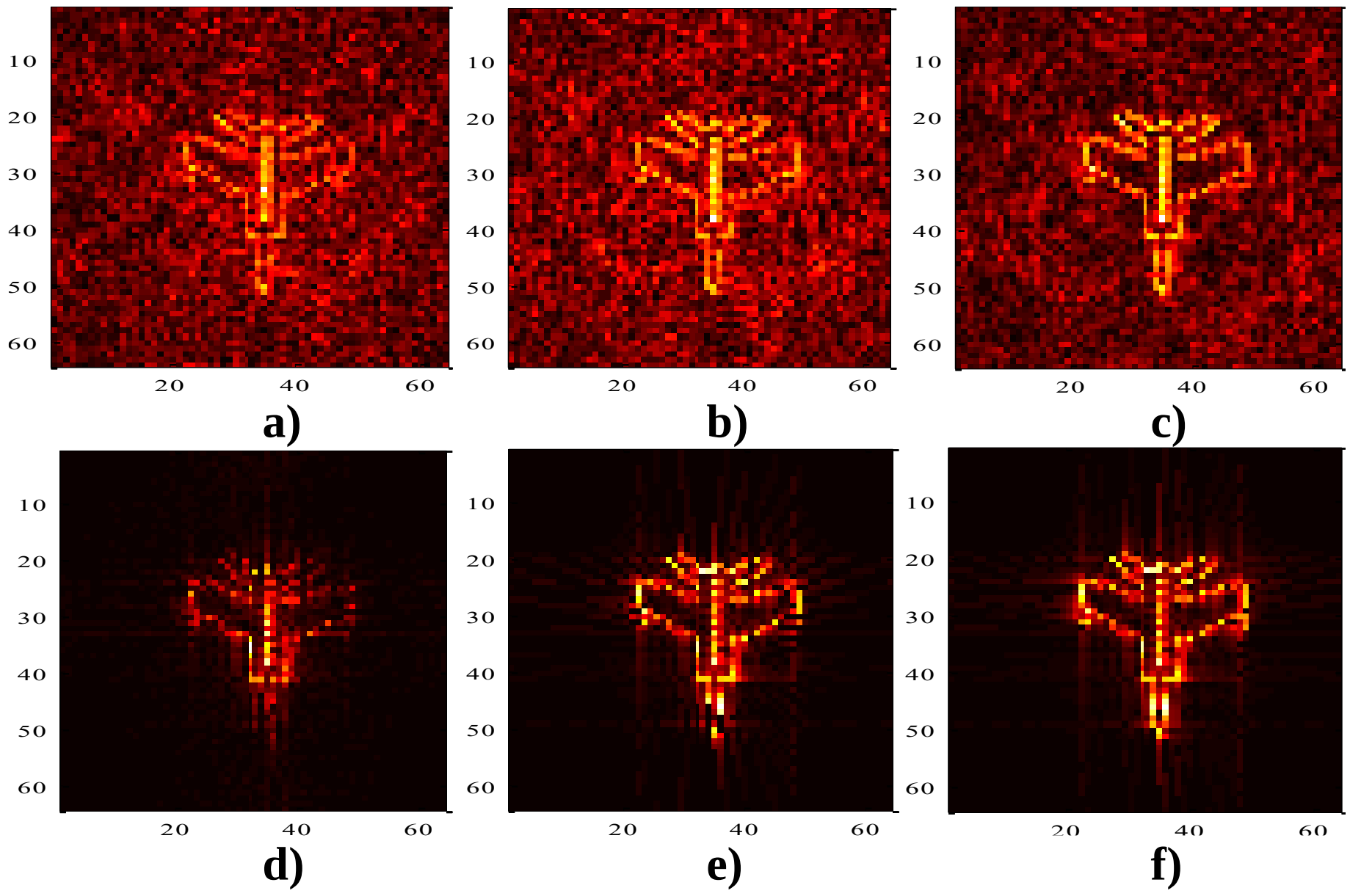

**Slika 3.6: a), b) i c) ISAR slike dobijene iz dostupnih odbiraka (33.4%, 38.5%, i 49.4%, respektivno), d), e) i f) Rekonstruisane ISAR slike (33.4%, 38.5%, i 49.4%, respektivno)**

Slike rekonstruisane primjenom rekonstrukcionog algoritma, prikazane su na slikama 3.6 d), e) i f), za 33.4%, 38.5%, i 49.4% dostupnih koeficijenata, respektivno. Postignut je značajno bolji kvalitet u poređenju sa slučajevima kada rekonstrukcioni algoritam nije primijenjen.

### 3.4.4. CS u rekonstrukciji slike i videa

Kompresivno očitavanje nalazi brojne aplikacije i u rekonstrukciji 2D signala. Procedura koja kombinuje CS pristup i rekonstrukciju slika u aplikacijama za detekciju objekata i monitoring teško pristupačnih terena, predložena je u [147]. Metod omogućava uspješnu rekonstrukciju čak i u slučajevima kada je 80% piksela slike nedostupno ili su zahvaćeni šumom. Rješenje je bazirano na 2D DCT domenu i gradijentnom algoritmu za rekonstrukciju 2D signala.

CS pristup za rekonstrukciju slika u astronomiji predložen je u [157]. Ispitivanje mogućnosti primjena CS-a u astronomiji je u poslednje vrijeme u porastu, imajući u vidu da konvencionalni metodi za kompresiju često ne daju zadovoljavajuće rezultate.

CS rekonstrukcija slike je primijenjena i u monitoringu nivoa ugljen-dioksida $CO_2$, prilikom transporta grožđa na udaljenu destinaciju na teritoriji Kine [158]. Precizno praćenje promjena različitih parametara, kao što su temperatura, relativna vlažnost, ili nivo različitih gasova u ovim transportnim lancima je od velikog značaja zbog očuvanja kvaliteta robe koja se prevozi. Međutim, konstantno praćenje parametara zahtijeva veliki broj senzora kao i velike memorijske kapacitete, što često dovodi do problema u komunikaciji između onog koji prati i uređaja koji šalju informacije. Imajući u vidu ove probleme, javila se ideja testiranja mogućnosti primjene CS-a u transportnim lancima. Posmatran je slučajno pododabrani signal koji opisuje nivo $CO_2$. Signal nema konciznu predstavu u posmatranim 1D transformacionim domenima, pa jedan od uslova koji je neophodan da bi se CS pristup mogao uspješno primijeniti, nije zadovoljen.

U cilju pronalaženja algoritma za uspješnu rekonstrukciju signala, umjesto da se ovaj 1D signal posmatra kao takav i da se primijenjuju algoritmi za rekonstrukciju 1D signala, korišćeni su algoritmi za rekonstrukciju 2D signala. Naime, signal je pretvoren u matricu i posmatran je kao 2D signal. Ako se vektor mjerenja signala označi sa $\mathbf{D}_a$, i ako je uzet iz 2D DCT domena (slučajnim rasporedom, nakon što je 2D DCT matrica cik-cak preuređenjem pretvorena u vektor), matrica 2D DCT transformacije označena sa $\mathbf{\Psi}$, a CS matrica sa $\mathbf{A}$, onda se optimizacioni problem može definisati kao:

$$\min_{\mathbf{F}} \mathrm{TV}(\mathbf{F}) \ \text{ uz uslov } \ \mathbf{f}_a = \mathbf{AF}, \tag{3.32}$$

gdje TV označava operator totalne varijacije, koji se definiše kao suma amplituda diskretnog gradijenta u svakoj tački, tj.:

$$\mathrm{TV}(\mathbf{F}) = \sum_{i,j} \left\| d_{i,j}\mathbf{F} \right\|_2 . \tag{3.33}$$

Aproksimacija gradijenta za piksel na poziciji *ij*, $d_{i,j}$, definisana je kao:

$$d_{i,j}\mathbf{F} = \begin{bmatrix} F(i+1,j) - F(i,j) \\ F(i,j+1) - F(i,j) \end{bmatrix} . \tag{3.34}$$

Rezultati dobijeni primjenom opisane metode na signal koji prati nivo $CO_2$ prikazani su na slici 3.7. Procenat dostupnih mjerenja je 45%. Kao što je napomenuto, signal je prvo transformisan u sliku, čije su originalna i rekonstruisana verzija prikazane na slici 3.7a, dok su originalna i rekonstruisana verzija 1D signala (koji je dobijen iz rekonstruisane slike), prikazani na slikama 3.7b.

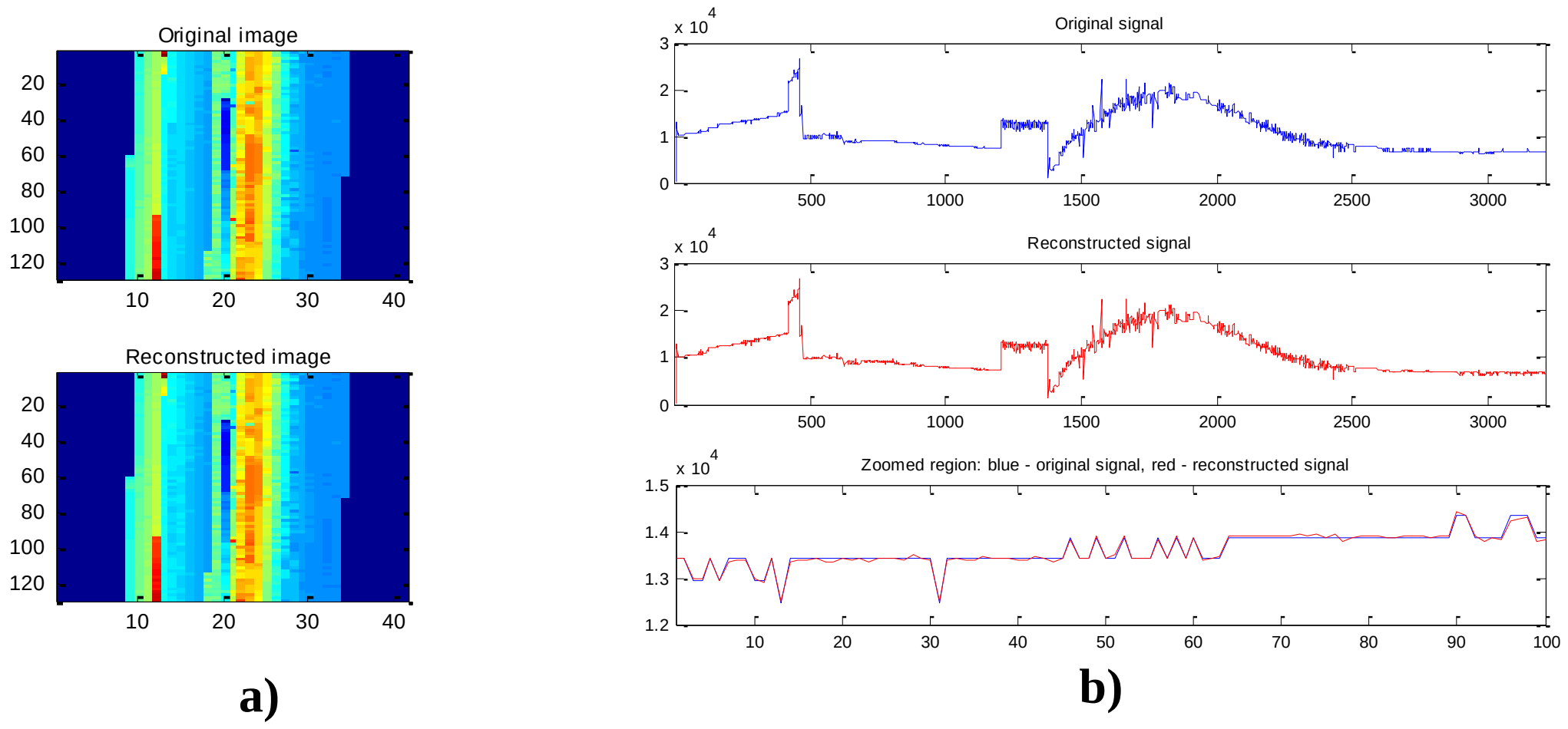

**Slika 3.7: a) Originalna i slika rekonstruisana korišćenjem 45 % od ukupnog broja odbiraka; b) Originalni signal (prvi red); Signal dobijen iz rekonstruisane slike prikazane na slici 3.7a (drugi red); Zumirani region originalnog signala – plavo, i rekonstruisanog signala - crveno (treći red)**

Osim aplikacija CS-a u rekonstrukciji slike, sve je veća primjena ovog metoda u rekonstrukciji video signala [119], [162]-[165]. Razvijeno je više algoritama za CS rekonstrukciju u video aplikacijama [165]-[171]. Pristup estimaciji parametara kretanja, baziran na CS-u, predložen je u [163]. Estimacija se radi u uslovima smanjenog broja frejmova video signala. Metod kombinuje algoritme za rekonstrukciju signala razrijeđenog spektra sa vremensko-frekvencijskom analizom.

Rekonstrukcija video podataka korišćenjem gradijentnog algoritma predložena je u [163]. Nedostajući odbirci su raspoređeni slučajno i javljaju se u svakom frejmu video signala. Frejmovi imaju kompaktnu predstavu u DCT domenu. Efikasnost algoritma testirana je u pogledu broja dostupnih odbiraka i nivoa razrijeđenosti spektra signala i pokazuje bolje rezultate u poređenju sa tzv. *greedy* algoritmima, kao i bolji kvalitet rekonstrukcije u slučajevima realnih video sekvenci.

### 3.4.5. Zaštita digitalnih podataka u uslovima komprimovanog očitavanja

Zaštita digitalnih podataka u uslovima CS-a obrađivana je kroz brojne radove u literaturi [166]-[171]. Tehnika umetanja i detekcije vodenog žiga bazirana na CS-u predložena je u [166] i testirana u uslovima MP3 kompresije. Ovdje se podrazumijeva da se i signal u koji se dodaje vodeni žig, i sam vodeni žig mogu kompaktno predstaviti u određenom domenu. Vodeni žig se dodaje u vektor mjerenja **y**: $\mathbf{y}=\mathbf{\Psi x}+\mathbf{a\Phi b}$, gdje je **x** signal, $\mathbf{\Psi}$ je transformaciona matrica a **b** vodeni žig dužine $L$.

Matrica $\mathbf{\Phi}_{M\times L}$ predstavlja slučajnu Gausovu matricu a $M$ je dužina vektora mjerenja. Jačina vodenog žiga, **a**, adaptira se za svaki frejm audio signala, u skladu sa relacijom: $\mathbf{a}=0.04\sqrt{\sum_{i=1}^{M}\mathbf{X}i^2}$, $\mathbf{X}=\mathbf{\Psi x}$. Da bi se rekonstruisao signal, mora se riješiti optimizacioni problem, što znači da matrica $\mathbf{\Phi}$ mora biti poznata.

Procedura watermarkiranja digitalne slike u uslovima CS-a je predložena u [168]. Pikseli, koji su odabrani slučajnim rasporedom i koji predstavljaju mjerenja signala, takođe služe kao koeficijenti u koje će biti umetnut watermark. Slika se prvo dijeli na blokove, a mjerenja se biraju iz blokova slike. Slika ima kompaktnu predstavu u domenu 2D DFT transformacije. Iz uzetih mjerenja, slika se rekonstruiše korišćenjem metoda totalne varijacije.

U [170] testirana je mogućnost uklanjanja watermark-a korišćenjem CS-a. Watermark $w$ je kreiran kao pseudoslučajna sekvenca i umetnut je u DCT koeficijente slike, u skladu sa relacijom: $DCT_w=DCT+\alpha w$, gdje parameter $\alpha$ označava jačinu watermarka. Kompresivno očitavanje je zatim primijenjeno kao atak na watermarkiranu sliku, sa ciljem uklanjanja watermarka. Uzet je samo mali procenat piksela slike, nakon čega je slika rekonstruisana iz uzetih mjerenja. Mjerenja su definisana sljedećom relacijom:

$$\mathbf{y}_{M\times 1}=\mathbf{\Phi}_{M\times N}\mathbf{x}_{N\times 1},\ \mathbf{y}=\mathbf{\Phi x}=\mathbf{\Phi\Psi}^{-1}\mathbf{X}=\mathbf{AX}, \tag{3.35}$$

gdje je $\mathbf{\Phi}$ matrica mjerenja, $\mathbf{\Psi}$ je transformaciona matrica, **A** je CS matrica, a **X** je signal u $\mathbf{\Psi}$ domenu. Mjerenja se sastoje od određenog procenta niskofrekvencijskih koeficijenata kao i određenog procenta slučajno odabranih koeficijenata iz ostaka DCT ravni (tj. koeficijenti koji pripadaju srednjim i visokim frekvencijama): **y**=[**y**1, **y**2], gdje su **y**1 niskofrekvencijski a **y**2 visokofrekvencijski DCT koeficijenti. Slika je rekonstruisana korišćenjem metoda totalne varijacije:

$$\min_{\mathbf{X}} \mathrm{TV}(\mathbf{X}) \text{ uz uslov } \mathbf{y} = \mathbf{AX}. \tag{3.36}$$

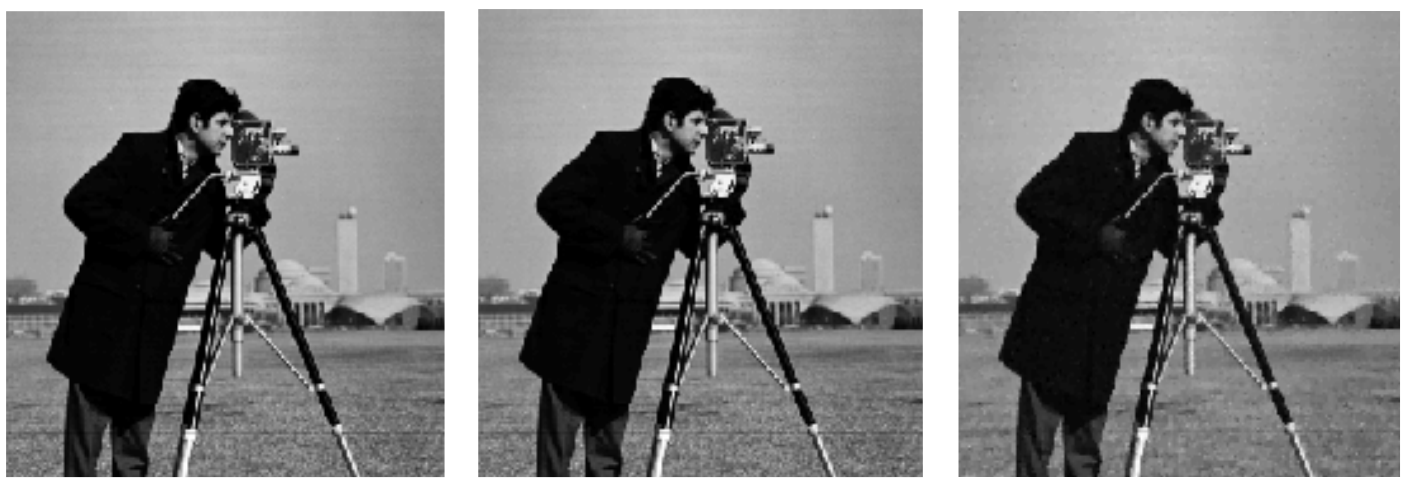

a) b) c)

**Slika 3.8: a) Originalna, b) watermarkirana i c) slika rekonstruisana korišćenjem 30% dostupnih odbiraka**

Operator totalne varijacije označen je sa TV i definisan je relacijom (3.33). Procedura detekcije watermarka je slijepa procedura a za detekciju koristi standardni korelator:

$$D_{ww} = \sum\nolimits_{i=L+1}^{L+K} w_i DCT_{w_i} > \sum\nolimits_{i=L+1}^{L+K} wrong_i DCT_{w_i} = D_{wr} \;, \tag{3.37}$$

gdje, za uspješnu detekciju watermarka, odziv detektora na watermark (pravi ključ), $D_{ww}$, treba da bude veći od odziva detektora na pogrešne ključeve, $D_{wr}$, generisane na isti način kao i pravi ključ.

Slika 3.8 prikazuje originalnu, watermarkinranu i sliku rekonstruisanu korišćenjem 30% mjerenja. Rezultati detekcije prikazani su na slici 3.9 i pokazuju da, iako je rekonstruisana slika dobrog kvaliteta, detekcija watermark-a nije bila uspješna.

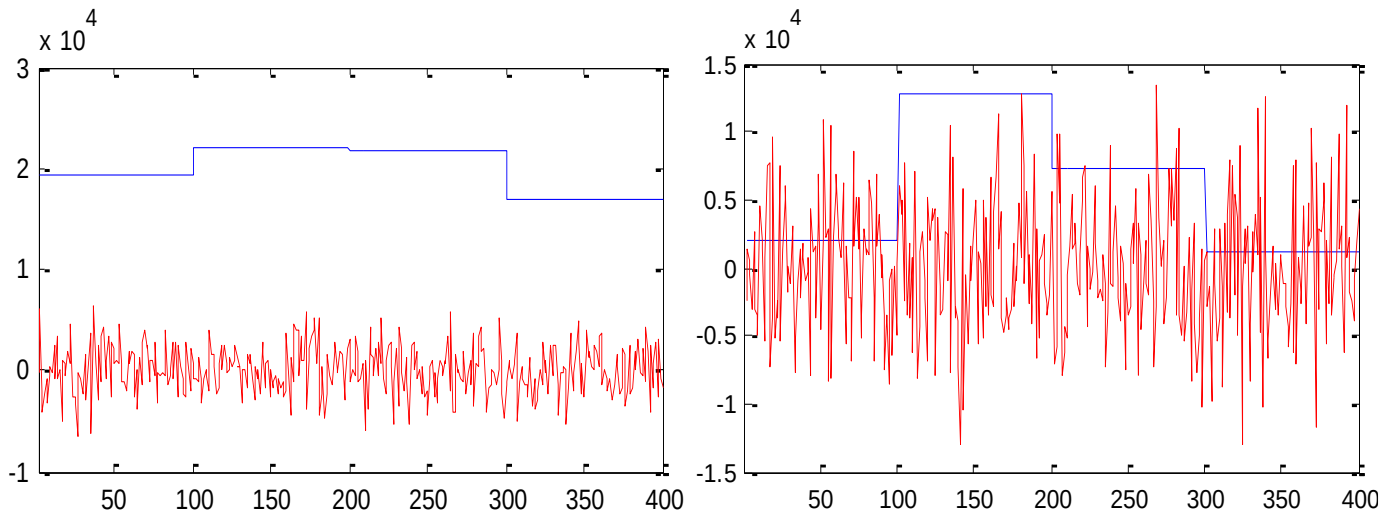

a) b)

**Slika 3.9: Rezultati detekcije za 4 prava ključa (plava linija) i 400 pogrešnih ključeva (crvena linija): a) bez ataka, b) CS rekonstrukcija sa 30% uzetih piksela**

# 4. Spars vremensko-frekvencijske reprezentacije sa primjenom na signale sa brzim promjenama trenutne frekvencije

Visoka rezolucija u vremensko-frekvencijskoj ravni zahtjev je za sve vremensko-frekvencijske distribucije. Međutim, visoka rezolucija i tačna estimacija trenutne frekvencije (*instantaneous frequency* - IF) zavisi kako od primijenjene distribucije tako i od kompleksnosti fazne funkcije signala. U cilju obezbjeđivanja što preciznijeg praćenja izuzetno brze promjene trenutne frekvencije signala, koriste se distribucije sa kompleksnim argumentom vremena. Ove distribucije omogućavaju efikasnu analizu takvih signala i u slučajevima kada standardne distribucije ne obezbjeđuju zadovoljavajuće rezultate. U ovoj glavi su posmatrane različite vremensko-frekvencijske distribucije u kontekstu CS-a [176]-[184]. U tom cilju, kompleksna distribucija 4-tog reda, bazirana na ambiguity funkciji, primijenjena je u analizi signala sa brzim promjenama trenutne frekvencije.

Signali sa brzom promjenom faze su npr. vibrirajući tonovi muzičkih instrumenata ili radarski signali. Različite informacije o radarskoj meti mogu se saznati na osnovu varijacija trenutne frekvencije signala, a interesantna je primjena vremensko-frekvencijskih distribucija u karakterizaciji *micro-Doppler* efekta kod radara. *Micro-Doppler* efekat proizvode brzo pokretni djelovi mete, a ovaj efekat se može koristiti u detekciji i prepoznavanju mete. Takođe, kada se posmatra kretanje ljudskog tijela, pokreti djelova tijela proizvode brze varijacije koje se teško mogu detektovati standardnim distribucijama, kao što su WD, S-metod ili distribucije iz Cohen-ove klase. Definisane su modifikacije S-metoda (adaptivni S-metod i S-metod sa višestrukim prozorima), koji poboljšavaju koncentraciju u vremensko-frekvencijskoj ravni. Međutim, i kod ovih distribucija postoje ograničenja zbog njihove kvadratne prirode, a javlja se i uticaj viših izvoda faze kod signala sa brzo promjenljivom faznom funkcijom (npr. izvodi sinusnih i kosinusnih modulacija prouzrokovanih rotirajućim djelovima mete). Zbog toga se, kao optimalan izbor distribucija u ovim slučajevima, nameću distribucije sa kompleksnim argumentom vremena.

U ovoj glavi CS pristup je primijenjen na signale sa brzim promjenama faze. Signali imaju kompaktnu (spars) predstavu u vremensko-frekvencijskom domenu, pa će i distribucije biti nazvane spars vremensko-frekvencijskim distribucijama. Cilj je što bolja lokalizacija trenutne frekvencije, korišćenjem malog broja mjerenja [172]-[176]. Kao mjerenja služe slučajno odabrani koeficijenti iz ambiguity domena. Mjerenja se mogu uzimati iz čitavog ambiguity domena, ili se oblast iz koje se uzimaju mjerenja može unaprijed definisati. Imajući u vidu da su koeficijenti koji pripadaju auto komponentama signala, u ambiguity domenu većinom koncentrisani oko koordinatnog početka, moguće je formirati masku oko koordinatnog početka i koeficijente birati slučajnim rasporedom iz regiona ograničenog maskom. Veličina maske mora biti dovoljna da uključi koeficijente koji pripadaju auto članovima signala a u isto vrijeme, da izbjegne kros komponente ili šum koji se takođe nalaze u ambiguity domenu, ali su dislocirani od koordinatnog početka.

U slučaju signala zahvaćenih šumom ovaj pristup se može kombinovati sa L-estimacijom, tako što se odbirci zahvaćeni šumom (ili dio ovakvih odbiraka) prvo otklone primjenom L-statistike, a zatim se otklonjeni odbirci rekonstruišu primjenom CS algoritama [177].

Opšti oblik distribucije $N$-tog reda sa kompleksnim argumentom vremena definisan je relacijom (1.42), tj.:

$$CTD_N(t,\omega)=\int_{-\infty}^{\infty}\prod_{\substack{i=-N/2,\\ i\neq 0}}^{N/2} x\left(t+\frac{1}{N}\frac{\tau}{sign(i)(a_i+jb_i)}\right)^{sign(i)(a_i+jb_i)} e^{-j\omega\tau}d\tau.$$

Distribucija četvrtog reda sa kompleksnim argumentom vremena dobija se za $N$=4 i biće primijenjena na signale korišćene u ovom dijelu. Opisana je sljedećom relacijom:

$$\begin{aligned}CTD_4(t,\omega)&=\int_{-\infty}^{\infty} R_4(t,\tau)e^{-j\omega\tau}d\tau=\\&=\int_{-\infty}^{\infty}\prod_{\substack{i=-2,\\ i\neq 0}}^{2} x\left(t+\frac{\tau}{4sign(i)(a_i+jb_i)}\right)^{sign(i)(a_i+jb_i)} e^{-j\omega\tau}d\tau.\end{aligned} \tag{4.1}$$

Parametri $a_i$ i $b_i$ označavaju tačke na jediničnom krugu. Posmatrajmo slučaj ($a_1$, $b_1$, $a_2$, $b_2$)=(1,0,0,1). Za ovakav odabir tačaka, distribucija 4-tog reda sa kompleksnim argumentom vremena biće oblika:

$$CTD_4(t,\omega)=\int_{-\infty}^{\infty} x\left(t+\frac{\tau}{4}\right)x^{-1}\left(t-\frac{\tau}{4}\right)x^{j}\left(t-j\frac{\tau}{4}\right)x^{-j}\left(t+j\frac{\tau}{4}\right)e^{-j\omega\tau}d\tau, \tag{4.2}$$

a njoj odgovarajući faktor rasipanja definisan je relacijom:

$$S(t,\tau)=\phi^{(5)}(t)\frac{\tau^5}{4^4 5!}+\phi^{(9)}(t)\frac{\tau^9}{4^8 9!}+... \tag{4.3}$$

Realni i imaginarni dio ambiguity funkcije opisani su sljedećim relacijama [42], [118]:

$$\begin{aligned} AF_{rt}(\theta,\tau)&=\int_{-\infty}^{\infty} x\left(t+\frac{\tau}{4}\right)x^*\left(t-\frac{\tau}{4}\right)e^{-j\theta t}dt,\\ AF_{ct}(\theta,\tau)&=\int_{-\infty}^{\infty} x^{-j}\left(t+j\frac{\tau}{4}\right)x^{j}\left(t-j\frac{\tau}{4}\right)e^{-j\theta t}dt, \end{aligned} \tag{4.4}$$

a obje funkcije se filtriraju funkcijom jezgra:

$$AF_{rt}^K(\theta,\tau)=K(\theta,\tau)AF_{rt}(\theta,\tau),\quad AF_{ct}^K(\theta,\tau)=K(\theta,\tau)AF_{ct}(\theta,\tau).$$

Rezultujuća ambiguity funkcija i distribucija sa kompleksnim argumentom definisane su relacijama:

$$AF_{CTD}(\theta,\tau)=\int_{-\infty}^{\infty}\int_{-\infty}^{\infty}\int_{-\infty}^{\infty} W(\varepsilon)e^{-j\varepsilon\tau_1}e^{j\varepsilon(\tau-\tau_1)}AF_{rt}^K(\theta_1,\tau_1)AF_{ct}^K(\theta-\theta_1,\tau-\tau_1)d\tau_1 d\theta_1 d\varepsilon,$$

$$CTD(t,\omega)=\frac{1}{2\pi}\int_{-\infty}^{\infty}\int_{-\infty}^{\infty} AF_{CTD}(\theta,\tau)e^{j\theta t-j\omega\tau}d\tau d\theta.$$

Pretpostavimo da imamo signal u vremenskom domenu **x**, čija je dužina $N$. Neka je sa $\mathbf{\Psi}^{-1}$ označen spars domen signala **x**. Tada se signal može predstaviti na sljedeći način:

$$\mathbf{x}=\mathbf{\Psi}_{N\times N}^{-1}\mathbf{X}\ . \tag{4.5}$$

Ako $\mathbf{\Psi}_{N\times N}^{-1}$ označava $N\times N$ inverznu Fourier-ovu transformacionu matricu, onda vektor **X** označava vektor koeficijenata Fourier-ove transformacije. U matričnom obliku prethodna relacija se može zapisati na sljedeći način:

$$\begin{bmatrix} x(0)\\ x(1)\\ ...\\ x(N-1)\end{bmatrix}=\frac{1}{N}\begin{bmatrix} 1 & 1 & ... & 1\\ 1 & e^{j\frac{2\pi}{N}} & ... & e^{j\frac{2(N-1)\pi}{N}}\\ ... & ... & ... & ...\\ 1 & e^{j\frac{2(N-1)\pi}{N}} & ... & e^{j\frac{2(N-1)(N-1)\pi}{N}}\end{bmatrix}\begin{bmatrix} X(0)\\ X(1)\\ ...\\ X(N-1)\end{bmatrix}, \tag{4.6}$$

gdje je svaki elemenat matrice $\mathbf{\Psi}_{N\times N}^{-1}$ eksponencijalni član $e^{j(2\pi kn/N)}$, $k$=0, …, $N$-1; $n$=0, …, $N$-1. Ako se uzme da je samo $M$ slučajno raspoređenih mjerenja signala **x** poznato, dok su ($N$-$M$) odbiraka signala nepoznati, Fourier-ova transformaciona matrica neće biti potpuna $N\times N$ matrica, već će biti slučajno pododabrana – sadržaće $M$ od ukupno $N$ slučajno raspoređenih vrsta originalne matrice. To se može matematički modelovati množenjem originalne matrice $\mathbf{\Psi}_{N\times N}^{-1}$ sa nekoherentnom matricom mjerenja $\mathbf{\Phi}_{M\times N}$.

Matrica formirana na ovaj način je slučajna parcijalna Fourier-ova matrica i može se definisati na sljedeći način:

$$\boldsymbol{\Psi}_P^{-1} = \boldsymbol{\Phi}_{M\times N}\boldsymbol{\Psi}_{N\times N}^{-1}. \tag{4.7}$$

Vektor mjerenja **y** se može definisati kao:

$$\mathbf{y} = \boldsymbol{\Psi}_P^{-1}\mathbf{X}. \tag{4.8}$$

Rješenje sistema definisanog prethodnom relacijom može se naći primjenom $\ell_1$ minimizacije:

$$\min_{\mathbf{X}} \|\mathbf{X}\|_{\ell_1} \text{ uz uslov } \mathbf{y} = \boldsymbol{\Psi}_P^{-1}\mathbf{X}. \tag{4.9}$$

### *CS problem u ambiguity domenu*

U slučaju da imamo dvodimenzioni problem, relacija između ambiguity funkcije i distribucije sa kompleksnim argumentum vremena se može zapisati na sljedeći način [118]:

$$AF(\theta,\tau) = \boldsymbol{\Psi}^{2D}\cdot CTD(t,\omega), \tag{4.10}$$

gdje je sa *CTD* označena distribucija sa kompleksnim argumentum vremena, $\boldsymbol{\Psi}^{2D}$ je dvodimenziona $N^2\times N^2$ matrica Fourier-ove transformacije, a domen u kom signal ima kompaktnu predstavu je distribucija sa kompleksnim argumentom vremena (4-tog reda). Matrica $\boldsymbol{\Psi}^{2D}$ je formirana kao Kronekerov proizvod jedinične matrice **I** i matrice jednodimenzione Fourier-ove transformacije $\boldsymbol{\Psi}_{N\times N}^{1D}$:

$$\boldsymbol{\Psi}_{N^2\times N^2}^{2D} = \mathbf{I}_{N\times N}\otimes\boldsymbol{\Psi}_{N\times N}^{1D}, \tag{4.11}$$

Simbol $\otimes$ označava Kronekerov proizvod. U matričnoj formi relacija (4.11) postaje:

$$\begin{bmatrix} \boldsymbol{\Psi}_{N\times N}^{1D} & 0 & ... & 0 \\ 0 & \boldsymbol{\Psi}_{N\times N}^{1D} & ... & 0 \\ ... & ... & ... & 0 \\ 0 & 0 & 0 & \boldsymbol{\Psi}_{N\times N}^{1D} \end{bmatrix} = \begin{bmatrix} \mathbf{1}_{N\times N} & 0 & ... & 0 \\ 0 & \mathbf{1}_{N\times N} & ... & 0 \\ ... & ... & ... & 0 \\ 0 & 0 & 0 & \mathbf{1}_{N\times N} \end{bmatrix} \otimes \begin{bmatrix} 1 & 1 & ... & 1 \\ 1 & e^{-j\frac{2\pi}{N}} & ... & e^{-j\frac{2(N-1)\pi}{N}} \\ ... & ... & ... & ... \\ 1 & e^{-j\frac{2(N-1)\pi}{N}} & ... & e^{-j\frac{2(N-1)^2\pi}{N}} \end{bmatrix} \tag{4.12}$$

Sada se CS problem može definisati u ambiguity domenu. Pretpostavka je da signal ima kompaktnu predstavu u vremensko-frekvencijskom domenu. Ako se za CS mjerenja uzme određeni procenat koeficijenata oko koordinatnog početka u ambiguity domenu, dobijena ambiguity funkcija naziva se mjernom ambiguity funkcijom. Mjerenja se biraju iz maske veličine $J \times J$ koja je postavljena oko koordinatnog početka. Nakon toga se, za potrebe rješavanja optimizacionog problema, od 2D maske formira vector $\mathbf{AF}^M$, veličine $J^2 \times 1$.

Spars vremensko-frekvencijska distribucija dobija se minimizacijom sljedeće funkcije:

$$\min_{\sigma}\left(F(\sigma)+G(\sigma)\right), \tag{4.13}$$

gdje $F$ predstavlja operator tzv. finog pragovanja (*soft thresholding*), u skladu sa relacijom: $F(\sigma)=\max(0,1-\frac{\lambda}{|\sigma|})\sigma$, λ je regularizacioni parameter a $G(\sigma)$ je definisano kao:

$$G(\sigma)=\mathbf{\Psi}_P^{'}(\mathbf{\Psi}_P\sigma-\mathbf{AF}^M). \tag{4.14}$$

Simbol $\mathbf{\Psi}_P$ označava parcijalnu 2D Fourier-ovu transformacionu matricu, $\mathbf{\Psi}_P^{'}$ je njena transponovana verzija a σ je vektor vrsta, koji se nakon rješavanja optimizacionog problema pretvara u matricu i formira spars vremensko-frekvencijsku reprezentaciju.

## 4.1. Robustni pristup estimaciji trenutne frekvencije

U slučaju da je ambiguity funkcija zahvaćena impulsnim šumom, ambiguity mjerenja će takođe biti zahvaćena šumom. U cilju obezbjeđivanja vektora mjerenja bez šuma, robustna statistika se može primijeniti na ambiguity funkciju u cilju otklanjanja odbiraka koji su zahvaćeni šumom. Ovo se može uraditi primjenom L-statistike. Vektor mjerenja uzetih iz ambiguity domena zahvaćenog šumom, se prvo sortira a zatim se određeni procenat najmanjih i najvećih koeficijenata odstranjuje. Robustna početna forma se može definisati na sljedeći način:

$$s_0=\sum_{i=P}^{Q}AF_{SORT},\ \text{gdje je}\ AF_{SORT}(\tau,\theta)=sort\{AF(\tau,\theta)e^{-j2\pi\tau k/M}e^{-j2\pi\theta l/M}\}, \tag{4.15}$$

a $P$ i $Q$ označavaju broj najvećih i najmanjih koeficijenata koji su odstranjeni, respektivno. Predloženi algoritam [118] za dobijanje spars vremensko-frekvencijske reprezentacije je prikazan na slici 4.1.

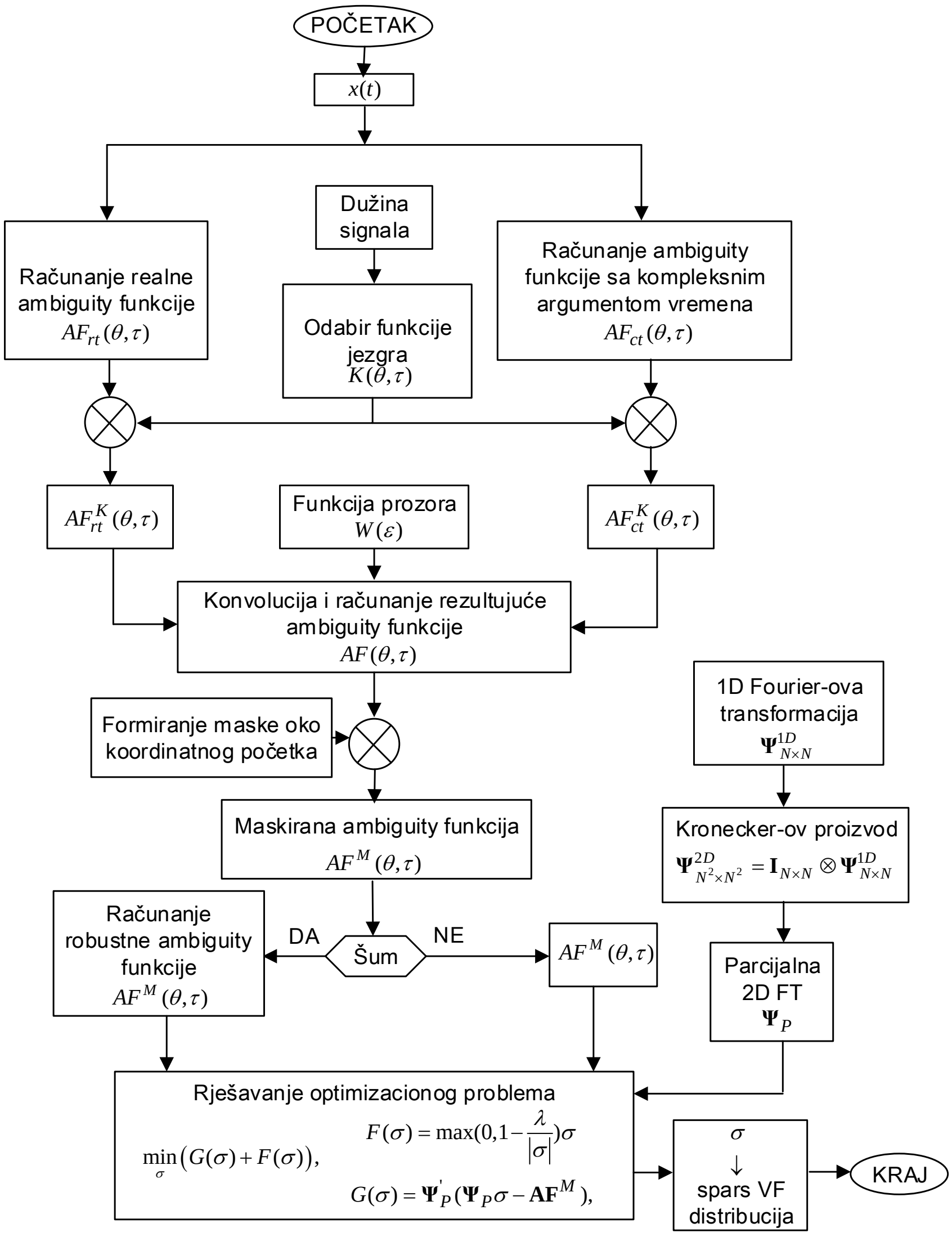

**Slika 4.1: Dijagram predloženog algoritma**

## 4.2. Eksperimentalni rezultati

### 4.2.1. Simulirani radarski signal sa brzim promjenama trenutnih frekvencija komponenti signala

Posmatran je multikomponentni signal sa nelinearnom faznom funkcijom:

$$x(t) = e^{j(2\cos(\pi t)+\cos(4\pi t)+4.5\pi t)/2} + e^{j(\cos(\pi t)+\cos(3\pi t)+\cos(4\pi t)-8\pi t)/2} . \qquad (4.16)$$

U cilju dobijanja optimalne vremensko-frekvencijske reprezentacije za posmatrani signal, testirano je nekoliko vremensko-frekvencijskih reprezentacija. Korišćene su matrice dimenzija 90×90. Vremensko-frekvencijske reprezentacije signala prikazane su na slici 4.2.

Prvo je posmatrana Wigner-ova distribucija. Sa slike se može primijetiti da ona ne prati promjene trenutne frekvencije komponenti signala (slika 4.2a i slika 4.2d), a dodatno unosi kros-članove. Zatim su testirane distribucije iz Cohen-ove klase. Kao funkcija jezgra upotrijebljena je Gausova funkcija, sa različitim vrijednostima parametra $\delta$ (distribucije dobijene korišćenjem $\delta$=80 i $\delta$=20 prikazane su na slikama 4.2b i 4.2c. Srednje kvadratne greške (MSE) i relativne srednje kvadratne greške estimacije (RMSE) za različite distribucije i različiti broj mjerenja date su u tabeli 4-1. Distribucije iz Cohen-ove klase eliminišu značajan broj kros članova koji se pojavljuju u vremensko-frekvencijskoj ravni, ali ni ove distribucije ne uspijevaju da obezbijede precizno praćenje promjena trenutne frekvencije nestacionarnih signala, što se može vidjeti sa slika 4.2e i 4.2f.

U cilju da potkrijepimo navedene tvrdnje, estimirane su trenutne frekvencije komponenti signala iz posmatranih distribucija, i prikazane su na slici 4.2 crvenom bojom, dok su tačne trenutne frekvencije komponenti prikazane na istoj slici, plavom bojom. Spars distribucije, računate počev od Wigner-ove distribucije preko distribucija iz Cohen-ove klase, prikazane su na slici 4.3. Kod spars, CS distribucija, maska u ambiguity domenu formirana je oko koordinatnog početka i veličine je 25×25. Ovaj region sadrži oko 7.7% od ukupnog broja odbiraka u ambiguity domenu. Iz tog regiona je 50% koeficijenata odabrano slučajnim rasporedom i predstavlja CS mjerenja. Drugim riječima, uzeto je 3.8% od ukupnog broja odbiraka.

Sa slike se može primijetiti da distribucije nemaju kompaktnu predstavu i da ne prate promjene trenutnih frekvencija pojedinačnih komponenti signala. Zbog toga smo prešli na računanje distribucija sa kompleksnim argumentom vremena. Standardna forma distribucije 4. reda sa kompleksnim argumentom, prikazana je na slici 4.4a. Filtriranje u ambiguity domenu Gausovim jezgrom je korišćeno prilikom računanja kompleksne distribucije. Rezultujuća, spars distribucija (slika 4.4b) dobijena je korišćenjem 60% odbiraka unutar maske, što predstavlja oko 5% od ukupnog broja odbiraka. Veoma slični rezultati dobijeni su korišćenjem manjeg broja mjerenja unutar maske, oko 40% (što je oko 3% od ukupnog broja odbiraka).

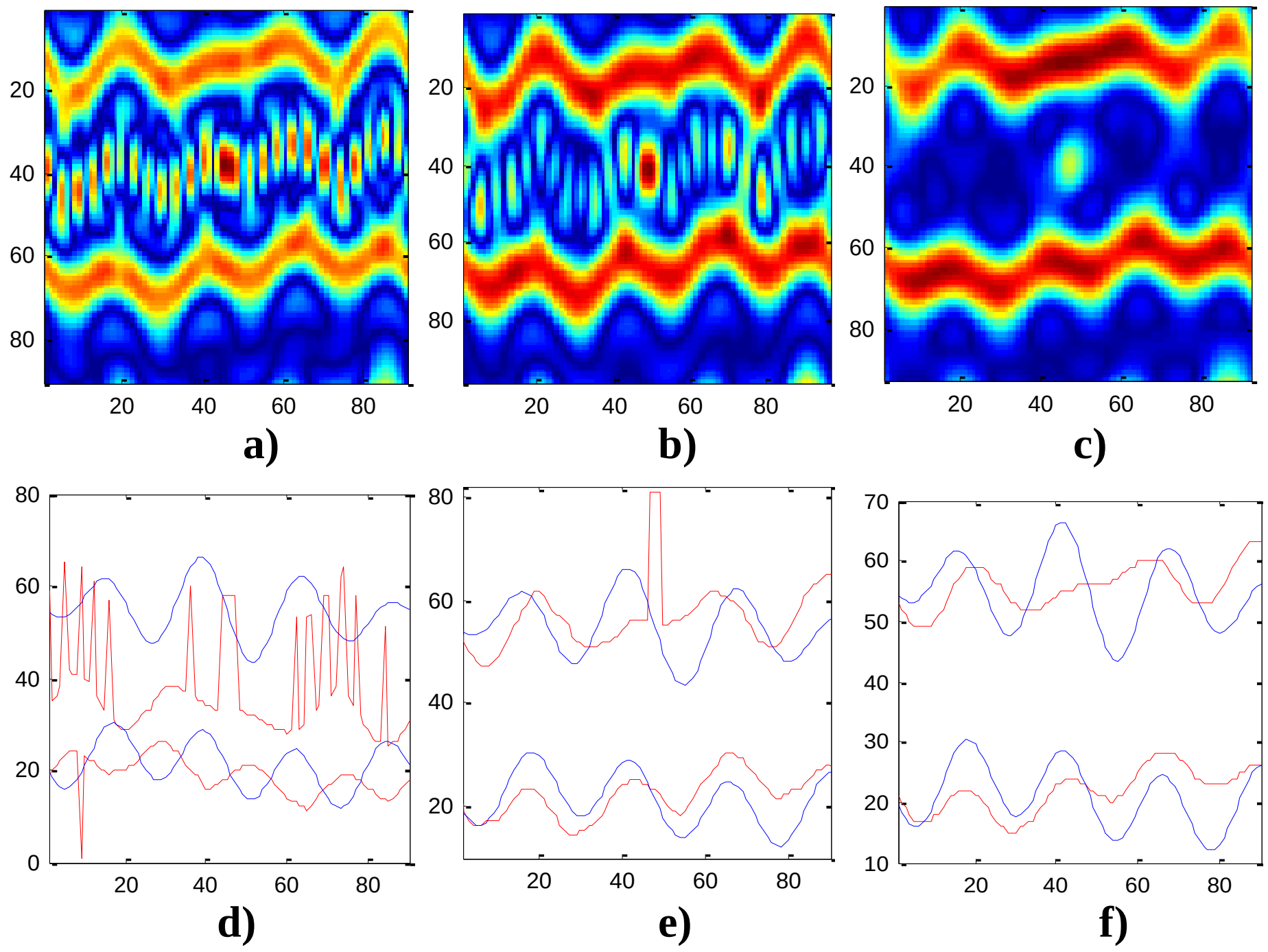

**Slika 4.2: a) Wignerova distribucija signala opisanog relacijom (4.16); b)-c) Distribucije iz Cohen-ove klase, dobijene korišćenjem Gausovog jezgra sa: b) δ =80 i c) δ =20; d) Trenutne frekvencije komponenti signala estimirane iz Wignerove distribucije; e) i f) Trenutne frekvencije komponenti signala estimirane iz Cohen-ove klase distribucija, sa Gausovim jezgrom i: e) δ=80 i f) δ =20. Tačne trenutne frekvencije označene su plavom linijom, dok su estimirane trenutne frekvencije označene crvenom linijom**

Trenutne frekvencije komponenti signala, estimirane iz spars distribucija (prikazanih na slikama 4.4a i 4.4b), veoma su bliske originalnim trenutnim frekvencijama komponenti signala (slika 4.4c i slika 4.4d).

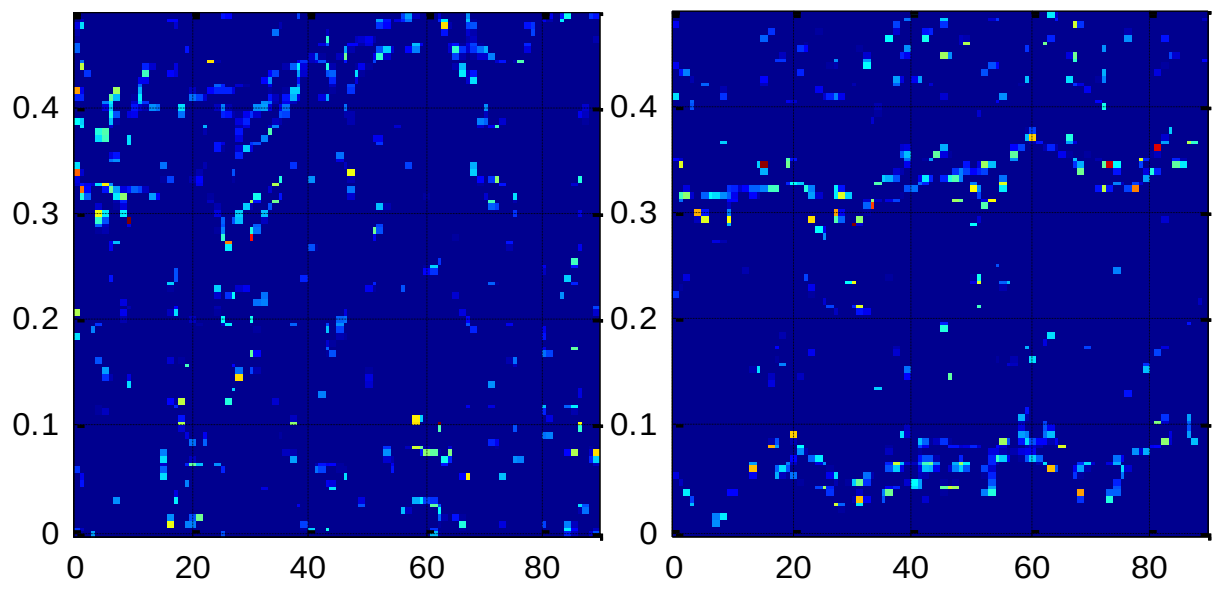

a) b)

**Slika 4.3: Spars vremensko-frekvencijske distribucije dobijene iz: a) Wignerove distribucije; b) Cohen-ove klase distribucija baziranih na Gausovom jezgru**

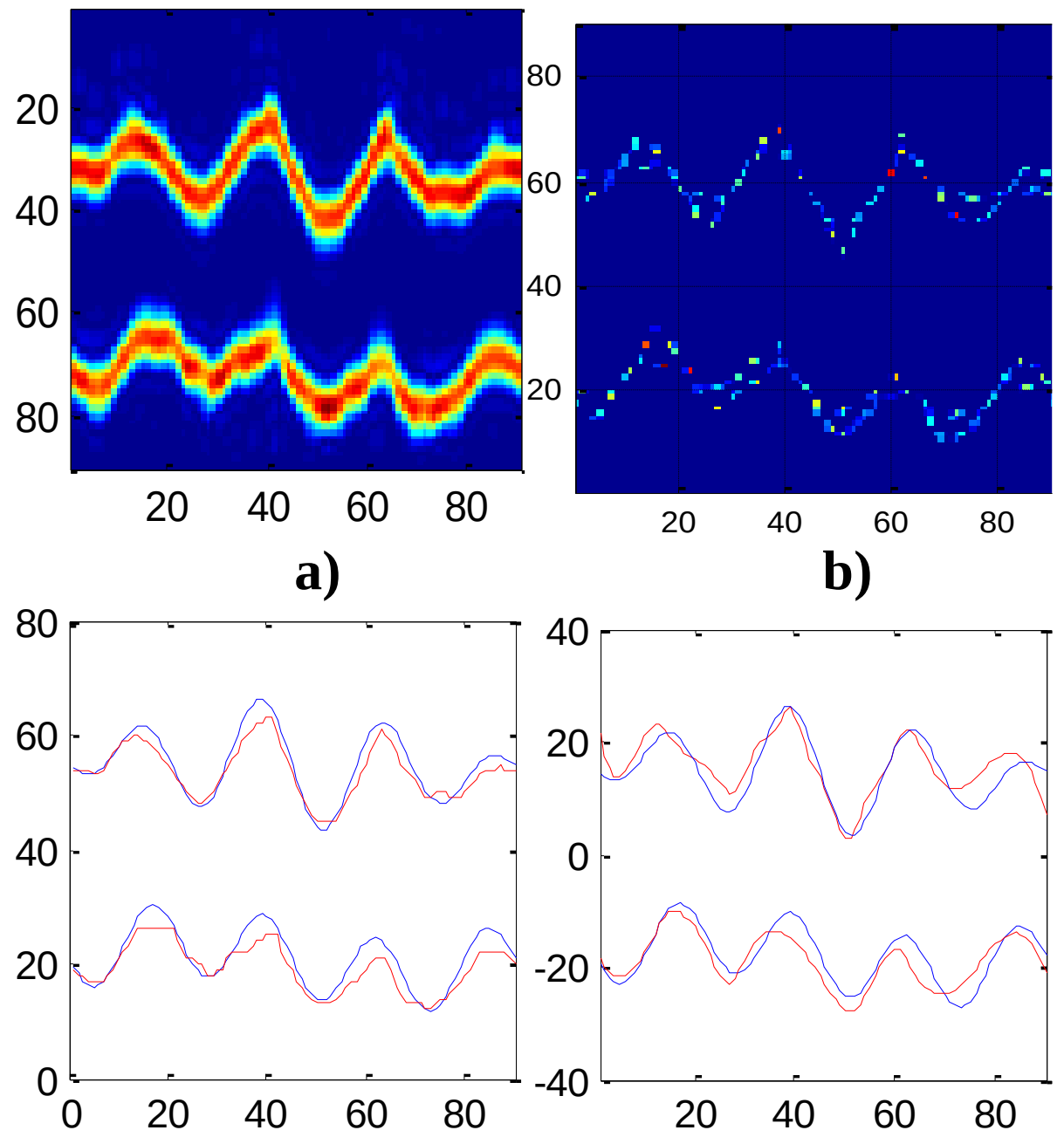

c) d)

**Slika 4.4: a) Distribucija sa kompleksnim argumentom vremena 4-tog reda; b) Spars vremensko-frekvencijska distribucija dobijena korišćenjem 60% slučajno odabranih mjerenja iz maske veličine 25×25; c) i d) Trenutne frekvencije komponenti signala estimirane iz distribucija prikazanih na slikama 4.4a i 4.4b, respektivno. Prave trenutne frekvencije komponenti prikazane su plavom linijom, a estimirane trenutne frekvencije prikazane su crvenom linijom**

Manje maske (veličine 15×15 i 20×20) takođe mogu da obezbijede tačno praćenje trenutnih frekvencija komponenti signala, ali zahtijevaju korišćenje većeg broja odbiraka koji će predstavljati CS mjerenja (npr. 70% ili 60% odbiraka iz maske su korišćeni u slučajevima maski veličine 15×15 i 20×20, respektivno).

**Tabela 4-1. Srednje kvadratne greške estimacije**

| *Distribucija* | *MSE* | | *Relativna MSE - RMSE (%)* | |
|---|---|---|---|---|
| | ***Komponenta 1*** | ***Komponenta 2*** | ***Komponenta 1*** | ***Komponenta 2*** |
| Wignerova distribucija | 3.3192×10$^3$ | 79.7761 | 67.81 | 5.79 |
| Distribucija iz Cohen-ove klase zasnovana na Gausovom jezgru $e^{-(\tau^2+\theta^2)/\delta^2}$ sa $\delta$=120 | 1.1563×10$^3$ | 3.1624×10$^3$ | 28.3 | 77.43 |
| Distribucija iz Cohen-ove klase zasnovana na Gausovom jezgru $e^{-(\tau^2+\theta^2)/\delta^2}$ sa $\delta$=80 | 1.3880×10$^3$ | 1.5905e×10$^3$ | 71.59 | 78.04 |
| Distribucija iz Cohen-ove klase zasnovana na Gausovom jezgru $e^{-(\tau^2+\theta^2)/\delta^2}$ sa $\delta$=20 | 1.6176e×10$^3$ | 1.5584e×10$^3$ | 61.02 | 54.82 |
| *Distribucija sa kompleksnim argumentom vremena, četvrtog reda* | | | | |
| Maska veličine 7×7, svi odbirci iz maske uzeti | neuspješno | neuspješno | / | / |
| Maska veličine 10×10, svi odbirci iz maske uzeti | neuspješno | neuspješno | / | / |
| Maska veličine 15×15 i 70% odbiraka uzeto | 81.3099 | 57.1962 | 11.08 | 0.02 |
| Maska veličine 20×20 i 60% odbiraka uzeto | 8.9314 | 20.5861 | 0.25 | 3.43 |
| *Maska veličine 25×25* | | | | |
| 40% odbiraka iz maske uzeto | 17.7834 | 32.7570 | 1.23 | 3.48 |
| 50% odbiraka iz maske uzeto | 10.7177 | 9.2834 | 1.09 | 0.36 |
| 60% odbiraka iz maske uzeto | 7.3854 | 8.1211 | 0.30 | 0.31 |

Posmatran je i uticaj veličine maske ($J\times J$) na broj mjerenja koji je neophodan za dobijanje spars vremensko-frekvencijske reprezentacije. Slika 4.5 pokazuje da su maske veličine 7×7 i 10×10 premale da obezbijede tačno praćenje trenutnih frekvencija komponenti signala, čak i ako su svi koeficijenti iz maske upotrijebljeni kao CS mjerenja.

### 4.2.2. Slučaj radarskih signala zahvaćenih impulsnim šumom

Posmatrajmo sada slučaj simuliranog radarskog signala. Signal je monokomponentni, sa brzim promjenama faze. Ovaj signal oponaša neuniformno rotaciono kretanje neke reflektujuće tačke u radarskim sistemima [118]. Trajanje signala je 96 sekundi. Signal $x(t)=e^{j(4cos(\pi t)+(2/3)cos(3\pi t)+(2/3)cos(5\pi t))}$ je sinusoidalno modulisan i odabran frekvencijom od 48 Hz.

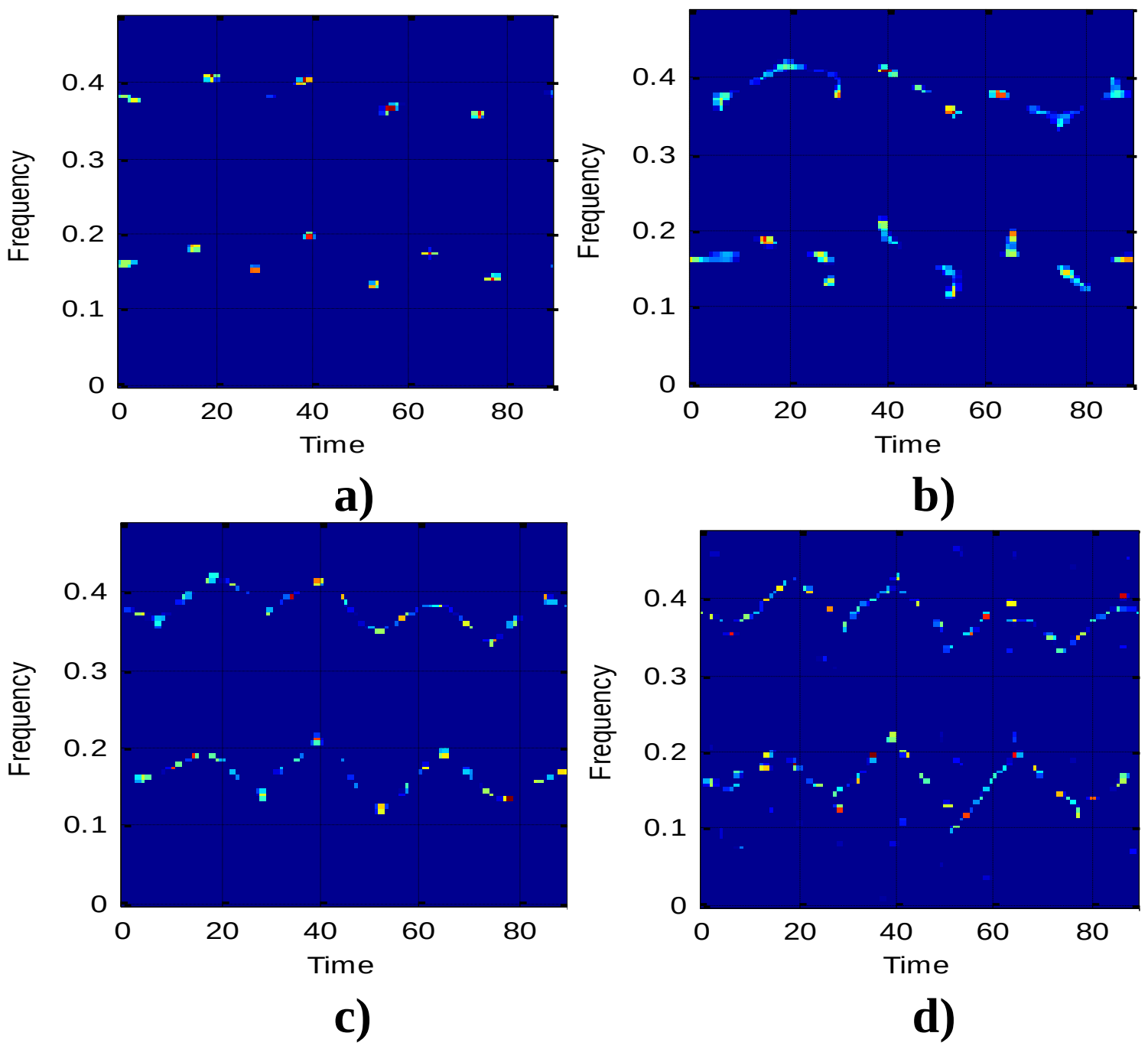

**Slika 4.5: Spars vremensko-frekvencijske distribucije dobijene korišćenjem maski različitih veličina i različitog procenta dostupnih odbiraka: a) maska veličine 7×7 sa 100% odbiraka uzetih iz maske; b) maska veličine 10×10 sa 100% odbiraka uzetih iz maske; c) maska veličine 15×15 sa 70% odbiraka uzetih iz maske; d) maska veličine 20×20 sa 60% odbiraka uzetih iz maske**

Pri analizi se polazi od pretpostavke da su mjerenja, uzeta iz ambiguity domena, zahvaćena impulsnim šumom. Vremensko-frekvencijska distribucija, dobijena korišćenjem 50% ambiguity mjerenja zahvaćenih šumom, unutar maske veličine 25×25, prikazana je na slici 4.6a. Impulsni šum kojim je zahvaćena ambiguity funkcija, onemogućava tačnu estimaciju trenutne frekvencije signala.

Vremensko-frekvencijska distribucija (slika 4.6a), dobijena korišćenjem ambiguity odbiraka zahvaćenih šumom, nema konciznu predstavu i ne obezbjeđuje precizno praćenje trenutne frekvencije. Zbog toga je na ambiguity funkciju zahvaćenu šumom primijenjena L-estimacija. Naime, 0.5% odbiraka koji imaju najveće vrijednosti u ambiguity domenu su odstranjeni korišćenjem robustne forme ambiguity funkcije. Pokazano je da je šum uspješno odstranjen pa je moguće je dobiti spars vremensko-frekvencijsku reprezentaciju (slika 4.6b). Estimirana trenutna frekvencija po obliku veoma dobro prati trenutnu frekvenciju (slika 4.6c).

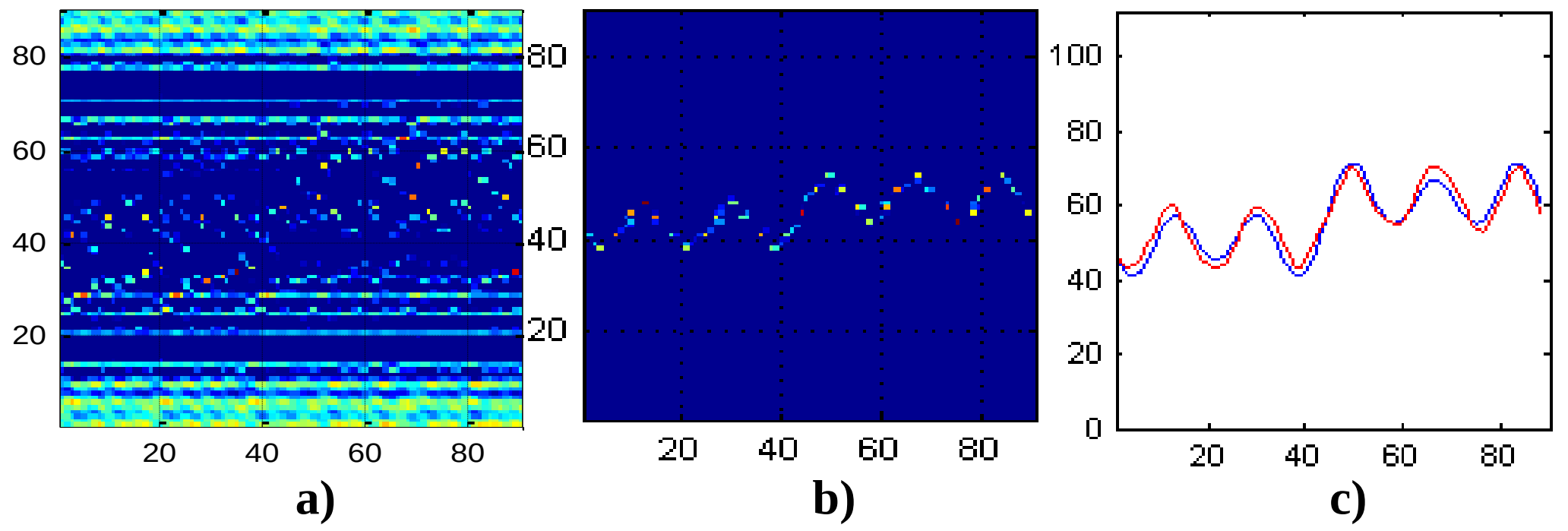

**Slika 4.6: a) Vremensko-frekvencijska distribucija dobijena iz mjerenja zahvaćenih šumom, uzetih iz ambiguity domena; b) Spars vremensko-frekvencijska distribucija dobijena korišćenjem mjerenja iz robustne ambiguity funkcije; c) Originalna (plavo) i estimirana (crveno) trenutna frekvencija signala, dobijena iz spars distribucije prikazane na slici 4.6b**

### 4.2.3. Realni radarski signal

Kao primjer realnog signala, snimljenog u prirodnom okruženju u prisustvu šuma, posmatran je dio radarskog signala koji odgovara pokretima ljudskog tijela. Pokrete koji se javljaju prilikom kretanja tijela, veoma je teško detektovati standardnim vremensko-frekvencijskim distribucijama, pa su, u cilju uspješne detekcije, korišćena visoko-rezoluciona rješenja, tj. vremensko-frekvencijske distribucije sa kompleksnim argumentom vremena.

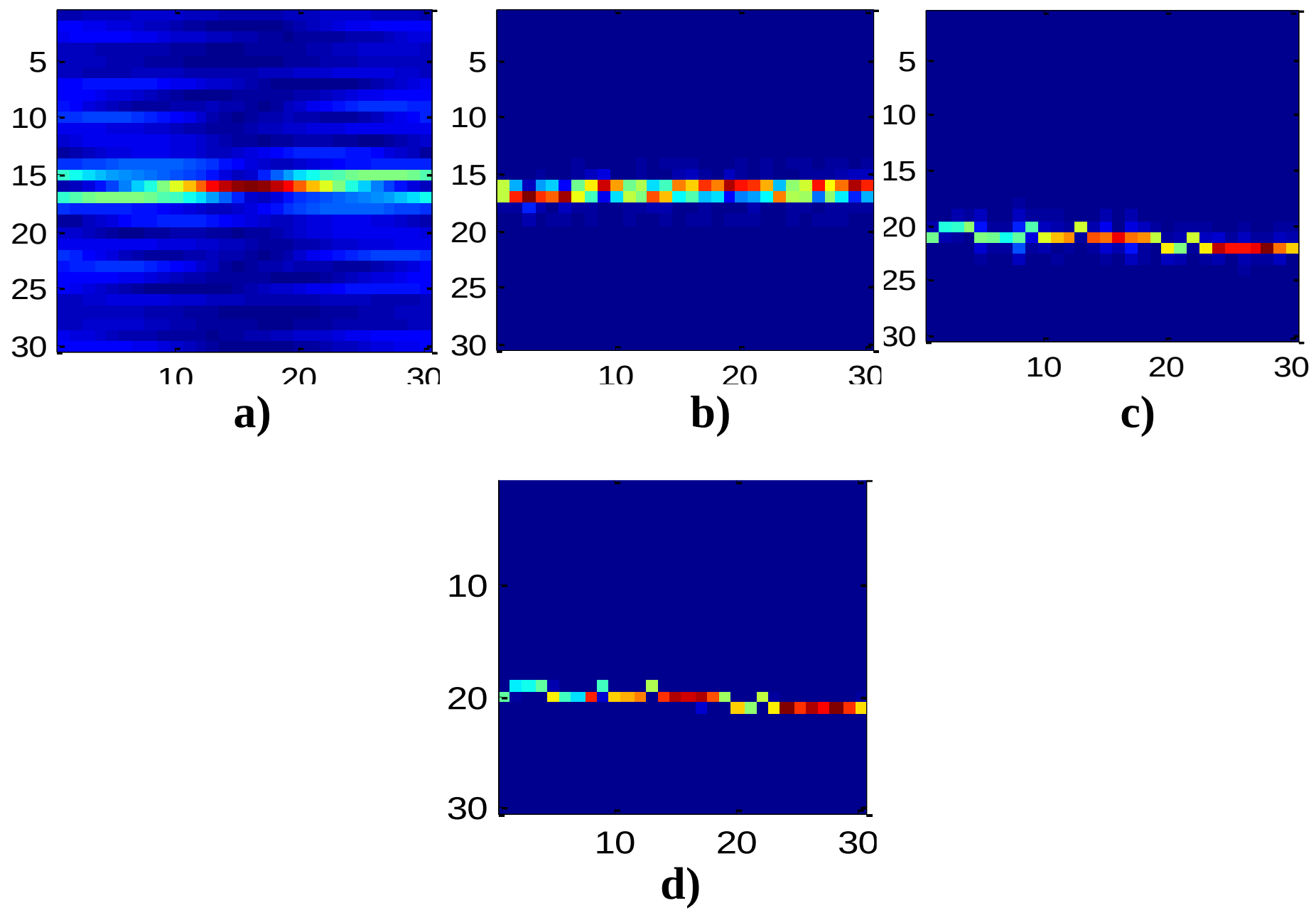

**Slika 4.7: a) Ambiguity funkcija realnog radarskog signala; b) Spektrogram; c) Distribucija sa kompleksnim argumentom vremena; d) Spars distribucija sa kompleksnim argumentom vremena**

Ove distribucije omogućavaju detekciju veoma malih varijacija trenutne frekvencije. Radar koji emituje signal radi na frekvenciji od 2.4 GHz, a nivo snage je 5 dBm. Signal je odabran frekvencijom od 1 KHz. Ambiguity funkcija signala prikazana je na slici 4.7a. Slika 4.7b prikazuje spektrogram signala, a distribucija sa kompleksnim argumentom vremena prikazana je na slici 4.7c. Pretpostavimo da je samo 11% odbiraka dostupno, od ukupnog broja odbiraka u vremensko-frekvencijskoj ravni. Na osnovu dostupnog broja odbiraka računata je spars vremensko-frekvencijska distribucija, i prikazana je na slici 4.7d. Trenutne frekvencije, estimirane iz originalne i iz spars vremensko-frekvencijske distribucije, prikazane su na slici 4.8.

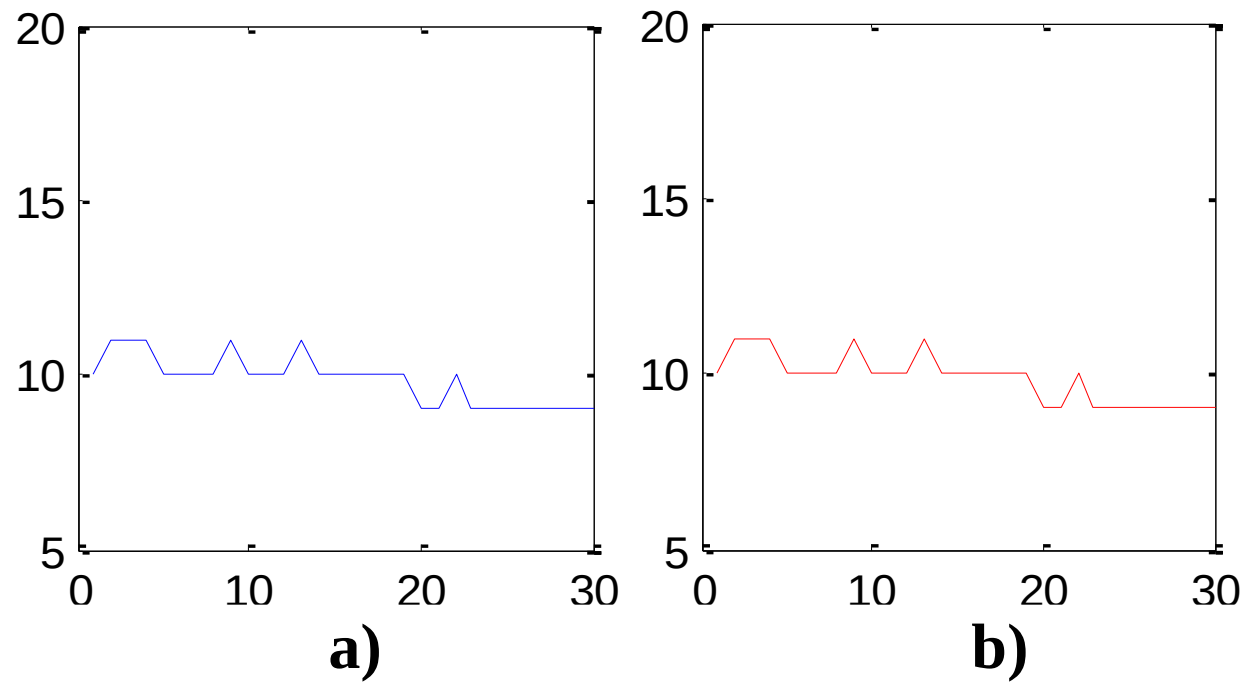

**Slika 4.8: Estimirane trenutne frekvencije: a) iz originalne distribucije sa kompleksnim argumentom vremena; b) iz spars distribucije sa kompleksnim argumentom vremena**

### 4.2.4. Primjena predloženog metoda u estimaciji trenutne frekvencije komponenti muzičkih signala

Predloženi metod je uspješno testiran i na muzičkim signalima. Ranije je napomenuto da su muzički signali multikomponentni i da se sastoje od većeg broja harmonika. Nelinearne promjene trenutne frekvencije komponenti ovih signala se ne mogu estimirati primjenom standardnih distribucija. Zato je u nastavku korišćena distribucija četvrtog reda sa kompleksnim argumentom vremena. Posmatran je signal flaute, koji se sastoji od ukupno 9 harmonika. U cilju pojednostavljenja procedure, estimacija trenutne frekvencije izvršena je samo za prva dva harmonika. Testirana je tačnost estimacije trenutne frekvencije komponenti signala ukoliko se koristi samo jedan dio koeficijenata iz ambiguity domena.

Znajući da auto komponente signala kod ambiguity funkcije nijesu u potpunosti koncentrisane oko koordinatnog početka, jedan dio mjerenja je uzet iz maske (10% koeficijenata), a veći dio je uzet iz okoline (50% koeficijenata). Maska je kvadratnog oblika, dimenzija 5×5, a dimenzija matrice vremensko-frekvencijske reprezentacije je 40×40.

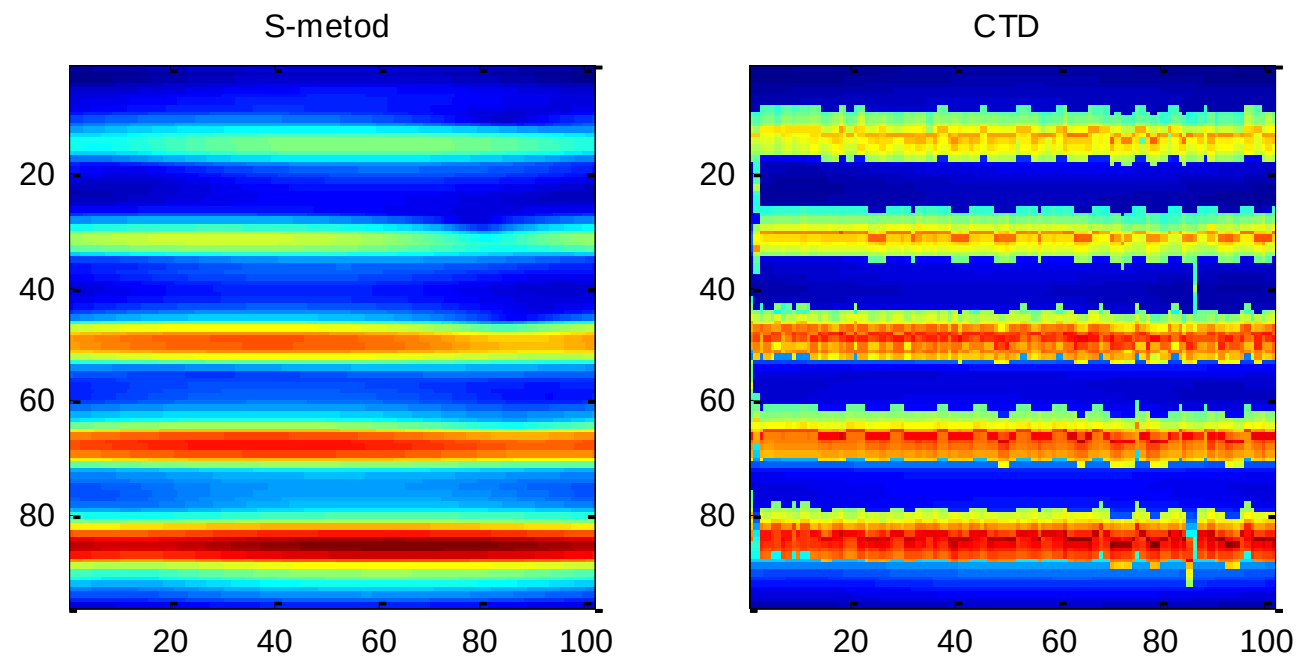

**a) b)**

**Slika 4.9: a) S-metod i b) distribucija sa kompleksnim argumentom vremena signala flaute (horizontalna osa - vrijeme, vertikalna osa - frekvencija)**

Poređenja radi, osim distribucije sa kompleksnim argumentom vremena računat je i S-metod signala. Za razliku od distribucije sa kompleksnim argumentom, S-metod ne može da prati brze promjene trenutnih frekvencija komponenti signala (Slika 4.9). To se može jasnije vidjeti sa slike 4.10, gdje su prikazane obje distribucije. Posmatrana su prva dva harmonika signala flaute. Ambiguity funkcija je prikazana na slici 4.11, dok je spars distribucija prikazana na slici 4.12a.

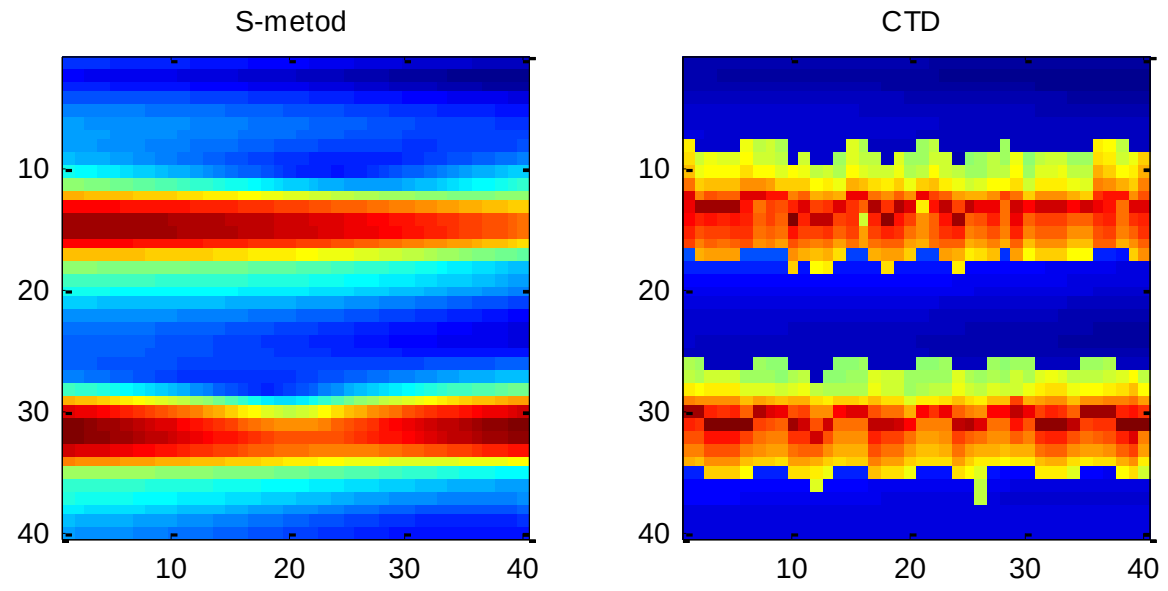

**a) b)**

**Slika 4.10: a) S-metod i b) distribucija sa kompleksnim argumentom vremena signala flaute, za prva dva harmonika (horizontalna osa - vrijeme, vertikalna osa - frekvencija)**

Estimirane trenutne frekvencije komponenti prikazane su na slici 4.12b. Plavom bojom prikazane su trenutne frekvencije komponenti estimirane iz standardne, a crvenom bojom trenutne frekvencije komponenti estimirane iz spars verzije distribucije. Kao što se vidi, obje linije se gotovo preklapaju, pa je srednja kvadratna greška estimacije veoma mala (za gornju komponentu MSE1=0, a za donju komponentu signala MSE2=0.0256).

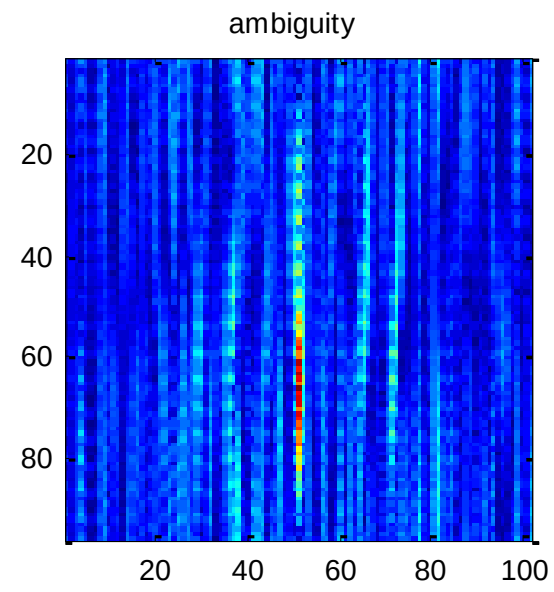

**Slika 4.11: Ambiguity funkcija signala flaute**

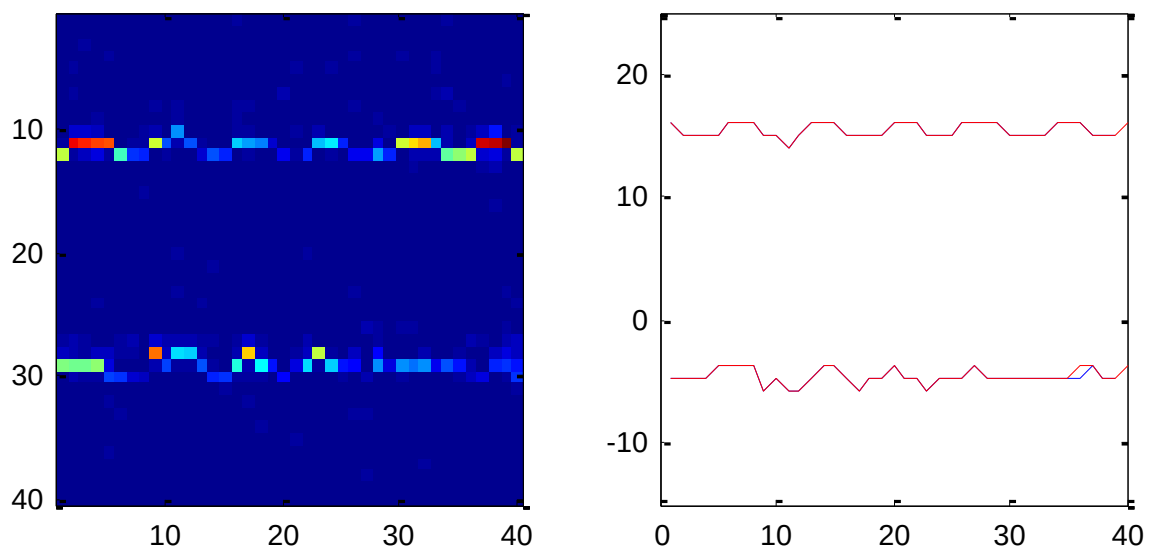

**a) b)**

**Slika 4.12: a) Spars vremensko-frekvencijska distribucija; b) Trenutne frekvencije prva dva harmonika signala flaute (trenutne frekvencije komponenti estimirane iz originalne distribucije sa kompleksnim argumentom vremena prikazane su plavom linijom, dok su trenutne frekvencije komponenti estimirane iz spars distribucije sa kompleksnim argumentom vremena prikazane crvenom linijom)**

# 5. Primjena Compressive Sensing tehnike na multikomponentne FHSS signale

U zavisnosti od zahtjeva sistema, standardi u bežičnim komunikacijama razlikuju se po brzini protoka, potrošnji energije, modulacionim tehnikama, distancama na kojima signali djeluju [121]. Niska potrošnja energije i brzi prenos poželjni su u svim standardima. Od velike važnosti je i siguran prenos podataka, zbog čega se kao tehnike modulacije signala koriste tehnike proširenog spektra (*spread spectrum modulations*), jer omogućavaju robusnost na šumove, zagušenja, intersimbolske interferencije (ISI), kao i na druge uticaje [121], [123], [129]. Tehnike proširenja spektra zasnivaju se na proširivanju spektra signala korišćenjem koda koji je jedinstven za svakog korisnika, i nekorelisan sa posmatranim signalom. Kao rezultat dobija se mnogo širi frekvencijski opseg od zahtijevanog.

U ovoj glavi biće dat prijedlog metoda za odvajanje i klasifikaciju signala koji pripadaju interferirajućim standardima: Bluetooth i IEEE 802.11b standardu. Ovi signali koriste isti opseg frekvencija – ISM opseg (*Industrial, Scientific and Medical*), zbog čega je česta pojava interferencija između signala koji pripadaju Bluetooth i IEEE 802.11b standardima [7], [121], [123]. Takođe, oba ova standarda koriste sinusoidalne signale, ali se komponente signala koje pripadaju različitim standardima razlikuju u pogledu fizičkih karakteristika. Bluetooth standard koristi FHSS, a IEEE 802.11b standard koristi DSSS modulaciju. Komponente FHSS signala imaju kratko trajanje, dok je trajanje komponenti DSSS signala duže. Fizičke razlike među komponentama signala biće iskorišćene prilikom njihove klasifikacije [123].

Procedura klasifikacije sastoji se od nekoliko koraka. U prvom koraku primijenjuje se odvajanje komponenti signala, korišćenjem dekompozicije na sopstvene vrijednosti. Kao rezultat, dobijaju se sopstveni vektori, koji odgovaraju komponentama signala.

U drugom koraku, odvojeni sopstveni vektori se pododabiraju, u skladu sa procedurom kompresivnog očitavanja, u cilju smanjenja broja odbiraka komponenti signala koje je potrebno prenijeti komunikacionim kanalom. Na strani prijema komponente signala treba da budu rekonstruisane iz malog broja poslatih odbiraka. To se postiže primjenom rekonstrukcionih algoritama. Bitan zahtjev za uspješnu rekonstrukciju sopstvenih vektora, jeste da oni imaju kompaktnu predstavu u nekom transformacionom domenu. U cilju obezbjeđivanja kompaktne predstave za oba tipa signala bira se pogodan transformacioni domen u kom odvojeni sopstveni vektor ima razrijeđenu, kompaktnu predstavu. Zbog različite prirode signala koji pripadaju različitim standardima, posmatrani su različiti transformacioni domeni, a kao najpogodniji pokazali su se DFT i HT domen.

Odabir pogodnog domena baziran je na računanju $\ell_1$-norme, koja se može iskoristiti za računanje mjere kompaktnosti signala. Nakon odabira domena u kom signali imaju kompaktnu predstavu, u posljednjem koraku procedure se, na prijemnoj strani, vrši klasifikacija komponenti signala. Klasifikacija se vrši posmatranjem karakteristika komponenti odvojenih u vremensko-frekvencijskom domenu.

Posmatrane su dvije tehnike proširenja spektra: *direct sequence spread spectrum* – DSSS i *frequency hopping spread spectrum* - FHSS, zbog njihove primjene u dva pomenuta bežična standarda. Prva modulaciona tehnika se koristi u IEEE 802.11b [7], [13], [129] i uzrokuje promjene faze signala nosioca, primjenom brzopromjenljive pseudoslučajne sekvence. Druga modulaciona tehnika se koristi u Bluetooth standardu i zasnovana je takođe na promjenama frekvencije po pseudoslučajnom ključu [7], [13], [129].

Procedura klasifikacije je zasnovana na EVD dekompoziciji i definisana je na sličan način kao procedura u Glavi 2, a ovdje će biti kratko sumirana.

Ako matricu kovarijanse označimo sa **R**, EVD se može opisati sljedećom relacijom:

$$\mathbf{R} = \mathbf{U}\Lambda\mathbf{U}^{\mathbf{T}} = \sum_{i=1}^{N+1} \lambda_i u_i(n) u^*_i(n), \tag{5.1}$$

gdje je sa **U** označena matrica sopstvenih vektora, $\Lambda$ je matrica sopstvenih vrijednosti sortiranih u opadajućem redosljedu, $\lambda_i$ su sopstvene vrijednosti a $u_i$ su sopstveni vektori. Matrica kovarijanse definisana je na osnovu autokorelacione matrice dobijene primjenom vremensko-frekvencijske analize.

Podsjetimo se na koji način je moguće definisati autokorelacionu matricu. Polazeći od inverzne Wigner-ove distribucije $i$-te komponente $x_i(n)$ multikomponentnog signala $x(n) = \sum_i x_i(n)$, definisane kao:

$$x_i(n+m)x_i^*(n-m) = \frac{1}{N+1} \sum_{k=-N/2}^{N/2} WD_i(n,k) e^{j\frac{2\pi}{N+1}2mk}, \tag{5.2}$$

definiše se autokorelaciona matrica $\mathbf{R}_i$ [15], [53], [123]:

$$R_i(n+m, n-m) = x_i\left(n+m\right)x_i^*\left(n-m\right) \ , \tag{5.3}$$

gdje je $x_i(n+m)$ vektor kolona a $x_i^*(n\text{-}m)$ vektor vrsta sa konjugovano kompleksnim vrijednostima. Za signal od $K$ komponenti važi:

$$R_K(n+m, n-m) = \sum_{i=1}^{K} x_i\left(n+m\right)x_i^*\left(n-m\right), \tag{5.4}$$

gdje se desna strana može definisati na osnovu inverzne Wigner-ove distribucije:

$$\sum_{i=1}^{K} x_i\left(n+m\right)x_i^*\left(n-m\right) = \frac{1}{N+1}\sum_{k=-N/2}^{N/2}\sum_{i=1}^{K} WD_i(n,k)e^{j\frac{2\pi}{N+1}2mk} \ . \tag{5.5}$$

Parametar $N$ označava dužinu signala, WD označava Wigner-ovu distribuciju $WD(n,k) = 2\sum_{m=-\infty}^{\infty} x(n+m)x^*(n-m)e^{-j2(2\pi/M)km}$, parameter $m$ je vremenski pomjeraj a $M$ je broj odbiraka signala u vremenskom ili frekvencijskom domenu koji bi bio potreban za računanje Wigner-ove distribucije [130].

Ukoliko nema preklapanja komponenti signala u vremensko-frekvencijskoj ravni, suma Wigner-ovih distribucija komponenti signala jednaka je S-metodu multikomponentnog signala. Stoga se prethodna relacija može napisati u obliku [15], [53], [123]:

$$\sum_{i=1}^{K} x_i(n+m)x_i^*(n-m) = \frac{1}{N+1}\sum_{k=-N/2}^{N/2} SM(n,k)e^{j\frac{4\pi}{N+1}mk} \ , \tag{5.6}$$

gdje SM označava S-metod, i važi:

$$R_K(n+m, n-m) = \frac{1}{N+1}\sum_{k=-N/2}^{N/2} SM(n,k)e^{j\frac{4\pi}{N+1}mk} \ , \tag{5.7}$$

gdje je $\mathbf{R_K}$ kvadratna autokorelaciona matrica, a EVD kvadratne matrice $\mathbf{R_K}$ definisana je na sljedeći način:

$$\mathbf{R_K} = \sum_{i=1}^{N+1} \lambda_i u_i(n) {u^*}_i(n) \, . \tag{5.8}$$

U cilju smanjenja broja odbiraka signala koji se šalju komunikacionim kanalom, sopstveni vektori su pododabrani na slučajan način i na prijemnoj strani rekonstruisani korišćenjem matematičkih algoritama za rekonstrukciju signala.

Ako sopstveni vektor $\mathbf{u}$, dužine $N$, ima kompaktnu predstavu u transformacionom domenu $\mathbf{\Psi}$, onda se može predstaviti korišćenjem baznih vektora na sljedeći način [123]:

$$\mathbf{u} = \sum_{i=1}^{N} \mathbf{U}_i \psi_i^{-1} = \mathbf{\Psi}^{-1}\mathbf{U}\,, \tag{5.9}$$

gdje je sa **U** označen transformacioni domen (za koji važi da je samo $S{<<}N$ koeficijenata nenulto), a $\mathbf{\Psi}^{-1}$ je inverzna transformaciona matrica. Ako se matrica koja modeluje selekciju odbiraka sopstvenih vektora na slučajan način označi sa **Φ**, vektor dostupnih mjerenja **y** (dužine $M_a$, $M_a<N$) može se definisati na sljedeći način:

$$\mathbf{y} = \mathbf{\Phi}\mathbf{\Psi}^{-1}\mathbf{U}. \tag{5.10}$$

Kao što je ranije napomenuto, sistem jednačina (5.10) je neodređen, jer je broj dostupnih mjerenja $M_a$ manji od ukupnog broja odbiraka sopstvenog vektora $N$, pa se u cilju dobijanja zadovoljavajućeg rješenja sistema jednačina, primjenjuju rekonstrukcioni algoritmi. U nastavku biće riječi o izboru transformacionog domena u kom komponente signala imaju kompaktnu predstavu.

## 5.1. Odabir domena

Imajući u vidu da DSSS modulisane komponente signala po svojoj prirodi imaju sinusoidalan oblik, DFT domen je pogodan za njihovo kompaktno predstavljanje. Sa druge strane, komponente FHSS signala su takođe sinusoidalne prirode ali je njihovo trajanje mnogo kraće u poređenju sa trajanjem DSSS komponenti, pa spektar takvog signala nema kompaktnu predstavu.

Posmatrajući oblik FHSS modulisanih komponenti i poređenjem sa oblikom Hermitskih funkcija, došlo se do zaključka da je Hermitski domen pogodniji za primjenu u FHSS signalima. Stoga je testirana mogućnost korišćenja Hermitske transformacije kao domena u kom FHSS komponente imaju kompaktnu predstavu.

Odabir domena je automatizovan i bazira se na mjerenju koncentracije u DFT i HT domenu, primjenom $\ell_1$-norme u skladu sa sljedećom relacijom [123]:

$$\left\|\mathbf{U}\right\|_{\ell_1} = \sum_{n=0}^{N-1} \left|U(n)\right|, \quad \mathbf{U} = \mathbf{\Psi}\mathbf{u}, \tag{5.11}$$

gdje **Ψ** može biti ili DFT ili HT matrica. Diskretna Hermitska funkcija $p$-tog reda se definiše kao $\psi_p(t_m) = \dfrac{e^{-t_m^2} H_p\left(t_m/\sigma\right)}{\sqrt{\sigma 2^p p! \sqrt{\pi}}}$ , a u matričnom obliku direktna i inverzna HT su definisane sljedećim relacijama, respektivno:

$$\mathbf{\Psi}=\frac{1}{M}\begin{bmatrix}\frac{\psi_0(t_1)}{\left(\psi_{M-1}(t_1)\right)^2} & \cdots & \frac{\psi_0(t_M)}{\left(\psi_{M-1}(t_M)\right)^2}\\ \frac{\psi_1(t_1)}{\left(\psi_{M-1}(t_1)\right)^2} & \cdots & \frac{\psi_1(tM)}{\left(\psi_{M-1}(t_M)\right)^2}\\ \cdots & \cdots & \cdots\\ \frac{\psi_{M-1}(t_1)}{\left(\psi_{M-1}(t_1)\right)^2} & \cdots & \frac{\psi_{M-1}(t_M)}{\left(\psi_{M-1}(t_M)\right)^2}\end{bmatrix},\quad \mathbf{\Psi}^{-1}=\begin{bmatrix}\psi_0(t_1) & \cdots & \psi_{M-1}(t_1)\\ \psi_0(t_2) & \cdots & \psi_{M-1}(t_2)\\ \cdots & \cdots & \cdots\\ \psi_0(t_M) & \cdots & \psi_{M-1}(t_M)\end{bmatrix}. \tag{5.12}$$

Parametar $M$ odgovara broju Hermitskih funkcija koje se koriste za predstavljanje signala, tj. $M$ označava red Hermitskog polinoma. Parametar $H_p$ označava Hermitski polinom čije su nule označene sa $t_m$. Faktor σ je skalirajući faktor i koristi se za kontrolu širine Hermitskih funkcija [39], [123]. To znači da se podešavanjem ovog parametra svaka Hermitska funkcija može prilagoditi posmatranom signalu, čime se obezbijeđuje dodatna sparsifikacija u Hermitskom domenu.

Treba napomenuti da je prilikom računanja Hermitske transformacije potrebno čak i preodabiranje sopstvenih vektora. Preodabiranje se vrši u neuniformnim tačkama koje su proporcionalne korjenima Hermitskog polinoma $N$-tog reda. U cilju preodabiranja koristi se *sinc* interpolacija kako bi se dobile vrijednosti u zahtijevanim, neuniformnim tačkama [39], [123]:

$$u(\lambda t_m)\approx\sum_{n=-N/2}^{N/2}u(n\Delta t)\frac{\sin\left(\pi(\lambda t_m-n\Delta t)/\Delta t\right)}{\pi(\lambda t_m-n\Delta t)/\Delta t}, \tag{5.13}$$

gdje je $m = 1, \dots, N$, $\Delta t$ je period odabiranja i skalira vremensku osu umjesto faktora σ. U cilju poboljšanja koncentracije i obezbjeđivanja što kompaktnije predstave signala u HT domenu, primijenjena je procedura za optimizaciju parametra λ. Mjera koncentracije bazirana na $\ell_1$-normi korišćena je za nalaženje optimalne vrijednosti parametra λ:

$$\lambda_{opt}=\min_{\lambda}\left\|U(\lambda t_m)\right\|_{\ell_1}=\left\|\mathrm{HT}\left\{\sum_{n=-N/2}^{N/2}u(n\Delta t)\frac{\sin\left(\pi(\lambda t_m-n\Delta t)/\Delta t\right)}{\pi(\lambda t_m-n\Delta t)/\Delta t}\right\}\right\|_1. \tag{5.14}$$

## 5.2. Procedura za klasifikaciju tipa signala u bežičnim komunikacijama i CS rekonstrukciju

Procedura za odvajanje i klasifikaciju komponenti Bluetooth i IEEE 802.11b standarda, data je u ovom poglavlju. Sastoji se od šest koraka, a bazirana je na prethodno opisanoj proceduri dekompozicije na sopstvene vrijednosti i vektore, i principima kompresivnog očitavanja [123].

**A.** Prvi korak u proceduri predstavlja računanje S-metoda polaznog signala. S-metod je odabran kao reprezentacija koja obezbjeđuje dobru koncentraciju auto komponenti signala i redukuje ili potpuno eliminiše pojavu kros-komponenti u vremensko-frekvencijskoj ravni multikomponentnih signala. Podsjetimo se, S-metod je računat polazeći od kratkotrajne Fourier-ove transformacije STFT:

$$STFT(n,k)=\sum\nolimits_{m=-N/2}^{N/2-1}w(m)x(n+m)e^{-j2\pi mk/N}\ . \tag{5.15}$$

Oznaka $x(n)$ je za ulazni signal, dok $w(n)$ označava prozor. S-metod se računa u skladu sa sljedećom relacijom:

$$SM(n,k)=\sum\nolimits_{i=-L}^{L}STFT(n,k+i)STFT^{*}(n,k-i)\,. \tag{5.16}$$

Parametar $L$ za ovaj tip signala može se birati tako da uzima vrijednosti između 3 i 6.

**B.** Sljedeći korak je računanje autokorelacione matrice $\mathbf{R}_K$. Autokorelaciona matrica se računa u skladu sa relacijom (5.7).

**C.** Izbor sparsifikacionog domena se vrši mjerenjem koncentracije u DFT i u HT domenu za svaki sopstveni vektor, na osnovu $\ell_1$-norme, u skladu sa relacijom (5.11).

**D.** Sljedeći korak podrazumijeva pododabiranje sopstvenih vektora, biranjem određenog procenta koeficijenata po slučajnom rasporedu.

**E.** Nakon prenosa samo onih koeficijenata sopstvenih vektora koji su odabrani po slučajnom rasporedu, vektori se rekonstruišu na prijemnoj strani primjenom optimizacionih algoritama. Ovdje je upotrijebljen algoritam baziran na minimizaciji $\ell_1$-norme:

$$\mathbf{U}=\min\left\|\mathbf{U}\right\|_{\ell_1}\ \text{ uz uslov }\ \mathbf{y}=\mathbf{\Phi}\mathbf{\Psi}^{-1}\mathbf{U}\ . \tag{5.17}$$

Rješenje $\ell_1$ minimizacionog problema bazira se na tzv. *basis pursuit primal-dual* algoritmu [15], a algoritam je izvršen primjenom softverskog paketa L1-magic.

**F.** U poslednjem koraku procedure se računa S-metod za svaki rekonstruisani sopstveni vektor. Zatim se osobine sopstvenih vektora estimiraju na bazi vremensko-frekvencijske reprezentacije. Na osnovu tih osobina vrši se klasifikacija posmatrane komponente signala.

Čitava procedura prikazana je na dijagramu na slici 5.1.

**Ulazi:**

- Signal $\boldsymbol{x}$
- Funkcija prozora za računanje STFT-a: $\boldsymbol{w}$
- Širina prozora S-metoda: $L$

**A.** Računanje STFT-a i S-metoda od signala $\boldsymbol{x}$:

$$STFT(n,k)=\sum_{m=-N/2}^{N/2-1} w(m)x(n+m)e^{-j2\pi mk/N}$$

$$SM(n,k)=\sum_{i=-L}^{L} STFT(n,k+i)STFT^{*}(n,k-i)$$

**B.** Računanje autokorelacione matrice i EVD-a:

$$\mathbf{R}_{\mathbf{K}}=inv(SM),\quad \text{EVD}\left\{\mathbf{R}_{\mathbf{K}}\right\}\rightarrow\sum_{i=1}^{N+1}\lambda_i u_i(n)u^{*}_{i}(n)$$

**C.** Izbor sparsifikacionog domena za svaki odvojeni sopstveni vektor $\mathbf{u}_i$.
Mjerenje $\ell_1$-koncentracija u HT i u DFT domenu:

$$\|\mathbf{U}\|_{\ell_1}^{H}=\sum_{n=0}^{N-1}|U_H(n)|,\quad \mathbf{U}_H=\mathbf{\Psi u} \qquad (\mathbf{\Psi}\text{ označava HT matricu})$$

$$\|\mathbf{U}\|_{\ell_1}^{F}=\sum_{n=0}^{N-1}|U_F(n)|,\quad \mathbf{U}_F=\mathbf{\Psi u} \qquad (\mathbf{\Psi}\text{ označava DFT matricu})$$

- Ako važi $\|\mathbf{U}\|_{\ell_1}^{H}>\|\mathbf{U}\|_{\ell_1}^{F}\Rightarrow\ \mathbf{\Psi}\rightarrow$ DFT matrica. Prelazi se na korak **D**.
- Ako važi $\|\mathbf{U}\|_{\ell_1}^{H}<\|\mathbf{U}\|_{\ell_1}^{F}\Rightarrow\ \mathbf{\Psi}\rightarrow$ HT matrica. HT optimizacija:

- *sinc* interpolacija sopstvenih vektora:

$$u(\lambda t_m)\approx\sum_{n=-N/2}^{N/2}u(n\Delta t)\frac{\sin\left(\pi(\lambda t_m-n\Delta t)/\Delta t\right)}{\pi(\lambda t_m-n\Delta t)/\Delta t}$$

- traženje optimalnog faktora skaliranja vremenske ose:

$$\lambda_{opt}=\min_{\lambda}\|U(\lambda t_m)\|_{\ell_1}=$$

$$=\left\|\text{HT}\left\{\sum_{n=-N/2}^{N/2}u(n\Delta t)\frac{\sin(\pi(\lambda t_m-n\Delta t)/\Delta t)}{\pi(\lambda t_m-n\Delta t)/\Delta t}\right\}\right\|_{\ell_1}$$

**D.** Slučajno pododabiranje sopstvenih vektora.

$$\mathbf{y}=\mathbf{\Phi\Psi}^{-1}\mathbf{U} \qquad (\text{domen rijetkosti } \mathbf{\Psi} \text{ odabran u koraku } \mathbf{C}.)$$

**E.** Rekonstrukcija pododabranih sopstvenih vektora:

$$\mathbf{U}=\min\|\mathbf{U}\|_{\ell_1}=\sum_{i=1}^{N}|\mathrm{U}_i| \ \text{ u skladu sa }\ \mathbf{y}=\mathbf{\Phi\Psi}^{-1}\mathbf{U}$$

**F.** Računanje S-metoda rekonstruisanih sopstvenih vektora i njihova klasifikacija

**Izlazi:**

- FHSS signal: $x_{FHSS}(t)=\sum u_{FHSS}(t)$
- DSSS signal: $x_{DSSS}(t)=\sum u_{DSSS}(t)$

**Slika 5.1: Algoritam za odvajanje i klasifikaciju komponenti signala, baziran na EVD-u i CS-u**

Predložena procedura ne zahtijeva da frekvencije komponenti signala budu unaprijed poznate. Ona omogućava slijepo razdvajanje komponenti različitih frekvencija, dok god se te komponente ne preklapaju i ne sijeku u vremensko-frekvencijskoj ravni. Važno je napomenuti činjenicu da FHSS i DSSS modulacije koriste različite frekvencijske kanale u ISM frekvencijskom opsegu. Komponente FHSS signala se pojavljuju unutar 79 kanala, dok DSSS modulacija koristi samo 14 kanala ISM frekvencijskog opsega. Imajući ovo u vidu može se zaključiti da se preklapanja i presijecanja komponenti ne dešavaju među komponentama koje pripadaju različitim tipovima signala.

Kao i kod dekompozicije muzičkih signala, i kod ovih signala se odvajaju prvo komponente najvećih energija, a zatim slijede komponente nižih energija. Najveća sopstvena vrijednost odgovaraće prvom sopstvenom vektoru, tj. po energiji najjačoj komponenti signala, druga sopstvena vrijednost odgovaraće drugoj po energiji komponenti, itd.

## 5.3. Eksperimentalni rezultati

Predložena procedura testirana je na sintetičkom signalu. Signal se sastoji od dva dijela koji pripadaju Bluetooth i IEEE 802.11b standardu. Dužina signala je 256 odbiraka a sastoji se od 7 komponenti: 4 komponente pripadaju FHSS signalu, dok 3 komponente pripadaju DSSS signalu.

S-metod signala prikazan je na slici 5.2. U ovom primjeru komponente koje pripadaju Bluetooth standardu imaju veću energiju u poređenju sa komponentama IEEE 802.11b signala.

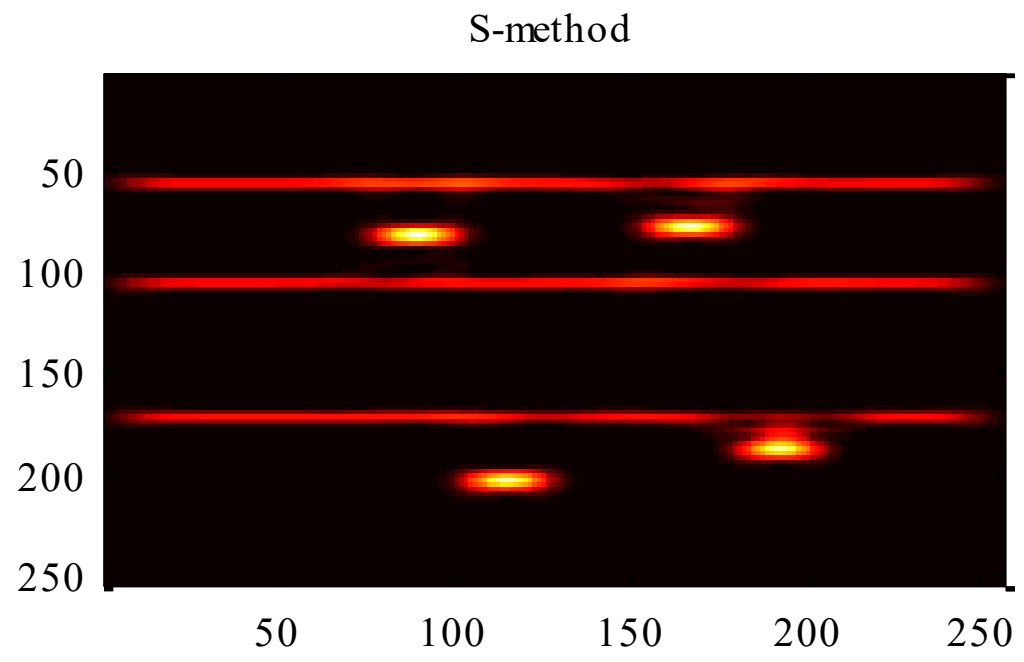

**Slika 5.2: S-metod polaznog signala; horizontalna osa predstavlja vrijeme, vertikalna osa predstavlja frekvenciju**

Zbog toga će prva 4 odvojena sopstvena vektora (koja generalno odgovaraju komponentama najvećih energija), pripadati Bluetooth signalu, dok će 5., 6. i 7. sopstveni vektor odgovarati komponentama IEEE 802.11b signala. Nakon odvajanja, prelazi se na traženje domena u kom će posmatrani sopstveni vektor imati kompaktnu predstavu. Mjere koncentracije za svih 7 sopstvenih vektora date su u tabeli 5-1.

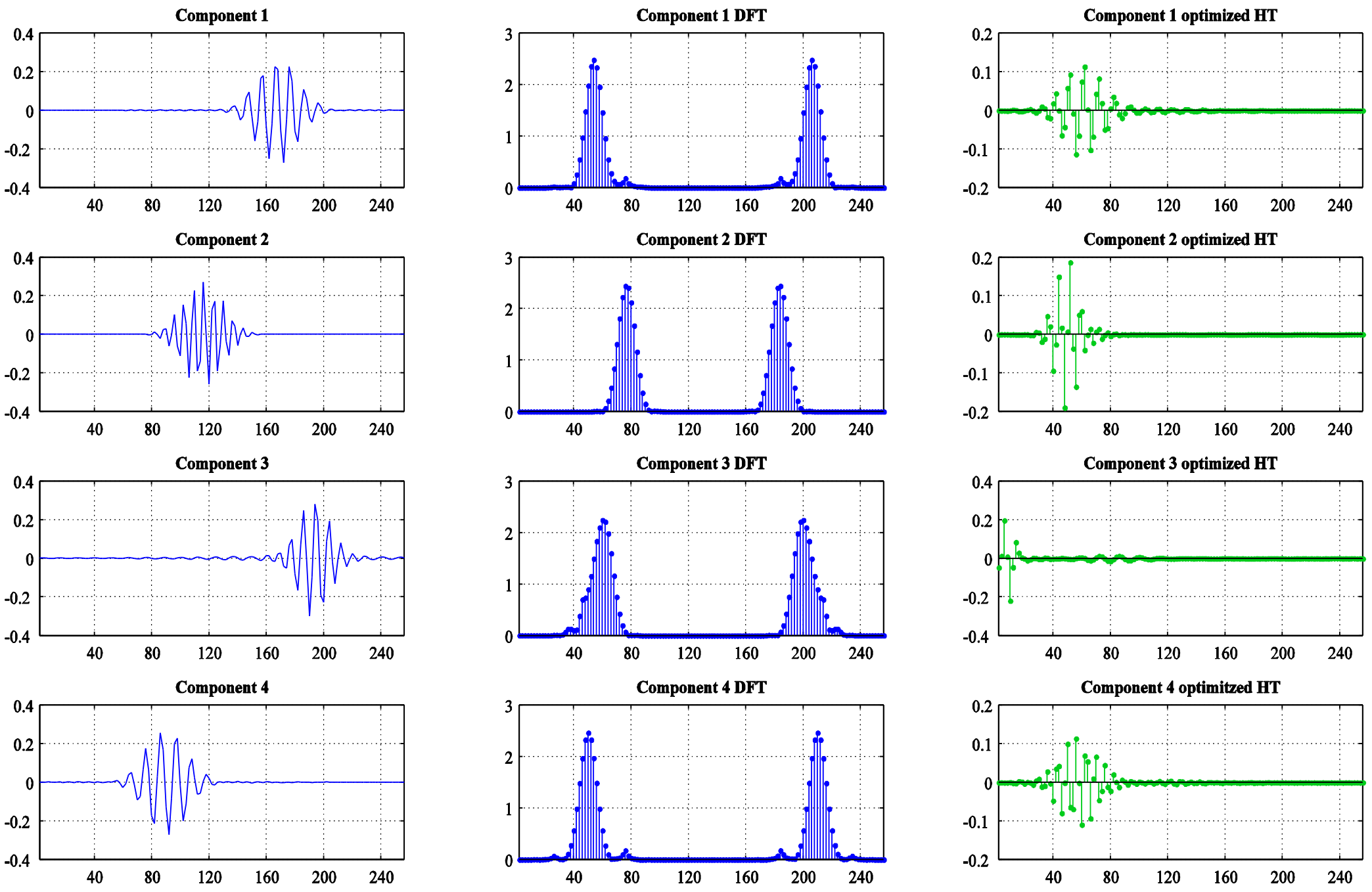

a)

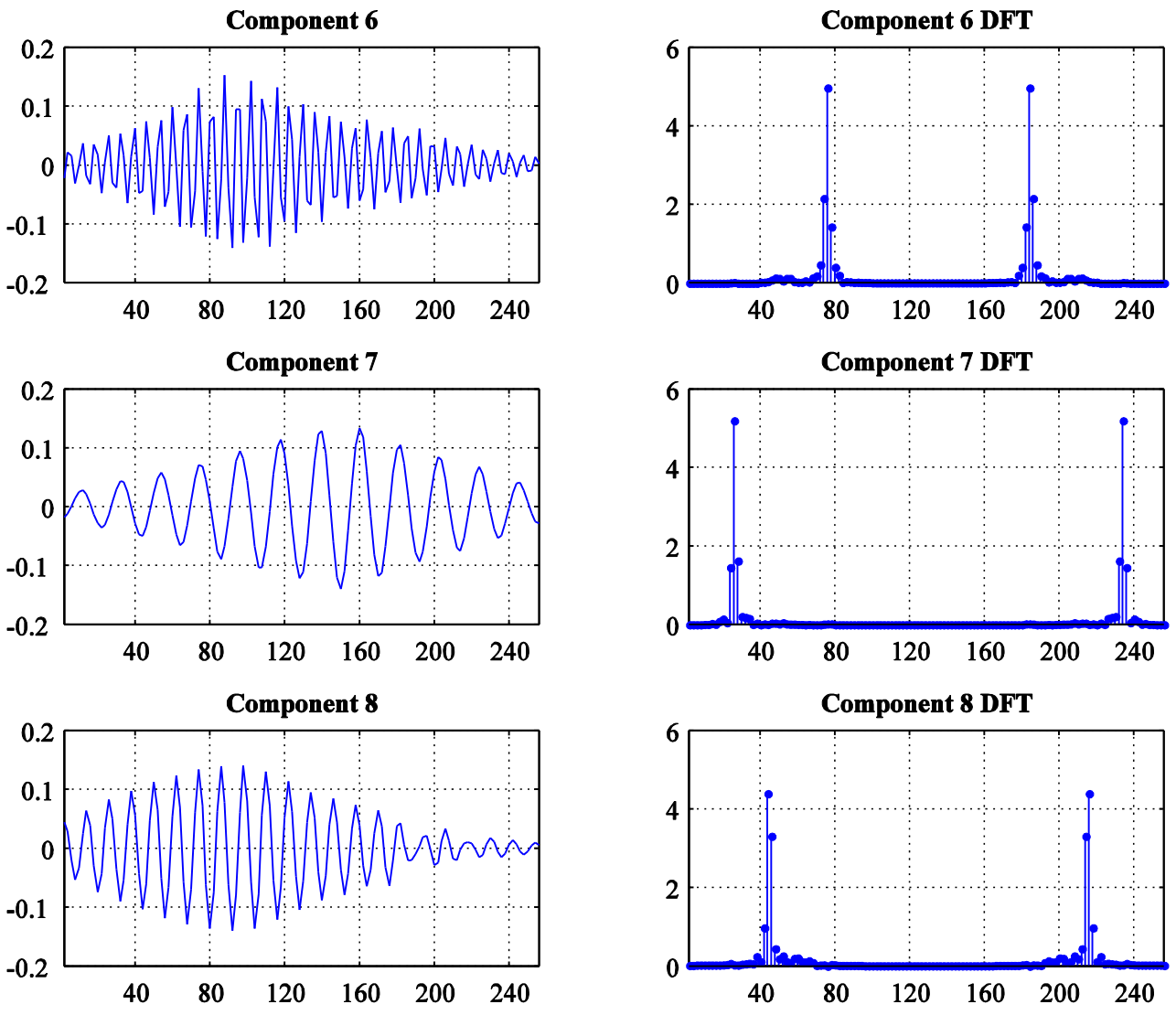

b)

**Slika 5.3: a) Prva kolona: Prva 4 sopstvena vektora (horizontalna osa je vrijeme, vertikalna osa je amplituda); Druga kolona: Odgovarajuće DFT (horizontalna osa je frekvencija, vertikalna osa je amplituda); Treća kolona: Optimizovane HT za prva 4 sopstvena vektora (horizontalna osa označava red Hermitskog koeficijenta, vertikalna osa označava amplitudu); b) Prva kolona: 5-ti, 6-ti i 7-mi sopstveni vektor (horizontalna osa je vrijeme, vertikalna osa je amplituda); Druga kolona: odgovarajuće DFT (horizontalna osa je frekvencija, vertikalna osa je amplituda)**

**Tabela 5-1: Mjere koncentracije sopstvenih vektora, računate u HT i DFT domenima**

| *Redni broj sopstvenog vektora* | *Mjera koncentracije-Hermitski domen* | *Mjera koncentracije-Fourier-ov domen* | *Odabrani domen u kom signal ima kompaktnu predstavu* |
|---|---|---|---|
| 1 | **27.9601** | 37.3387 | ***HT*** |
| 2 | **28.4100** | 36.2452 | ***HT*** |
| 3 | **32.7542** | 41.4611 | ***HT*** |
| 4 | **27.8796** | 37.4397 | ***HT*** |
| 5 | 68.2208 | **22.9848** | ***DFT*** |
| 6 | 67.7154 | **19.7623** | ***DFT*** |
| 7 | 94.0820 | **25.4799** | ***DFT*** |

Za prva 4 sopstvena vektora, mjera koncentracije je bila manja u HT domenu, pa je taj domen odabran kao domen u kom prva 4 sopstvena vektora imaju kompaktnu predstavu. Takođe, odrađena je i dodatna sparsifikacija za vektore koji imaju kompaktnu predstavu u HT domenu. Optimizovane HT prikazane su na slici 5.3. Na istoj slici prikazane su i DFT i HT, u cilju da se, pored vrijednosti mjere koncentracije, i ilustrativno prikaže da HT obezbjeđuje kompaktnije predstavljanje komponente signala, za slučaj FHSS signala. Poslednja 3 sopstvena vektora prikazana su u DFT domenu, kao domenu koji, za ovaj tip vektora, obezbjeđuje kompaktno predstavljanje, slika 5.3b.

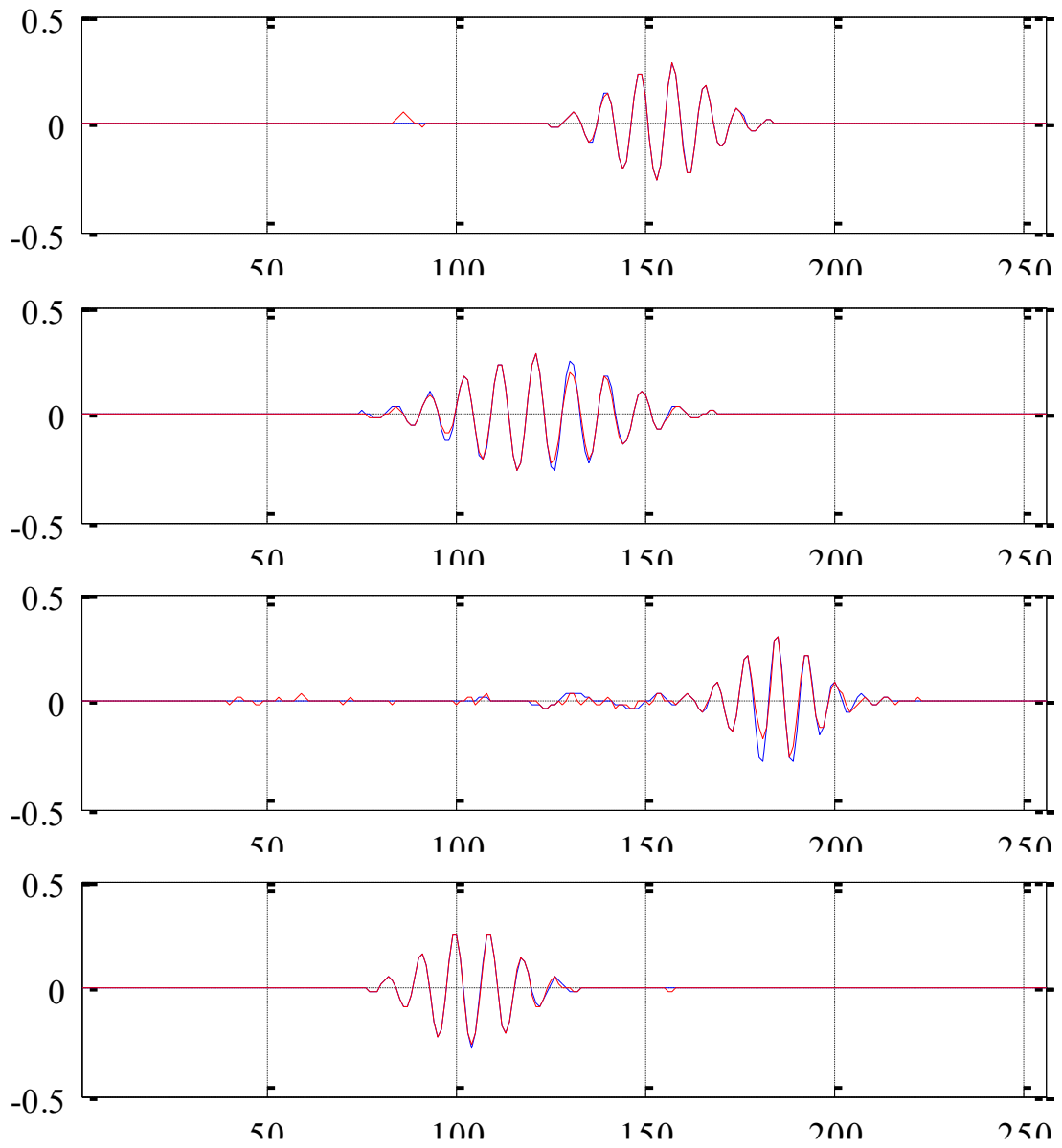

**Slika 5.4: Odvojene komponente Bluetooth signala; originalna komponenta označena je plavom bojom, dok je CS rekonstruisana komponenta označena crvenom bojom (45% od ukupnog broja odbiraka signala je dostupno); horizontalna osa označava vrijeme, vertikalna osa označava amplitudu**

Dobijeni sopstveni vektori su pododabrani (po slučajnom rasporedu), u cilju prenosa što je moguće manjeg broja odbiraka signala. Samo 45% od ukupnog broja odbiraka je uzeto od svakog sopstvenog vektora.

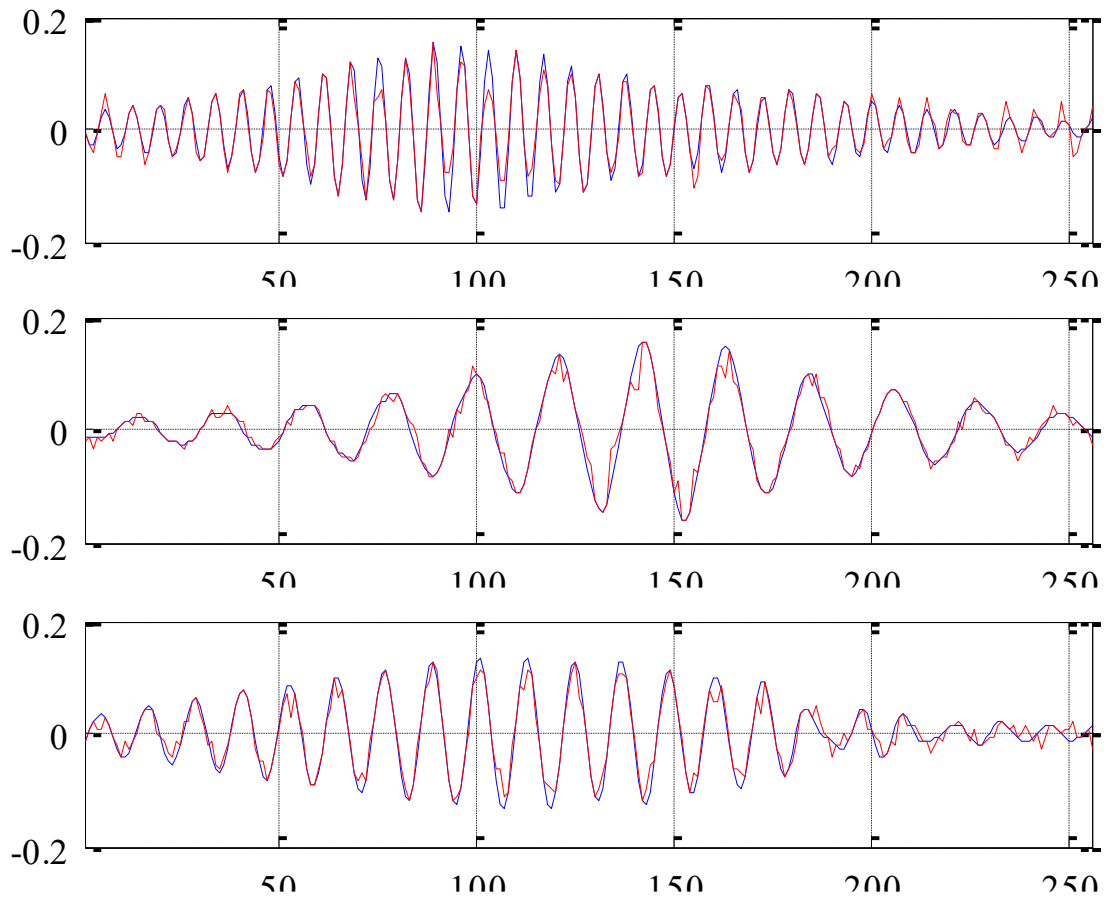

**Slika 5.5: Odvojene komponente IEEE 802.11b signala – originalne komponente označene su plavom bojom a CS rekonstruisane komponente označene su crvenom bojom. Procenat slučajno odabranih odbiraka od svakog sopstvenog vektora je 45% (horizontalna osa je vrijeme, vertikalna osa je amplituda)**

Na prijemnoj strani izvršena je rekonstrukcija pododabranih vektora korišćenjem *basis pursuit primal-dual* rekonstrukcionog algoritma. Originalni (plava linija) i rekonstruisani (crvena linija) sopstveni vektori prikazani su na slikama 5.4 i 5.5.

Komponente FHSS signala prikazane su na slici 5.4, dok su DSSS modulisane komponente prikazane na slici 5.5. Poslednji korak u proceduri je računanje vremensko-frekvencijskih reprezentacija rekonstruisanih komponenti. S-metodi pojedinačnih sopstvenih vektora prikazani su na slici 5.6.

Klasifikacija komponenti signala vrši se posmatranjem vremensko-frekvencijske reprezentacije sopstvenih vektora i estimacijom parametara signala iz vremensko-frekvencijske ravni. Na prvom mjestu posmatra se vremensko trajanje komponenti. Estimiranje parametara signala i klasifikacija na osnovu vremensko-frekvencijske reprezentacije signala predložena je u [7]. Estimirana trajanja komponenti dobijena iz S-metoda, prikazana su u tabeli 5.2. Iz tabele 5.2 se može vidjeti da su trajanja FHSS komponenti manja u poređenju sa trajanjima DSSS komponenti. Srednje kvadratne greške rekonstrukcije, računate između originalnog i rekonstruisanog sopstvenog vektora, date su u tabeli 5.3.

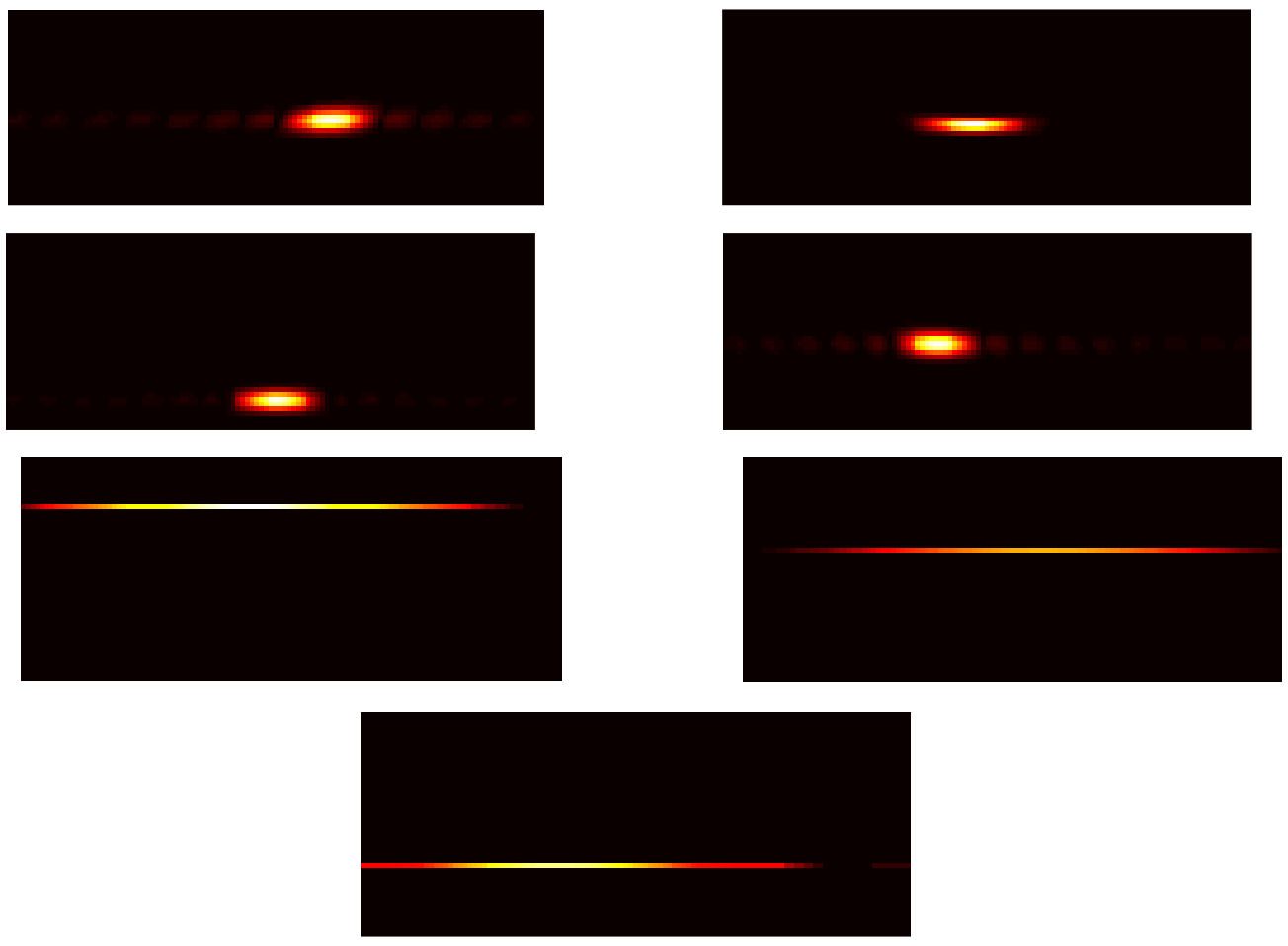

**Slika 5.6: S-metod odvojenih komponenti signala: prve 4 figure odgovaraju S-metodu FHSS signala, dok preostale 3 figure odgovaraju S-metodu DSSS signala (horizontalna osa - vrijeme, vertikalna osa - frekvencija)**

**Tabela 5-2**: **Trajanja komponenti signala, estimirana iz vremensko-frekvencijske reprezentacije odvojenih komponenti**

| *Redni broj komponente* | *Trajanje (broj odbiraka)* | *Signal* |
|---|---|---|
| 1 | 21 | FHSS |
| 2 | 20 | FHSS |
| 3 | 18 | FHSS |
| 4 | 20 | FHSS |
| 5 | 108 | IEEE 802.11b |
| 6 | 112 | IEEE 802.11b |
| 7 | 102 | IEEE 802.11b |

**Tabela 5-3**: **Srednje kvadratne greške (MSE) računate između originalnog i rekonstruisanog sopstvenog vektora. Broj dostupnih odbiraka signala je 45% od ukupnog broja**

| *Redni broj komponente* | *MSE* |
|---|---|
| 1 | 0.0103 |
| 2 | 0.0087 |
| 3 | 0.0077 |
| 4 | 0.0122 |
| 5 | $2.3406\times10^{-4}$ |
| 6 | $3.06781\times10^{-4}$ |
| 7 | $2.5467\times10^{-4}$ |

### 5.3.1. Signal u prisustvu šuma

U ovom poglavlju će biti testirane performanse predložene procedure u slučaju da je posmatrani signal zahvaćen šumom. Analiziran je broj odvojenih sopstvenih vektora, koji odgovaraju komponentama signala, za različiti odnos-signal šum (SNR). Pretpostavljeno je da je signal zahvaćen Gausovim šumom.

Rezultati za različite vrijednosti SNR-a prikazani su u tabeli 5-4, gdje je prikazan broj odvojenih komponenti, od ukupno 7 komponenti signala. Sve komponente signala mogu biti uspješno odvojene dok god je SNR iznad -4dB. Treba napomenuti da bi broj iteracija EVD–a, u slučaju da je prisutan šum u signalu, trebalo da bude malo veći od očekivanog broja komponenti (u našem slučaju, broj EVD iteracija je 8 za SNR = -3.7823 dB, dok je broj komponenti signala 7).

**Tabela 5-4**: **Broj odvojenih sopstvenih vektora koji odgovaraju komponentama signala, u prisustvu šuma. Posmatran je Gausov šum, za različiti odnos SNR**

| *SNR (dB)* | *Broj odvojenih sopstvenih vektora* | *Broj EVD iteracija* |
|---|---|---|
| -4.6599 | 5 | 15 |
| -3.7823 | 7 | 8 |
| -2.0454 | 7 | 8 |
| -0.3898 | 7 | 7 |
| 0.4935 | 7 | 7 |
| 1.9965 | 7 | 7 |

Da zaključimo, eksperimentalno je pokazano da predložena procedura uspješno klasifikuje komponente signala koji pripadaju interferirajućim standardima u bežičnim komunikacijama, dok god se komponente signala ne preklapaju ili ne sijeku u vremensko-frekvencijskoj ravni. Takođe, moguće je pododabrati signale i komunikacionim signalom prenositi redukovani broj odbiraka signala (manje od 50%) a na prijemnoj strani rekonstruisati signal bez gubitka informacije.

Procedura uspješno odvaja i klasifikuje komponente čak i u prisustvu šuma. Predložena su dva sparsfikaciona domena, u zavisnosti od prirode komponente signala – DFT i HT.

# 6. Arhitektura algoritma za rekonstrukciju spars signala

U ovoj glavi je dat prijedlog hardverske realizacije algoritma za rekonstrukciju signala koji imaju kompaktnu predstavu u nekom domenu [61]. Algoritam pretpostavlja DFT domen kao domen u kom se signal može kompaktno predstaviti, ali se ovaj pristup može prilagoditi i drugim transformacionim domenima. Algoritam se bazira na analizi šuma koji se javlja kao posljedica nedostajućih odbiraka u signalu. Vrijednost praga, koji razdvaja šum i komponente signala, zavisiće od broja dostupnih odbiraka u signalu. Prag služi da u DFT domenu, pod određenim okolnostima, u jednom koraku (iteraciji) odvoji komponente koje se javljaju kao posljedica šuma od komponenti koje pripadaju signalu.

Pretpostavićemo da signal $x(n)$ ima $K$ nenultih komponenti u DFT domenu, gdje važi da je $K<<N$ (a $N$ je ukupna dužina signala). Signal se, u skladu sa teorijom CS-a, može rekonstruisati koristeći značajno manji broj odbiraka nego što to zahtijeva Teorema o odabiranju. Kao što je rečeno, nedostajući odbirci signala uzrokovaće izobličenja u vidu šuma u spektralnom domenu. Varijansa tog šuma se može dobiti na osnovu sljedeće relacije [14], [61], [73]:

$$\sigma^2 = M\frac{N-M}{N-1}\sum_{a=1}^{M}\frac{x^2(n_a)}{M} = \mu\sum_{a=1}^{M} y^2(a), \tag{6.1}$$

gdje su sa $n_a$ označene pozicije dostupnih odbiraka, $y(a)=x(n_a)$ je vektor dostupnih odbiraka, a μ je konstanta: $\mu=(N-M)/(N-1)$. Varijansa šuma zavisi od broja nedostajućih odbiraka $M$, tj. broja dostupnih odbiraka $N-M$. Ova varijansa je ključna za računanje praga $T$ [14], [61]:

$$T = \frac{1}{N}\sqrt{-\sigma^2 \ln(1-P^{1/(N-K)})} \approx \frac{1}{N}\sqrt{-\sigma^2 \ln(1-P^{1/N})} \ . \tag{6.2}$$

Prag $T$ se postavlja tako da omogućava da su sve $(N-K)$ komponente šuma ispod vrijednosti praga, sa odgovarajućom vjerovatnoćom $P$. Diskretna Fourier-ova transformacija, DFT, računata samo na osnovu dostupnih odbiraka signala, naziva se početna DFT i označena je sa **X**, a dobija se na osnovu sljedeće relacije:

$$X(k) = \sum_{a=1}^{M} y(a) e^{-j\frac{2\pi}{N}n_a k}, k = 1, ..., N \ .$$

Nakon računanja početne DFT, vrijednosti koje su iznad praga $T$ definišu pozicije frekvencijskih komponenti signala i dobijaju se u skladu sa relacijom:

$$\mathbf{k} = \arg\{|\mathbf{X}| > T\} \; . \tag{6.3}$$

Na ovaj način određene su frekvencije komponenti signala. Amplitude ovih komponenti su i dalje nepoznate. Za određivanje tačnih vrijednosti amplituda potrebno je riješiti optimizacioni problem.

Od pune matrice DFT, $\mathbf{A}_{N\times N}$, formira se tzv. CS matrica $\mathbf{A}_{CS}$. Vrste ove matrice odgovaraju pozicijama dostupnih odbiraka signala, $n_a$, dok kolone matrice odgovaraju frekvencijama komponenti signala $\mathbf{k}$, koje su određene na osnovu relacije (6.3). Minimizacioni problem je definisan sljedećom relacijom:

$$\mathbf{X} = (\mathbf{A}_{CS}^{*}\mathbf{A}_{CS})^{-1}(\mathbf{A}_{CS}^{*}\mathbf{y}) \; . \tag{6.4}$$

Vektor $\mathbf{y}$ je vektor dostupnih odbiraka: $\mathbf{y}=\mathbf{x}(n_a)$, $a=1,\ldots,M$, a $\mathbf{A}_{CS}^{*}$ je transponovana matrica matrice $\mathbf{A}_{CS}$. Matrica $\mathbf{A}_{CS}$ se naziva još i slučajnom, parcijalnom matricom Fourier-ove transformacije, imajući u vidu da je nastala selekcijom samo određenih vrsta i kolona matrice Fourier-ove transformacije, $\mathbf{A}$. Rješenje sistema (6.4) obezbjeđuje tačne amplitude za $K$ komponenti signala, u DFT domenu.

U ovoj glavi je dat prijedlog modifikacije originalnog algoritma, sa ciljem smanjenja kompleksnosti sistema i omogućavanja jednostavnije hardverske realizacije. Umjesto da se u optimizacionom problemu koristi parcijalna matrica Fourier-ove transformacije, minimizacioni problem je preformulisan i sveden na upotrebu trougaone matrice $\mathbf{R}$ [72]. Optimizacioni problem je sveden na trougaonu matricu korišćenjem QR dekompozicije, o čemu će kasnije biti više riječi. Za rješavanje osnovne forme optimizacionog problema, bilo je neophodno invertovati proizvod parcijalne DFT matrice i njene transponovane verzije. Realizacija inverzne matrice je hardverski veoma zahtjevna, pa se u cilju redukcije kompleksnosti svodi na računanje inverzne trougaone matrice. Inverzna trougaona matrica je hardverski mnogo lakša za implementaciju u poređenju sa inverznom osnovnom matricom.

U radu je predložen fleksibilan i skalabilan sistem za realizaciju inverzije trougaone matrice $\mathbf{R}$. Dimenzije ove matrice mogu se mijenjati u skladu sa promjenama broja dostupnih odbiraka signala, kao i broja frekvencijskih komponenti signala. Takođe, numerički zahtjevan dio algoritma, koji se odnosi na računanje praga, je modifikovan i sveden na konstantu.

U narednom poglavlju će biti opisane blok šeme za hardversku realizaciju, sa predloženim modifikacijama algoritma.

## 6.1. Blok šeme hardverske arhitekture

Arhitektura sistema za hardversku realizaciju algoritma za rekonstrukciju kompresivno očitavanih signala prikazana je na slici 6.1. *Blok 1* služi za određivanje pozicija spektralnih komponenti koje su iznad praga *T*. Za računanje početne Fourier-ove transformacije koristi se blok FFT. *Blok 2* služi za računanje parcijalne Fourier-ove matrice, $\mathbf{A}_{CS}$.

*Blok 3* predstavlja dio za računanje trougaone matrice **R**, tj. taj dio vrši QR dekompoziciju nad ulaznom matricom $\mathbf{A}_{CS}$. U ovom bloku nalazi se i dio za rješavanje optimizacionog problema.

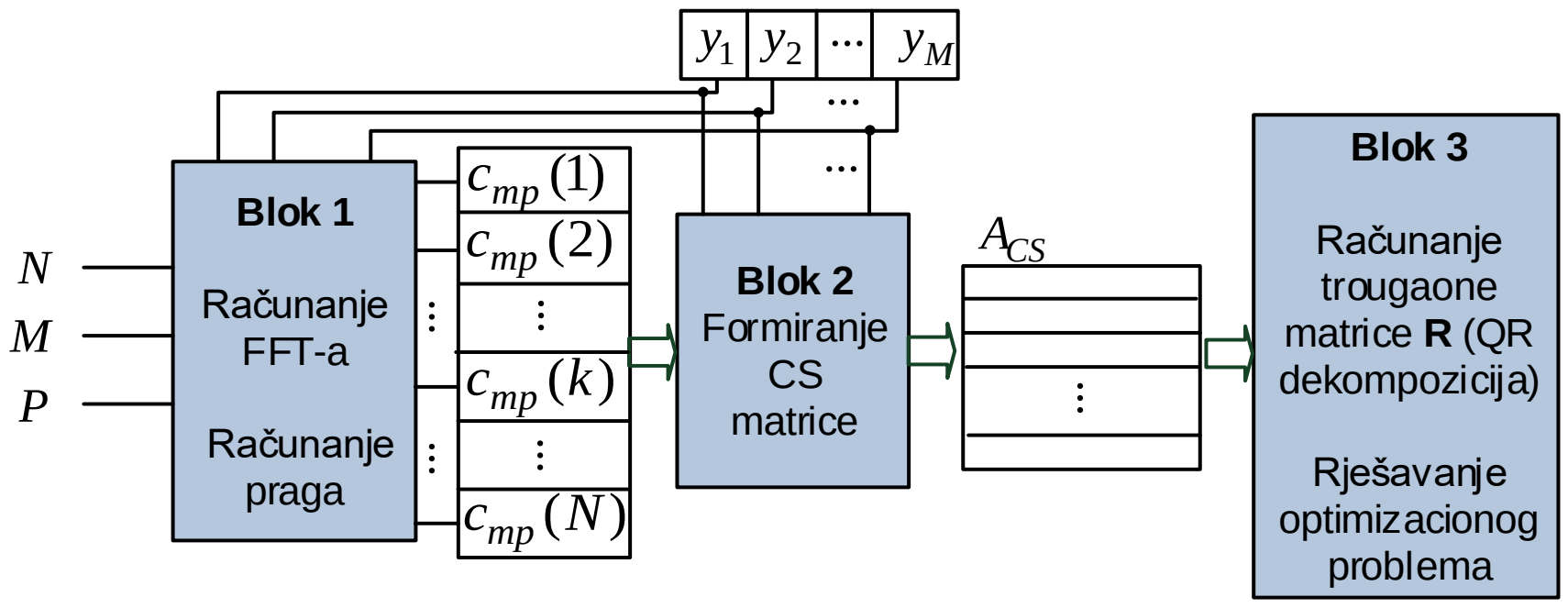

**Slika 6.1: Blok šema za hardversku implementaciju algoritma za rekonstrukciju kompresivno očitavanih signala, koji radi u jednom koraku**

***Blok 1*** prikazan na slici 6.1 sastoji se od dijela za računanje početne Fourier-ove transformacije, korišćenjem sklopova koji računaju brzu Fourier-ovu transformaciju (FFT) [72]. Paralelna grana ovog bloka računa prag, zajedno sa računanjem apsolutnih vrijednosti brze Fourier-ove transformacije FFT, slika 6.2. Važno je primijetiti da su FFT koeficijenti kompleksne vrijednosti i zbog toga ABS kola uključuju računanje kvadratnog korijena od zbira kvadrata realnog i imaginarnog dijela.

Apsolutne vrijednosti početne Fourier-ove transformacije kao i prag *T* dovode se na ulaz komparatorskog bloka – COMP (slika 6.2). Na izlazu ovog bloka dobija se vektor $\boldsymbol{c}_{mp}$. Vrijednosti vektora $\boldsymbol{c}_{mp}$ su ili logička nula ili logička jednica. Vrijednost „1“ dobija se na poziciji na kojoj je apsolutna vrijednost FFT koeficijenta iznad praga *T*, a u suprotnom, na izlazu je vrijednost „0“.

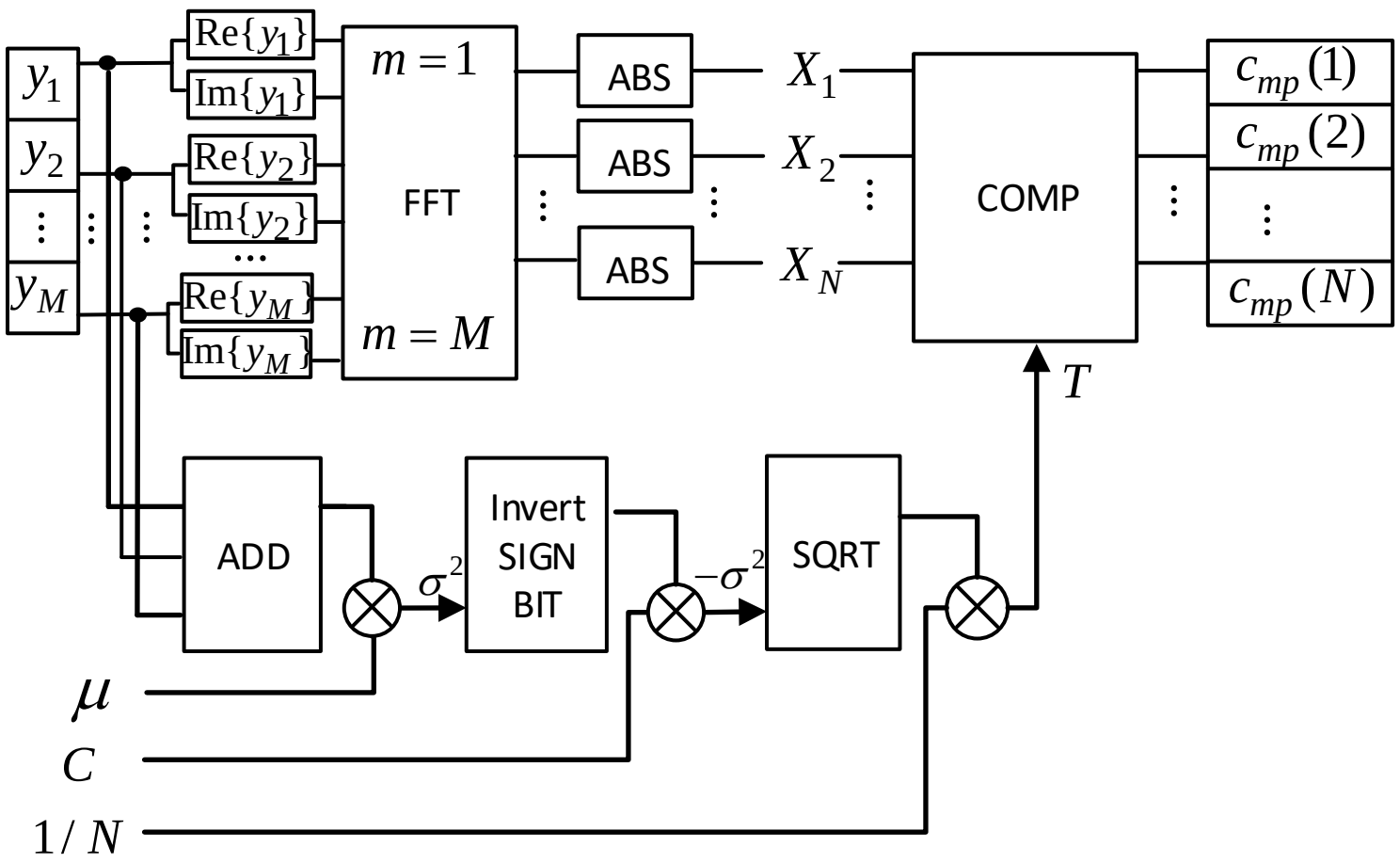

**Slika 6.2: Blok 1 hardverske arhitekture – dio za računanje FFT-a, dio za računanje praga i komparatorski blok**

Najzahtjevniji dio ovog bloka je realizacija funkcije za računanje praga. Naime, u skladu sa relacijom (6.2) potrebno je računati logaritam, stepenu funkciju i kvadratni korjen, a sve to je hardverski zahtjevno. Umjesto računanja praga u skladu sa originalnim algoritmom, ovdje je predložen efikasniji dizajn sistema koji bi bio jednostavan za realizaciju i primjenljiv u aplikacijama koje rade u realnom vremenu. U tom cilju, uvedene su aproksimacije koje su eksperimentalno testirane nad različitim signalima.

- Na prvom mjestu, vrijednost vjerovatnoće *P*, koja je generalno definisana od strane korisnika, može biti fiksirana na vrijednost *P*=0.99. Eksperimentalno je pokazano da se sa ovom vrijednošću vjerovatnoće dobijaju zadovoljavajući rezultati [61].
- Druga modifikacija se odnosi na logaritamski član. Za različite, značajnije vrijednosti parametra *N* važi da je:

$$\ln(1-P^{1/N}) \approx const. = C. \tag{6.5}$$

- Posljedično sa prethodnim modifikacijama, i izraz za računanje praga postaje jednostavniji:

$$T = \frac{1}{N}\sqrt{-\sigma^2 C}. \tag{6.6}$$

Varijansa $\sigma^2$ se računa korišćenjem relacije (6.1). Ovo dalje znači da je najzahtijevnija operacija prilikom računanja praga kvadratni korijen. U literaturi postoje različiti metodi za realizaciju kvadratnog korijena: Vavilonski metod, Newton-Raphson-ov metod, algoritam zasnovan na razvoju u Taylor-ov red, obnovljivi (*restoring*) i ne-obnovljivi (*non-restoring*) algoritam [72].

Za hardversku implementaciju kvadratnog korijena potrebni su sabirač/oduzimač, kombinaciona logika, registri i pomjerački registri.

***Blok 2*** Drugi dio sistema, prikazanog na slici 6.1 odnosi se na dobijanje parcijalne Fourier-ove transformacije. Iz originalne, kompletne matrice Fourier-ove transformacije, potrebno je selektovati kolone koje odgovaraju komponentama signala, i vrste koje odgovaraju pozicijama dostupnih odbiraka. Drugim riječima, CS matrica $\mathbf{A}_{CS}$ formira se od elemenata matrice DFT koji se nalaze na presjeku odgovarajućih vrsta i kolona. Ako pretpostavimo da je DFT matrica složena u memoriji po vrstama, onda se elementi matrice $\mathbf{A}_{CS}$ selektuju korišćenjem memorijskih adresa na sljedeći način:

$$address = (n_a - 1)N + b(j), \quad n_a \in \{n_1, ..., n_M\}, \quad j \in \{1, ..., K\}. \tag{6.7}$$

U vektoru $\boldsymbol{b}$ nalaze se pozicije DFT koeficijenata koji su iznad praga, tj. pozicije vektora na izlazu kola komparatora COMP koje su jednake jedinici, $\boldsymbol{c}_{mp}(k)=1$, $k=1, \ldots, N$.

***Blok 3*** predstavlja arhitekturu za rješavanje optimizacionog problema. Najveći izazov u ovom dijelu jeste realizacija inverzne matrice $\mathbf{A}_{CS}$. Najčešće se inverzna matrica realizuje korišćenjem QR dekompozicije. U narednom poglavlju će biti više riječi o metodama za računanje QR dekompozicije [73]-[75], [77]-[81].

## 6.2. QR dekompozicija i metode za njeno računanje

QR dekompozicija matrice **A** se može zapisati kao:

$$\mathbf{A} = \mathbf{QR}, \tag{6.8}$$

gdje je **Q** ortogonalna matrica ($\mathbf{Q}^T\mathbf{Q}=\mathbf{Q}\mathbf{Q}^T=\mathbf{I}$, $\mathbf{Q}^T= \mathbf{Q}^{-1}$), a matrica **R** je gornja trougaona matrica, i važi da je:

$$\mathbf{R} = \mathbf{Q}^T\mathbf{A}. \tag{6.9}$$

Postoji više metoda za računanje QR dekompozicije. Najčešće korišćeni su:

1. Householder transformacija;
2. Gram-Schmidt dekompozicija;
3. Givens-ove rotacije.

Za svaki od ovih metoda postoje hardverske realizacije, i svaki od metoda ima prednosti i nedostatke. Householder transformacija je hardverski komplikovanija za implementaciju u poređenju sa ostalim algoritmima i teška je za paralelizaciju. U pogledu kompleksnosti, Gram-Schmidt metoda je ekvivalentna metodi Givens-ovih rotacija.

***Householder transformacija***

Householder transformacija [72], [75], [77], [78] je jedna od metoda koja se koristi za računanje QR dekompozicije matrice. Ova transformacija se primjenjuje na kolone matrice i anulira sve elemente u posmatranoj koloni, osim prvog elementa. Householder transformacija se označava sa $\mathbf{H}_i$ i računa se prema relaciji:

$$\mathbf{H}_i = \mathbf{I} - 2\mathbf{v}\mathbf{v}^T, \tag{6.10}$$

gdje je **I** matrica identiteta, a vektor **v** je definisan kao:

$$\mathbf{v} = \frac{\mathbf{u}}{\|\mathbf{u}\|_{\ell_2}}, \quad \mathbf{u} = \mathbf{a} - \alpha\mathbf{e}. \tag{6.11}$$

Vektor **a** je kolona matrice čije elemente (osim prvog) anuliramo, vektor **e** je $\mathbf{e}=(1,0,\ldots,0)^T$, a parameter α je $|\alpha| = \|\mathbf{a}\|_{\ell_2}$ i važi da je $\|\mathbf{u}\|_{\ell_2} = 1$. Trougaona matrica **R** dobija se nakon $i$ Householder transformacija kao:

$$\mathbf{R} = \mathbf{H}_i \ldots \mathbf{H}_1 \mathbf{A}, \tag{6.12}$$

a ortogonalna matrica **Q** je definisana:

$$\mathbf{Q} = \mathbf{H}_1^T \ldots \mathbf{H}_i^T \ . \tag{6.13}$$

***Primjer računanja Q i R matrica primjenom Householder transformacije:***

Neka je data matrica $\mathbf{A} = \begin{bmatrix} 1 & 3 & 1 \\ 0 & 2 & 1 \\ 1 & 2 & 3 \end{bmatrix}$.

- Prva Householder transformacija. Primjenjuje se na prvu kolonu matrice **A** i kao rezultat anulira elemenate ispod glavne dijagonale u prvoj koloni matrice **A**:

Prva kolona matrice **A** je $\mathbf{a} = (1,0,1)^T$. Računa se vektor **u**:

$$\mathbf{u} = \mathbf{a} - \alpha\mathbf{e},\ \text{gdje je}\ \alpha = \|\mathbf{a}\|_{\ell_2} = \sqrt{2},\ \mathbf{e} = (1,0,0)^T,\ \text{pa je}$$

$$\mathbf{u} = (1,0,1)^T - \sqrt{2}\,(1,0,0)^T = (1-\sqrt{2},0,1)^T$$

Sljedeći korak je računanje vektora **v**:

$$\mathbf{v} = \frac{\mathbf{u}}{\|\mathbf{u}\|_{\ell_2}} = \frac{(1-\sqrt{2},0,1)^T}{\sqrt{(1-\sqrt{2})^2+1^2}} = \frac{(1-\sqrt{2},0,1)^T}{\sqrt{4-2\sqrt{2}}}.$$

Householder transformacija $\mathbf{H}_1$ je:

$$\mathbf{H}_1 = \mathbf{I} - 2\mathbf{v}\mathbf{v}^T = \mathbf{I} - 2\frac{1}{4-2\sqrt{2}}\begin{pmatrix}1-\sqrt{2}\\0\\1\end{pmatrix}\left(1-\sqrt{2},0,1\right) = \begin{bmatrix}1/\sqrt{2} & 0 & 1/\sqrt{2}\\ 0 & 1 & 0\\ 1/\sqrt{2} & 0 & -1/\sqrt{2}\end{bmatrix}.$$

Množenjem matrice $\mathbf{H}_1$ sa polaznom matricom **A** dobija se nova matrica $\mathbf{A}_1$:

$$\mathbf{A}_1 = \mathbf{H}_1\mathbf{A} = \left[\begin{array}{c|cc}2/\sqrt{2} & 5/\sqrt{2} & 4/\sqrt{2}\\ \hline 0 & 2 & 1\\ 0 & 1/\sqrt{2} & -2/\sqrt{2}\end{array}\right],\quad \mathbf{A}_1' = \begin{bmatrix}0 & 0 & 0\\ 0 & 2 & 1\\ 0 & 1/\sqrt{2} & -2/\sqrt{2}\end{bmatrix}$$

- Drugu Householder transformaciju primjenjujemo na drugu kolonu matrice $\mathbf{A}_1$':

$$\mathbf{u} = \mathbf{a} - \alpha\mathbf{e},\ \text{gdje je}\ \mathbf{a}\ \text{sada:}\ \mathbf{a} = \left(0,2,1/\sqrt{2}\right)^T,$$

$$\alpha = \|\mathbf{a}\|_{\ell_2} = \sqrt{9/2},\ \mathbf{e} = (0,1,0)^T,\ \text{pa je}$$

$$\mathbf{u} = \left(0,2,1/\sqrt{2}\right)^T - \sqrt{9/2}\,(0,1,0)^T = \left(0,2-\sqrt{9/2},1/\sqrt{2}\right)^T$$

Sljedeći korak je računanje vektora **v**:

$$\mathbf{v} = \frac{\mathbf{u}}{\|\mathbf{u}\|_{\ell_2}} = \frac{\left(0,2-\sqrt{9/2},1/\sqrt{2}\right)^T}{\sqrt{9-6\sqrt{2}}}.$$

Matrica $\mathbf{H}_2$ je:

$$\mathbf{H}_2 = \mathbf{I} - 2\mathbf{v}\mathbf{v}^T = \mathbf{I} - 2\frac{1}{9-6\sqrt{2}}\begin{pmatrix}0\\2-\sqrt{9/2}\\ \frac{1}{\sqrt{2}}\end{pmatrix}\left(0,2-\sqrt{9/2},\frac{1}{\sqrt{2}}\right) = \begin{bmatrix}1 & 0 & 0\\ 0 & 0.943 & 0.333\\ 0 & 0.333 & -0.943\end{bmatrix}.$$

Sada se matrice **R** i **Q** mogu dobiti kao:

$$\mathbf{R} = \mathbf{H}_2\mathbf{H}_1\mathbf{A} = \begin{bmatrix}1.414 & 3.536 & 2.828\\ 0 & 2.121 & 0.471\\ 0 & 0 & 1.667\end{bmatrix},\quad \mathbf{Q} = \mathbf{H}_1^T\mathbf{H}_2^T = \begin{bmatrix}0.707 & 0.236 & -0.667\\ 0 & 0.943 & 0.333\\ 0.707 & -0.236 & 0.667\end{bmatrix}.$$

### *Gram-Schmidt metod*

Gram-Schmidt metod omogućava paralelno računanje, sa velikom brzinom obrade, pa se često koristi u aplikacijama gdje je poželjna paralelizacija [73], [75], [79]. Postoje dvije varijante Gram-Schmidt metoda:

-Klasični Gram-Schmidt metod;

-Modifikovani Gram-Schmidt metod.

U nastavku je opisan klasični Gram-Schmidt algoritam. Označimo kolone matrice **A** kao $\mathbf{a}_1,\ldots,\mathbf{a}_n$ , tj. :

$$\mathbf{A} = \left[\mathbf{a}_1 \middle| \mathbf{a}_2 \middle| \ldots \middle| \mathbf{a}_n\right]. \tag{6.14}$$

Definišu se vektori $\mathbf{u}_1,\ldots,\mathbf{u}_n$ i $\mathbf{e}_1,\ldots,\mathbf{e}_n$ na sljedeći način :

$$\begin{aligned}
&\mathbf{u}_1 = \mathbf{a}_1, && \mathbf{e}_1 = \frac{\mathbf{u}_1}{\|\mathbf{u}_1\|_2} \\
&\mathbf{u}_2 = \mathbf{a}_2 - \langle \mathbf{a}_2, \mathbf{e}_1 \rangle \mathbf{e}_1, && \mathbf{e}_2 = \frac{\mathbf{u}_2}{\|\mathbf{u}_2\|_2} \\
&\mathbf{u}_3 = \mathbf{a}_3 - \langle \mathbf{a}_3, \mathbf{e}_1 \rangle \mathbf{e}_1 - \langle \mathbf{a}_3, \mathbf{e}_2 \rangle \mathbf{e}_2, && \mathbf{e}_3 = \frac{\mathbf{u}_3}{\|\mathbf{u}_3\|_2} \quad , \\
&\vdots \\
&\mathbf{u}_n = \mathbf{a}_n - \langle \mathbf{a}_n, \mathbf{e}_1 \rangle \mathbf{e}_1 - \ldots - \langle \mathbf{a}_n, \mathbf{e}_{n-1} \rangle \mathbf{e}_{n-1}, && \mathbf{e}_n = \frac{\mathbf{u}_n}{\|\mathbf{u}_n\|_2}
\end{aligned} \tag{6.15}$$

gdje je $\ell_2$ norma definisana kao:

$$\|\mathbf{u}\|_2 = \sqrt{\sum_{k=1}^{n} u_k^2} \ ,$$

a proizvod :

$$\langle \mathbf{a}, \mathbf{e} \rangle = \langle (a_1,\ldots,a_n), (e_1,\ldots,e_n) \rangle = a_1 e_1 + \ldots + a_n e_n \,.$$

Ortogonalna matrica **Q** i trougaona matrica **R** sada se definišu na sljedeći način:

$$\mathbf{Q} = \left[\mathbf{e}_1 \middle| \mathbf{e}_2 \middle| \ldots \middle| \mathbf{e}_n\right], \quad \mathbf{R} = \begin{bmatrix} \mathbf{a}_1\mathbf{e}_1 & \mathbf{a}_2\mathbf{e}_1 & \ldots & \mathbf{a}_n\mathbf{e}_1 \\ 0 & \mathbf{a}_2\mathbf{e}_2 & \ldots & \mathbf{a}_n\mathbf{e}_2 \\ \vdots & \vdots & \ddots & \vdots \\ 0 & 0 & \ldots & \mathbf{a}_n\mathbf{e}_n \end{bmatrix} \tag{6.16}$$

***Primjer računanja Q i R matrica primjenom Gram-Schmidt metoda:***

Neka je data matrica $\mathbf{A}=\begin{bmatrix}1&3&1\\0&2&1\\1&2&3\end{bmatrix}$, gdje je $\mathbf{a}_1=\begin{bmatrix}1\\0\\1\end{bmatrix}$, $\mathbf{a}_2=\begin{bmatrix}3\\2\\2\end{bmatrix}$, $\mathbf{a}_3=\begin{bmatrix}1\\1\\3\end{bmatrix}$. Prvi korak je računanje vektora $\mathbf{u}_1$ i $\mathbf{e}_1$:

$$\mathbf{u}_1=\mathbf{a}_1,\quad \mathbf{e}_1=\frac{\mathbf{u}_1}{\|\mathbf{u}_1\|_2}=\frac{(1,0,1)}{\sqrt{1^2+0^2+1^2}}=\left(\frac{1}{\sqrt{2}},0,\frac{1}{\sqrt{2}}\right), \tag{6.17}$$

U sljedećem koraku računaju se vektori $\mathbf{u}_2$ i $\mathbf{e}_2$:

$$\mathbf{u}_2=\mathbf{a}_2-\langle\mathbf{a}_2,\mathbf{e}_1\rangle\mathbf{e}_1=(3,2,2)-\left\langle(3,2,2)\left(\frac{1}{\sqrt{2}},0,\frac{1}{\sqrt{2}}\right)\right\rangle\left(\frac{1}{\sqrt{2}},0,\frac{1}{\sqrt{2}}\right)=$$
$$=(3,2,2)-\frac{5}{\sqrt{2}}\left(\frac{1}{\sqrt{2}},0,\frac{1}{\sqrt{2}}\right)=\left(\frac{1}{2},2,-\frac{1}{2}\right),$$

$$\mathbf{e}_2=\frac{\mathbf{u}_2}{\|\mathbf{u}_2\|_2}=\frac{\left(\frac{1}{2},2,-\frac{1}{2}\right)}{\sqrt{\frac{18}{4}}}=\left(\frac{1}{3\sqrt{2}},\frac{4}{3\sqrt{2}},\frac{-1}{3\sqrt{2}}\right), \tag{6.18}$$

a zatim se računaju $\mathbf{u}_3$ i $\mathbf{e}_3$:

$$\mathbf{u}_3=\mathbf{a}_3-\langle\mathbf{a}_3,\mathbf{e}_1\rangle\mathbf{e}_1-\langle\mathbf{a}_3,\mathbf{e}_2\rangle\mathbf{e}_2=(1,1,3)-\left\langle(1,1,3)\left(\frac{1}{\sqrt{2}},0,\frac{1}{\sqrt{2}}\right)\right\rangle\left(\frac{1}{\sqrt{2}},0,\frac{1}{\sqrt{2}}\right)-$$
$$-\left\langle(1,1,3)\left(\frac{1}{3\sqrt{2}},\frac{4}{3\sqrt{2}},\frac{-1}{3\sqrt{2}}\right)\right\rangle\left(\frac{1}{3\sqrt{2}},\frac{4}{3\sqrt{2}},\frac{-1}{3\sqrt{2}}\right)=$$
$$=(1,1,3)-\frac{4}{\sqrt{2}}\left(\frac{1}{\sqrt{2}},0,\frac{1}{\sqrt{2}}\right)-\frac{2}{3\sqrt{2}}\left(\frac{1}{3\sqrt{2}},\frac{4}{3\sqrt{2}},\frac{-1}{3\sqrt{2}}\right)=\left(\frac{-10}{9},\frac{5}{9},\frac{10}{9}\right)$$

$$\mathbf{e}_3=\frac{\mathbf{u}_3}{\|\mathbf{u}_3\|_2}=\frac{\left(\frac{-10}{9},\frac{5}{9},\frac{10}{9}\right)}{\sqrt{\frac{(-10)^2}{9^2},\frac{5^2}{9^2},\frac{10^2}{9^2}}}=\left(\frac{-2}{3},\frac{1}{3},\frac{2}{3}\right) \tag{6.19}$$

Na osnovu $\mathbf{e}_1$, $\mathbf{e}_2$ i $\mathbf{e}_3$ računa se matrica $\mathbf{Q}$:

$$\mathbf{Q} = \left[\mathbf{e}_1 \middle| \mathbf{e}_2 \middle| \mathbf{e}_3\right] = \begin{bmatrix} 1/\sqrt{2} & 1/\left(3\sqrt{2}\right) & -2/3 \\ 0 & 4/\left(3\sqrt{2}\right) & 1/3 \\ 1/\sqrt{2} & -1/\left(3\sqrt{2}\right) & 2/3 \end{bmatrix} = \begin{bmatrix} 0.707 & 0.236 & -0.667 \\ 0 & 0.943 & 0.333 \\ 0.707 & -0.236 & 0.667 \end{bmatrix}$$

(6.20)

Matrica **Q** je ortogonalna matrica pa treba da važi da je $\mathbf{Q}^T\mathbf{Q}=\mathbf{I}$. Uvrštavanjem konkretnih vrijednosti dobija se da je taj uslov zadovoljen: $\mathbf{Q}^T\mathbf{Q} = \begin{bmatrix} 1 & 0 & 0 \\ 0 & 1 & 0 \\ 0 & 0 & 1 \end{bmatrix}$.

Matrica **R** je gornja trougaona matrica čije su vrijednosti:

$$\mathbf{R} = \begin{bmatrix} \mathbf{a}_1\mathbf{e}_1 & \mathbf{a}_2\mathbf{e}_1 & \mathbf{a}_3\mathbf{e}_1 \\ 0 & \mathbf{a}_2\mathbf{e}_2 & \mathbf{a}_3\mathbf{e}_2 \\ 0 & 0 & \mathbf{a}_3\mathbf{e}_3 \end{bmatrix} = \begin{bmatrix} 2/\sqrt{2} & 5/\sqrt{2} & 4/\sqrt{2} \\ 0 & 3/\sqrt{2} & 2/\left(3\sqrt{2}\right) \\ 0 & 0 & 5/3 \end{bmatrix} = \begin{bmatrix} 1.414 & 3.536 & 2.828 \\ 0 & 2.121 & 0.471 \\ 0 & 0 & 1.667 \end{bmatrix}.$$

(6.21)

***Givens-ove rotacije***

Metod zasnovan na Givens-ovim rotacijama u poslednje vrijeme privlači dosta pažnje, zbog dobrih numeričkih osobina i mogućnosti paralelizacije [73], [80], [81]. Givens-ove rotacije baziraju se na skupu rotacija posmatrane matrice **A**. Matrica **G**, koja se koristi za implementaciju rotacija, definisana je na sljedeći način:

$$G(i,j,\theta) = \begin{bmatrix} 1 & \cdots & 0 & \cdots & 0 & \cdots & 0 \\ \vdots & \ddots & \vdots & \cdots & \vdots & \cdots & \vdots \\ 0 & \cdots & C & \cdots & S & \cdots & 0 \\ \vdots & \cdots & \vdots & \ddots & \vdots & \cdots & \vdots \\ 0 & \cdots & -S & \cdots & C & \cdots & 0 \\ \vdots & \cdots & \vdots & \cdots & \vdots & \ddots & \vdots \\ 0 & \cdots & 0 & \cdots & 0 & \cdots & 1 \end{bmatrix} \begin{matrix} \\ \\ i \\ \\ j \\ \\ \\ \end{matrix}, \quad \text{(columns } i, j\text{)} \tag{6.22}$$

a koeficijenti $C$ i $S$ se definišu kao:

$$\begin{aligned} C &= \cos(\theta_{i,j}) \\ S &= -\sin(\theta_{i,j}) \end{aligned} \quad . \tag{6.23}$$

Množenjem matrice **A** matricom **G** anulira se onaj element matrice **A** koji se nalazi na poziciji (*i,j*). Matrica **Q** i gornja trougaona matrica **R** dobijaju se u skladu sa sljedećim relacijama:

$$\begin{aligned} \mathbf{Q} &= \mathbf{G}_1\mathbf{G}_2 \ldots \mathbf{G}_i, \\ \mathbf{Q}^{\mathrm{T}}\mathbf{A} &= \mathbf{R}, \end{aligned} \tag{6.24}$$

gdje $\mathbf{G}_i$ predstavlja matricu **G** u *i*-toj Givens-ovoj rotaciji.

## 6.3. Modifikacija optimizacionog problema i implementacija QR dekompozicije korišćenjem sistoličkog niza

Optimizacioni problem, opisan relacijom (6.4), moguće je modifikovati primjenom QR dekompozicije, čime se svodi na sljedeći oblik [73]:

$$\begin{aligned} &1)\ \mathbf{A_{CS}} = \mathbf{Q_{CS}}\mathbf{R_{CS}}, \\ &2)\ \mathbf{X} = \left(\left(\mathbf{Q_{CS}}\mathbf{R_{CS}}\right)^{\mathrm{T}}\left(\mathbf{Q_{CS}}\mathbf{R_{CS}}\right)\right)^{-1}\left(\mathbf{A_{CS}^{T}}\mathbf{y}\right), \\ &3)\ \mathbf{X} = \left(\mathbf{R_{CS}^{T}}\mathbf{R_{CS}}\right)^{-1}\left(\mathbf{A_{CS}^{T}}\mathbf{y}\right). \end{aligned} \tag{6.25}$$

Konačno, modifikovana verzija optimizacionog problema ima oblik [73]:

$$\mathbf{X} = \mathbf{R_{CS}^{-1}}\left(\mathbf{R_{CS}^{-1}}\right)^{\mathrm{T}}\left(\mathbf{A_{CS}^{T}}\mathbf{y}\right). \tag{6.26}$$

Na ovaj način, inverzija proizvoda matrice $\mathbf{A}_{CS}$ i njene transponovane verzije svedena je na inverziju trougaone matrice $\mathbf{R}_{\mathrm{CS}}$. Važno je napomenuti da ortogonalnu matricu $\mathbf{Q}_{\mathrm{CS}}$ nije potrebno koristiti, imajući u vidu da ona ne figuriše u relaciji za modifikovani optimizacioni problem. U poređenju sa inverzijom proizvoda polazne CS matrice i njene transponovane verzije, računanje inverzne trougaone matrice $\mathbf{R}_{\mathrm{CS}}$ je hardverski mnogo manje zahtjevno - za matricu dimenzija $N \times N$, računska kompleksnost je $N^2$.

U nastavku će biti opisani blokovi za realizaciju QR dekompozicije i inverzije gornje trougaone matrice. Za realizaciju QR dekompozicije korišćen je metod zasnovan na Givens-ovim rotacijama. Arhitekture su bazirane na tzv. sistoličkom nizu [75]-[77], koji predstavlja često korišćeni dizajn za sisteme koji uključuju inverziju trougaone matrice.

Sistolički niz sastoji se od procesorskih elementa, koji su međusobno povezani i omogućavaju paralelno izvršavanje operacija. Ove arhitekture su obično trougaone ali mogu biti i linearne, a postoje i arhitekture sa jednim procesorskim elementom [75]. Ovdje će biti korišćene i opisane trougaone arhitekture bazirane na sistoličkom nizu, i za QR dekompoziciju a kasnije i za inverziju trougaone matrice.

Šema za računanje matrice $\mathbf{R}_{CS}$ na osnovu ulazne matrice $\mathbf{A}_{CS}$, prikazana je na slici 6.3. Ulazna matrica $\mathbf{A}_{CS}$ se prvo reorganizuje na način kako je to prikazano na slici, a zatim se kolone matrice vode na ćelije koje vrše QR dekompoziciju (ćelije označene sa *r*), a koje ulaze u sastav sistoličkog niza [73]. Svaka kolona reorganizovane matrice ima kašnjenje za po jedan element, u poređenju sa prethodnom kolonom. To je na slici predstavljeno kao nulti ulaz u sistolički niz (slika 6.3). Kao posljedica toga, za matricu veličine $M\times K$, potrebno je $M+(K\text{-}1)$ koraka za računanje trougaone matrice $\mathbf{R}_{CS}$. Dijagonalne ćelije sistoličkog niza (ćelije u obliku kruga na slici) služe za računanje koeficijenata $c_i$ i $s_i$, tj. koeficijenata Givens-ovih rotacija. Elementi matrice $\mathbf{A}_{CS}$ koji se nalaze u ćelijama, označeni su sa $r_{in}$. Koeficijenti rotacije se računaju na sljedeći način [73]:

$$c_i = \frac{r_{i,j}}{\sqrt{r_{i,j}^2 + r_{in}^2}}, \quad s_i = \frac{r_{in}}{\sqrt{r_{i,j}^2 + r_{in}^2}} , \tag{6.27}$$

gdje su $c_i$ i $s_i$ koeficijenti rotacije računati u posmatranoj ćeliji, $r_{i,j}$ je trenutna vrijednost u ćeliji koja se ažurira nakon što se izračunaju koeficijenti $c_i$ i $s_i$, i to prema sljedećoj formuli:

$$r_{i,j} = \sqrt{r_{i,j}^2 + r_{in}^2} . \tag{6.28}$$

Unutrašnje (kvadratne) ćelije ažuriraju svoje vrijednosti u skladu sa relacijom:

$$r_{i,j} = c_{in} r_{i,j} + s_{in} r_{in} , \tag{6.29}$$

gdje su $c_{in}$ i $s_{in}$ koeficijenti na ulazima kvadratnih ćelija. Važno je napomenuti da $c_{in}$ i $s_{in}$ koeficijenti, koji se nalaze na ulazima kvadratnih ćelija, ne odgovaraju $c_i$ i $s_i$ koeficijentima koji se računaju u okruglim ćelijama u odgovarajućoj vrsti sistoličkog niza. Na ulaze kvadratnih ćelija dolaze zakašnjeli koeficijenti rotacija, a kašnjenje se povećava povećanjem distance između kvadratnih i kružnih ćelija, što će biti prikazano u primjeru u Poglavlju 6.4.

Izlazi kvadratnih ćelija $w_{out}$ računaju se kao [73]:

$$w_{out} = -s_{in} r_{i,j} + c_{in} r_{in} \; . \tag{6.30}$$

Kao rezultat, u poslednjem koraku u svakoj ćeliji će se naći po jedna vrijednost, koja odgovara vrijednosti elementa gornje trougaone matrice $\mathbf{R}_{CS}$, a vrijednosti u matrici koje se nalaze ispod glavne dijagonale biće jednake nuli. Imajući u vidu da sistem treba da radi za matrice različitih dimenzija, kao i za signale sa različitim brojem komponenti i za različiti broj dostupnih mjerenja signala, treba obezbijediti fleksibilnost arhitekture. Naime, treba obezbijediti sistemu da može da radi sa različitim brojem komponenti signala (različitim parametrom $K$) kao i različitim brojem dostupnih odbiraka $M$. Drugim riječima, treba omogućiti da sistolički niz ima adaptibilnu veličinu. Inicijalno, sistolički niz ima $N$ okruglih i $(N^2\text{-}N)/2$ kvadratnih ćelija, i važi da je $N>M$ i $N>K$. Ovo je maksimalna veličina sistoličkog niza. U cilju obezbjeđivanja fleksibilnosti u pogledu veličine niza, uvedeni su kontrolni signali $E_1$:$E_N$ (Slika 6.3).

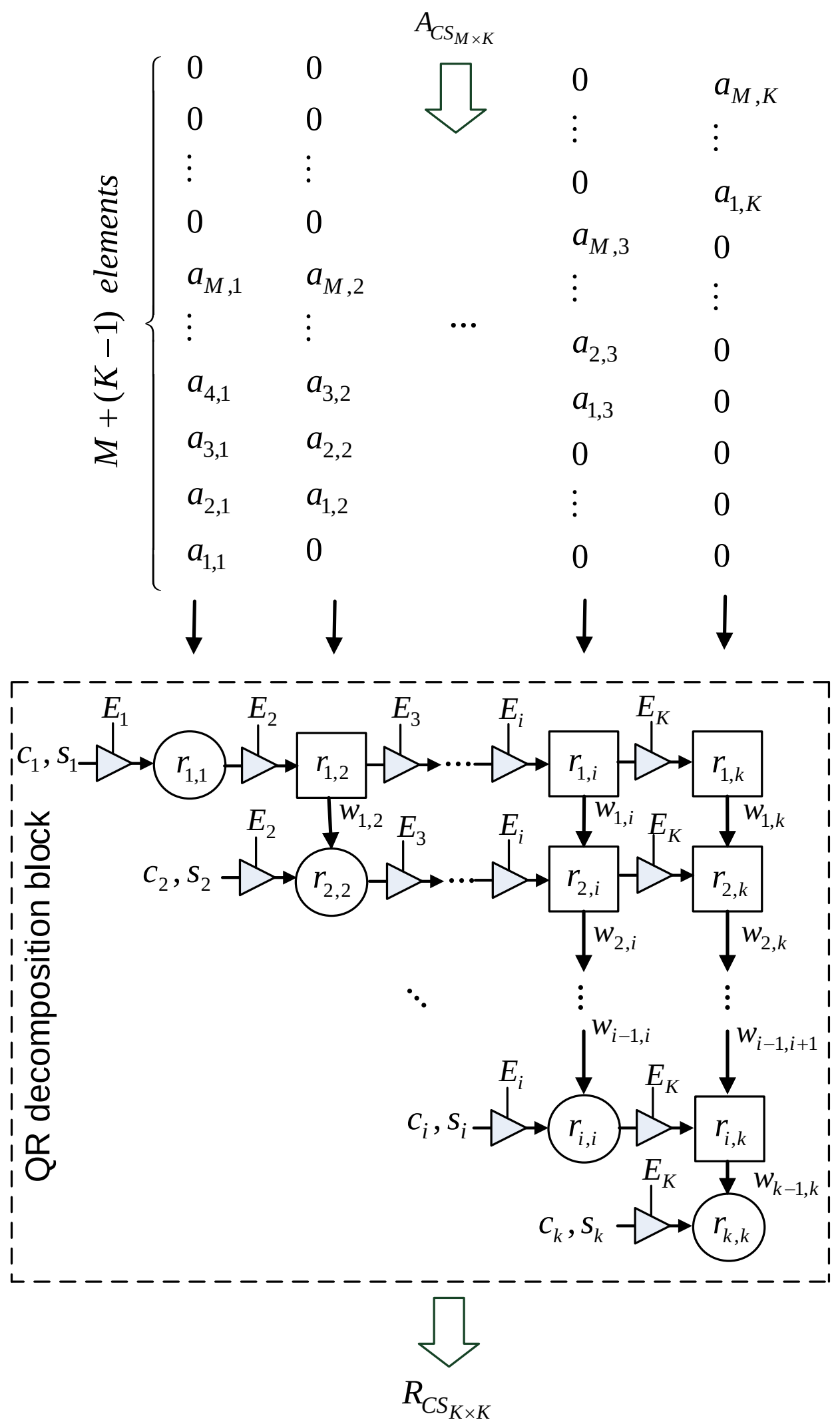

**Slika 6.3: Blok za QR dekompoziciju**

Kontrolni signali djeluju na kolone matrice i “isključuju” one kolone niza čiji su odgovarajući kontrolni signali postavljeni na “0”. Npr. za matricu $\mathbf{A}_{CS}$ dimenzija $M\times K$, ($M>K$) sistolički niz sa slike 6.3 imaće aktivno prvih $K$ kružnih ćelija i $(K^2-K)/2$ kvadratnih ćelija. Ostatak ćelija sistoličkog niza biće isključen, što se postiže postavljanjem odgovarajućih kontrolnih signala na nulu, tj. $E_i=0$, za $i=K+1,\ldots,N$. Ovim je realizovan dio koji se odnosi na QR dekompoziciju, tj. na računanje trougaone matrice $\mathbf{R}_{CS}$. Napominjemo da je u ovom bloku računata samo trougaona matrica jer ortogonalna matrica $\mathbf{Q}_{CS}$ nije neophodna pri rješavanju modifikovanog optimizacionog problema. Nakon dobijanja matrice $\mathbf{R}_{CS}$, ona se reorganizuje i dopunjava nulama, u skladu sa slikom 6.4. Zatim se ova matrica dovodi na blok za inverziju trougaone matrice, koji je prikazan na istoj slici. Za matricu dimenzija $K\times K$, inverzna matrica dobija se u $2K$-1 koraka. I za realizaciju inverzne matrice koristi se sistolički niz, sastavljen od kružnih i kvadratnih ćelija. Međutim, funkcije ćelija ovog sistoličkog niza razlikuju se od funkcija kružnih i kvadratnih ćelija sistoličkog niza za QR dekompoziciju, i opisane su u nastavku. Ulaz u kvadratne (kružne) ćelije može biti ili vrijednost “1” (ili “0” ako je u pitanju kružna ćelija), ili element trougaone matrice, ili izlaz susjedne ćelije. Slično kao kod prethodnog bloka, sistolički niz prvobitno ima $N$ kvadratnih i $(N^2-N)/2$ kružnih ćelija. Kontrolni signali $E_i=0$ isključuju ćelije čiji indeks $i$ je $i>K$ (slika 6.4).

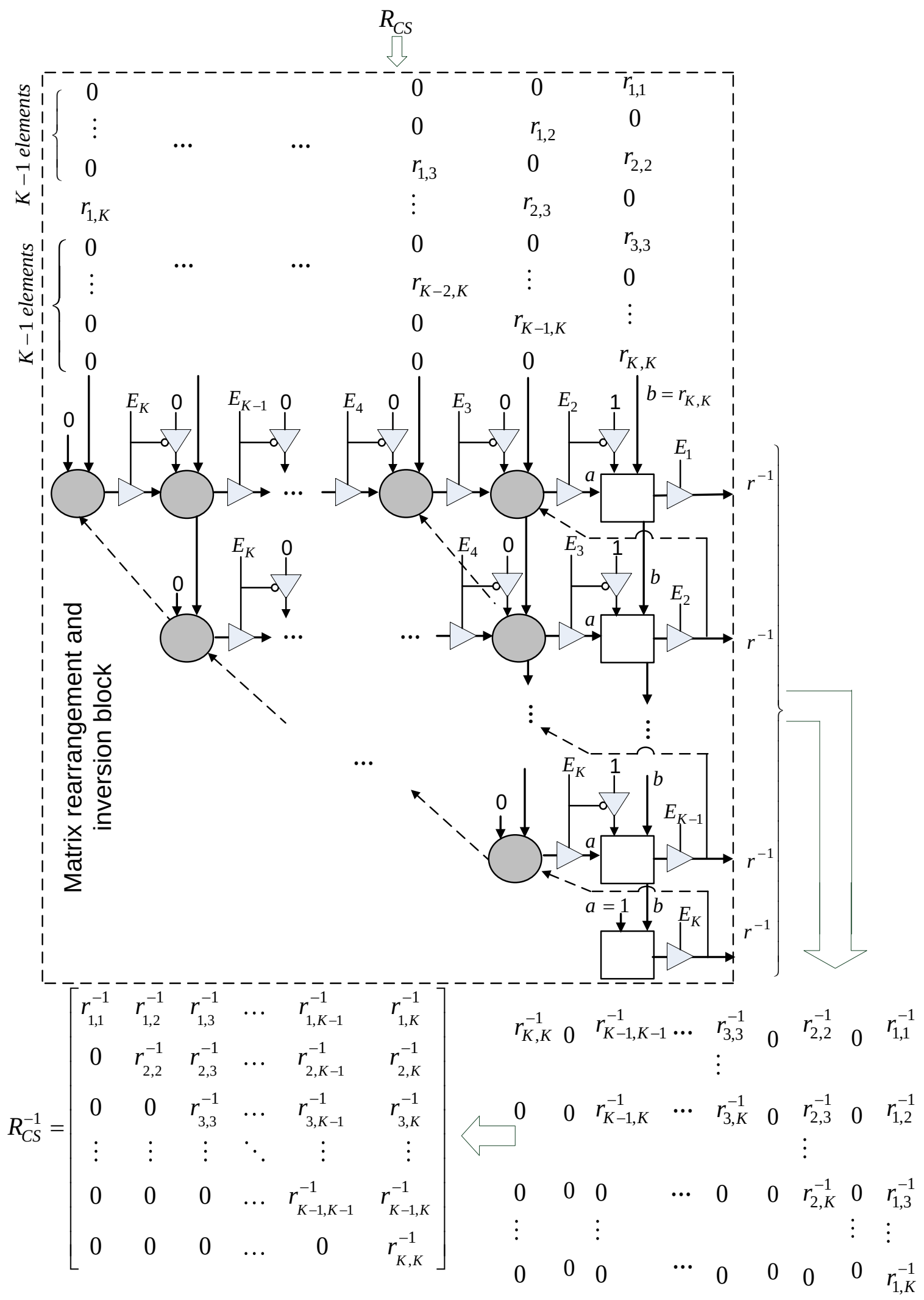

**Slika 6.4: Blok za računanje inverzne matrice**

Na izlazima kvadratnih ćelija nalaze se elementi invertovane matrice. Izlazi iz kvadratnih ćelija dobijaju se kao $r^{-1}$=-$a$ /$r_{i,i}$ (slika 6.4):

$$r_{i,i}^{-1} = -1/r_{i,i}, \quad r_{i,j}^{-1} = -u_{i,j}/r_{i,i} \tag{6.31}$$

gdje je $r_{i,i}$ dijagonalni element, parametar $a$ može biti -1 ili jednak $u_{i,j}$, za $j$=$i$+1,…, $K$. Vrijednost $u_{i,j}$ računa se kao:

$$u_{i,j} = \sum_{k=1}^{j-1} r_{i,k} r_{k,j}^{-1} \ . \tag{6.32}$$

a $r_{i,k}$ je vrijednost unutar ćelije, dok je $r_{k,j}^{-1}$ jednak elementu invertovane matrice koji ulazi u ćeliju od dijagonale (slika 6.4).

## 6.4. Primjer dobijanja trougaone matrice i njene inverzne forme korišćenjem sistoličkih nizova

U nastavku je dat primjer funkcionisanja sistoličkih nizova za QR dekompoziciju i inverziju matrice. Metod je opisan korišćenjem matrice dimenzija 3×2 : **A**=[1, 0.8; 0.8, 2; 2 1].

### 6.4.1. QR dekompozicija

Prvi korak procedure je reorganizacija matrice **A**:

$$\mathbf{A}=\begin{bmatrix}1 & 0.8\\ 0.8 & 2\\ 2 & 1\end{bmatrix}=\begin{bmatrix}a_{11} & a_{12}\\ a_{21} & a_{22}\\ a_{31} & a_{32}\end{bmatrix}\rightarrow\begin{matrix}a_{31} & a_{32}\\ a_{21} & a_{22}\\ a_{11} & a_{12}\end{matrix}$$

Inverzna matrica se dobija kroz 5 koraka. Koraci su numerisani od 0 do 4.

**Korak 0:**

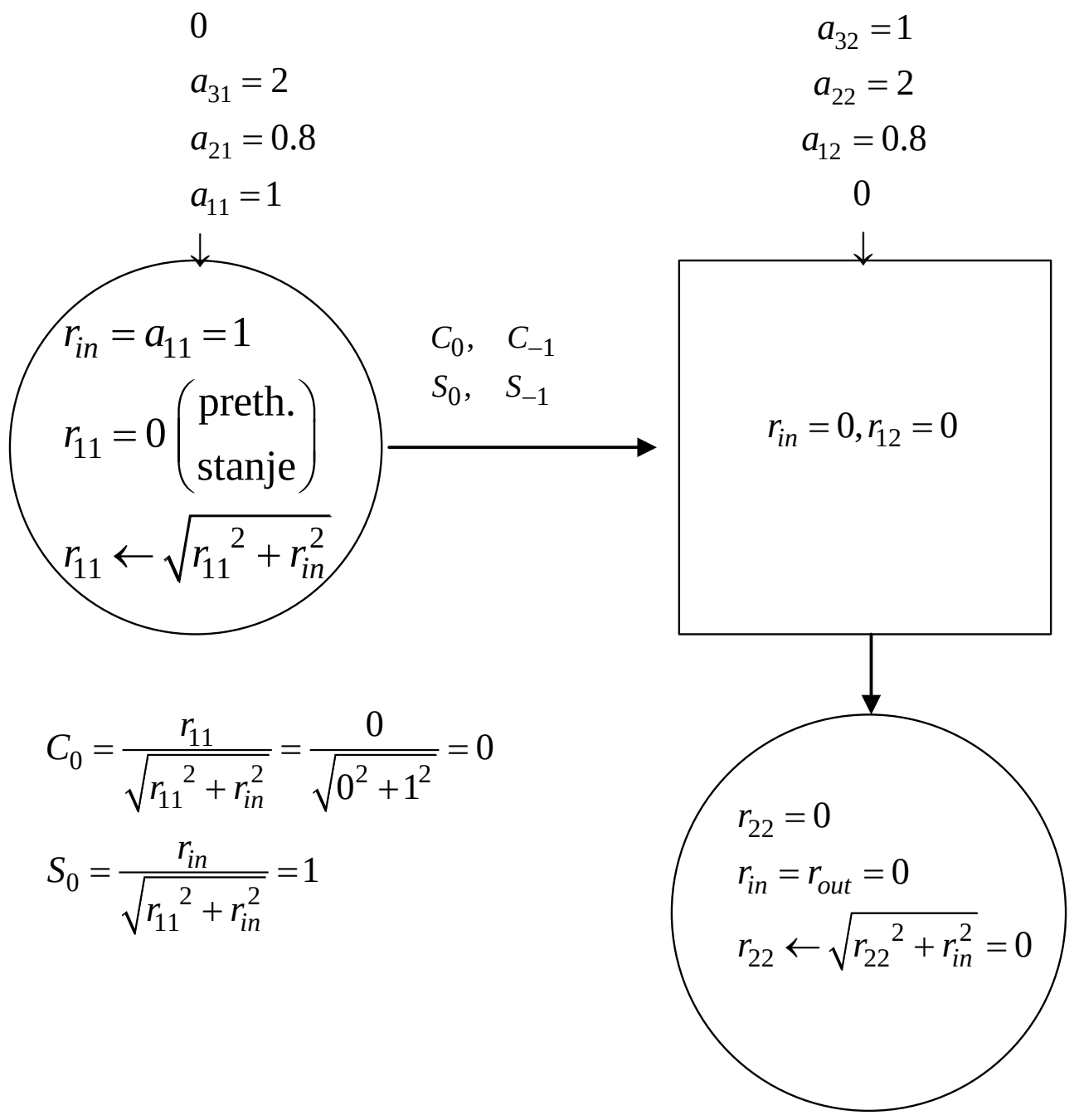

## Korak 1:

$$0,\quad a_{31} = 2,\quad a_{21} = 0.8 \quad \downarrow$$

$$r_{in} = a_{21} = 0.8$$
$$r_{11} = 1$$
$$r_{11} \leftarrow \sqrt{{r_{11}}^2 + r_{in}^2} = 1.2806$$

$$C_1, C_0 \rightarrow \qquad S_1, S_0 \rightarrow$$

$$a_{32} = 1,\quad a_{22} = 2,\quad a_{12} = 0.8 \quad \downarrow$$

$$r_{in} = a_{12} = 0.8$$
$$r_{12} = 0$$
$$r_{12} \leftarrow C_0 r_{12} + S_0 r_{in}$$
$$r_{12} = 0 + 1 \cdot 0.8 = 0.8$$

$$C_1 = \frac{r_{11}}{\sqrt{{r_{11}}^2 + r_{in}^2}} = \frac{1}{\sqrt{1^2 + 0.8^2}} = 0.7809$$

$$S_1 = \frac{r_{in}}{\sqrt{{r_{11}}^2 + r_{in}^2}} = \frac{0.8}{\sqrt{1^2 + 0.8^2}} = 0.6247$$

$$w \leftarrow -S_0 r_{12} (\text{u koraku } n = 0) + C_0 r_{in}$$
$$w = -1 \cdot 0 + 0 \cdot 0.8 = 0$$

$$r_{22} = 0$$
$$r_{in} = w(n = 0) = 0$$
$$r_{22} \leftarrow \sqrt{{r_{22}}^2 + r_{in}^2} = 0$$

## Korak 2:

$$0,\quad a_{31} = 2 \quad \downarrow$$

$$r_{in} = a_{31} = 2$$
$$r_{11} = 1.2806$$
$$r_{11} \leftarrow \sqrt{{r_{11}}^2 + r_{in}^2} = 2.3749$$

$$C_2, C_1 \rightarrow \qquad S_2, S_1 \rightarrow$$

$$a_{32} = 1,\quad a_{22} = 2 \quad \downarrow$$

$$r_{in} = a_{22} = 2$$
$$r_{12} = 0.8$$
$$r_{12} \leftarrow C_1 r_{12} + S_1 r_{in}$$
$$r_{12} = 0.7809 \cdot 0.8 + 0.6247 \cdot 2$$
$$r_{12} = 1.8741$$

$$C_2 = \frac{r_{11}}{\sqrt{{r_{11}}^2 + r_{in}^2}} = \frac{1.2806}{\sqrt{1.2806^2 + 2^2}} = 0.5392$$

$$S_2 = \frac{r_{in}}{\sqrt{{r_{11}}^2 + r_{in}^2}} = \frac{2}{\sqrt{1.2806^2 + 2^2}} = 0.8422$$

$$w \leftarrow -S_1 r_{12} (\text{u koraku } n = 1) + C_1 r_{in}$$
$$w = -0.6247 \cdot 0.8 + 0.7809 \cdot 2$$
$$w = 1.062$$

$$r_{22} = 0$$
$$r_{in} = w(n = 1) = 0$$
$$r_{22} \leftarrow \sqrt{{r_{22}}^2 + r_{in}^2} = 0$$

## Korak 3:

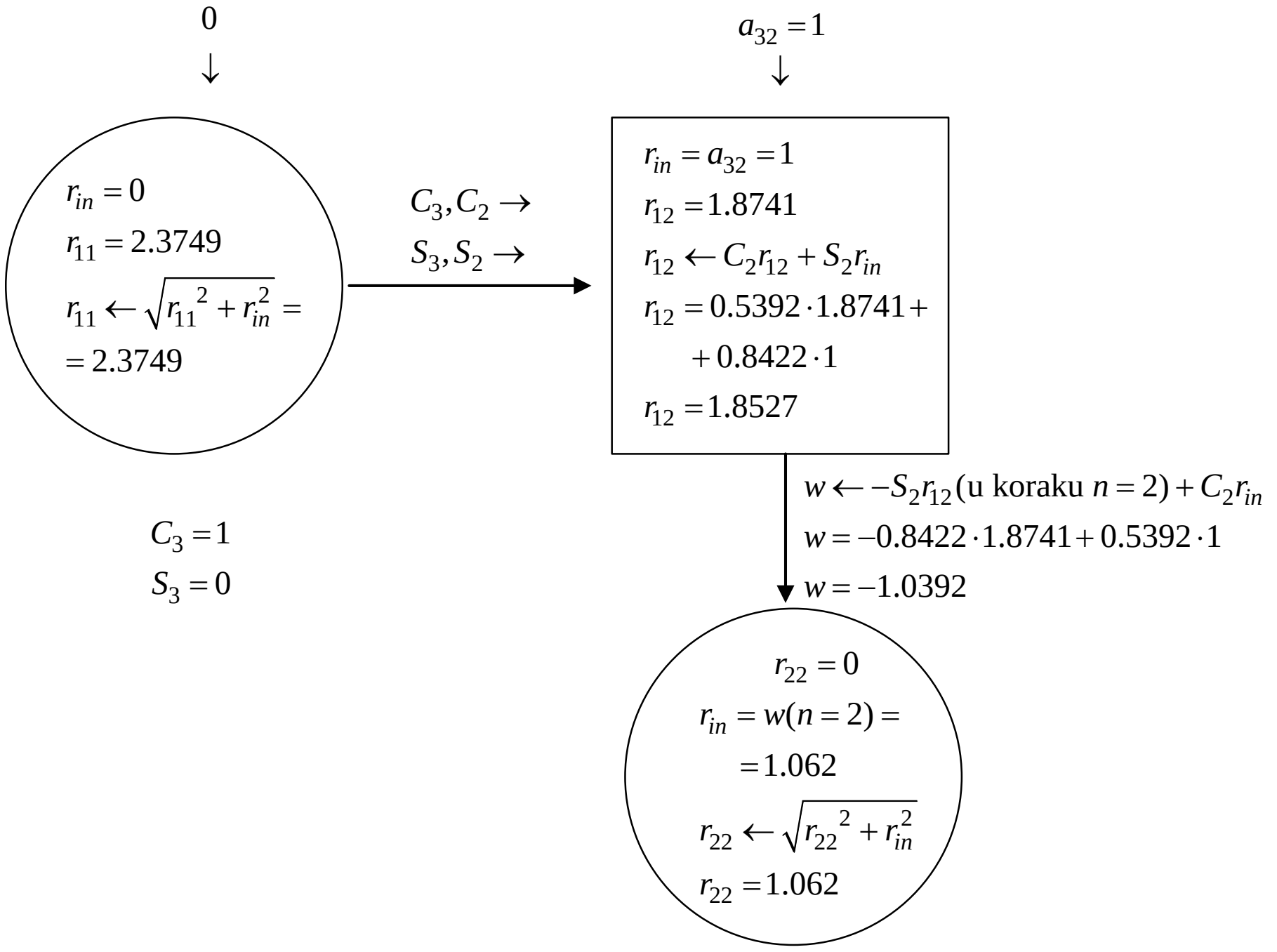

## Korak 4:

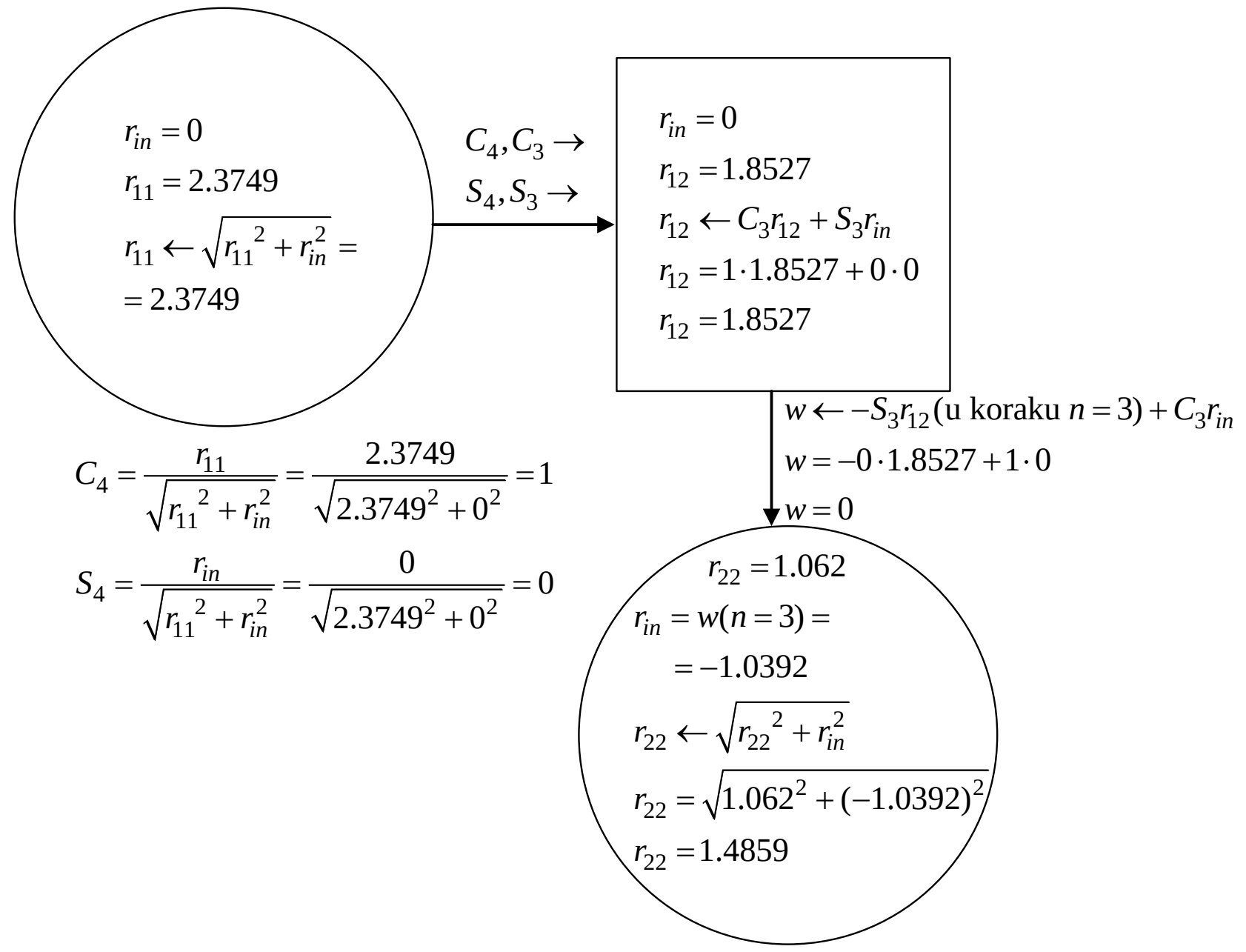

Konačno, vrijednosti elemenata trougaone matrice jednake su vrijednostima koje su se zatekle u ćelijama nakon izvršavanja Koraka 4 procedure, tj.:

$$\mathbf{R} = \begin{bmatrix} 2.3749 & 1.8527 \\ 0 & 1.4859 \end{bmatrix}.$$

Opšta šema QR dekompozicije se može prikazati grafikom na slici 6.5.

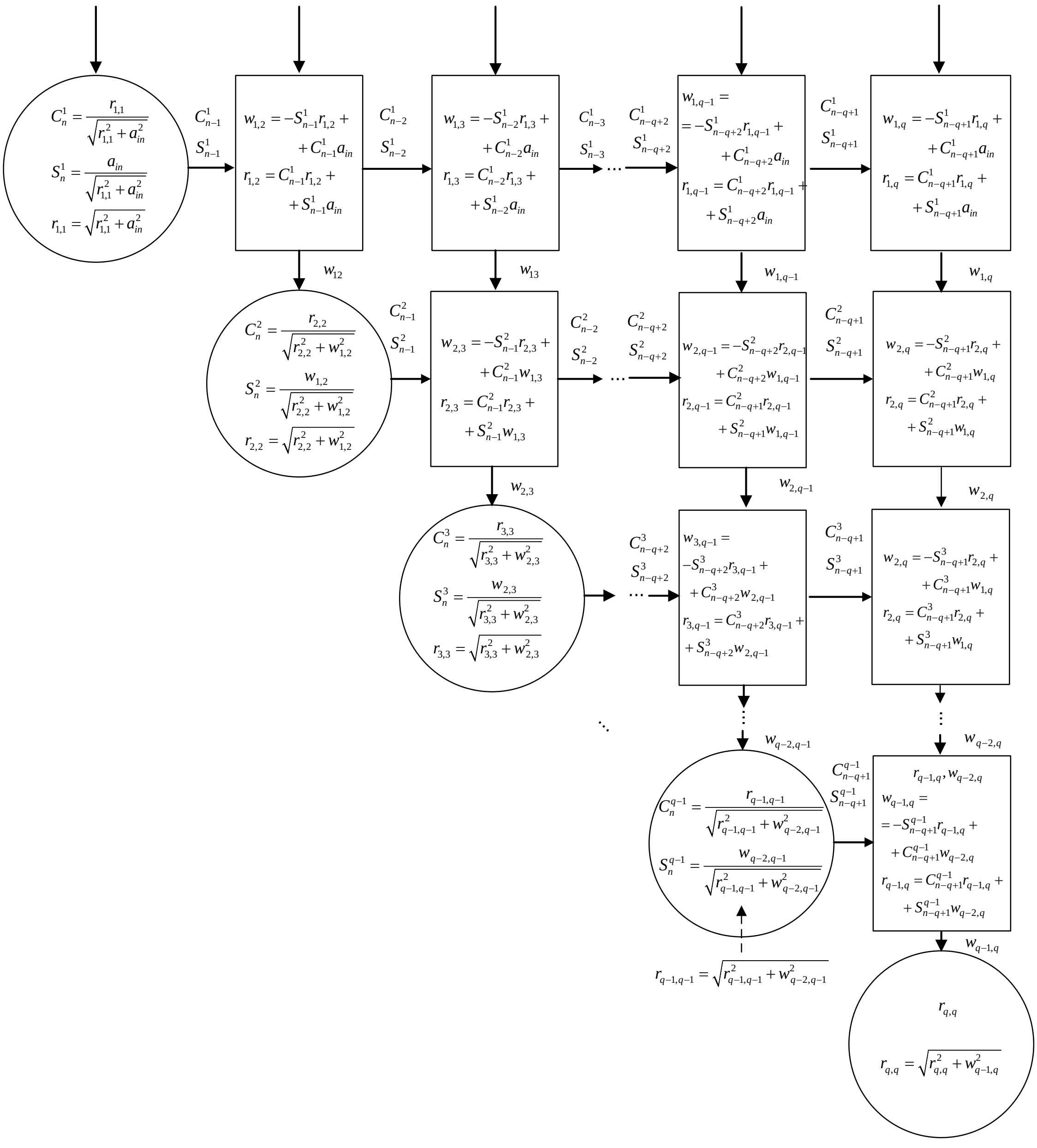

Slika 6.5: Opšta šema QR dekompozicije

### 6.4.2. Inverzna trougaona matrica R

U nastavku će, kroz primjer, biti opisano funkcionisanje sistoličkog niza za dobijanje inverzne matrice od polazne trougaone matrice. Posmatrana je matrica **R** dobijena u prethodnom primjeru. Matrica je veličine $K \times K = 2 \times 2$.

$$\mathbf{R} = \begin{bmatrix} 2.3749 & 1.8527 \\ 0 & 1.4859 \end{bmatrix} = \begin{bmatrix} r_{1,1} & r_{1,2} \\ r_{2,1} & r_{2,2} \end{bmatrix}$$

**Korak 0:**

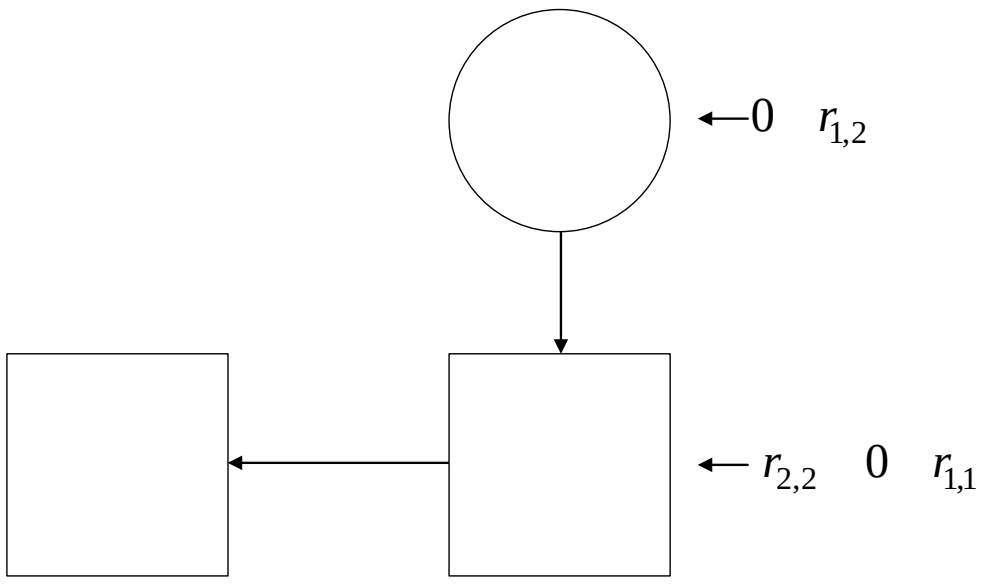

**Korak 1:**

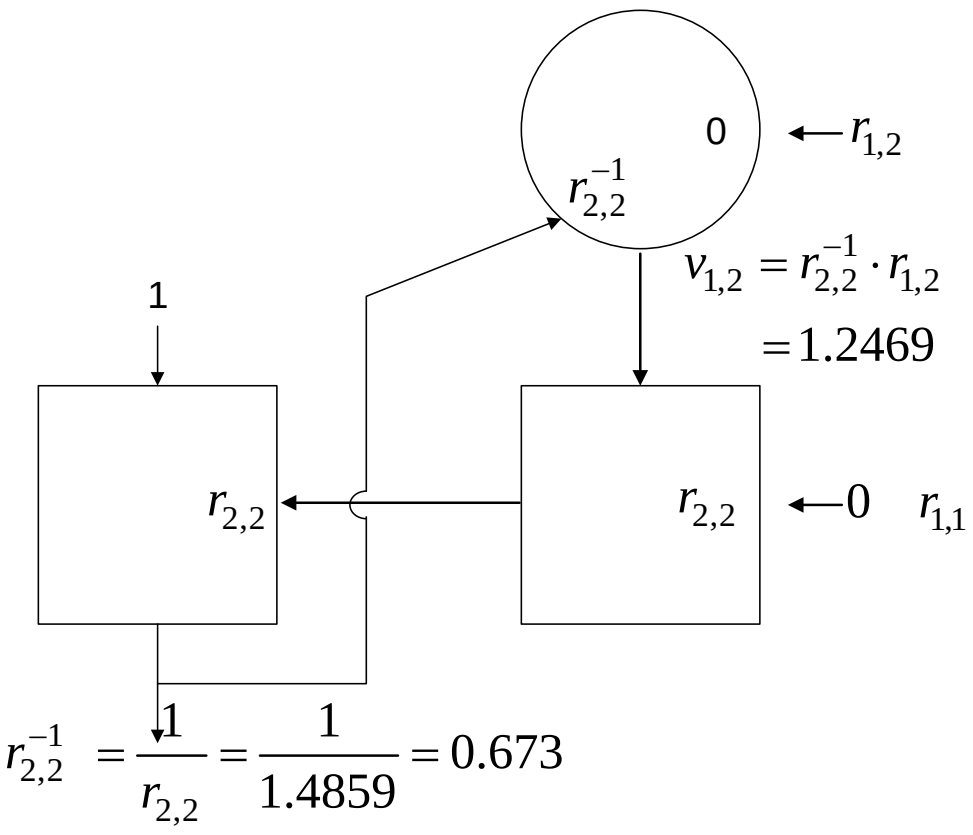

**Korak 2:**

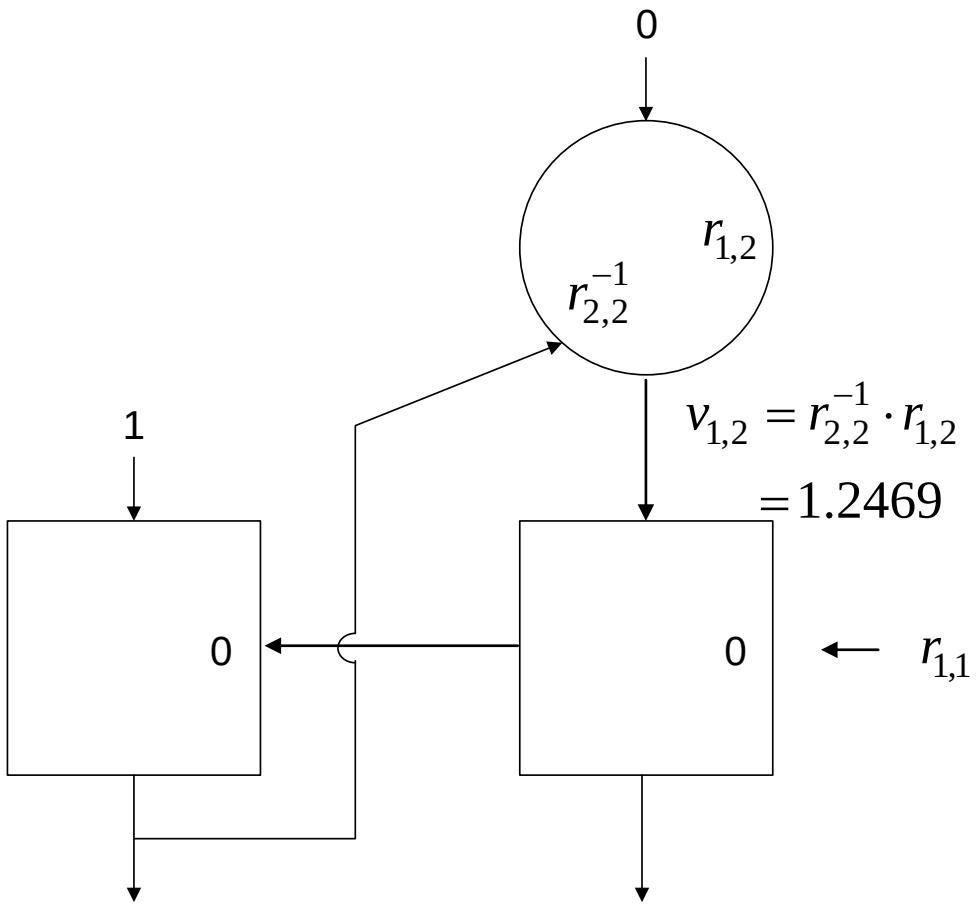

**Korak 3:**

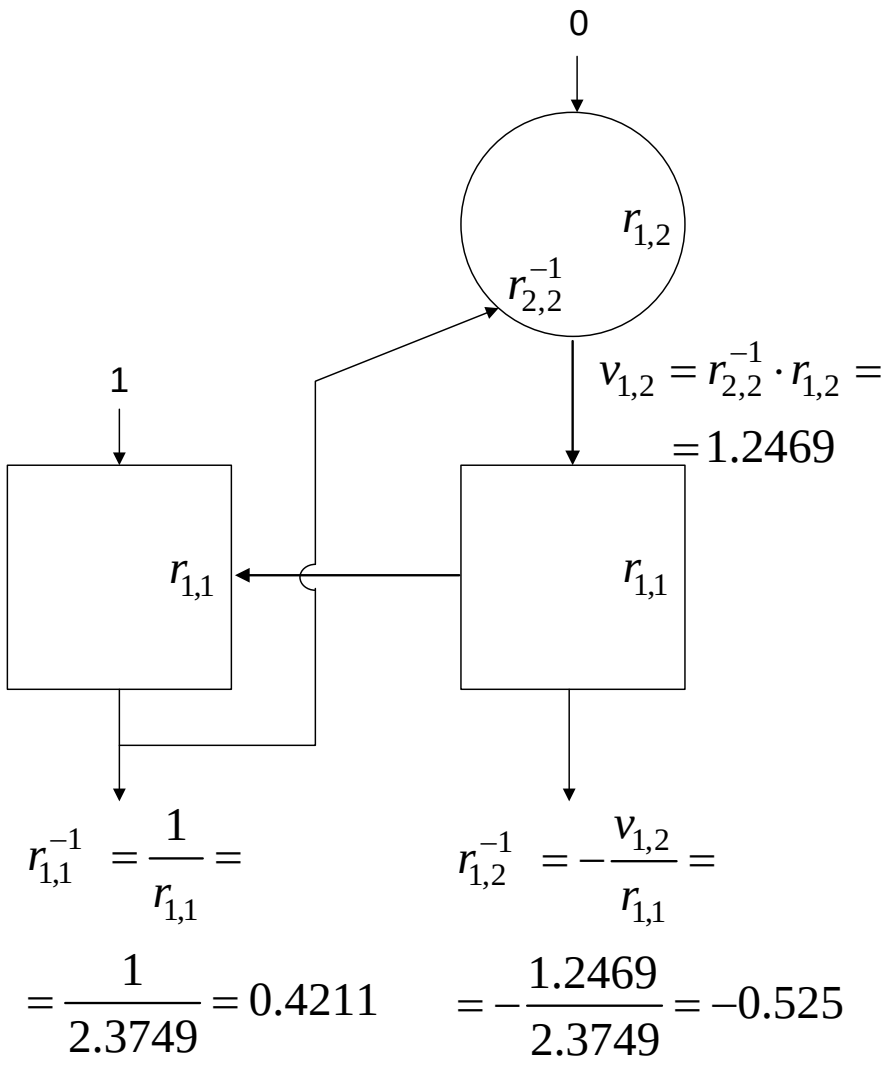

Dobijena inverzna matrica $\mathbf{R}^{-1}$ ima vrijednosti : $\mathbf{R}^{-1} = \begin{bmatrix} 0.421 & -0.525 \\ 0 & 0.673 \end{bmatrix}$.

## 6.5. Kompleksnost predložene arhitekture

U ovom poglavlju je dat kratak osvrt na računsku kompleksnost predloženog metoda zasnovanog na Givens-ovim rotacijama. Za pravougaonu matricu dimenzija $M \times N$, $\mathbf{A}_{M \times N}$, gdje važi da je $M>N$, Givens-ove rotacije zahtijevaju $3N^2(M-N/3)$ operacija (*floating point operations* – flops) za računanje matrice **R**, dok je za računanje ortogonalne matrice **Q** neophodno $4N(M^2-N^2/3)$ operacija. U predloženoj proceduri rješavanja optimizacionog problema nije neopodno računanje ortogonalne matrice **Q**, pa je moguće izbjeći $4N(M^2-N^2/3)$ operacija koje bi bile potrebne za njeno izračunavanje.

Osvrnućemo se sada na računsku kompleksnost predloženog rješenja za rješavanje optimizacionog problema. Kako u relaciji figuriše množenje matrica, polazimo od toga. Množenje dvije matrice čiji su elementi kompleksne vrijednosti, a koje su dimenzija $P \times N$ i $N \times M$ zahtijeva $4PMN-2PM$ sabiranja i $4PMN$ množenja [73].

Ukoliko pretpostavimo da su dimenzije matrica i vektora koji figurišu u relaciji (6.26) sljedeće: dimenzije matrice $\mathbf{R}^{-1}_{\mathrm{CS}}$ $K \times K$, dimenzije parcijalne matrice Fourier-ove transformacije $\mathbf{A}_{\mathrm{CS}}$ $M \times K$ (tj. $K \times M$ za transponovanu matricu) a dimenzije vektora mjerenja **y** $M \times 1$, onda rješenje optimizacionog problema datog relacijom (6.26) zahtijeva $2K(K^2+K+2M-2)$ sabiranja i $4K(K^2+M+K)$ množenja.

Poređenja radi, originalni optimizacioni problem, definisan relacijom (6.4), zahtijeva $2K(2MK-K+2M-1)$ sabiranja i $4MK(K+1)$ množenja, a računska kompleksnost je dodatno povećana računanjem inverzne forme proizvoda parcijalne matrice Fourier-ove transformacije i njene transponovane verzije ($\mathbf{A}^*_{CS}\mathbf{A}_{CS}$) za $K^2$.

Primjera radi, posmatrajmo realni slučaj signala koji ima $K$=15 komponenti, ako je broj dostupnih odbiraka $M$=250. Predloženo rješenje zahtijeva 22140 sabiranja i 29400 množenja. Korišćenjem prvobitnog rješenja optimizacionog problema dobili smo da je zahtijevani broj sabiranja 239520, a broj množenja 240000, što je oko 11 (za operacije sabiranja) odnosno 8 (za operacije množenja) puta više od predloženog rješenja.
Broj takt ciklusa neophodan za izvršavanje algoritma je:

a) $3M$+61 za blok koji računa prag,

b) O(logM) za kolo komparatora,

c) 126$K$-8 za računanje matrice **R**,

d) 34$K$-4 za inverziju matrice, i

e) O($KM$+$K^2$) za rješavanje optimizacionog problema.

# Zaključak

Analizu multikomponentnih signala umnogome olakšava analiza svake komponente signala pojedinačno, pa je u radu poseban akcenat stavljen na procedure za razdvajanje komponenti signala, zasnovane na njihovom vremensko-frekvencijskom predstavljanju. Posmatrani su harmonijski i neharmonijski signali kao i signali koji se javljaju u bežičnim komunikacijama. U zavisnosti od posmatrane aplikacije, korišćene su različite vremensko-frekvencijske distribucije, a posebna pažnja je posvećena S-metodu i distribucijama sa kompleksnim argumentom vremena. Definisan je algoritam za dekompoziciju multikomponentnih signala, kod kojih se energije komponenti signala značajno razlikuju. Algoritam je baziran na dekompoziciji na sopstvene vrijednosti i vektore, a koristi iterativnu proceduru za odvajanje komponenti. Predloženi algoritam testiran je na harmonijskim i neharmonijskim signalima. Predložena procedura poređena je sa standardnim pristupom kao što je MUSIC (*Multiple Signal Classification*) algoritam. Pokazalo se da:

- MUSIC algoritam zahtijeva da je broj komponenti u signalu unaprijed poznat, što u realnim aplikacijama obično nije slučaj;
- Primjenom MUSIC algoritma nije moguće odvojiti komponente signala čije su energije veoma male;
- U poređenju sa predloženim algoritmom, MUSIC pruža lošu rezoluciju odvojenih komponenti u vremensko-frekvencijskoj ravni.

U radu je posmatran problem signala koji pripadaju različitim standardima u bežičnim komunikacijama, a djeluju u istom opsegu frekvencija - Bluetooth i IEEE 802.11b signali. Definisana je procedura za klasifikaciju komponenti signala. Nakon razdvajanja komponenti moguće je zasebno posmatrati svaku od njih i klasifikaciju izvršiti na osnovu njenih fizičkih osobina u vremensko-frekvencijskoj ravni (npr. na osnovu centralne frekvencije komponente signala, njenog vremenskog trajanja i frekvencijskog opsega komponente). Na ovaj način moguće je razdvojiti signale koji pripadaju različitim standardima.

Procedura dekompozicije je kombinovana sa pristupom komprimovanog očitavanja, koji omogućava rekonstrukciju koeficijenata signala koji su, ili namjerno izostavljeni prilikom akvizicije signala, ili su originalne vrijednosti signala uništene prisustvom šuma u signalu. U ovom radu, posmatran je slučaj namjernog izostavljanja određenog procenta koeficijenata razdvojenih komponenti signala, u cilju prenosa što je moguće manje podataka komunikacionim kanalom, a da se na prijemu dobije kompletna informacija o signalu.

Definisan je dvodimenzioni CS pristup estimaciji trenutne frekvencije nestacionarnih signala, korišćenjem ambiguity funkcije i distribucije sa kompleksnim argumentom vremena. Pristup je adaptiran i za signale koji su zahvaćeni šumom, pa je definisana i robustna forma algoritma.

Dat je prijedlog hardverske arhitekture za realizaciju Single Iteration Reconstruction Algorithm-a za rekonstrukciju kompresivno očitavanih signala, koji je zasnovan na pragu a kojim se odvajaju komponente signala od komponenti šuma u transformacionom domenu. Pokazano je da se prilikom implementacije algoritma može izbjeći računanje kompleksnih izraza koji su sastavni dio algoritma, kao što su logaritamska i stepena funkcija, što u velikoj mjeri utiče na smanjenje računske kompleksnosti. Takođe, modifikovan je optimizacioni problem koji se javlja u originalnoj verziji algoritma, na način da se izbjegne računanje inverzne matrice relativno velikih dimenzija i inverzija svede na računanje inverzne trougaone matrice. Inverzna trougaona matrica se relativno jednostavno hardverski implementira, što je u literaturi i poznato. Sistem je skalabilan pa se može koristiti za različite dimenzije ulazne matrice, različite tipove signala i za različit broj dostupnih odbiraka signala. Skalabilnost predloženog sistema postignuta je uvođenjem kontrolnih signala kod elemenata sistoličkog niza, koji uključuju/isključuju odgovarajuće kolone matrice.

# Lista referenci Anđele Draganić

## Radovi publikovani u međunarodnim časopisima

## Radovi publikovani u regionalnim časopisima

## Radovi publikovani na međunarodnim konferencijama

### Radovi publikovani na regionalnim konferencijama